\documentclass[useAMS,usenatbib]{mn2e}
\usepackage{amsmath, amssymb, mathtools}
\usepackage{graphicx}
\usepackage{subfig}

\usepackage[morefloats=25]{morefloats}

\usepackage{times}
\usepackage{mathrsfs}
 
\setcitestyle{round,semicolon,aysep={},yysep={,}}
\defcitealias{Malareckietal2013}{Paper I}

\newcommand{\RadioAppName}[1]{{\sc#1}}

\newcommand{\subfigletter}[1]{\emph{#1}}
\newcommand{\bJ}{b_{\mathrm{J}}}

\usepackage{rotating}
\title[GRGs -- II. Tracers of LSS]{Giant radio galaxies -- II. Tracers of large-scale structure}

\author[J. M. Malarecki et al.]
  {J. M.~Malarecki,$^1$\thanks{E-mail: jurek.malarecki@icrar.org}
   D.~H.~Jones,$^{2,3}$ L.~Saripalli,$^4$ L.~Staveley-Smith,$^{1,5}$ R.~Subrahmanyan$^{4,5,6}$ \\ 
  $^1$International Centre for Radio Astronomy Research, M468, The University of Western Australia, Crawley, WA 6009, Australia\\
  $^2$Department of Physics and Astronomy, Macquarie University, Sydney, NSW 2109, Australia\\
  $^3$School of Physics, Faculty of Science, Monash University, Clayton, Victoria 3800, Australia\\
  $^4$Raman Research Institute, C V Raman Avenue, Sadashivanagar, Bangalore 560080, India\\
  $^5$ARC Centre of Excellence for All-Sky Astrophysics (CAASTRO)\\
  $^6$National Radio Astronomy Observatory, Socorro, NM 87801, USA
  }
\date{Accepted 2015 January 30. Received 2015 January 14; in original form 2014 September 25}

\pagerange{\pageref{firstpage}--\pageref{lastpage}} \pubyear{2015}

\begin{document}

\label{firstpage}

\maketitle

\begin{abstract}
We have carried out optical spectroscopy with the Anglo-Australian Telescope for $24{,}726$ objects surrounding a sample of 19 Giant Radio Galaxies (GRGs) selected to have redshifts in the range 0.05 to 0.15 and projected linear sizes from 0.8 to 3.2 Mpc. Such radio galaxies are ideal candidates to study the Warm-Hot Intergalactic Medium (WHIM) because their radio lobes extend beyond the ISM and halos of their host galaxies, and into the tenuous IGM. We were able to measure redshifts for $9{,}076$ galaxies. Radio imaging of each GRG, including high-sensitivity, wideband radio observations from the Australia Telescope Compact Array for 12 GRGs and host optical spectra~\citep[presented in a previous paper,][]{Malareckietal2013}, is used in conjunction with the surrounding galaxy redshifts to trace large-scale structure.

We find that the mean galaxy number overdensity in volumes of ${\sim}700$ Mpc$^3$ near the GRG host galaxies is ${\sim}70$ indicating an overdense but non-virialized environment. A Fourier component analysis is used to quantify the anisotropy in the surrounding galaxy distribution. For GRGs with radio components offset from the radio axis, there is a clear influence of the environment with lobes appearing to be deflected away from overdensities in the surrounding medium. Furthermore, the GRG lobes tend to be normal to the plane defined by the galaxy neighbourhood close to the host. This indicates the tendency for lobes to grow to giant sizes in directions that avoid dense regions on both small and large scales.
\end{abstract}

\begin{keywords}
galaxies: active -- galaxies: distances and redshifts -- intergalactic medium -- galaxies: jets -- radio continuum: galaxies.
\end{keywords}

\section{Introduction}
Giant radio galaxy lobes expand into the intergalactic medium, well beyond the confines of the host galaxy and its thermal halo. As such, they are ideal probes of the Warm-Hot Intergalactic Medium (WHIM), the tenuous 10$^5$ to 10$^7$~K gas permeating large-scale structures in which these galaxies reside. Big-bang nucleosynthesis theory predicts more baryons than are seen, suggesting that the WHIM harbours most not yet accounted for by observations~\citep*[e.g.][]{Shulletal2012}. Giant radio galaxies (GRGs) are therefore ideal objects to investigate the ambient medium and the missing baryons by virtue of their large physical extent ({\scriptsize $\gtrsim$} 0.7 Mpc). The morphology of their radio lobes may be influenced by their proximity to large-scale structures, such as the filaments in which the WHIM gas is thought to be found, interacting via ram pressure, buoyancy or gravitational effects. The distribution of the IGM follows that of galaxies and so these large-scale structures may be traced by the local galaxy distribution and the properties of the WHIM gas inferred by studying the giant lobes in dynamic equilibrium with the IGM.

We have embarked on a large program to probe the baryon content of the WHIM using 19 GRGs~\citep[][hereafter Paper I]{Malareckietal2013}. This follows earlier work on two individual sources: MSH 05--22~\citep[B0503--286;][]{Subrahmanyanetal2008} and MRC B0319--454~\citep{Safourisetal2009}. A major part of our study is spectroscopy of field galaxies within a degree (${\sim}6$ Mpc) of the GRG hosts. These give redshift distances and allow us to relate the local galaxy distribution with the growth and morphology of the extended radio structure. The interaction between the large-scale structure WHIM and the GRGs can be used to derive the properties of the WHIM gas~\citep[e.g.][]{Subrahmanyanetal2008}, which we present in a companion paper (Malarecki et al., in prep.; hereafter Paper III).

In this paper we present clustering and density measures using $10{,}316$ optically selected galaxies surrounding the 19 GRGs. We use this data set to show a relationship between the radio morphologies of the GRGs and the anisotropy in the surrounding galaxy distribution. The paper is laid out as follows: in \textsection\ref{sec:AAT} we describe the spectroscopic observations, and outline the data reduction and redshift determination. In \textsection\ref{sec:Analysis} we describe estimation of galaxy stellar masses for galaxies in the neighbourhood of GRGs as well as the methods used to quantify overdensity and anisotropy in those regions. In \textsection\ref{sec:Results} we present the results of a Fourier component analysis of the GRG environments and also provide the projected galaxy distribution around each GRG. Finally, in \textsection\ref{sec:Discussion} we discuss our results and explore the population properties of the sample of GRGs. We adopt a flat cosmology with parameters $\Omega_m = 0.27$, $\Omega_{\Lambda} = 0.73$ and a Hubble constant of $H_0 = 71\mbox{ km s}^{-1}\mbox{Mpc}^{-1}$.

\section{Target selection, observations and reduction}
\label{sec:AAT}

The full radio-selected input sample consists of 19 giant radio galaxies. The sources were selected from a number of complete radio source surveys on the basis of size and proximity~\citep{JonesMcAdam1992,Subrahmanyanetal1996,Saripallietal2005,Subrahmanyanetal2010}. In particular, projected linear sizes exceed 700 kpc and host galaxy redshifts are less than 0.15, permitting spectroscopic observations of the surrounding galaxies. \citetalias{Malareckietal2013} contains further details of the radio sample and the optical spectra of their hosts. Table~\ref{tbl:GRG_radio_properties} lists the radio properties of the sample, including a measure of asymmetry between the longer and shorter lobes. We define radio lobe length as the distance between the host and the emission peak in each lobe using the highest resolution, 2.1 GHz wideband, ATCA radio map, where available. The criterion for asymmetry is when the ratio between the lengths of the longer lobe and shorter lobe exceeds 1.3. The emission peaks largely represent the termination points of the jets and indicate the extent to which the lobes have reached in their growth. With this choice of lobe-extent measurement however, for some sources (J0034--6639, J0400--8456, B0511--305 and J0843--7007) prominent but faint lobe regions beyond the emission peaks were excluded. In relating the galaxy distribution around the GRGs and the radio structures, we have given due attention to the presence and location of these faint lobe emission regions (see Section~\ref{subsubsec:Individual_sources}).

\begin{table}
 \caption{GRG sample properties including longer lobe PA and length, and lobe length ratio. PA are measured over the range [-180, 180]. For sources with non-collinear lobes, a pair of angles representing the average radio axis are provided in parentheses after the longer lobe PA. Sources are considered asymmetric if the lobe length ratio exceeds 1.30.}\label{tbl:GRG_radio_properties}
 \begin{tabular}{llll}
  \hline
   Source                            & Longer lobe PA$^a$ & Longer lobe & Lobe ratio \\
                                      & (degrees)          & (Mpc)       &            \\
  \hline
  \multicolumn{4}{l}{(a) \emph{Asymmetric radio lobes}}\\
  \\
  J0116--473$^b$  & $ -33$ $(-22, 158)$ & $0.86$ & $1.41$ \\
  B0319--454      & $-131$ & $1.30$ & $2.20$ \\
  J0331--7710     & $-176$ & $1.75$ & $2.71$ \\
  B0503--286      & $-171$ $(-179, 1)$ & $1.00$ & $1.76$ \\
  B0511--305      & $  34$ $(-156, 24)$ & $0.48$ & $1.83$ \\
  B0703--451$^c$  & $  65$ & $0.46$ & $1.41$ \\
  B0707--359      & $ 121$ & $0.59$ & $1.48$ \\
  B1308--441      & $ -70$ $(-60, 120)$ & $0.58$ & $2.12$ \\
  \hline
  \multicolumn{4}{l}{(b) \emph{Symmetric radio lobes}}\\
  \\
  J0034--6639     & $ 169$ & $0.76$ & $1.20$ \\
  J0400--8456     & $ -12$ $(-180, 0)$ & $0.22$ & $1.02$ \\
  J0459--528      & $-108$ & $0.52$ & $1.09$ \\
  J0515--8100     & $ -36$ $(-25, 155)$ & $0.35$ & $1.07$ \\
  J0746--5702     & $-124$ & $0.40$ & $1.17$ \\
  J0843--7007     & $ 111$ & $0.37$ & $1.27$ \\
  B1302--325      & $-146$ & $0.54$ & $1.20$ \\
  B1545--321      & $ 146$ & $0.49$ & $1.04$ \\
  J2018--556$^d$  & $ -22$ & $0.79$ & $1.10$ \\
  J2159--7219     & $ -39$ & $0.56$ & $1.23$ \\
  B2356--611      & $ -48$ & $0.34$ & $1.02$ \\
  \hline
 \end{tabular} 
 
 \medskip
 Notes -- $^a$ [N=$0^{\circ}$; E=$90^{\circ}$]; $^b$ J0116--473 = B0114--476; $^c$ the host galaxy of B0703--451 is in doubt; $^d$ J2018--556 = B2014--558
\end{table}

We obtained optical spectroscopic redshifts for $24{,}726$ objects, including $9{,}076$ galaxies, in fields around 17 GRGs using the 2dF/AAOmega spectrograph on the Anglo-Australian Telescope (AAT) between 2010 and 2013. For the additional case study GRGs, B0319--454 and B0503--286, we used the data of~\citet{Safourisetal2009} and~\citet{Subrahmanyanetal2008}, respectively. 2dF/AAOmega is a multi-fibre spectrograph delivering up to 392 simultaneous spectra across a $2^\circ$-diameter field~\citep{Sharpetal2006}. Our spectra employ the lowest spectral resolution mode of AAOmega ($R=\lambda/\Delta\lambda=1300$) with 580V and 385R lines mm$^{-1}$ gratings used in the respective blue and red arms. The spectra cover 400--900~nm with a spectral dispersion of 0.1--0.16~nm~pix$^{-1}$. A 570 nm dichroic was used throughout the observing campaign with the exception of a set of observations in 2011 for which a 670 nm dichroic was used. Table~\ref{tbl:AATObservations} summarizes the spectroscopic observations and Table~\ref{tbl:Optical_field_properties} gives properties as well as details of the measurements for the 19 GRG optical fields.

\begin{table}
\centering
 \caption{Journal of AAT observations with the AAOmega spectrograph. A 560 nm dichroic was used unless otherwise indicated. Optical data reduction was performed with the version of {\sc 2dfdr} noted.}\label{tbl:AATObservations}
 \begin{tabular}{llc}
  \hline
   Field                            & Observation & {\sc 2dfdr} \\
                                        &                        & version     \\
  \hline
  \multicolumn{3}{l}{(a) \emph{Asymmetric radio lobes}}\\
  J0116--473        & 2013 Nov                           & 5.35 \\
                               & 2013 Dec                           & 5.35 \\
  B0319--454        & Archival data only \\
  J0331--7710       & 2010 Jun                         & 4.66 \\
                                & 2011 Jan$^a$                           & 4.66 \\
  B0503--286        & Archival data only \\
  B0511--305        & 2011 Jan$^a$                           & 4.66 \\
  B0703--451        & 2010 Jun                           & 4.66 \\
                                & 2011 Jan$^a$                           & 4.66 \\
                                & 2012 Apr                           & 4.66 \\
  B0707--359       & 2010 Jun                            & 4.66 \\
                               & 2011 Jan$^a$                          & 4.66 \\
                               & 2013 Dec                           & 5.35 \\
  B1308--441       & 2010 Jun                           & 4.66 \\
  \hline
  \multicolumn{3}{l}{(b) \emph{Symmetric radio lobes}}\\
  J0034--6639      & 2010 Jun                           & 4.66 \\
                               & 2013 Nov                           & 5.35 \\
  J0400--8456      & 2010 Jun                           & 4.66 \\
                                & 2011 Jan$^a$                           & 4.66 \\
  J0459--528         & 2011 Jan$^a$                           & 4.66 \\
  J0515--8100      & 2013 Dec                           & 5.35 \\
  J0746--5702      & 2012 Apr                           & 5.35 \\
                               & 2013 Dec                           & 5.35 \\
  J0843--7007      & 2012 Apr                           & 4.66 \\
                                & 2013 Apr                           & 5.35 \\
                                 & 2013 Dec                           & 5.35 \\
  B1302--325       & 2012 Apr                           & 4.66 \\
                               & 2013 Dec                           & 5.35 \\
  B1545--321       & 2010 Jun                          & 4.66 \\
                                & 2012 Apr                          & 4.66 \\
  J2018--556       & 2010 Jun                           & 4.66 \\
                              & 2012 Apr                           & 4.66 \\
                              & 2013 Apr                           & 5.35 \\
                              & 2013 Jul                           & 5.35 \\
  J2159--7219      & 2010 Jun                           & 4.66 \\
                               & 2013 Jul                           & 5.35 \\
                               & 2013 Nov                           & 5.35 \\
  B2356--611       & 2010 Jun                           & 4.66 \\
                                & 2013 Jul                          & 5.35 \\
                               & 2013 Nov                           & 5.35 \\
  \hline
 \end{tabular} 
 
 \medskip
$^a$ The indicated configuration used a 670 nm dichroic
\end{table}

Field galaxy targets were selected from the $\bJ$ SuperCOSMOS catalogue~\citep{Hamblyetal2001a,Hamblyetal2001b} within a 1$^{\circ}$ radius of each GRG host. The brightest stars ($b_{\mathrm{J}}<17$) flagged by SuperCOSMOS were filtered out but everything fainter than this was included. The remaining objects were cut in apparent magnitude to produce subsets comprising 600 to 700 objects per field at increasingly fainter apparent magnitude. Of these, 350 to 390 were allocated fibres along with 6 to 8 guide stars and 30 sky positions, all of which were chosen to distribute the fibres as uniformly as possible across the field. Unobserved targets were rolled into subsequent configurations. Guide stars and sky fibre locations were manually inspected to exclude cases of contamination and multiple objects. Some of the GRGs lie at low Galactic latitude ($|b| < 15^\circ$), resulting in significant Galactic stellar contamination. We removed some of these stars through a colour cut ($b_{\mathrm{J}}-r_{\mathrm{F}}< 1.8$) to exclude the bluest contaminants.

The targets in all fields were selected as uniformly as possible across each 2$^{\circ}$-diameter field. In fields with bright saturating stars, the same bright star masks applied by SuperCOSMOS have been applied here. This saturation prevents SuperCOSMOS from finding faint adjacent objects in the vicinity of these bright stars. We account for this masking in the analysis described in the following sections.

Observations for each configuration included flat field and arc calibration frames. Each GRG field had several configurations. The raw spectral data was reduced with the software package {\sc 2dfdr} (version 5.35) using the standard default options, with the red and blue halves being reduced separately and spliced together at the end. Several of the 2011 fields included master bias and master dark frames for reduction of the blue half-spectrum. A Cu-Ar-Ne arc lamp was used for wavelength calibration throughout.

To determine redshifts we used the program {\sc runz} (modified from the original version developed by W. Sutherland). It displays each science spectrum and assigns a redshift based on the best-fitting cross-correlation and Gaussian fit to emission line features. The user evaluates the correctness of these redshifts and assigns a value classifying its reliability: $Q = 6$ for stars, 4 for reliable extragalactic spectra, 3 for acceptable redshifts and 1 for unreliable spectra, \citep[see e.g.][]{Jonesetal2009}. Typical redshifts of $z \sim 0.16$ were found for those fields selected down to $b_{\mathrm{J}}\sim 19.9$. Stellar contamination rates varied from 13.7 per cent to 85.1 per cent. Redshifts derived from this procedure typically have an error of \mbox{$60$ km s$^{-1}$} for high signal-to-noise sources and \mbox{${\sim}100$ km s$^{-1}$} for lower quality spectra~\citep[e.g.][]{Driveretal2011}. These are conservative estimates based on similar spectroscopic observations by the GAMA survey with the AAT/AAOmega using fainter galaxies.

Unambiguous identification of the GRG with a clear optical counterpart was possible in all cases but two. The sources B1308--441 and B1545--321 each have two optical components close to the central region of the respective GRG, at similar redshifts. For B1308--441, SuperCOSMOS assigned two positions for one object and so we designated the candidate three magnitudes brighter as the host optical galaxy. In the case of B1545--321, one object was located at a slightly more central position in agreement with the previously accepted host while the other had features consistent with a spiral galaxy.

We supplemented our observations with redshifts from all available redshift catalogues including the 6dF Galaxy Survey~\citep[6dFGS;][]{Jonesetal2009} and the 2MASS Redshift Survey~\citep[2MRS;][]{Huchraetal2012}. The Millennium Galaxy Catalogue~\citep[MGCz;][]{Driveretal2005} and GAMA survey~\citep{Driveretal2009} regions do not coincide with any of our fields. We also applied the automated redshift algorithm {\sc autoz}~\citep{Baldryetal2014} to verify our redshift catalogues for any misclassifications.

Our final redshift catalogues were inspected for intersecting fibre positions and we excluded duplicate observations attributed to spurious input catalogue targets. The best available photometry was used for objects with multiple overlapping positions from SuperCOSMOS.

In summary, $9{,}076$ galaxy redshifts are newly observed by us, $486$ are from \citet{Safourisetal2009}, $323$ are from \citet{Subrahmanyanetal2008}, $412$ are from 6dFGS and $19$ are from 2MRS giving a total of $10{,}316$ galaxies.

\begin{sidewaystable*}
 \caption{Details of optical observations of the GRG sample. The number of cuts denotes the number of independent observations obtained for each field. Completeness refers to the percentage of SuperCOSMOS objects (galaxies and contaminating stars) with $b_{\mathrm{J}}<19.0$ for which redshifts were determined.}\label{tbl:Optical_field_properties}
 \begin{tabular}{llccccccccc}
  \hline
Source            &  Redshift  &  Number   &  Objects   &  Galaxies    &  Stellar contamination  &  \multicolumn{2}{c}{Limiting magnitude}  &  Galactic latitude  &  Completeness  &  Input catalogue masked  \\
                         &                   &  of cuts      &  observed & identified   &  (per cent)                       &  $b_{\mathrm{J}}$ & $M_{\mathrm{B}}$   &  (degrees)              &  (per cent)           &  (per cent)               \\
  \hline
  \multicolumn{10}{l}{(a) \emph{Asymmetric radio lobes}} \\
  J0116--473       & $0.1470$ & $2$ &  $719$ &  $605$ &  $13.8$ & $19.95$ & $-19.5$ & $-69.2$ & $98.5$ & $7.9$ \\
  B0319--454$^{a}$ & $0.0622$ & $3$ &  $567$ &  $486$ &  $14.3$ & $19.51$ & $-17.7$ & $-55.3$ & $66.1$ & $11.5$ \\
  J0331--7710      & $0.1447$ & $3$ &  $1037$ & $261$ &  $73.2$ & $19.91$ & $-19.5$ & $-36.5$ & $94.2$ & $11.5$ \\
  B0503--286$^{b}$ & $0.0383$ & $2$ &  $564$ &  $323$ &  $13.7$ & $18.24$ & $-17.8$ & $-34.5$ & $59.4$ & $11.1$ \\
  B0511--305       & $0.0577$ & $5$ &  $1745$ & $554$ &  $66.1$ & $20.11$ & $-16.9$ & $-33.3$ & $92.9$ & $15.6$ \\
  B0703--451       & $0.1241$ & $6$ &  $2049$ & $425$ &  $75.8$ & $20.01$ & $-19.0$ & $-16.6$ & $99.9$ & $24.5$ \\
  B0707--359       & $0.1109$ & $4$ &  $1409$ & $193$ &  $85.1$ & $19.87$ & $-18.8$ & $-12.3$ & $97.2$ & $25.8$ \\
  B1308--441       & $0.0507$ & $12$ & $4156$ & $778$ &  $80.8$ & $20.51$ & $-16.2$ & $+18.3$ & $99.8$ & $15.1$ \\
  \hline
  \multicolumn{10}{l}{(b) \emph{Symmetric radio lobes}} \\
  J0034--6639      & $0.1102$ & $3$ &  $1063$ & $772$ &  $21.9$ & $20.11$ & $-18.5$ & $-50.4$ & $98.4$ & $10.3$ \\
  J0400--8456      & $0.1033$ & $3$ &  $1033$ & $330$ &  $64.4$ & $20.01$ & $-18.5$ & $-30.5$ & $96.7$ & $10.8$ \\
  J0459--528       & $0.0957$ & $3$ &  $1047$ & $782$ &  $22.0$ & $20.21$ & $-18.1$ & $-38.1$ & $92.1$ & $13.9$ \\
  J0515--8100      & $0.1050$ & $3$ &  $1081$ & $462$ &  $52.4$ & $20.21$ & $-18.3$ & $-30.4$ & $98.4$ & $12.1$ \\
  J0746--5702      & $0.1301$ & $4$ &  $1365$ & $190$ &  $84.8$ & $19.41$ & $-19.7$ & $-15.5$ & $96.8$ & $19.1$ \\
  J0843--7007      & $0.1389$ & $3$ &  $1041$ & $211$ &  $78.4$ & $19.01$ & $-20.3$ & $-16.7$ & $81.1$ & $17.6$ \\
  B1302--325       & $0.1528$ & $2$ &  $678$ &  $323$ &  $47.1$ & $19.23$ & $-20.3$ & $+30.0$ & $83.6$ & $13.3$ \\
  B1545--321       & $0.1081$ & $4$ &  $1354$ & $203$ &  $83.5$ & $19.51$ & $-19.1$ & $+17.1$ & $93.5$ & $36.9$ \\
  J2018--556       & $0.0605$ & $4$ &  $1391$ & $750$ &  $43.4$ & $20.01$ & $-17.1$ & $-34.2$ & $98.3$ & $16.6$ \\
  J2159--7219      & $0.0967$ & $5$ &  $1751$ & $1117$ & $34.2$ & $20.21$ & $-18.1$ & $-39.2$ & $99.2$ & $9.6$ \\
  B2356--611       & $0.0962$ & $5$ &  $1807$ & $1120$ & $33.9$ & $21.01$ & $-17.3$ & $-55.1$ & $98.5$ & $13.2$ \\
  \hline
 \end{tabular} 
  
  \medskip
 Notes -- $^a$redshift data from~\citet{Safourisetal2009}; $^b$redshift data from~\citet{Subrahmanyanetal2008}
\end{sidewaystable*}

\section{Analysis}
\label{sec:Analysis}

\subsection{Galaxy distributions in the GRG fields}
\label{sec:Galaxy_distributions}

\begin{figure}
\centering
\includegraphics[width=\hsize]{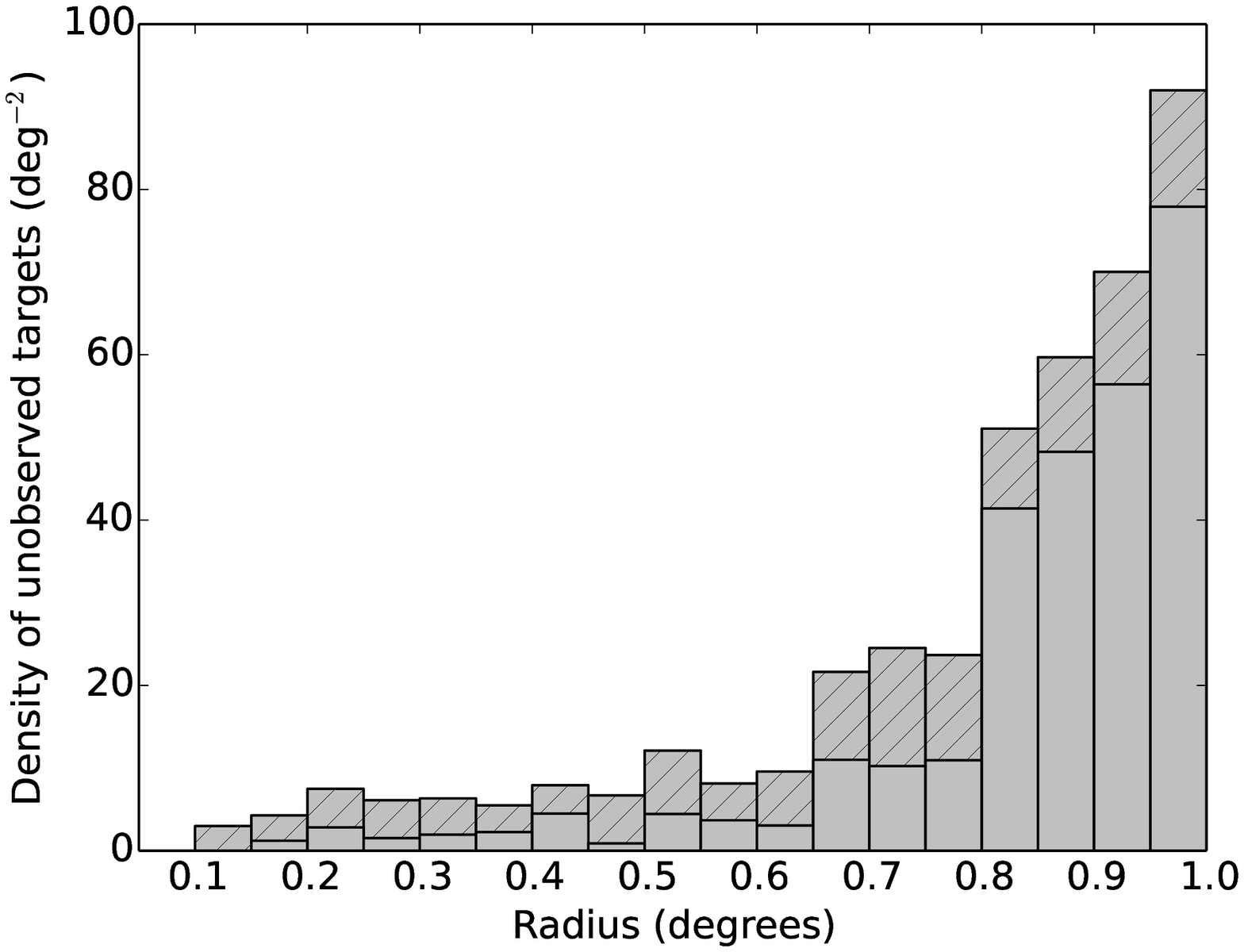}
\caption{Density of remaining SuperCOSMOS targets in annular regions (intervals of $0.05^{\circ}$) for the set of 17 fields that we observed with the AAT/AAOmega, applying a faint $b_{\mathrm{J}}$ apparent magnitude limit of 19.0. The hatched portion represents the effect of the two most incomplete fields, J0843-7007 and B1302-325.}
\label{fig:unobserved_target_density}
\end{figure}

The redshifts derived for galaxies in the 2$^{\circ}$ fields surrounding each GRG have been used to create plots of the projected galaxy distribution about the radio structure. For most fields, redshift completeness within 0.8$^{\circ}$ approaches 100 per cent and so only galaxies within a 0.8$^{\circ}$ radius of the field centre have been considered due to a $\sim$10 to 20~per cent drop in completeness at the field edges. {\sl Completeness} in this context is the fraction of spectroscopically confirmed objects (both stars and galaxies) with respect to potentially observable targets. Fig.~\ref{fig:unobserved_target_density} shows the density of unobserved SuperCOSMOS objects with apparent $\bJ$ magnitudes less than 19.0 for the 17 fields we observed, divided into annular rings.

Fig.~\ref{fig:J0034} to~\ref{fig:B2356} show radio-optical overlays of each GRG field, with field galaxies ($|\Delta z|\leq0.003$ represented as + signs) overlaid on the radio contour map of the GRG (to scale). The location of the host optical galaxy is also shown. To gauge the quality and depth to which each volume was sampled we computed the completeness as a function of both radius and apparent magnitude, presented in panels (c) and (d) respectively. The field galaxy completeness in GRG fields at low Galactic latitude ($\left|b\right|$ {\scriptsize $\lesssim$} $30^{\circ}$) is lower due to the preponderance of bright stars that mask regions of the SuperCOSMOS field. Also, for the most part, the innermost regions of these GRG fields ({\scriptsize $\lesssim$} $1$~Mpc) are unaffected. Additional completeness plots are provided in Appendix A showing distributions of $\bJ$ apparent magnitude for all SuperCOSMOS targets, observed objects and galaxies with redshifts in each field. Inset plots show radial completeness with respect to the field centre for all fields. Appendix A is available online in a supplementary document.

\subsection{Galaxy stellar masses}
\label{sec:StellarMass}

We have estimated the stellar masses of all galaxies in our sample en route to deriving the stellar mass density of the local environment. This is used to characterise the galaxy distribution in relation to the orientation of the GRG radio axes. Stellar masses are determined from $g$-band luminosities ($M_g$) and the corresponding $M/L$ ratio, derived from the $(g-r)$ calibration of~\citet[][Table 7]{Belletal2003} adopting \mbox{$M_{\odot,g} = 5.15$}. We converted existing $b_{\mathrm{J}}$ and $r_{\mathrm{F}}$ SuperCOSMOS photometry to $g$ and $r$ using relations given in~\citet{DeProprisetal2004}.

\subsection{Overdensities}
Our observations of the optical fields surrounding the GRGs allow us to determine the density of the regions in which they grow as well as how the galaxies in their neighbourhood are distributed and related to the radio structures. Galaxies have a distribution of optical luminosities; in order to measure environmental richness, our observations need to probe the lower end of the galaxy luminosity distribution. For each of the GRGs we have estimated the average galaxy number density and luminosity density for a given minimum luminosity by assuming the luminosity function of~\citet{Jonesetal2006}. The minimum luminosity is determined from the limiting apparent magnitude of each field, applying a $k$-correction appropriate to the redshift of the source. The galaxy number and luminosity density for each field is computed using cylindrical volumes centred on the host galaxy, applying a faint absolute magnitude limit of $-19.49$~mag. Galaxy number and luminosity overdensities are determined using these density values and a corresponding average for the Universe. In particular, the average galaxy number density down to limiting luminosity $L$ is given by

\begin{flalign}\label{eq:average_galaxy_number_density}
n(L) &= \int_L^\infty \phi(L)\,\mathrm{d}L\\
 &= \frac{\ln10}{2.5}\phi^*\int_{\left(\frac{L}{L^*}\right)}^\infty \left(\cfrac{L}{L^*}\right)^\alpha\mathrm{e}^{-\left(\frac{L}{L^*}\right)}\,\mathrm{d}\left(\frac{L}{L^*}\right)
\end{flalign}

\noindent where we have used $h=0.71$, $\phi^*=10^{-1.983} h^3$ Mpc$^{-3}$, \mbox{$M_{\odot,B} = 5.48$},\\
$M^* = -19.91 + 5\log_{10}(h)$,\\
$L^* = -0.4(M^*-M_{\odot,B})$\\
and $\alpha=-1.21$ as given in~\citet{Jonesetal2006}. The corresponding average galaxy luminosity density is given by
\begin{flalign}\label{eq:average_galaxy_luminosity_density}
\mathscr{L}(L) = \frac{\ln10}{2.5}\phi^*L^*\int_{\left(\frac{L}{L^*}\right)}^\infty \left(\cfrac{L}{L^*}\right)^{\alpha+1}\mathrm{e}^{-\left(\frac{L}{L^*}\right)}\,\mathrm{d}\left(\frac{L}{L^*}\right).
\end{flalign}

Stellar mass overdensity is determined from the stellar mass density in volumes around each host and the average stellar mass density for the Universe. The latter is estimated with the $g$-band stellar mass function of \citet{Belletal2003}. We apply the limiting $b_{\mathrm{J}}$ apparent magnitude for a given field to the galaxies found in the volume enclosing the host galaxy. Also, a lower stellar mass limit is applied to both densities, which is computed using the limiting $b_{\mathrm{J}}$ apparent magnitude for the given field and an estimated $r_{\mathrm{F}}$ value assuming an average Sab galaxy $b-r$ colour of $1.34$ from Table 3(e) in~\citet{Fukugitaetal1995}.

\subsection{Fourier Components}
\label{sec:Fourier_component_analysis}
The Fourier Component analysis technique detailed here is the method we use to quantify anisotropy in the galaxy distribution surrounding each GRG host~\citep[e.g.][]{Thoratetal2013}. We define five parameters that between them represent galaxy overdensities within large regions surrounding the GRG hosts in locations relative to the GRG structure. The Fourier components $a_k$ are given by
\begin{equation}\label{eq:Fourier_components}
a_k = \cfrac{\sum\limits_{i=1}^{n}(w_i.f_k(\theta_i))}{\bar{\alpha}_1},
\end{equation}
where $n$ is the number of galaxies excluding the host, $w_i$ is a weight factor, $f_k(\theta_i)$ are Fourier terms equal to $1$, $\sin(\theta_i)$, $\cos(\theta_i)$, $\sin(2\theta_i)$, and $\cos(2\theta_i)$ for $k$ = $1$, $2$, $3$, $4$ and $5$, respectively, and $\bar{\alpha}_1$ is a normalisation factor applied to each Fourier component term. The angles $\theta_i$ between the longer lobe vector (here defined as $0^{\circ}$) and the vector made by each surrounding galaxy with the host (as projected on the plane of the sky) are measured clockwise from $0$ to $-180^{\circ}$ and anticlockwise from $0$ to $+180^{\circ}$. The normalisation factor $\bar{\alpha}_1$ is equal to the mean of the numerator of equation~\eqref{eq:Fourier_components}, with $k=1$, for volumes (equal in size to that centred at the host) placed randomly in the surrounding $2^{\circ}$ field.

In Fig.~\ref{fig:Fourier_components_schematic} we present a visual representation of the sectors corresponding to each Fourier component parameter. The $a_1$ parameter represents the average overdensity in the field, while the $a_2$ to $a_5$ parameters provide a way to quantify the dipole and quadrupole angular distribution of objects surrounding the host. For a non-uniform distribution of galaxies around the radio source, these parameters have non-zero values and the sign of each parameter indicates the direction of the overdensity with respect to the longer lobe. Specifically, the $a_2$ parameter provides a measure of the density difference between the two sides of the radio axis, and the sign of $a_2$ indicates which side is overdense. The $a_3$ parameter is a measure of the overdensity along the radio axis and a positive sign indicates an overdensity on the longer side, whereas a negative $a_3$ value corresponds to an overdensity on the shorter side. The $a_4$ and $a_5$ parameters measure the quadrupole anisotropy in the galaxy density distribution about the longer radio lobe axis. A positive $a_4$ value corresponds to an overdensity at angles (to the radio axis) of $45^{\circ}$ and $225^{\circ}$, whereas a negative $a_4$ value corresponds to an overdensity along $135^{\circ}$ and $315^{\circ}$. A positive $a_5$ term indicates that there is an overdensity along the longer lobe axis compared to a negative $a_5$ value indicating an overdensity orthogonal to this radio axis.

We consider only galaxies within cylindrical volumes of radius $R$ and length $2\ell$ around each host galaxy. Hence, equation~\eqref{eq:Fourier_components} becomes
\begin{equation}\label{eq:Fourier_components_cylinder}
a_k(r,l) = \cfrac{\sum\limits_{i=1}^{n}(w_i.f_k(\theta_i,r,l))}{\bar{\alpha}_1},
\end{equation}
where $f_k(\theta_i,r,l)$ is equal to 0 for objects with offsets from the centre of the given cylinder either projected on the sky with $r > R$ or offset orthogonal to the plane of the sky with $l > \ell$. Error estimates are computed for each dipole and quadrupole Fourier component parameter using Jackknife resampling of the galaxies in the volume surrounding the host. The Jackknife method consists of removing a galaxy, re-computing each $a_k$ parameter, replacing the galaxy and repeating these steps for each galaxy found in the cylinder. The error for each Fourier component is then given by the Jackknife standard error for the respective set of re-computed $a_k$ values. The standard deviation of the corresponding Fourier components computed for a set of `reference volumes' randomly placed in the surrounding field is instead used as the error of the monopole parameter $a_1$ and for the other parameters if there are fewer than three galaxies detected in the host volume. These error terms are also normalised by the factor $\bar{\alpha}_1$.

The reference volumes are cylinders, equal in size to the host volume, placed throughout the field with the aim of filling the surrounding three-dimensional region without intersection. Each reference volume is centred on a dummy GRG host with a position determined by first randomly selecting an angle from $0$ to $360^{\circ}$. A displacement $r$ from the field centre on the plane of the sky is taken from the interval $r^2\sim\left[0, (0.8^{\circ}-R)^2\right]$ to avoid bias towards small radii, where 0.8$^{\circ}$ is the completeness boundary (see Section~\ref{sec:Galaxy_distributions}) and $R$ is the radius of the given reference volume. The largest physical radius is 3 Mpc (included for the extreme case of J0331--7710), whereas 2-Mpc radii are sufficient for the remaining sources. For B0503--286 and B0511--305 the host galaxies lies within 2~Mpc of the 0.8$^{\circ}$ completeness boundary, due to an offset in host position from each AAT/AAOmega field centre, and so in these two cases the host volume extends beyond the $0.8^{\circ}$ threshold. Finally, a random offset in redshift within $\pm 0.02$ from the host is applied. Also, it is necessary to vary the random configuration of reference volumes should the denominator $\bar{\alpha}_1$ equal zero for a sparsely sampled field. However, it is not possible to compute Fourier components in cases where no other objects are detected in the volume enclosing the host; such GRGs would be considered isolated with respect to objects above the limiting apparent magnitude for which the field has been sampled.

We introduce a sixth Fourier component parameter $a_0$ that provides a measure of overdensity of each host volume with respect to the mean of the surrounding regions of all fields given by\\
\begin{equation}\label{eq:a0_definition}
a_0=\cfrac{a^*_1}{\bar{a}_0},
\end{equation}
where $a^*_1$ is the numerator of equation~\eqref{eq:Fourier_components_cylinder} for $k=1$, i.e. the number of galaxies in the volume surrounding a given GRG host galaxy; $\bar{a}_0 = \frac{1}{N}\sum\limits_{j=1}^N a_{1_j}$; and $N$ is the number of randomly placed reference volumes in the given set of fields. The quantities $a^*_1$ and $a_{1_j}$ are normalised by the volumes of the cylindrical regions for which they are computed. This $a_0$ parameter permits comparison of the local environments of each GRG, whereas the $a_1$ parameter represents the relative density of a host volume compared to the rest of the given field.

As a test of the effect of stellar contamination on the derived Fourier component values, we constructed uniform random distributions about each host galaxy and noted any change in the Fourier component values when introducing the masks from the SuperCOSMOS input catalogues attributed to bright stars. We find that the masks in general have negligible effect on the significance of the Fourier component results, except those of $a_2$ and $a_3$ (the dipole moments) in the cases of the two most heavily masked fields, B0707--359 and B1545--321.

\section{Results}
\label{sec:Results}

Our multi-object spectroscopic observations of GRG fields are analysed in two ways. In one approach we use the Fourier components that we computed for each GRG field and in the other we use the overlays of GRG radio images and the projected optical galaxy distribution. Table~\ref{tbl:host_volume_density} lists galaxy numbers and densities in the regions around each GRG host galaxy, and Table~\ref{tbl:host_volume_overdensity} provides the corresponding number, luminosity and stellar mass overdensities. These quantities have errors due to redshift measurement uncertainties and peculiar velocities of individual galaxies; however in most cases these are dominated by Poisson errors, which are given as error estimates in both tables. In Table~\ref{tbl:Fourier_components_equal_weight} we present the results of our Fourier Component analysis (see Section~\ref{sec:Fourier_component_analysis} for details). The five parameters calculated for each GRG field, which quantify the surrounding galaxy distribution in terms of dipole and quadrupole moments, are presented together with their respective errors. In the second method we give overlays of the radio maps and respective optical fields that show the distribution of galaxies within a certain redshift range of the host redshift.

The Fourier component analysis method has been carried out using cylindrical volumes with $|\Delta z| \leq 0.003$, where $\Delta z$ is the redshift difference between a neighbouring galaxy and the GRG host. This corresponds to ${\sim}2$ correlation lengths and is the median extent in the redshift distributions about the host for which the number of galaxies approaches zero. This redshift range corresponds to physically large distances ($\pm11.8$ to $\pm12.5$~Mpc) but is similar to that used in the previous two case studies, B0503--286~\citep{Subrahmanyanetal2008} and B0319--454~\citep{Safourisetal2009}, and allows for a complete picture to be discerned for any galaxy concentrations seen in the neighbourhood of the GRG host. The extent of these cylinders encloses the majority of galaxies with redshifts about the host. The corresponding velocity offset of $\pm900$~km~s$^{-1}$ is greater than the velocity dispersion expected for the GRG host optical galaxies in group environments~\citep[as compared to clusters, e.g.][]{Rueletal2014}. Appendix B, available online in a supplementary document, presents redshift distributions about each GRG host redshift.

The observations we present in this paper are focused on studying the surrounding WHIM gas environment in which the GRGs are expected to form and evolve. The underlying presumption is that galaxy concentrations also indicate the presence of gas. A point to note is that, given the physically large sizes of GRGs, the host galaxy environment is not necessarily the same as the environment into which the lobes grow. The median projected linear size of the GRG sample is ${\sim}1$~Mpc. The Fourier component analysis is based on volumes (typically cylinders of 4-Mpc diameter) surrounding the giant radio sources on scales that encompass their immediate environment. The $2^{\circ}$ regions observed with the AAT/AAOmega, discussed with respect to the radio-optical overlays, are on the order of 10~Mpc in diameter. The previous two case studies have been able to clearly relate the GRG morphologies with the galaxy distribution on Mpc scales. Our sample study presented here is an opportunity to examine the influences of ambient medium on formation, growth and evolution of GRGs, and their morphologies.

In the course of our study, we assessed the robustness of the results we obtained by considering the extent of stellar contamination, completeness of the data and uniformity over the 2$^{\circ}$-diameter field. As can be seen from panels (c) and (d) in Fig.~\ref{fig:J0034} to~\ref{fig:B2356}, the data are highly complete within a radius of 0.8$^{\circ}$ centred on the GRG host. For J0843--7007 and B1302--325, where the completeness levels are more affected, it is still only down to ${\sim70}$ per cent at the far edges of the region. Within 2~Mpc of the GRG host, most fields are nearly 100 per cent complete. As mentioned, data used for B0503--286 and B0319--454 were obtained by~\citet{Subrahmanyanetal2008} and \citet{Safourisetal2009} and do not share the same observational set up as the rest in the sample. Only one GRG, B0707--359, is severely affected by stellar contamination due to its low Galactic latitude and as a result we do not include it in our analysis of the results. Separately, B0703--451 has been excluded from analysis and discussion pending unambiguous identification of the host galaxy.

\subsection{Environmental characteristics of the GRG population}

\subsubsection{Overdensities}
\label{subsubsec:Overdensities}

In Table~\ref{tbl:host_volume_density} the mean galaxy number density is 0.07 Mpc$^{-3}$ for a radius of 1~Mpc. On average this is typical for the density in groups and poor clusters, and a factor of 10 less dense than galaxy clusters~\citep{Bahcall1999}. The number overdensities corresponding to the field luminosity function, shown in Table~\ref{tbl:host_volume_overdensity}, are in the range 6 to 299 with a mean of 73. The corresponding mean luminosity and stellar mass overdensities are 46 and 14, respectively. There is no clear difference between the asymmetric and symmetric GRGs. However, there is considerable variation (a factor of ${\sim}50$) in the overdensities indicating that GRGs are found in a range of overdense environments.

\begin{table}
 \caption{The number of galaxies (N) and the corresponding galaxy number density ($n$; Mpc$^{-3}$) within cylindrical volumes containing the host with 1, 2 and 3-Mpc radii, and 24-Mpc length along the line of sight. A faint absolute magnitude threshold of $-19.49$~mag was applied to avoid bias towards better sampled fields and required the exclusion of the three most poorly sampled fields (J0746--5702, J0843--7007 and B1302--325). Galaxies are weighted to account for the observational completeness of SuperCOSMOS objects at the corresponding apparent magnitude. Error estimates are given in parentheses beside each number density ($\Delta n$; \mbox{$\times 10^{-3}$ Mpc$^{-3}$}).}\label{tbl:host_volume_density}
 \begin{tabular}{l@{\hspace{0.8em}}r@{\hspace{0.9em}}r@{\hspace{1em}}r@{\hspace{0.9em}}r@{\hspace{1em}}r@{\hspace{0.9em}}r}
  \hline
   Source & \multicolumn{6}{c}{Radius} \\
                 & \multicolumn{2}{c}{1 Mpc} & \multicolumn{2}{c}{2 Mpc} & \multicolumn{2}{c}{3 Mpc} \\
                 & N & \multicolumn{1}{c}{$n$} & N & \multicolumn{1}{c}{$n$} & N & \multicolumn{1}{c}{$n$} \\
  \hline
  \multicolumn{7}{l}{(a) \emph{Asymmetric radio lobes}}\\
  \\
  J0116--473   & $3 $ & $0.046(26)$ & $4 $ & $0.015(7)$ & $10$ & $0.017(5)$ \\
  B0319--454   & $4 $ & $0.064(32)$ & $7 $ & $0.028(11)$ & $9 $ & $0.016(5)$ \\
  J0331--7710  & $1 $ & $0.014(14)$ & $1 $ & $0.003(3)$ & $3 $ & $0.005(3)$ \\
  B0503--286$^a$ & $3 $ & $0.041(24)$ & $10$ & $0.034(11)$ & \multicolumn{2}{c}{$-$} \\
  B0511--305$^a$ & $4 $ & $0.057(28)$ & $5 $ & $0.018(8)$ & \multicolumn{2}{c}{$-$} \\
  B1308--441   & $14$ & $0.186(50)$ & $26$ & $0.086(17)$ & $29$ & $0.043(8)$ \\
  \\
  Mean         & $4.83$ & $0.068(13)$ & $8.83$ & $0.031(4)$ & $12.75$ & $0.020(3)$ \\
  \hline
  \multicolumn{7}{l}{(b) \emph{Symmetric radio lobes}}\\
  \\
  J0034--6639  & $5 $ & $0.067(30)$ & $9 $ & $0.030(10)$ & $15$ & $0.022(6)$ \\
  J0400--8456  & $4 $ & $0.054(27)$ & $5 $ & $0.017(8)$ & $11$ & $0.017(5)$ \\
  J0459--528   & $6 $ & $0.084(34)$ & $18$ & $0.064(15)$ & $22$ & $0.035(7)$ \\
  J0515--8100  & $1 $ & $0.014(14)$ & $1 $ & $0.003(3)$ & $4 $ & $0.006(3)$ \\
  B1545--321   & $3 $ & $0.047(27)$ & $3 $ & $0.012(7)$ & $5 $ & $0.008(4)$ \\
  J2018--556   & $5 $ & $0.066(30)$ & $7 $ & $0.023(9)$ & $11$ & $0.016(5)$ \\
  J2159--7219  & $8 $ & $0.107(38)$ & $11$ & $0.037(11)$ & $18$ & $0.027(6)$ \\
  B2356--611   & $6 $ & $0.080(33)$ & $20$ & $0.067(15)$ & $29$ & $0.043(8)$ \\
  \\
  Mean         & $4.75$ & $0.065(11)$ & $9.25$ & $0.032(4)$ & $14.38$ & $0.022(2)$ \\
  \hline
  Overall mean & $4.79$ & $0.066(8)$ & $9.07$ & $0.031(3)$ & $13.83$ & $0.021(2)$ \\
 \end{tabular} 
 
 \medskip
 $^a$ Excluding the 3-Mpc radius case for which the volume centred on the host would extend beyond the boundaries of the observed region.
\end{table}

\begin{table}
 \caption{Measurements of galaxy number, luminosity and stellar mass overdensity in the environments of the GRG sample from cylindrical volumes of 3-Mpc radius and 24-Mpc length (unless otherwise noted) centred on the GRG host. The dimensionless quantities listed here are a measure of the overdensity of each field with respect to an estimated average density for the Universe. The $b_{\mathrm{J}}$ luminosity function of~\citet{Jonesetal2006} was used to estimate the average number and luminosity densities. The $g$-band stellar mass function of~\citet{Belletal2003} was used to determine the corresponding average stellar mass density. The integration limits for both functions are based on the limiting $b_{\mathrm{J}}$ apparent magnitude of the respective field. The GRGs B0703--451 and B0707--359 are excluded due to uncertainty in the host redshift and problems with observational completeness due to low Galactic latitude, respectively. Error estimates are given for each quantity.}\label{tbl:host_volume_overdensity}
 \begin{tabular}{lrrr}
  \hline
   Source & Number & Luminosity & Stellar mass \\
                 & overdensity        & overdensity            & overdensity \\
  \hline
  \multicolumn{4}{l}{(a) \emph{Asymmetric radio lobes}}\\
  \\
  J0116--473     & $23\pm7$ & $14\pm5$ & $37\pm15$ \\
  B0319--454     & $67\pm14$ & $42\pm9$ & $1\pm1$ \\
  J0331--7710    & $6\pm3$ & $3\pm2$ & $1\pm1$ \\
  B0503--286$^a$ & $134\pm26$ & $84\pm16$ & $13\pm4$ \\
  B0511--305$^a$ & $84\pm19$ & $53\pm12$ & $4\pm1$ \\
  B1308--441     & $299\pm25$ & $191\pm16$ & $16\pm2$ \\
  \\
  Mean           & $102\pm7$ & $65\pm5$ & $12\pm3$ \\
  \hline
  \multicolumn{4}{l}{(b) \emph{Symmetric radio lobes}}\\
  \\
  J0034--6639    & $62\pm12$ & $38\pm7$ & $9\pm2$ \\
  J0400--8456    & $30\pm8$ & $19\pm5$ & $6\pm4$ \\
  J0459--528     & $106\pm15$ & $67\pm10$ & $8\pm2$ \\
  J0515--8100    & $16\pm6$ & $10\pm4$ & $3\pm1$ \\
  J0746--5702    & $10\pm5$ & $6\pm3$ & $1\pm1$ \\
  J0843--7007    & $7\pm4$ & $4\pm2$ & $9\pm9$ \\
  B1302--325     & $30\pm9$ & $18\pm6$ & $23\pm10$ \\
  B1545--321     & $16\pm6$ & $9\pm4$ & $1\pm1$ \\
  J2018--556     & $94\pm14$ & $59\pm9$ & $2\pm1$ \\
  J2159--7219    & $68\pm12$ & $43\pm8$ & $8\pm2$ \\
  B2356--611     & $190\pm20$ & $121\pm13$ & $94\pm15$ \\
  \\
  Mean           & $57\pm3$ & $36\pm2$ & $15\pm2$ \\
  \hline
  Overall mean   & $73\pm3$ & $46\pm2$ & $14\pm2$ \\
 \end{tabular} 
 
 \medskip
 $^a$ Using a 2-Mpc radius to ensure that the volume does not extend beyond the boundaries of the observationally complete region.
\end{table}

In Fig.~\ref{fig:GRG_SDSS_histogram} we show the distribution of galaxy number densities for cylindrical volumes of 24-Mpc length and 2-Mpc radius centred on each GRG host. This is overlaid on a corresponding distribution for $1.4\times10^4$ equally sized volumes centred on Sloan Digital Sky Survey~\citep[SDSS;][]{Abazajianetal2009} galaxies with absolute magnitudes within $0.5$~mag of the median GRG host absolute $B$-band magnitude of $-21.4$~mag. We apply a faint absolute magnitude threshold of $-19.49$~mag to all GRG and SDSS volumes in order to permit comparison of the galaxy number densities without bias due to varying observational completeness between fields. This limit corresponds to the brightest limiting absolute magnitude for the set of 14 GRGs fields displayed. We exclude the three most poorly sampled GRG fields (J0746--5702, J0843--7007 and B1302--325), which do not reach this completeness limit, in addition to the two previously mentioned fields B0703--451 and B0707--359. The SDSS sample has a mean (median) galaxy number density of 0.012 (0.010) Mpc$^{-3}$ and the overlaid sample of 14 GRGs has a mean (median) galaxy number density of 0.031 \mbox{(0.026) Mpc$^{-3}$}, a factor of ${\sim}3$ higher. We compare these samples using the non-parametric Kolmogorov-Smirnov test and find that they are not derived from a common distribution at a 95 per cent level of confidence \mbox{($p$-value = $4.6\times10^{-4}$)}. The average galaxy number density for all GRGs in our sample, of 0.03 Mpc$^{-3}$ as already noted, is similar to the densities in groups and poor clusters~\citep{Bahcall1999}.

\subsubsection{Fourier component analysis}
\label{subsubsec:Fourier_component_analysis}

Table~\ref{tbl:Fourier_components_equal_weight} presents the Fourier component parameters computed for the individual GRG fields. The GRGs are divided into two groups based on lobe length ratio. Eight GRGs in our sample are in the asymmetric group (Table~\ref{tbl:GRG_radio_properties}), for which lobes in one direction have grown at least 30 per cent more than those in the opposite direction, and the rest form the symmetric GRGs group. For each group, and for the whole sample, we also list `stacked' Fourier components which are derived from the co-added fields after aligning to an average redshift of 0.1 and rotating such that the longer lobe radio axis lies at PA $=0^{\circ}$. For reasons given above, regarding uncertainty in the host and low Galactic latitude respectively, we exclude two GRGs B0703--451 and B0707--359 from our discussion. Stacked parameters are listed for an apparent $b_{\mathrm{J}}$ magnitude-limited sample, as well as for an absolute magnitude-limited sample. The latter prevents bias towards fields with lower redshift GRGs and better-sampled fields with fainter limiting magnitudes. However, adopting this limit also further reduces the sample size and significantly reduces the total number of galaxies in the stacked field analysis. As in Section~\ref{subsubsec:Overdensities}, a threshold of $-19.49$~mag is used. Three fields with limiting absolute magnitudes brighter than this threshold were excluded (J0746--5702, J0843--7007, and B1302--325). We also investigated stellar mass-weighting of galaxies to avoid bias towards better-sampled fields. However, we find a faint absolute magnitude limit accounts for this bias sufficiently without introducing additional uncertainties. Tables of stellar mass-weighted Fourier component results are presented in Appendix C for comparison (available online in a supplementary document).

In Table~\ref{tbl:Fourier_components_equal_weight}, the $a_1$ parameters, representing the density ratio around GRG hosts compared with the mean of the rest of the field surrounding that GRG, are high for 9 of the 17 GRGs, suggesting that there is clustering of galaxies in the vicinity of the GRG hosts. The overall mean of ${\sim}10$ is lower than the overdensities calculated in Table~\ref{tbl:host_volume_overdensity}, implying that the fields beyond the cylinder enclosing the GRGs are themselves overdense.

Two individual dipole $a_2$ parameters (density gradient perpendicular to the radio axis) are significant but the stacked values are not significant. Similarly, the overall stacked dipole $a_3$ parameter (density gradient parallel with the radio axis) is not significant. However, for the asymmetric sources, we find that the majority of the sources have a negative $a_3$ dipole parameter. One individual source (B0503--286) is strongly negative, and the stacked value, compiled from 207 galaxies, is significant at the $2.3\sigma$ level. The negative sign of $a_3$ implies an increased overdensity on the side of the shorter lobes compared to longer lobes. That is, the jets appear to have grown to shorter extents on the side of larger galaxy overdensity. In the case of the asymmetric GRG, B0319--454, which does not have a negative $a_3$, the galaxies on the longer lobe side are widely dispersed and the jet appears to have taken a path devoid of galaxies (see Section~\ref{subsubsec:Individual_sources} for additional notes on this GRG). Additionally for this GRG, several bright galaxies have not been catalogued on the shorter lobe side (see Section~\ref{subsubsec:Individual_sources}). Local environments over scales of few hundred kpc, to which the Fourier component method may not be sensitive, may also play a more influential role in impeding jet propagation as in the case of B1308--441 (see Section~\ref{subsubsec:Individual_sources}).

Some of the individual quadrupole $a_4$ and $a_5$ parameters in Table~\ref{tbl:Fourier_components_equal_weight} are also significant. The stacked values for $a_4$ in particular are significant, though of opposite signs for the two groups. For the asymmetric-GRG subset, the highly significant negative component suggests an overdensity at angles to the radio axis of $135^{\circ}$ and $315^{\circ}$ (see Section~\ref{subsec:QuadrupoleMoment} for a separate discussion). For symmetric GRGs, the positive $a_4$ value shows that the galaxy overdensities are along $45^{\circ}$ and $225^{\circ}$ position angle. For the latter group, the $a_4$ value (as with $a_3$) is dominated by the galaxy distribution surrounding GRG B2356--611. The stacked quadrupole $a_5$ values, which measure galaxy overdensity orthogonal to the GRG lobes, are not as significant.

Fig.~\ref{fig:Stacked_Fourier_component_parameters_vs_radius} shows plots of Fourier component values for a stacked field, comprising the fields from entry (c) in Table~\ref{tbl:Fourier_components_equal_weight}, for cylinders of various radii and indicates that sensitivity decreases with increasing radius.

\begin{table*} 
 \caption{Fourier components and estimates of their errors for the complete sample of giant radio galaxies. Optical galaxies detected in cylindrical volumes of 2-Mpc radius (unless otherwise noted) are given equal weighting. A redshift range of $|\Delta z|\leq0.003$ is used to define the extent of the cylindrical volume surrounding the host galaxy. The corresponding physical length is determined using the Hubble constant at the given host galaxy redshift and this length is used for all reference volumes placed throughout the field. Only galaxies brighter than the limiting magnitude for each field are included in the Fourier component analysis. Entries for stacked fields denoted `Abs. mag. lim.' as well as all $a_0$ parameters are computed with a faint absolute magnitude limit of $-19.49$~mag applied. The number of observed galaxies within the host volume and all reference volumes are also listed. Measurements that are significant at the 3$\sigma$ level appear in bold typeface. Error estimates for $a_2$ to $a_5$ are computed using Jackknife resampling of galaxies in the host volume (see Section~\ref{sec:Fourier_component_analysis} for details).}\label{tbl:Fourier_components_equal_weight}
 \begin{tabular}{l@{}c@{ }c@{}rrrrrrr}
  \hline
                                            &\multicolumn{2}{c}{Number of galaxies}&     &                                                                     & \multicolumn{2}{c}{Dipole} & \multicolumn{2}{c}{Quadrupole}\\
   Field                            & host vol. & random vols & \multicolumn{1}{c}{$a_0$} &  \multicolumn{1}{c}{$a_1$}        &     \multicolumn{1}{c}{$a_2$}             & \multicolumn{1}{c}{$a_3$}                       & \multicolumn{1}{c}{$a_4$}       & \multicolumn{1}{c}{$a_5$} \\
\hline
\multicolumn{6}{l}{(a) \emph{Asymmetric radio lobes}}\\
J0116--473 & $3$ & $40$ & $2.84$ & $2.10\pm1.50$ & $-0.82\pm0.54$ & $-1.16\pm0.49$ & $-0.02\pm0.64$ & $0.20\pm0.75$ \\
B0319--454 & $15$ & $55$ & $5.46$ & $1.09\pm1.17$ & $-0.09\pm0.20$ & $0.40\pm0.16$ & $-0.14\pm0.21$ & $-0.05\pm0.16$ \\
J0331--7710$^a$ & $2$ & $7$ & $-$ & $1.43\pm0.73$ & $-0.23\pm0.41$ & $\pmb{-1.41\pm0.25}$ & $0.45\pm0.49$ & $\pmb{1.35\pm0.41}$ \\
B0503--286 & $27$ & $2$ & $8.11$ & $\pmb{54.00\pm1.00}$ & $-8.31\pm5.04$ & $\pmb{-30.93\pm6.48}$ & $11.55\pm5.81$ & $\pmb{25.03\pm6.56}$ \\
B0511--305$^b$ & $10$ & $8$ & $-$ & $\pmb{25.00\pm2.00}$ & $-6.26\pm5.20$ & $1.95\pm5.03$ & $4.34\pm5.09$ & $-2.19\pm5.31$ \\
B0511--305 & $19$ & $6$ & $3.63$ & $\pmb{12.67\pm1.00}$ & $-1.47\pm1.75$ & $-2.44\pm2.13$ & $-0.67\pm1.90$ & $2.64\pm2.01$ \\
B0703--451$^c$ & $2$ & $41$ & $-$ & $1.12\pm1.45$ & $0.42\pm0.75$ & $0.54\pm0.76$ & $-0.28\pm0.62$ & $-0.07\pm0.76$ \\
B0707--359$^d$ & $0$ & $3$ & $-$ & $-$ & $-$ & $-$ & $-$ & $-$ \\
B1308--441 & $146$ & $49$ & $23.55$ & $\pmb{11.92\pm0.75}$ & $1.53\pm0.72$ & $-0.78\pm0.66$ & $\pmb{-3.83\pm0.64}$ & $-1.20\pm0.67$ \\
\\
Asymmetric stacked$^e$ & $207$ & $122$ & $-$ & $\pmb{6.79\pm0.65}$ & $0.30\pm0.33$ & $-0.77\pm0.33$ & $\pmb{-1.41\pm0.32}$ & $0.09\pm0.33$ \\
Abs. mag. lim. stack$^{ef}$ & $47$ & $42$ & $-$ & $\pmb{4.48\pm0.79}$ & $0.20\pm0.44$ & $-0.77\pm0.46$ & $-0.95\pm0.43$ & $0.27\pm0.46$ \\
\hline
\multicolumn{6}{l}{(b) \emph{Symmetric radio lobes}}\\
J0034--6639 & $15$ & $21$ & $7.44$ & $\pmb{10.00\pm1.28}$ & $1.65\pm2.13$ & $-1.29\pm1.20$ & $3.03\pm1.83$ & $\pmb{-5.07\pm0.84}$ \\
J0400--8456 & $6$ & $19$ & $3.71$ & $2.53\pm1.59$ & $0.58\pm0.62$ & $-0.35\pm0.66$ & $-0.30\pm0.76$ & $0.08\pm0.54$ \\
J0459--528 & $36$ & $12$ & $15.71$ & $\pmb{21.00\pm1.11}$ & $3.02\pm2.65$ & $-1.28\pm2.14$ & $1.55\pm2.61$ & $-4.64\pm2.12$ \\
J0515--8100 & $2$ & $8$ & $0.00$ & $2.00\pm1.32$ & $0.70\pm0.91$ & $-1.53\pm0.63$ & $-0.62\pm0.81$ & $0.54\pm0.34$ \\
J0515--8100$^a$ & $6$ & $17$ & $-$ & $1.41\pm1.11$ & $0.26\pm0.29$ & $-0.51\pm0.39$ & $0.12\pm0.29$ & $0.47\pm0.40$ \\
J0746--5702 & $2$ & $9$ & $-$ & $5.56\pm2.71$ & $0.06\pm2.53$ & $0.44\pm1.90$ & $-1.52\pm1.65$ & $-5.27\pm1.93$ \\
J0843--7007 & $2$ & $19$ & $-$ & $2.74\pm1.34$ & $0.60\pm1.25$ & $2.66\pm1.49$ & $1.16\pm1.61$ & $2.44\pm0.99$ \\
B1302--325 & $4$ & $16$ & $-$ & $\pmb{7.75\pm1.70}$ & $-3.17\pm2.10$ & $-0.56\pm2.21$ & $2.83\pm1.00$ & $-0.92\pm2.93$ \\
B1545--321 & $3$ & $24$ & $1.86$ & $1.12\pm1.52$ & $\pmb{0.61\pm0.12}$ & $0.17\pm0.42$ & $0.58\pm0.35$ & $0.35\pm0.23$ \\
J2018--556 & $26$ & $11$ & $5.46$ & $\pmb{9.45\pm1.73}$ & $-3.06\pm1.21$ & $1.69\pm1.18$ & $-0.80\pm1.24$ & $-0.90\pm1.31$ \\
J2159--7219 & $18$ & $16$ & $9.24$ & $\pmb{7.88\pm1.16}$ & $0.84\pm1.17$ & $0.20\pm1.36$ & $0.57\pm1.33$ & $1.10\pm1.19$ \\
B2356--611 & $63$ & $13$ & $17.56$ & $\pmb{33.92\pm0.78}$ & $\pmb{11.01\pm2.71}$ & $7.09\pm2.82$ & $\pmb{11.64\pm3.10}$ & $-0.92\pm2.49$ \\
\\
Symmetric stacked & $176$ & $69$ & $-$ & $\pmb{10.20\pm0.67}$ & $1.37\pm0.56$ & $0.76\pm0.51$ & $1.69\pm0.57$ & $-0.97\pm0.49$ \\
Abs. mag. lim. stack$^{fg}$ & $66$ & $32$ & $-$ & $\pmb{8.25\pm0.74}$ & $1.48\pm0.71$ & $1.00\pm0.68$ & $\pmb{2.30\pm0.74}$ & $-0.58\pm0.62$ \\
\hline
\multicolumn{6}{l}{(c) \emph{All sources}}\\
Abs. mag. lim. stack$^{efg}$ & $113$ & $70$ & $-$ & $\pmb{6.46\pm0.35}$ & $0.80\pm0.42$ & $-0.00\pm0.42$ & $0.48\pm0.45$ & $-0.10\pm0.40$ \\
  \hline
 \end{tabular}
 
  \medskip
  \begin{flushleft}
  Notes -- $^a$ using 3-Mpc radius cylinders; $^b$ using 1-Mpc radius cylinders; $^c$ the host galaxy of B0703--451 is in doubt - we assume the redshift of the optical galaxy at the GRG centroid ($z=0.1242$) for the individual field Fourier component values but exclude it from the stacked fields; $^d$ no galaxies were detected in the host volume and so Fourier component values could not be computed; $^e$ excluding B0703--451, B0707--359; $^f$a faint absolute magnitude threshold of $M_{\mathrm{B}} = -19.49$ was applied to all constituent fields; $^g$ the three most poorly sampled GRG fields (J0746--5702, J0843--7007 and B1302--325) were excluded as their observational completeness did not reach the faint absolute magnitude limit.
  \end{flushleft}
\end{table*}

\subsubsection{Environments of GRG subsets}
\label{subsubsec:GRG_subsets}

In Table~\ref{tbl:stacked}, we compute Fourier components for several sets of stacked fields grouped according to GRG properties. As before, we use a faint absolute magnitude limit of $-19.49$~mag and 2-Mpc radius reference volumes. Each stacked field is constructed by co-adding the constituent fields as described in Section~\ref{sec:Fourier_component_analysis}. It was necessary to invert certain fields east-west (as indicated in the Table~\ref{tbl:stacked} footnotes) for sources with non-collinear components (categories \emph{d}, \emph{e} and \emph{f}) to align radio structures offset from the central axis.

The significant $a_1$ parameters for all stacked categories indicate that the host volumes for the respective categories of GRGs are more densely populated than the rest of the region in the fields. In the case of sources grouped according to evidence of restarted nuclear activity, see entries (a) to (c) in Table~\ref{tbl:stacked}, there is no clear correlation between the environment and restarted nuclear activity. The dipole and quadrupole parameters are marginally significant at best without any discernible pattern.

The fields associated with GRGs exhibiting offset radio components were combined as a stacked field of 8 GRGs, listed in Table~\ref{tbl:stacked} as entry (d). Four of these fields (J0400--8456, B0503--286, B0511--305 and B2356--611) were inverted about PA $=0^{\circ}$ such that their offset radio components were aligned along a common axis around PA $\sim 45^{\circ}$ and $225^{\circ}$. The significant negative $a_4$ value reflects a distribution of galaxies spanning the stacked field along PA $\sim 135^{\circ}$ and $315^{\circ}$ orthogonal to the offset components. Fig.~\ref{fig:Stacked_Fourier_components_wing_like} represents the significant, negative $a_4$ Fourier component for the first subset of these non-collinear sources consisting GRGs with an extension on one side of the radio axis and no detected counterpart (see entry \emph{e} in Table~\ref{tbl:stacked}). In Fig.~\ref{fig:Stacked_Fourier_components_noncollinear} where we present the other subset comprising GRGs with non-collinear lobes (entry \emph{f} in Table~\ref{tbl:stacked}), which were combined in the same way, the shorter lobes are aligned with PA $\sim 225^{\circ}$ and the significant negative $a_4$ parameter indicates that this is in a direction orthogonal to an overdensity along PA $\sim 135^{\circ}$ and $315^{\circ}$.

\begin{table*}
 \caption{Fourier component values for co-added optical fields surrounding giant radio galaxies grouped according to morphology, including varying evidence of restarted nuclear activity, extended radio components and non-collinear lobes. The stacked fields are centred on the respective GRG host galaxy, aligned longer lobe to north and fields indicated with $\dagger$ are inverted east-west to align offset radio components. Optical galaxies detected in cylindrical volumes of 2-Mpc radius are given equal weighting. A redshift range of $|\Delta z|\leq0.003$ is used to define the extent of the cylinder surrounding the stacked field host galaxy with a redshift of 0.1 (the mean GRG host optical galaxy redshift). The corresponding physical length is used for all reference volumes placed throughout the field. The stacked fields are restricted to galaxies brighter than an absolute magnitude of $M_B=-19.49$~mag to address differences in observational completeness between fields. We exclude the fields J0746--5702, J0843--7007 and B1302--325 with limiting magnitudes brighter than this absolute magnitude limit, as well as B0703--451 due to uncertainty in the host ID and B0707--359 due to excess masking of bright stars at low Galactic latitude. Measurements that are significant at the 3$\sigma$ level appear in bold typeface.}\label{tbl:stacked}
 \begin{tabular}{l@{}c@{ }c@{}rrrrrr}
  \hline
                                            &\multicolumn{2}{c}{Number of galaxies}&                                                                     & \multicolumn{2}{c}{Dipole} & \multicolumn{2}{c}{Quadrupole}\\
   Stacked field                            &host vol.&random vols&  \multicolumn{1}{c}{$a_1$}        &     \multicolumn{1}{c}{$a_2$}             & \multicolumn{1}{c}{$a_3$}                       & \multicolumn{1}{c}{$a_4$}       & \multicolumn{1}{c}{$a_5$} \\
  \hline
(a) strong evidence of restarted nuclear activity$^a$ & $33$ & $25$ & $\pmb{5.28\pm0.69}$ & $-0.17\pm0.68$ & $0.26\pm0.59$ & $0.71\pm0.64$ & $-0.72\pm0.62$ \\
(b) some evidence of restarted nuclear activity$^b$ & $30$ & $6$ & $\pmb{20.00\pm0.33}$ & $2.78\pm2.30$ & $-5.52\pm2.53$ & $5.26\pm2.52$ & $2.84\pm2.32$ \\
(c) no evidence of restarted nuclear activity$^c$ & $50$ & $39$ & $\pmb{5.13\pm0.91}$ & $1.11\pm0.49$ & $0.67\pm0.49$ & $-0.40\pm0.55$ & $-0.15\pm0.45$ \\
(d) lobe structure offset from the jet axis$^d$ & $70$ & $43$ & $\pmb{6.51\pm0.90}$ & $-0.41\pm0.53$ & $-0.21\pm0.55$ & $\pmb{-3.00\pm0.45}$ & $0.22\pm0.52$ \\
\hspace*{3mm}(e) an extension on only one side of the jet axis$^e$ & $25$ & $29$ & $3.45\pm1.24$ & $-0.70\pm0.47$ & $0.93\pm0.43$ & $\pmb{-1.80\pm0.44}$ & $-0.18\pm0.37$ \\
\hspace*{3mm}(f) non-collinear lobes$^f$ & $45$ & $17$ & $\pmb{10.59\pm0.98}$ & $0.16\pm1.06$ & $-2.12\pm1.10$ & $\pmb{-4.52\pm0.82}$ & $0.86\pm1.14$ \\
(g) collinear lobes$^g$ & $43$ & $29$ & $\pmb{5.93\pm0.90}$ & $0.63\pm0.65$ & $0.31\pm0.60$ & $0.84\pm0.68$ & $-0.57\pm0.56$ \\
\hline
 \end{tabular}
  
  \medskip
  \begin{flushleft}
  $^a$ J0034--6639, J0116--473, B0511--305, B1545--321, J2018--556 and J2159--7219\\
  $^b$ J0331--7710, J0400--8456, J0459--528 and B0503--286 \\
  $^c$ B0319--454, J0515--8100, B1308--441 and B2356--611\\
  $^d$ J0116--473, B0319--454, J0400--8456$\dagger$, B0503--286$\dagger$, B0511--305$\dagger$, J0515--8100, B1308--441 and B2356--611$\dagger$\\
  $^e$ B0319--454 and B2356--611$\dagger$\\
  $^f$ J0116--473, J0400--8456$\dagger$, B0503--286$\dagger$, B0511--305$\dagger$, J0515--8100 and B1308--441\\
  $^g$ J0034--6639, J0331--7710, J0459--528, B1545--321, J2018--556 and J2159--7219
  \end{flushleft}
\end{table*}

\subsubsection{Limitations}

We note that limitations with the Fourier component method (due to the use of a fixed cylindrical region of radius 2~Mpc) can affect the individually estimated Fourier component parameters for the galaxy overdensity. This is discussed for each GRG in Section~\ref{subsubsec:Individual_sources}, where we present notes on each field. GRG fields J0116--473, B0503--286, B0511--305, B1302--325 and J2159--7219 are cases where either the imposed radial limit of 2~Mpc, over which the parameters have been computed, excludes the bulk of the galaxy filament (e.g. J2159--7219 and in the sparse fields of J0116--473 and B1302--325) or has incorporated well-separated and distinct galaxy concentrations and thereby offsets the effect of the galaxy distribution in the host vicinity (e.g. B0511--305, also see the notes on this GRG). In the case of B0503--286 there is a strong asymmetry in the galaxy distribution about the host, particularly in the form of a strong clustering of galaxies seen to the NE of the northern lobe at a distance of ${\sim}500$~kpc from the lobe. This results in a large negative $a_3$ parameter but also a positive $a_5$ parameter. For this GRG, the jets having formed in a particular position angle (almost north-south) are affected by the asymmetric galaxy distribution where the northern jet extends to just half the extent of the southern jet.

\subsection{Notes on individual sources}
\label{subsubsec:Individual_sources}

\begin{figure*}
  \centering
  \begin{tabular}{ll}
      \includegraphics[width=0.5\hsize]{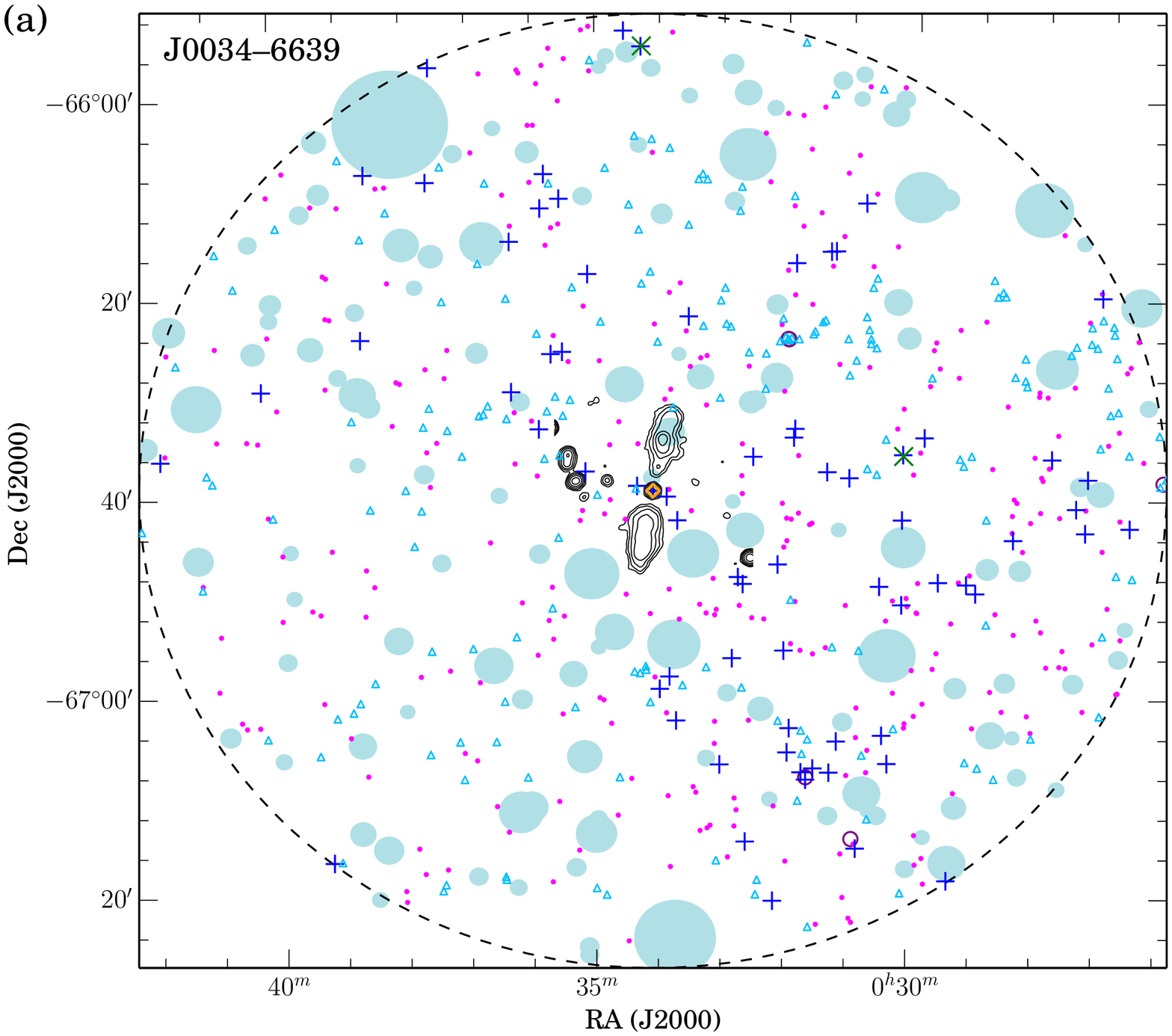} &
      \includegraphics[width=0.5\hsize]{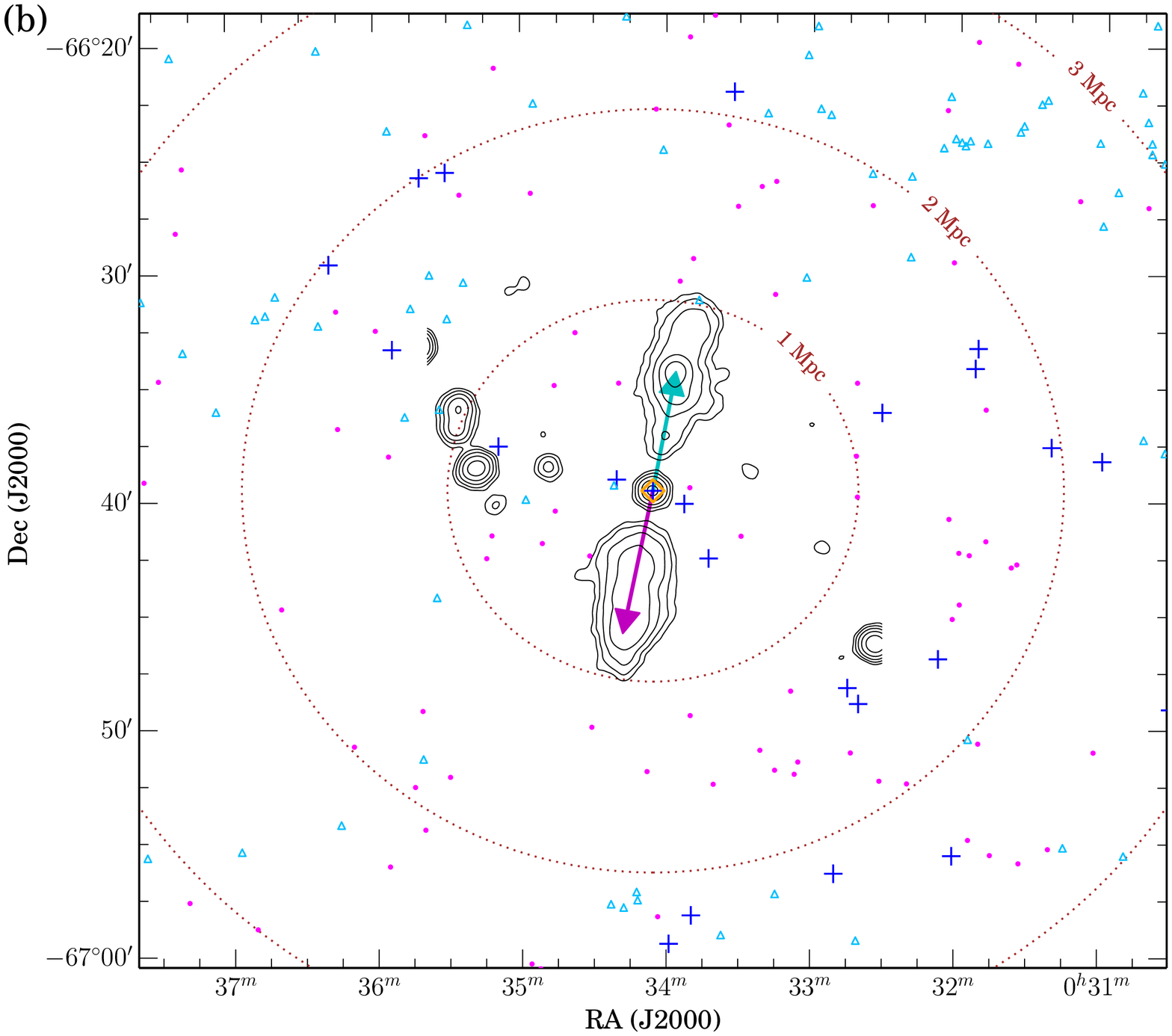}\\
      \includegraphics[width=0.5\hsize]{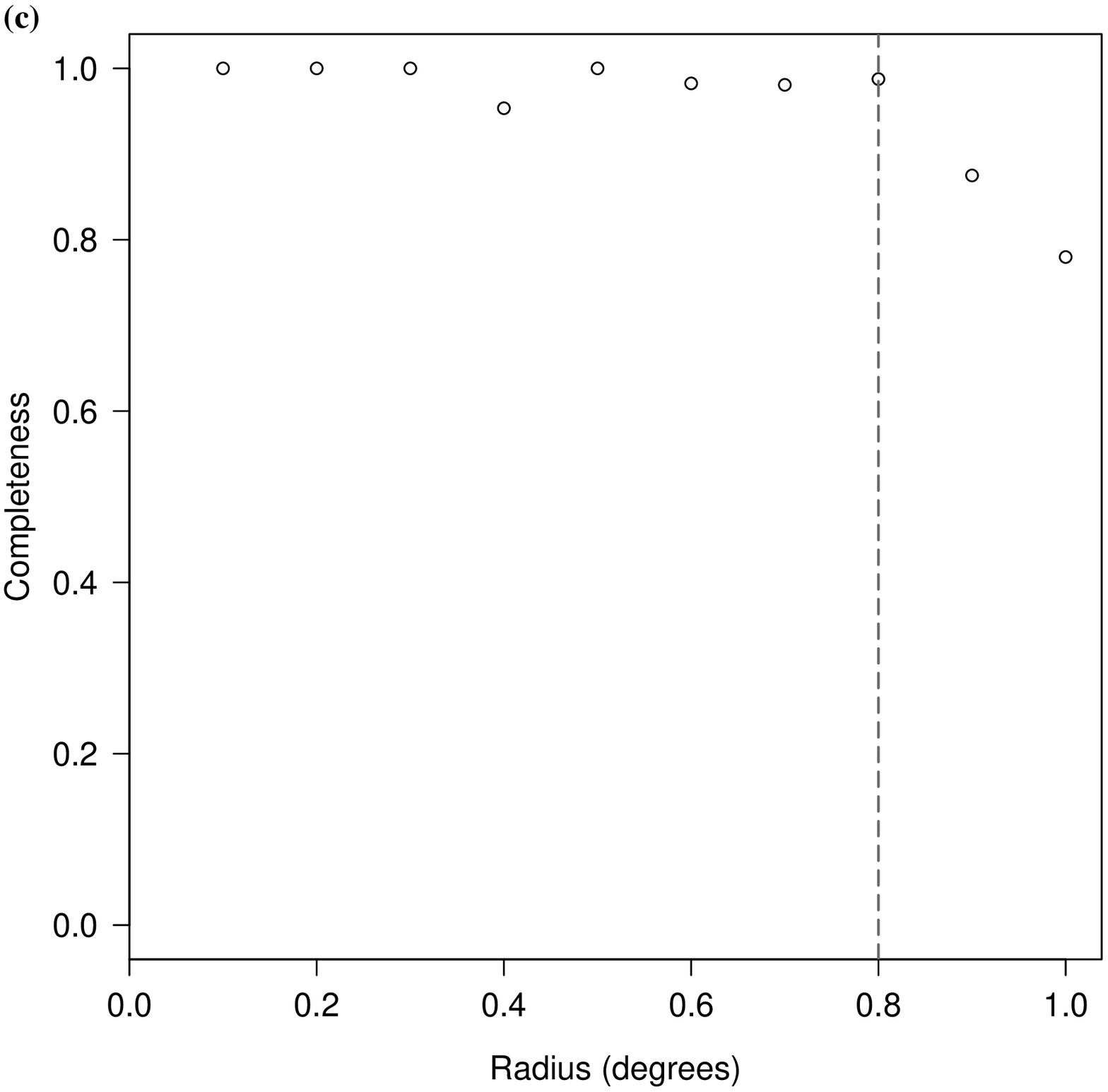} &
      \includegraphics[width=0.5\hsize]{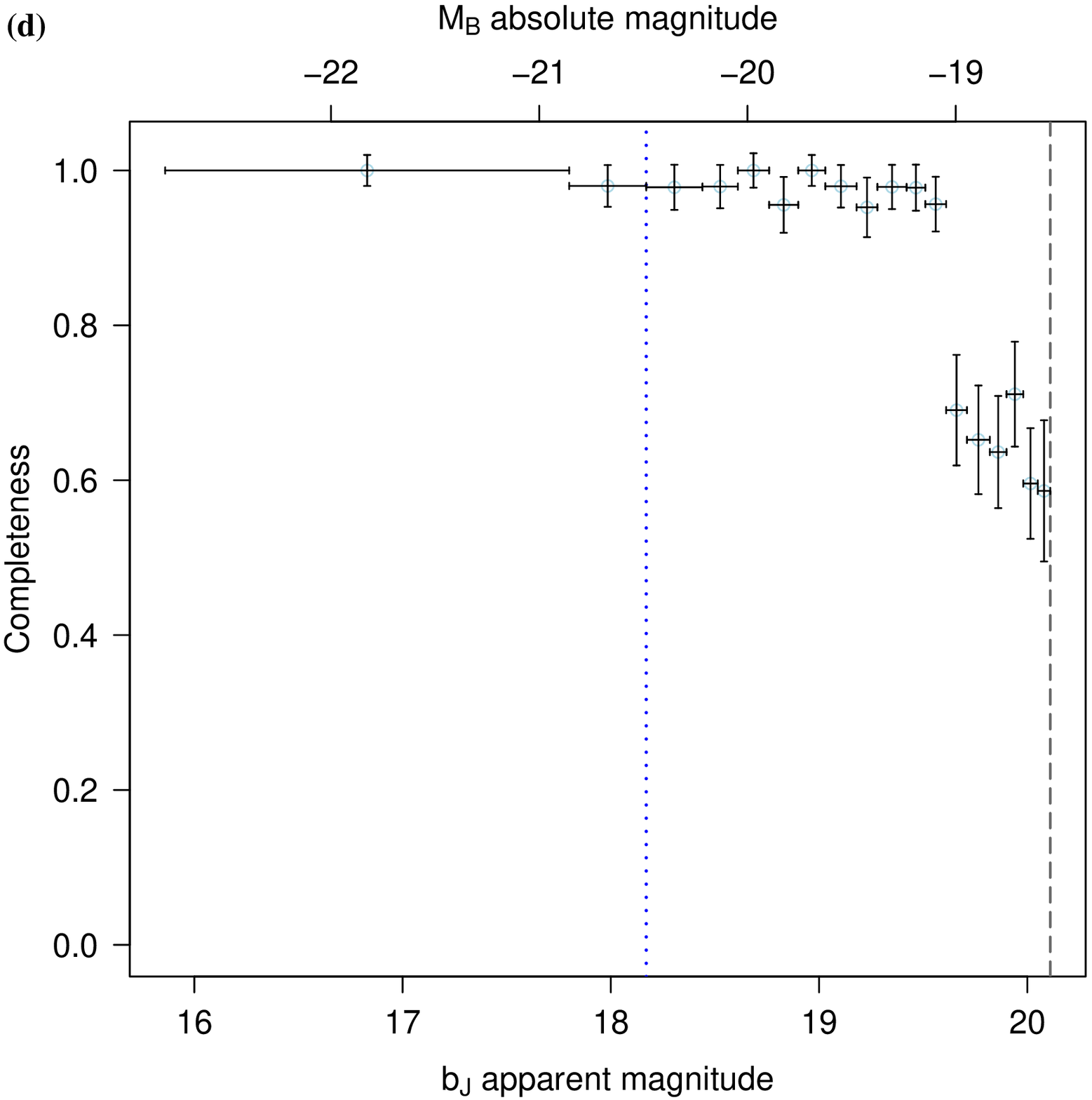}
\end{tabular}
\caption{(\subfigletter{a}) Distribution of galaxies around the giant radio galaxy J0034--6639 within the 2$^{\circ}$ field of the AAT/AAOmega; a 20 arcmin angular scale corresponds to a linear size of 2.38 Mpc. The dashed, black circle represents a 0.8$^{\circ}$ radial boundary chosen to account for observational completeness. The following symbols are used to represent optical objects in this plot and all subsequent radio-optical overlay plots. Plus symbols represent galaxies within $\pm0.003$ of the host redshift, in this case z = 0.1102. Multiple galaxies within 10 arcsec of each other are also marked with a green cross. Redshift-confirmed galaxies outside of this redshift range are marked with cyan triangles and magenta dots to indicate foreground and background galaxies, respectively. Objects with $b_{\mathrm{J}}$ photographic apparent magnitudes brighter than 19.0 that were not observed due to time constraints are indicated by open violet circles (hexagons if more than 1 mag fainter than the host). The GRG host is marked with an orange diamond. The contours displayed in all radio-optical overlay plots represent the low-resolution, wideband, total intensity ATCA radio image of the given GRG at 2.1 GHz with a symmetric beam of FWHM 48 arcsec, unless otherwise noted. Here the contours represent (6, 12, 24, 48, and 96) $\times$ 40 $\mu$Jy beam$^{-1}$. SuperCOSMOS input catalogue masks around bright stars are shaded in light blue. (\subfigletter{b}) An enlarged view of the optical field, centred at the host galaxy, with dotted circles at 1, 2 and 3-Mpc radii. Arrows indicate the P.A. of the longer and shorter lobes, shown in magenta and cyan respectively. (\subfigletter{c}) Radial completeness shown as the fraction of observed SuperCOSMOS targets [including both galaxies and contaminating stars] in intervals of 0.1$^{\circ}$ measured from the host optical galaxy. The dashed line indicates the completeness radius of 0.8$^{\circ}$ used in the Fourier component analysis. (\subfigletter{d}) A completeness plot showing the fraction of observed targets against $B$-band magnitude. The absolute magnitude scale assumes the host redshift, the blue dotted line indicates the $b_{\mathrm{J}}$ apparent magnitude (18.17) of the host galaxy and the dashed line indicates the field's limiting apparent magnitude.}
\label{fig:J0034}
\end{figure*}

\begin{figure*}
\begin{minipage}[t][\textheight]{\textwidth}
  \centering
  \begin{tabular}{ll}
      \includegraphics[width=0.5\hsize]{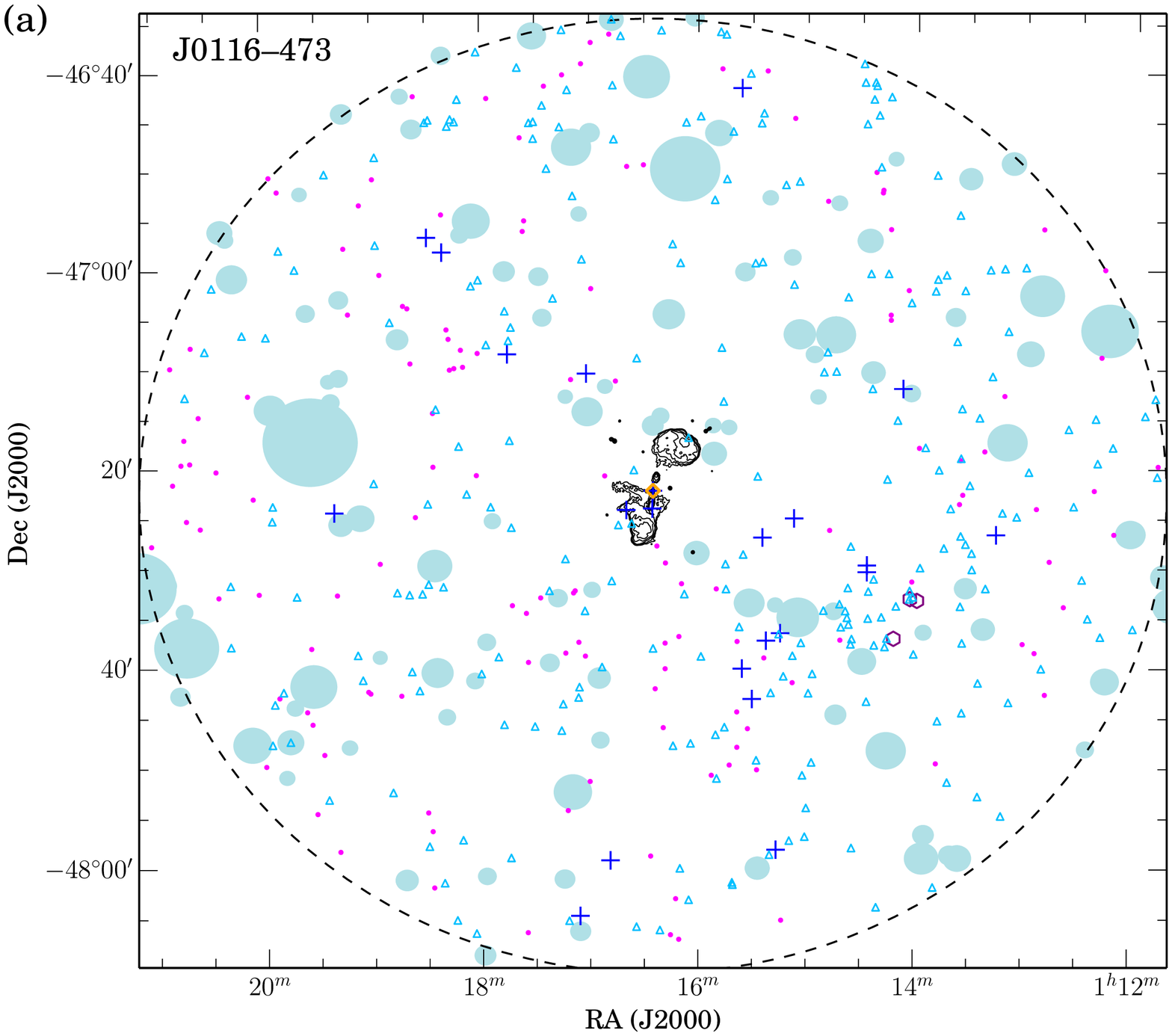} &
      \includegraphics[width=0.5\hsize]{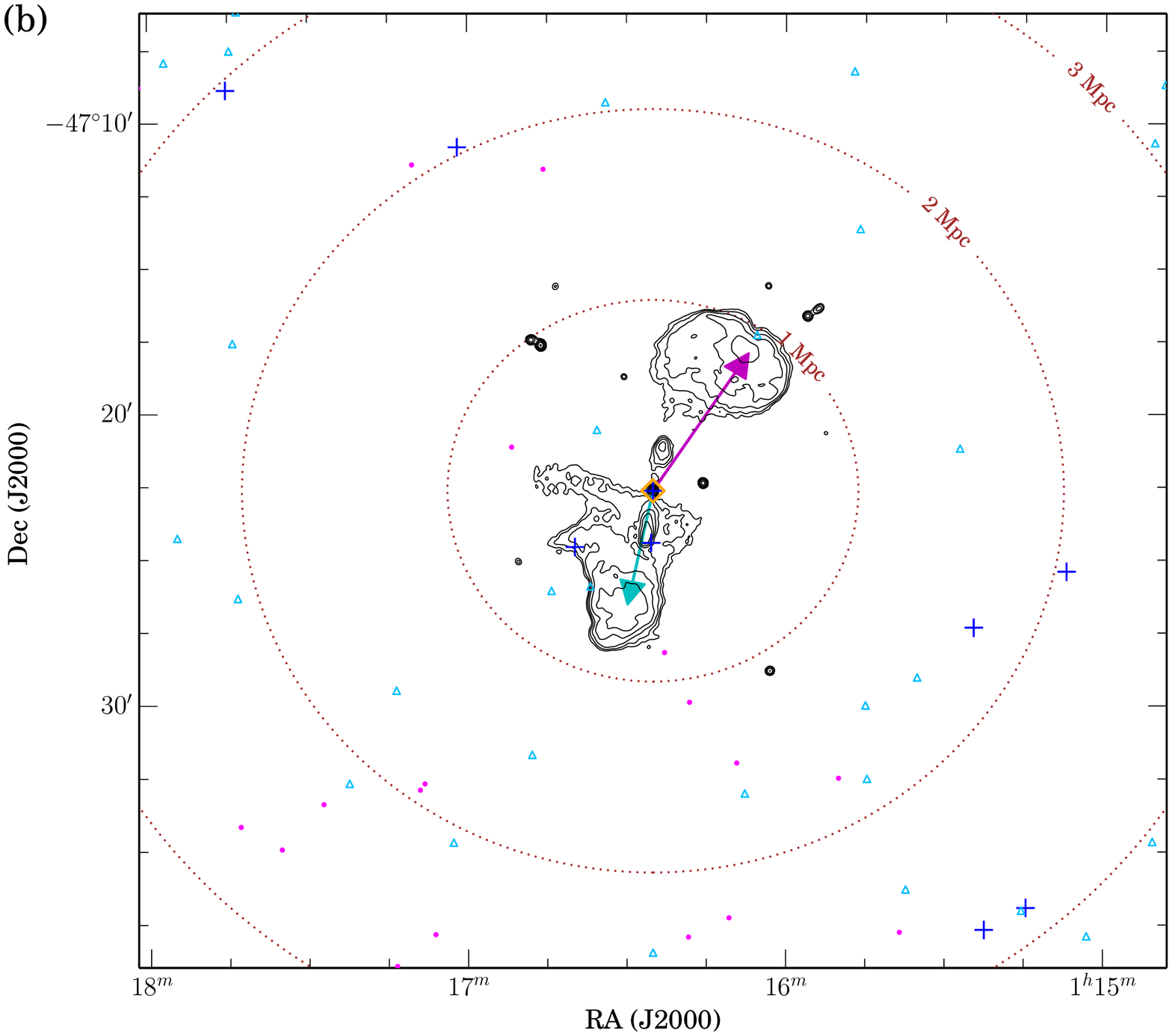}\\
      \includegraphics[width=0.5\hsize]{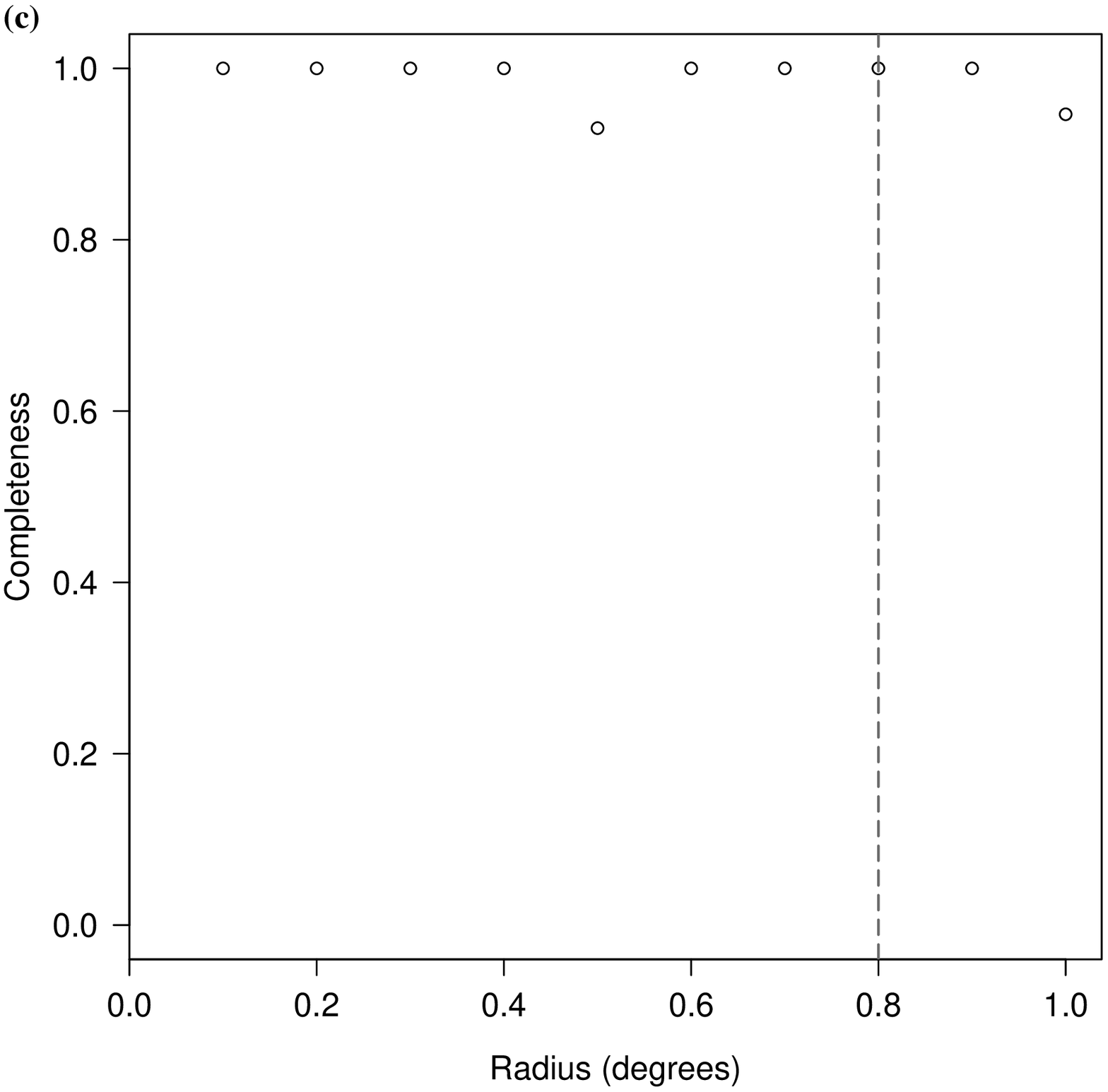} &
      \includegraphics[width=0.5\hsize]{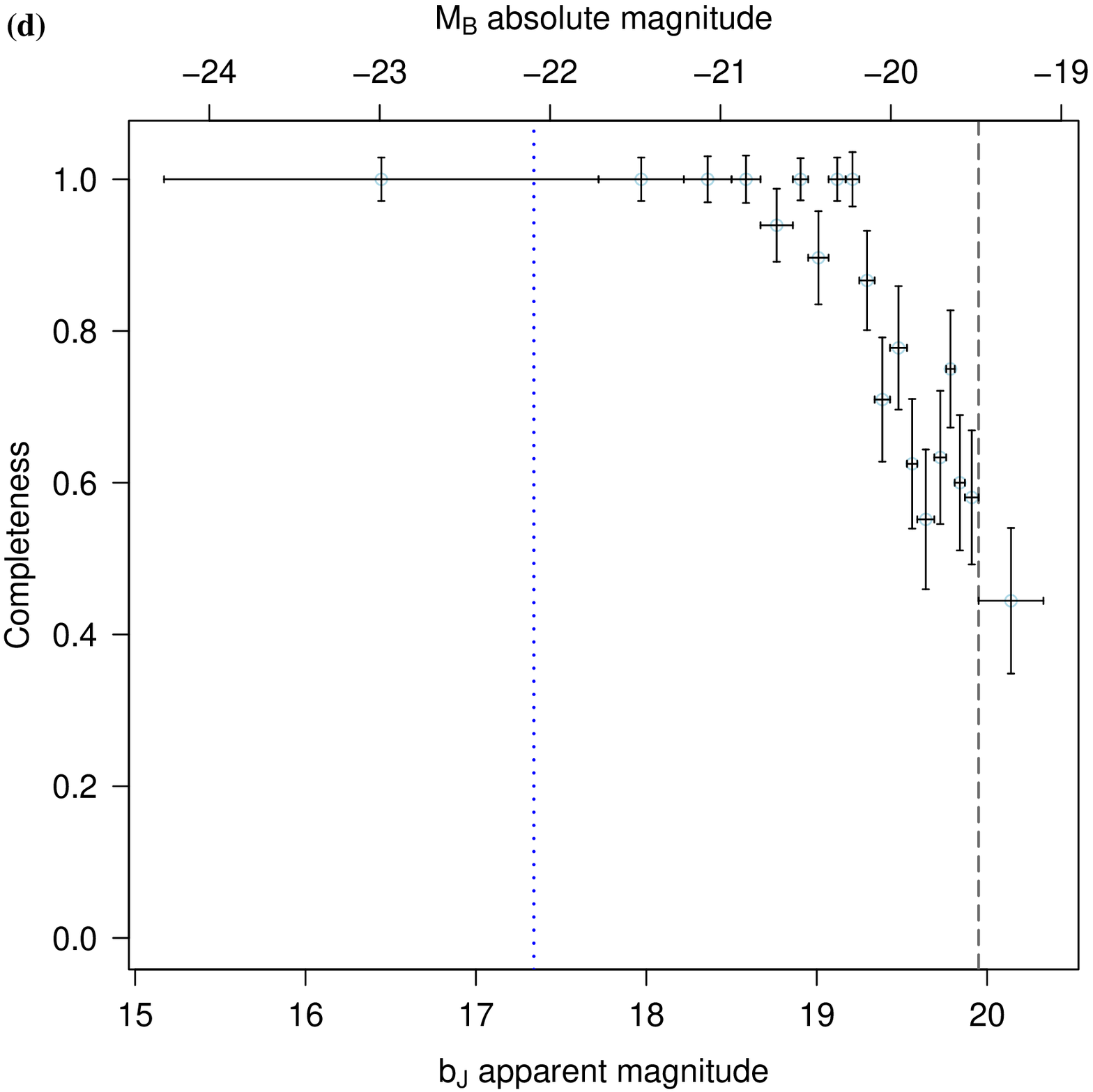}
\end{tabular}
\caption{(\subfigletter{a}) Distribution of galaxies around J0116--473; a 20 arcmin angular scale corresponds to a linear size of 3.05 Mpc. Plus symbols represent galaxies within $\pm0.003$ of the host redshift at z = 0.1470. Radio contours are shown at (6, 12, 24, 48, 96, and 192) $\times$ 85 $\mu$Jy beam$^{-1}$. (\subfigletter{b}) An enlarged view of the optical field, centred at the host galaxy, with dotted circles at 1, 2 and 3-Mpc radii. (\subfigletter{c}) A completeness plot showing the fraction of observed SuperCOSMOS targets against radius, measured in intervals of 0.1$^{\circ}$ from the field centre. (\subfigletter{d}) A completeness plot showing the fraction of observed targets against $B$-band magnitude with the $b_{\mathrm{J}}$ apparent magnitude (17.34) of the host galaxy indicated by a blue dotted line.}
\label{fig:J0116}
\end{minipage}
\end{figure*}

\begin{figure*}
\begin{minipage}[t][\textheight]{\textwidth}
  \centering
  \begin{tabular}{ll}
      \includegraphics[width=0.5\hsize]{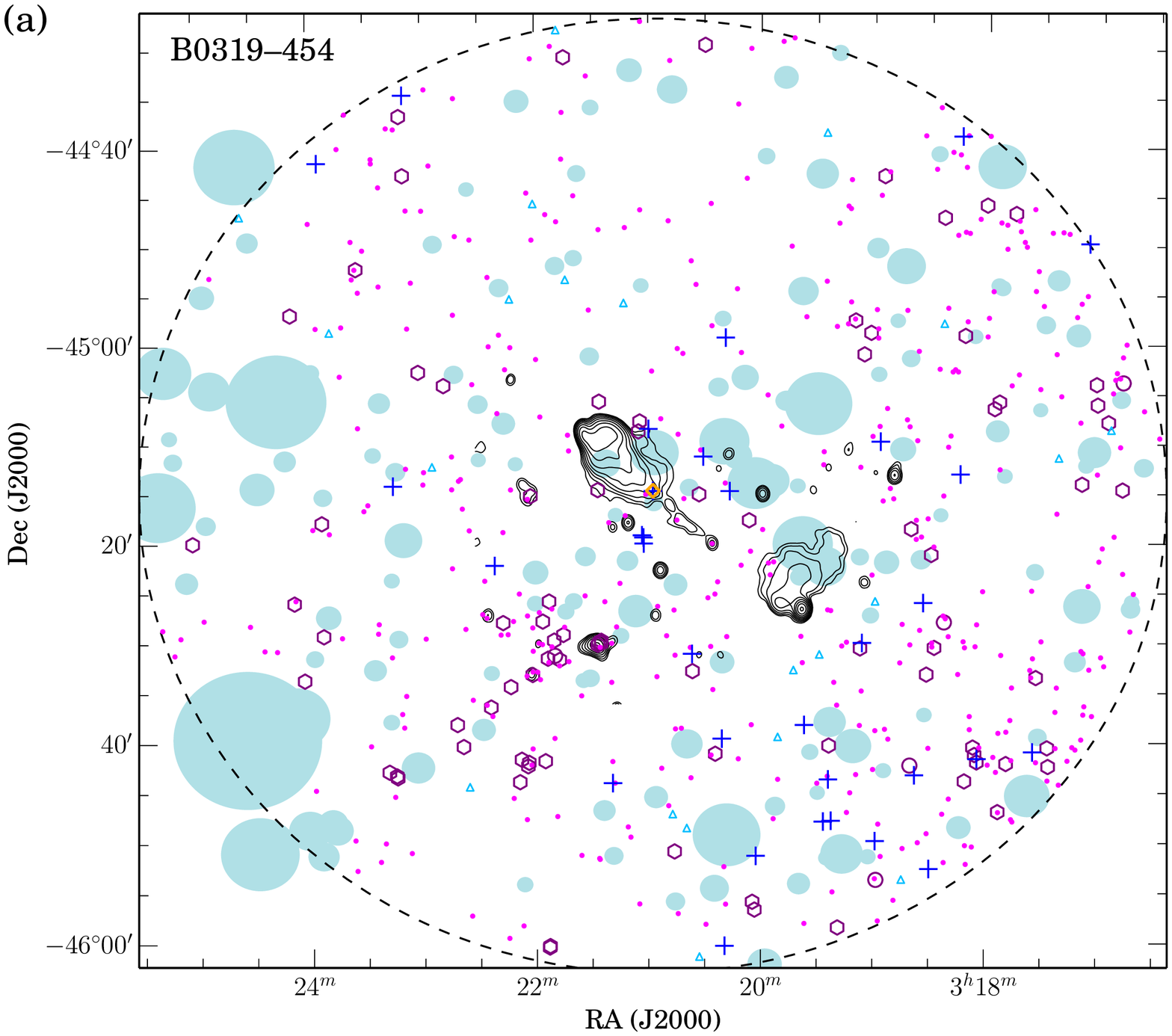} &
      \includegraphics[width=0.5\hsize]{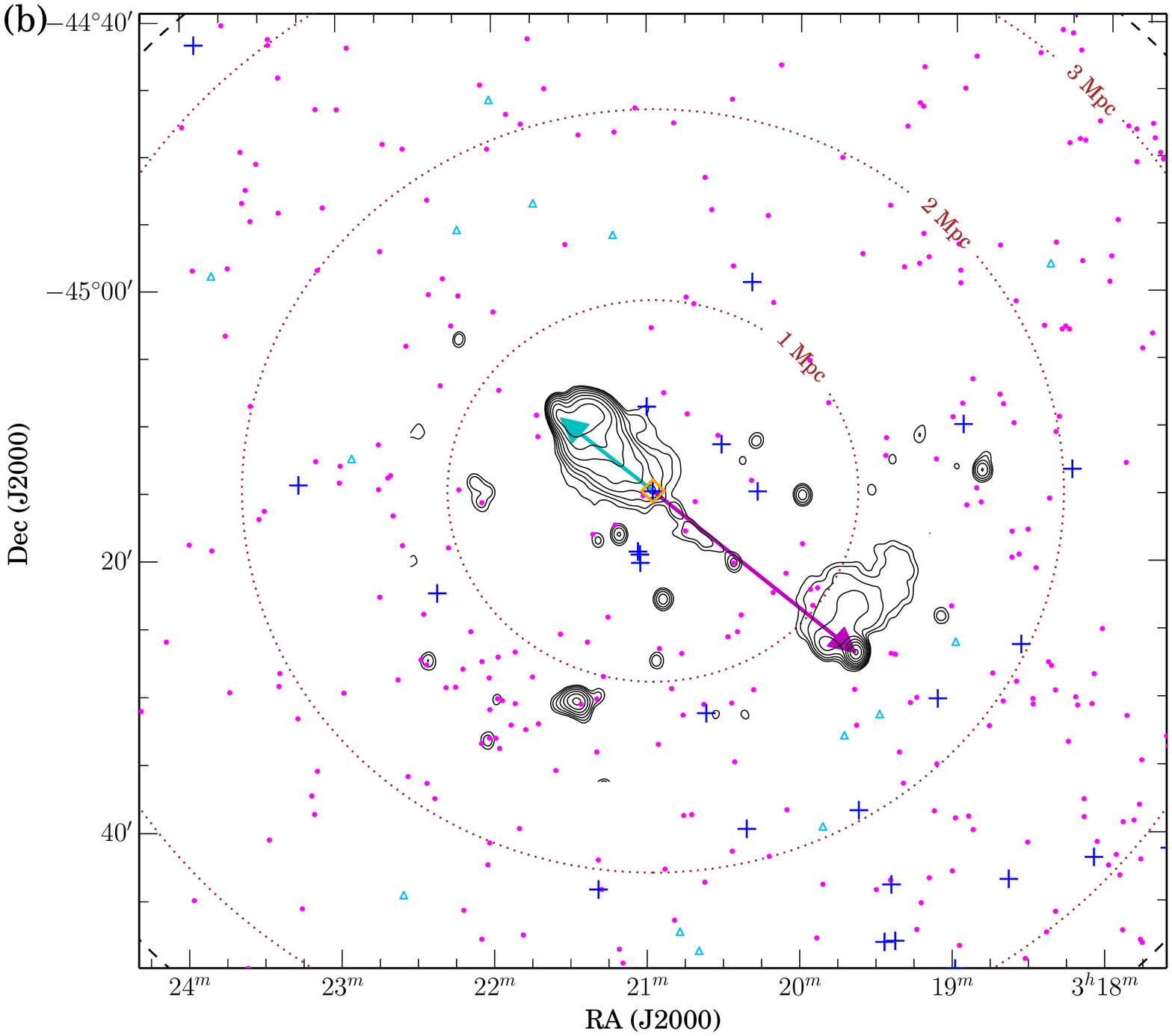}\\
      \includegraphics[width=0.5\hsize]{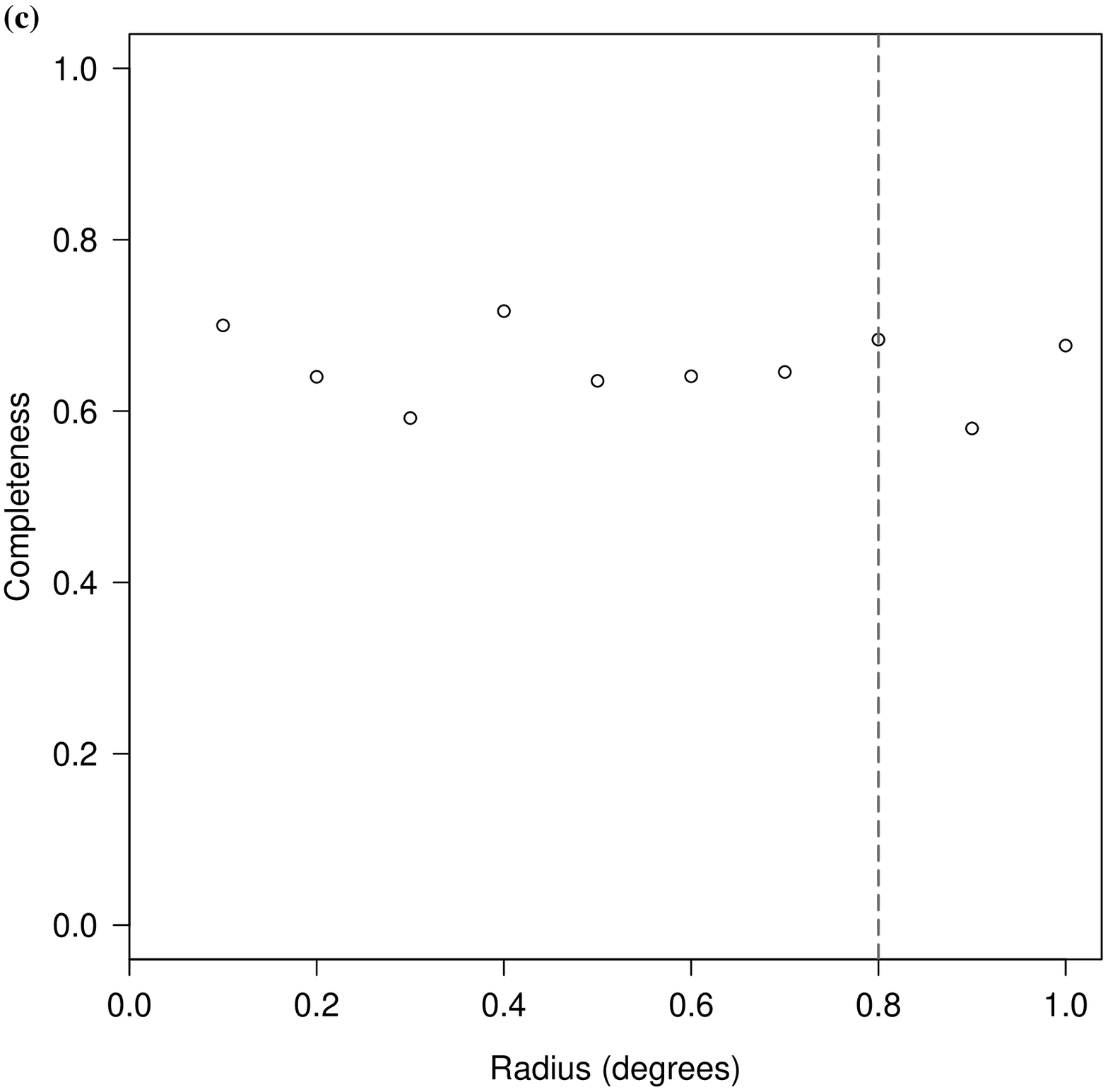} &
      \includegraphics[width=0.5\hsize]{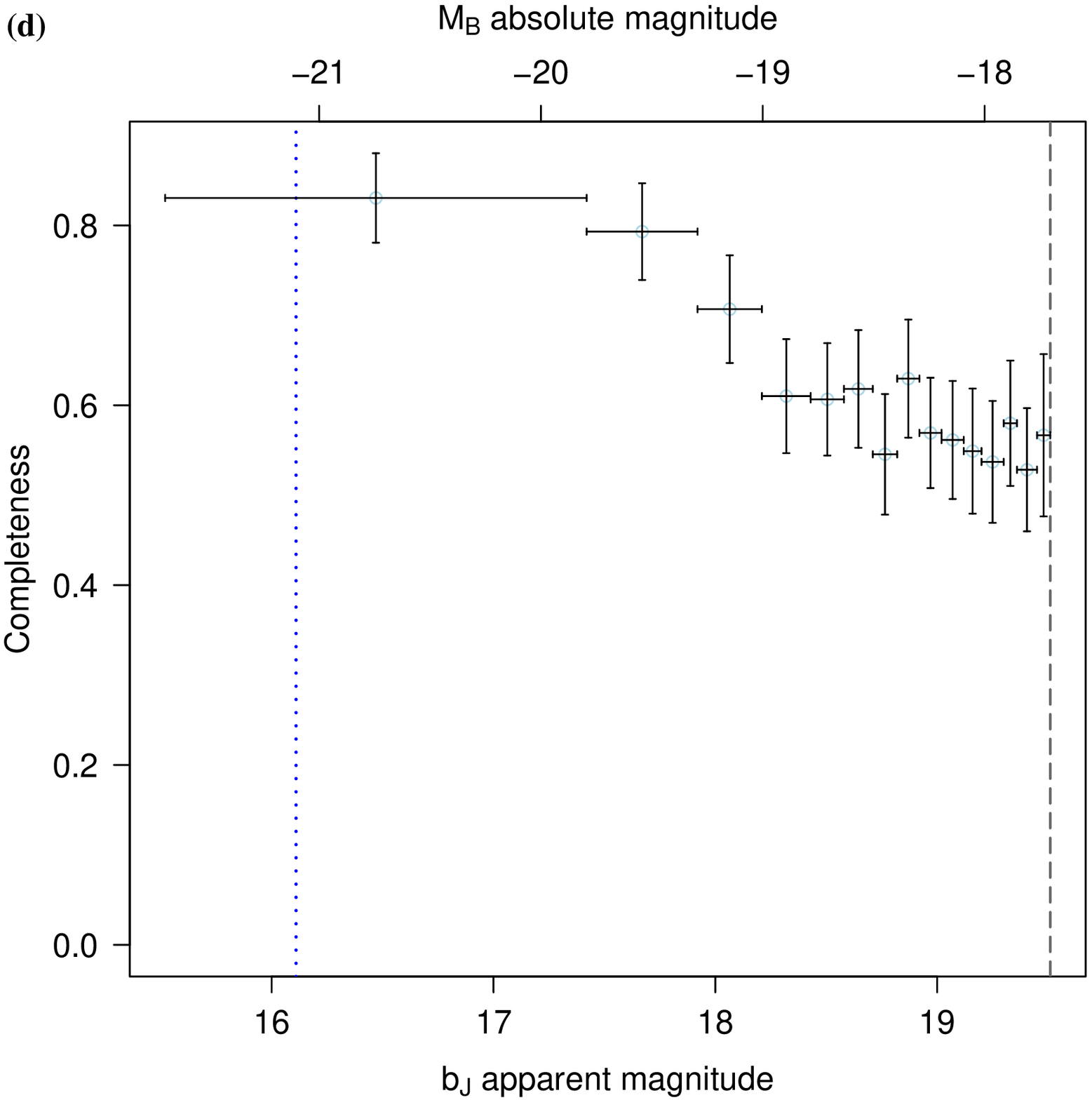}
\end{tabular}
\caption{(\subfigletter{a}) Distribution of galaxies around B0319--454 within the 2-degree field of the AAT/AAOmega (using redshift data from~\citet{Safourisetal2009}); a 20 arcmin angular scale corresponds to a linear size of 1.42 Mpc. Plus symbols represent galaxies within $\pm0.003$ of the host redshift at z = 0.0622. Radio contours represent 1, 2, 4, 8, 16, 32, 64, 128 and 256 mJy beam$^{-1}$. (\subfigletter{b}) An enlarged view of the optical field, centred at the host galaxy, with dotted circles at 1, 2 and 3-Mpc radii. (\subfigletter{c}) A completeness plot showing the fraction of observed SuperCOSMOS targets against radius, measured in intervals of 0.1$^{\circ}$ from the field centre. (\subfigletter{d}) A completeness plot showing the fraction of observed targets against $B$-band magnitude with the $b_{\mathrm{J}}$ apparent magnitude (16.11) of the host galaxy indicated by a blue dotted line.}
\label{fig:B0319}
\end{minipage}
\end{figure*}

\begin{figure*}
\begin{minipage}[t][\textheight]{\textwidth}
  \centering
  \begin{tabular}{ll}
      \includegraphics[width=0.5\hsize]{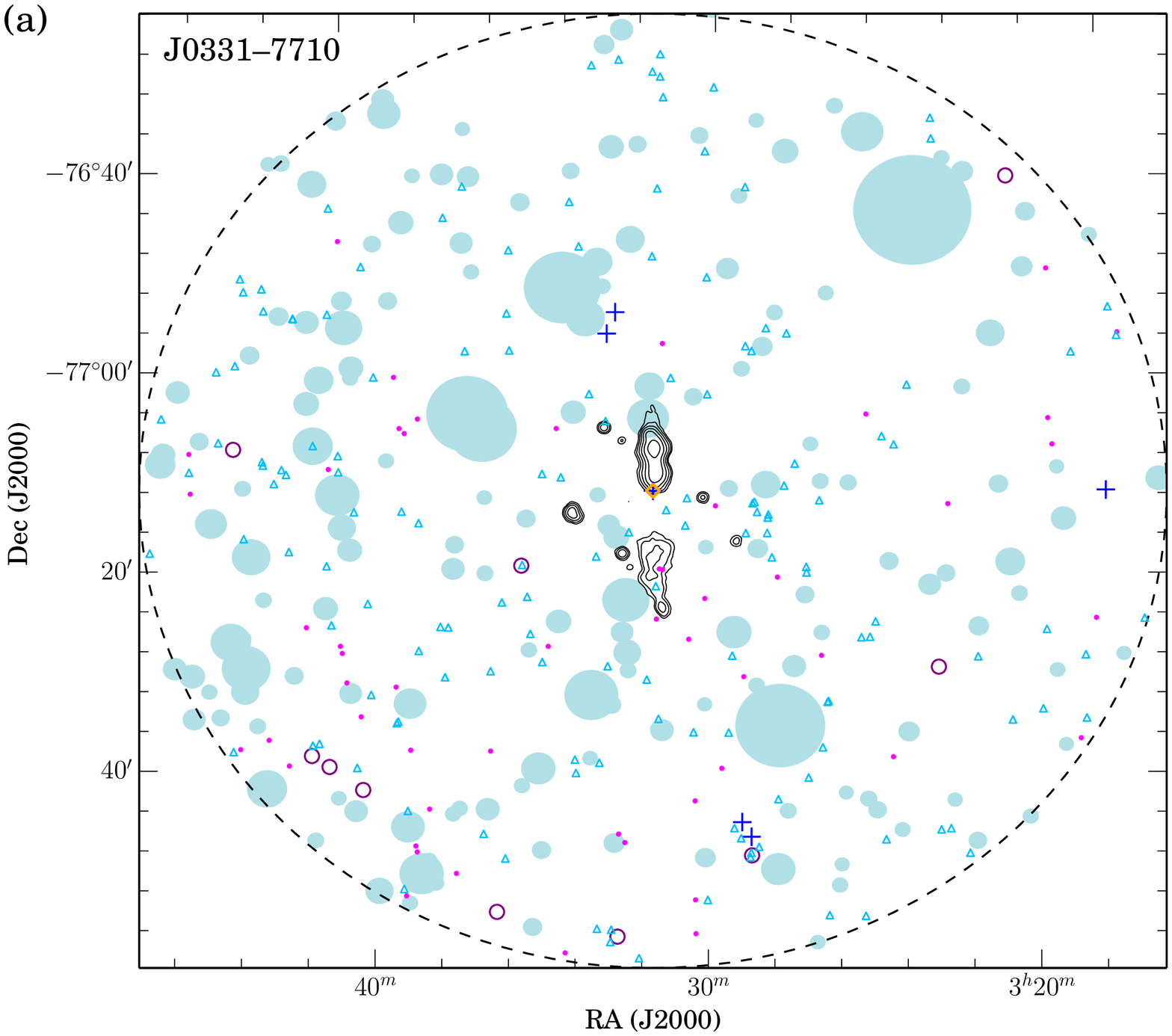} &
      \includegraphics[width=0.5\hsize]{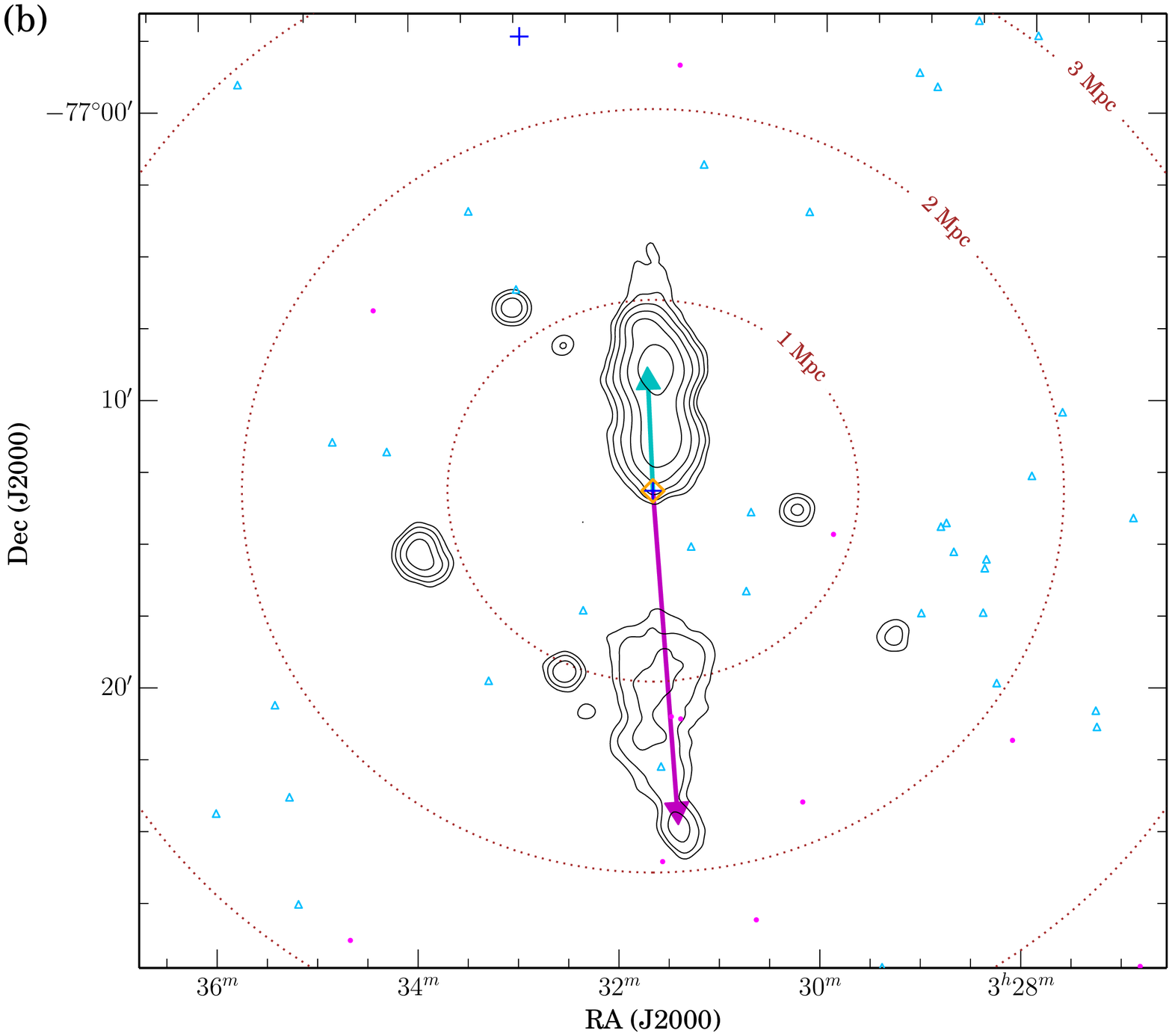}\\
      \includegraphics[width=0.5\hsize]{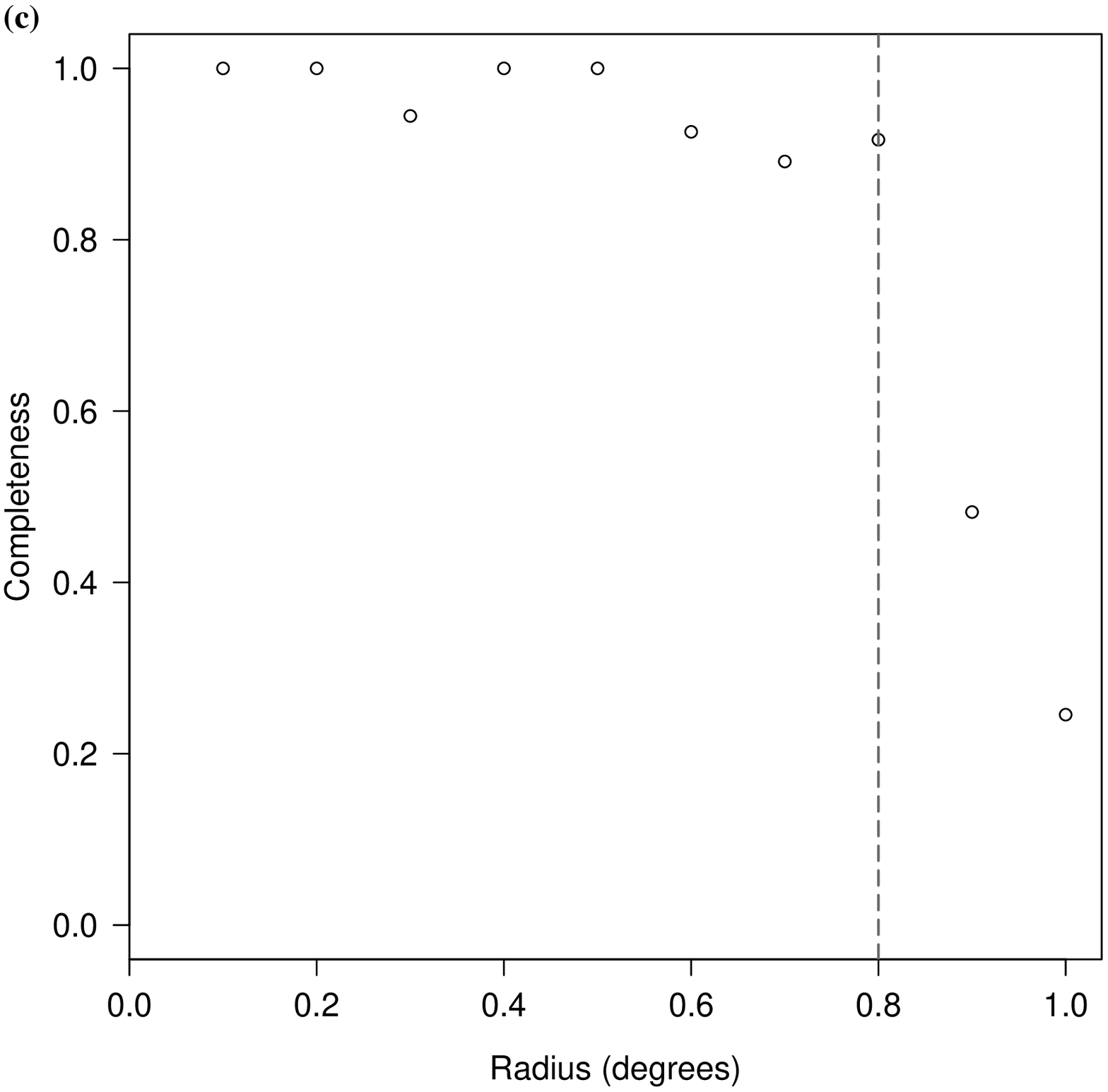} &
      \includegraphics[width=0.5\hsize]{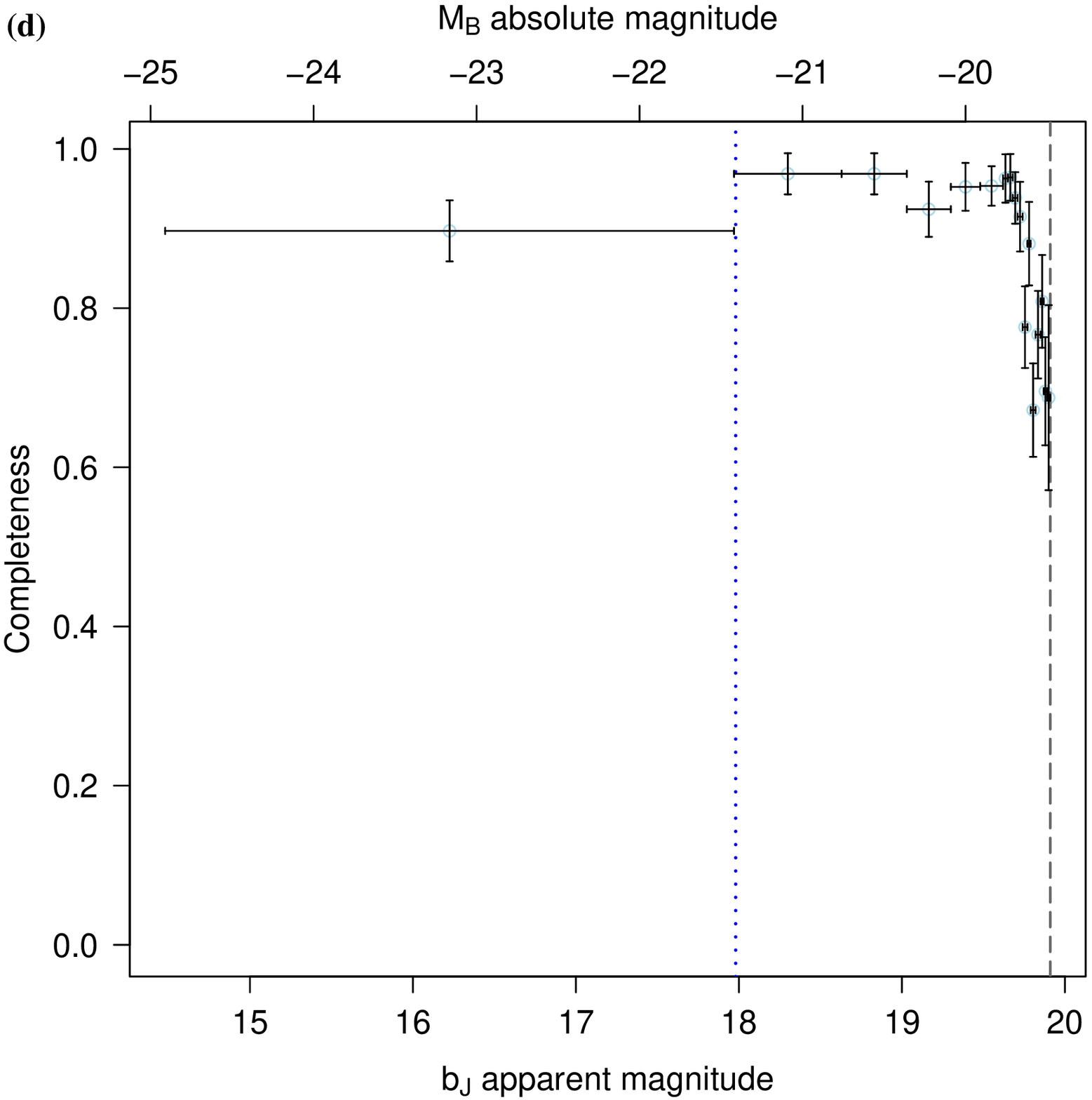}
\end{tabular}
\caption{(\subfigletter{a}) Distribution of galaxies around J0331--7710; a 20 arcmin angular scale corresponds to a linear size of 3.01 Mpc. Plus symbols represent galaxies within $\pm0.003$ of the host redshift at z = 0.1447. Radio contours are shown at (6, 12, 24, 48, 96, and 192) $\times$ 85 $\mu$Jy beam$^{-1}$. (\subfigletter{b}) An enlarged view of the optical field, centred at the host galaxy, with dotted circles at 1, 2 and 3-Mpc radii. (\subfigletter{c}) A completeness plot showing the fraction of observed SuperCOSMOS targets against radius, measured in intervals of 0.1$^{\circ}$ from the field centre. (\subfigletter{d}) A completeness plot showing the fraction of observed targets against $B$-band magnitude with the $b_{\mathrm{J}}$ apparent magnitude (17.98) of the host galaxy indicated by a blue dotted line.}
\label{fig:J0331}
\end{minipage}
\end{figure*}

\begin{figure*}
\begin{minipage}[t][\textheight]{\textwidth}
  \centering
  \begin{tabular}{ll}
      \includegraphics[width=0.5\hsize]{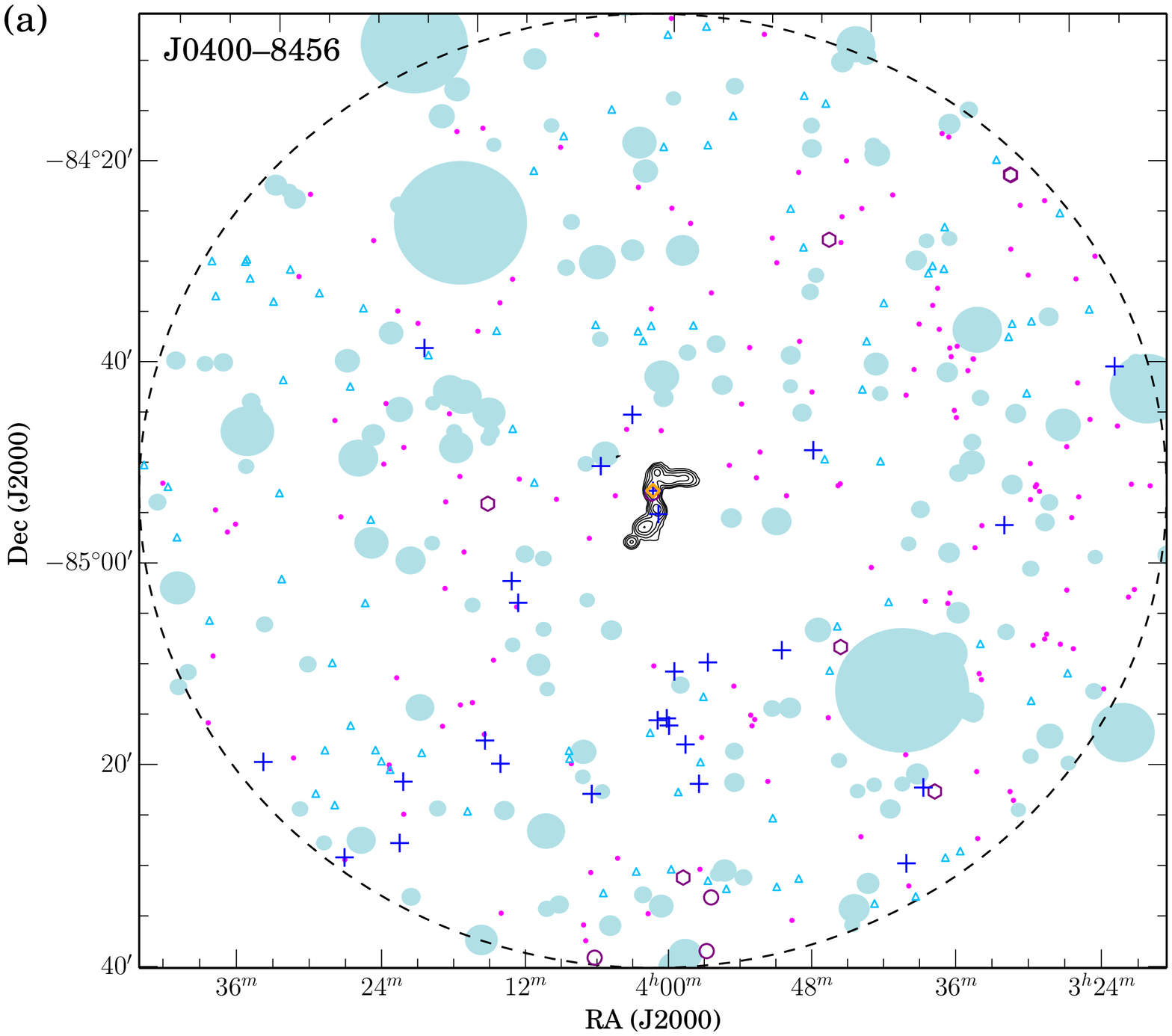} &
      \includegraphics[width=0.5\hsize]{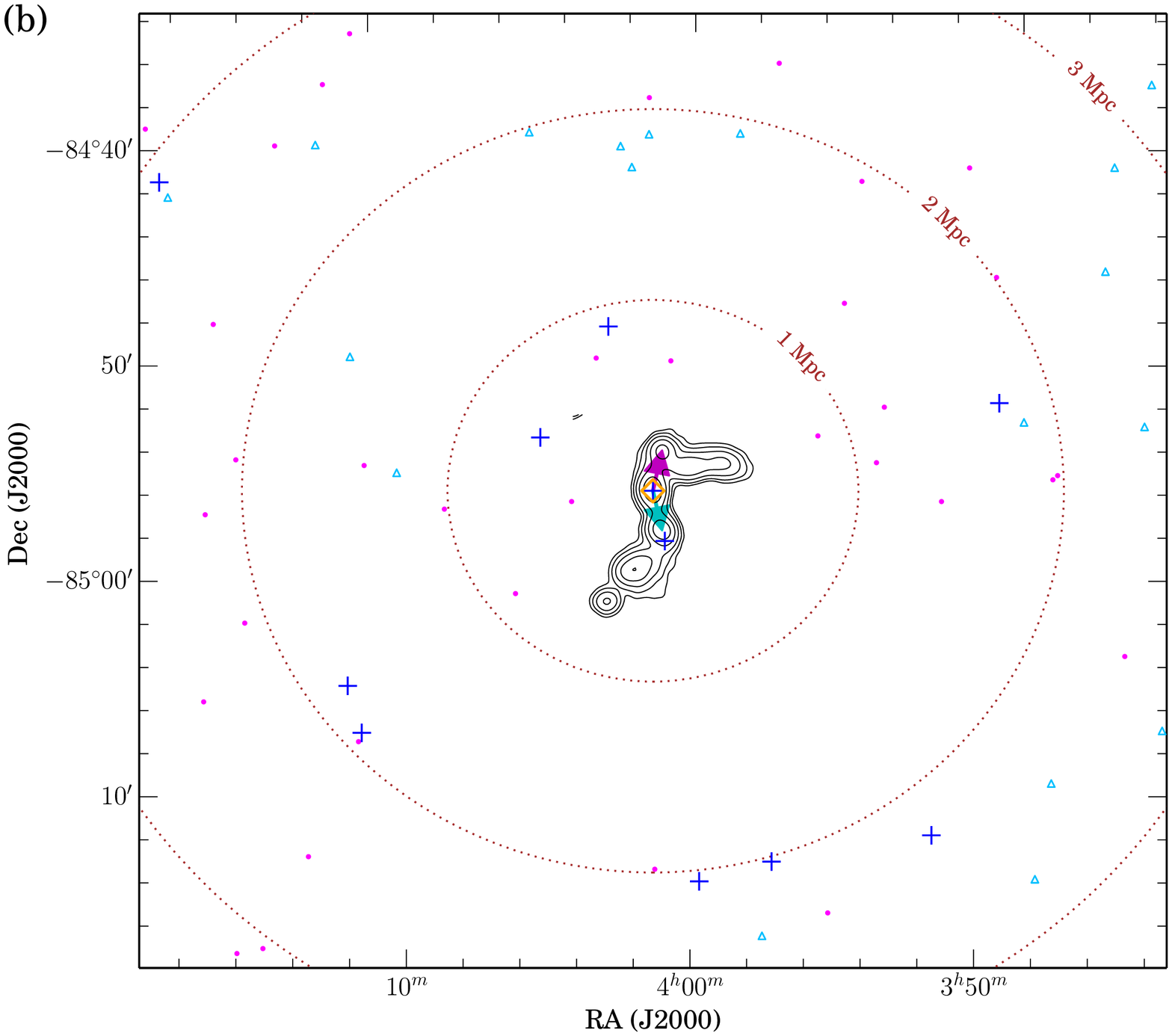}\\
      \includegraphics[width=0.5\hsize]{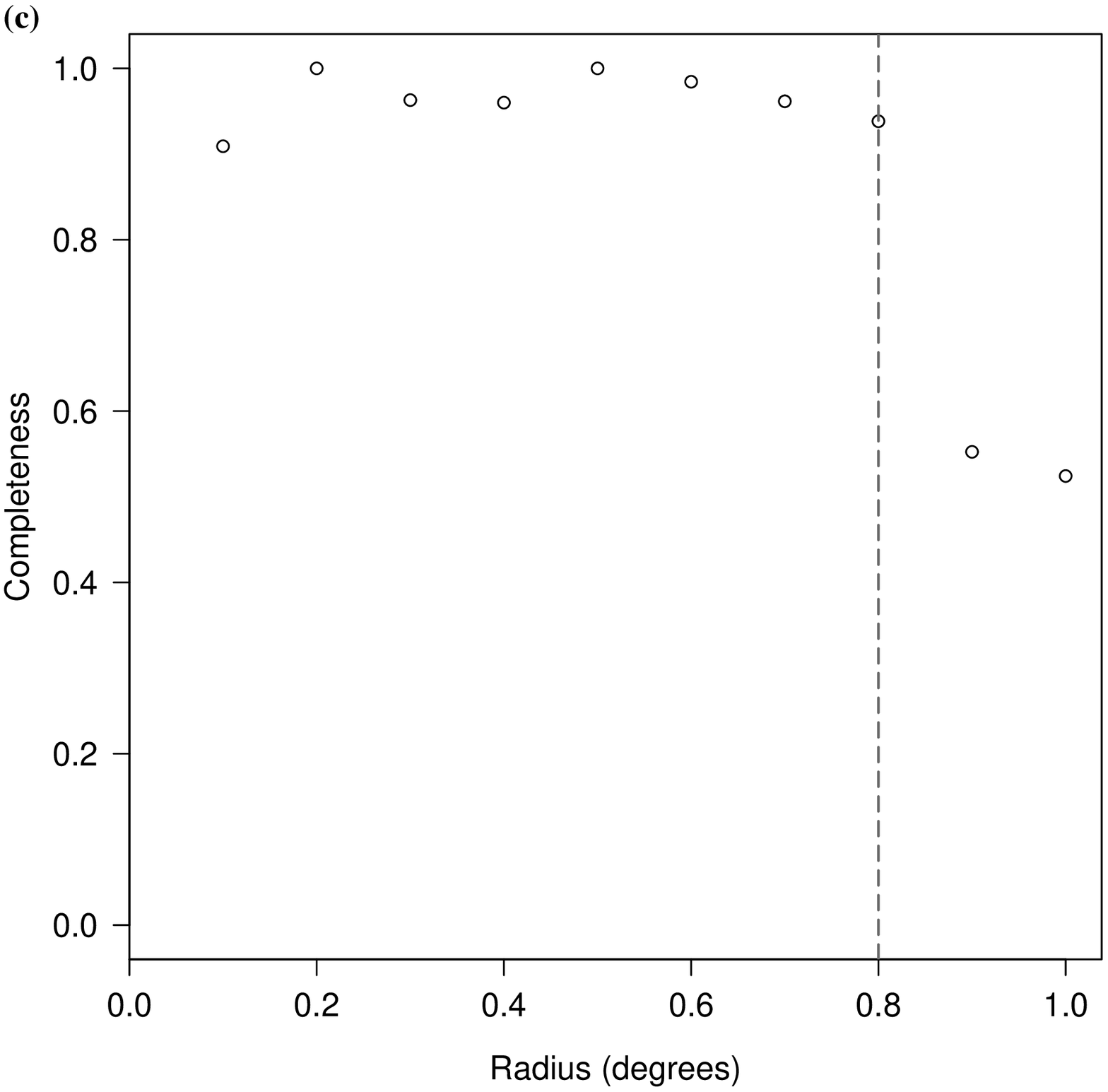} &
      \includegraphics[width=0.5\hsize]{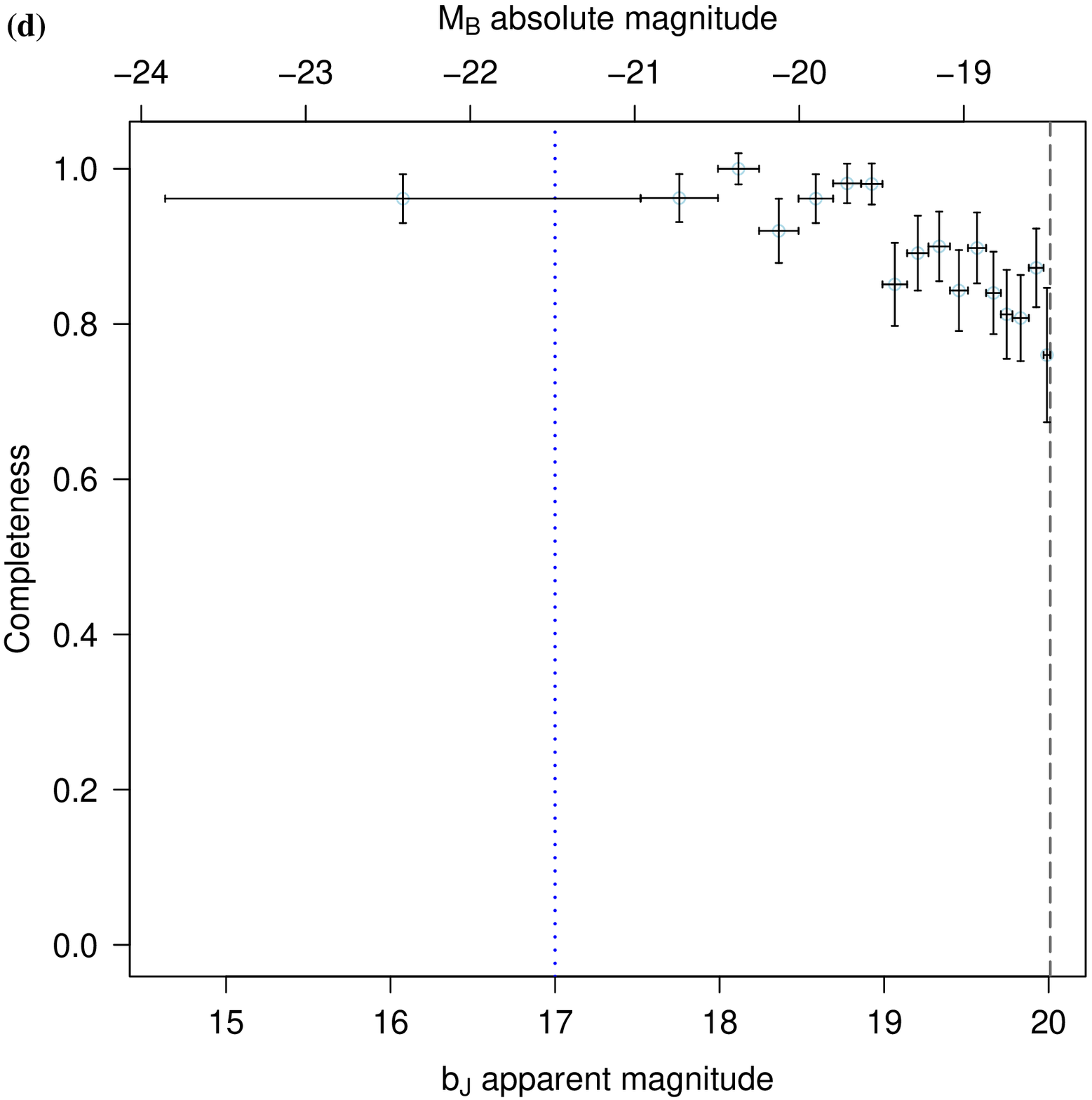}
\end{tabular}
\caption{(\subfigletter{a}) Distribution of galaxies around J0400--8456; a 20 arcmin angular scale corresponds to a linear size of 2.25 Mpc. Plus symbols represent galaxies within $\pm0.003$ of the host redshift at z = 0.1033. Radio contours are shown at (6, 12, 24, 48, and 96) $\times$ 0.1 mJy beam$^{-1}$. (\subfigletter{b}) An enlarged view of the optical field, centred at the host galaxy, with dotted circles at 1, 2 and 3-Mpc radii. (\subfigletter{c}) A completeness plot showing the fraction of observed SuperCOSMOS targets against radius, measured in intervals of 0.1$^{\circ}$ from the field centre. (\subfigletter{d}) A completeness plot showing the fraction of observed targets against $B$-band magnitude with the $b_{\mathrm{J}}$ apparent magnitude (17.00) of the host galaxy indicated by a blue dotted line.}
\label{fig:J0400}
\end{minipage}
\end{figure*}

\begin{figure*}
\begin{minipage}[t][\textheight]{\textwidth}
  \centering
  \begin{tabular}{ll}
      \includegraphics[width=0.5\hsize]{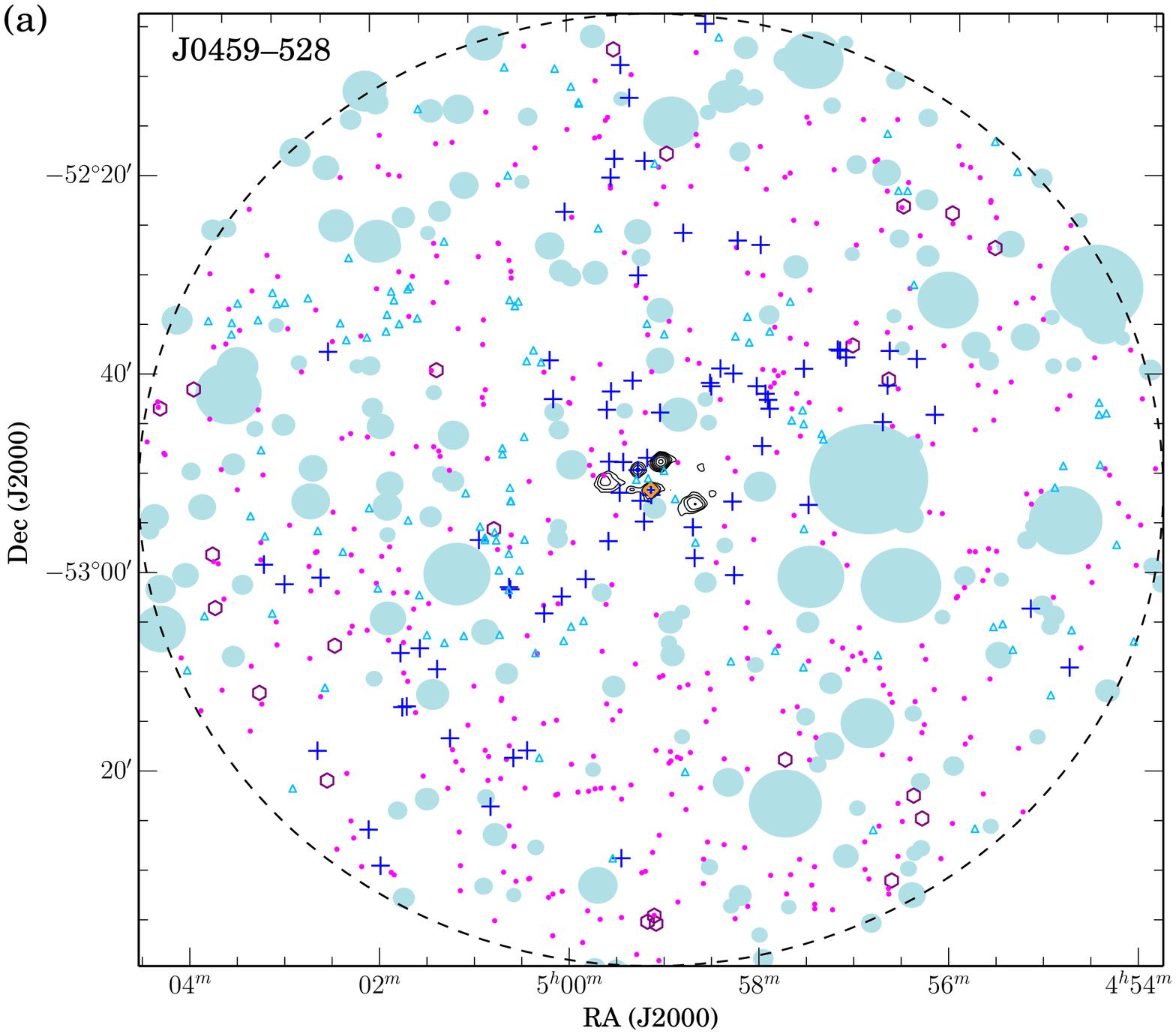} &
      \includegraphics[width=0.5\hsize]{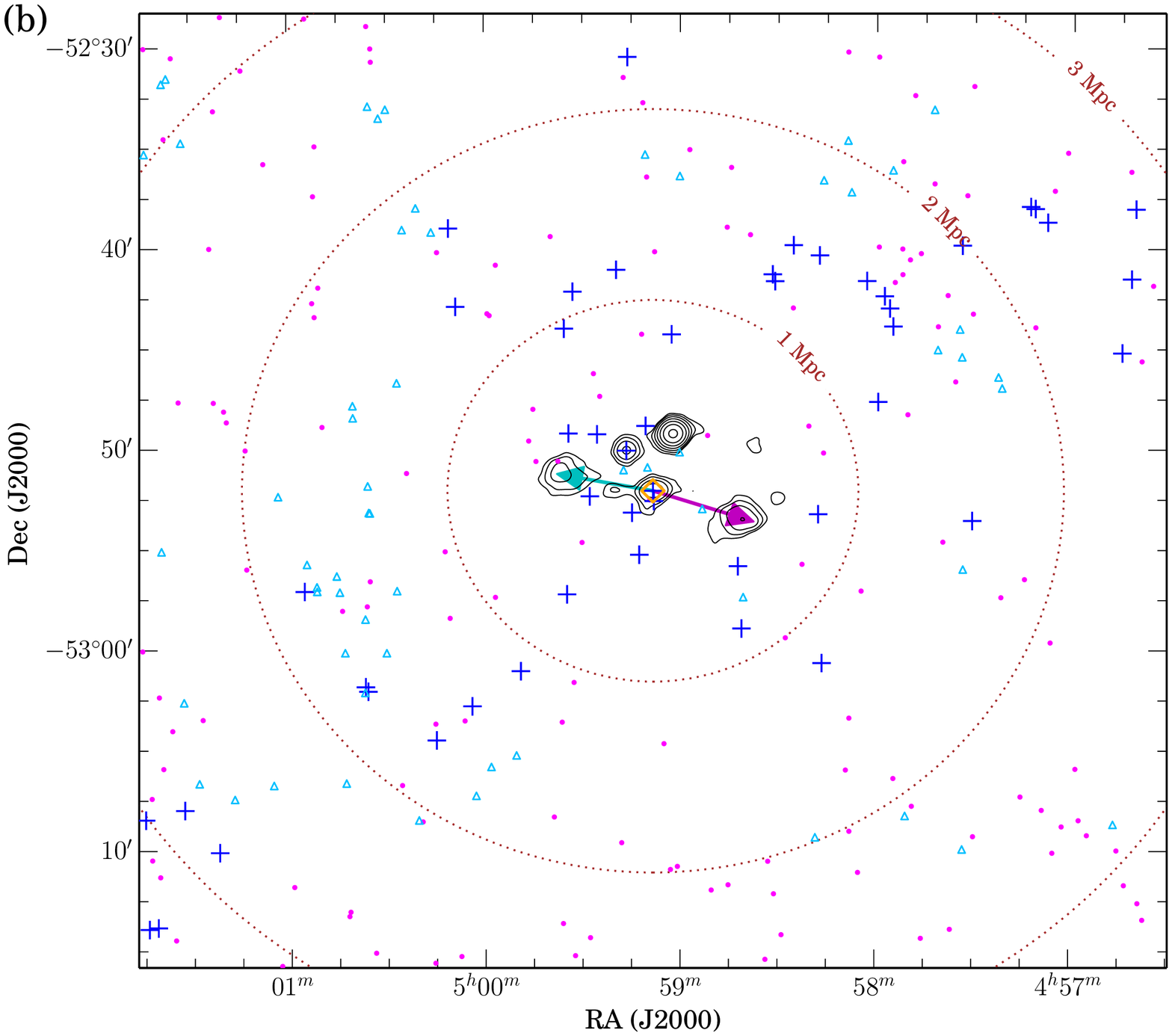}\\
      \includegraphics[width=0.5\hsize]{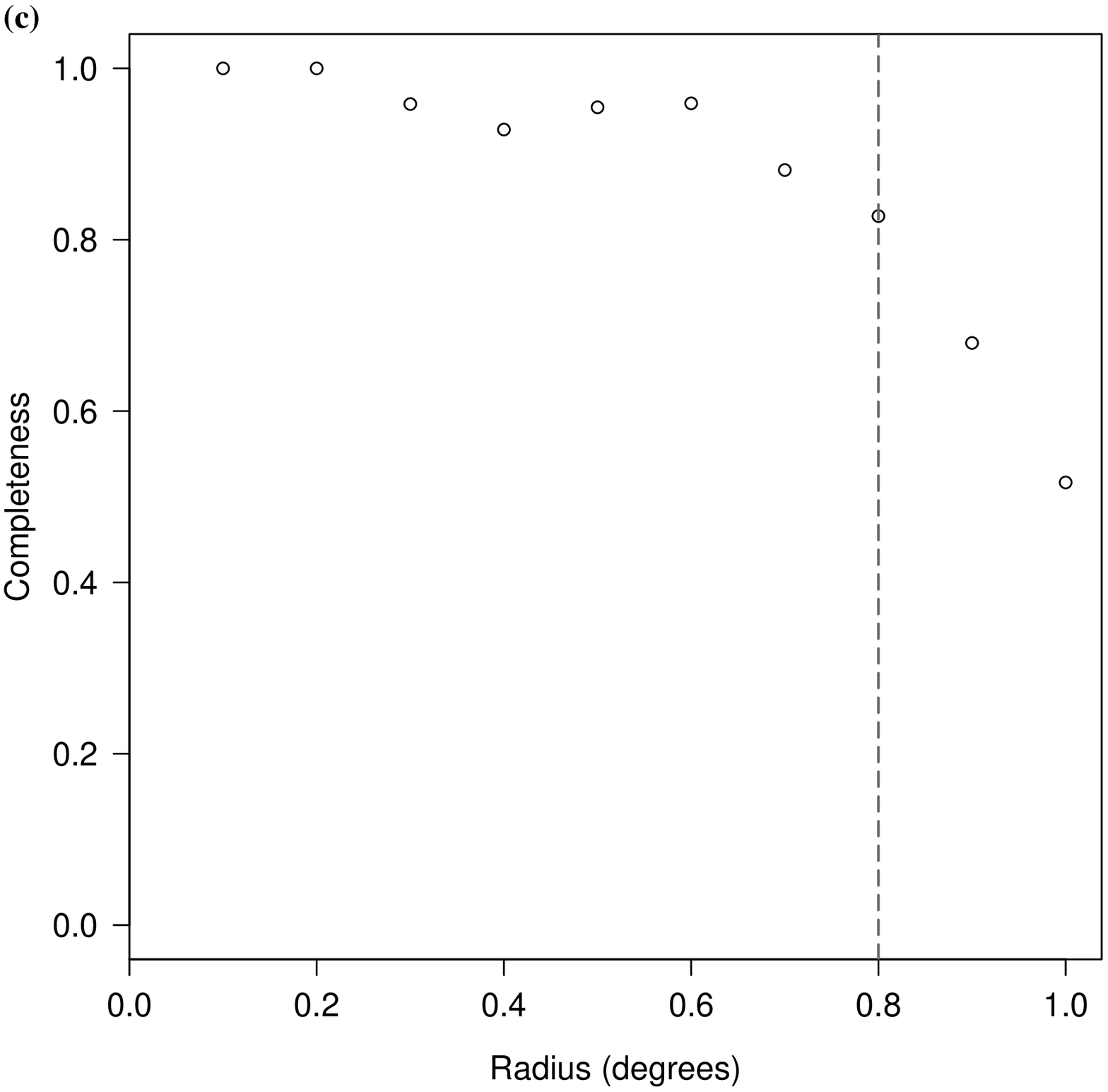} &
      \includegraphics[width=0.5\hsize]{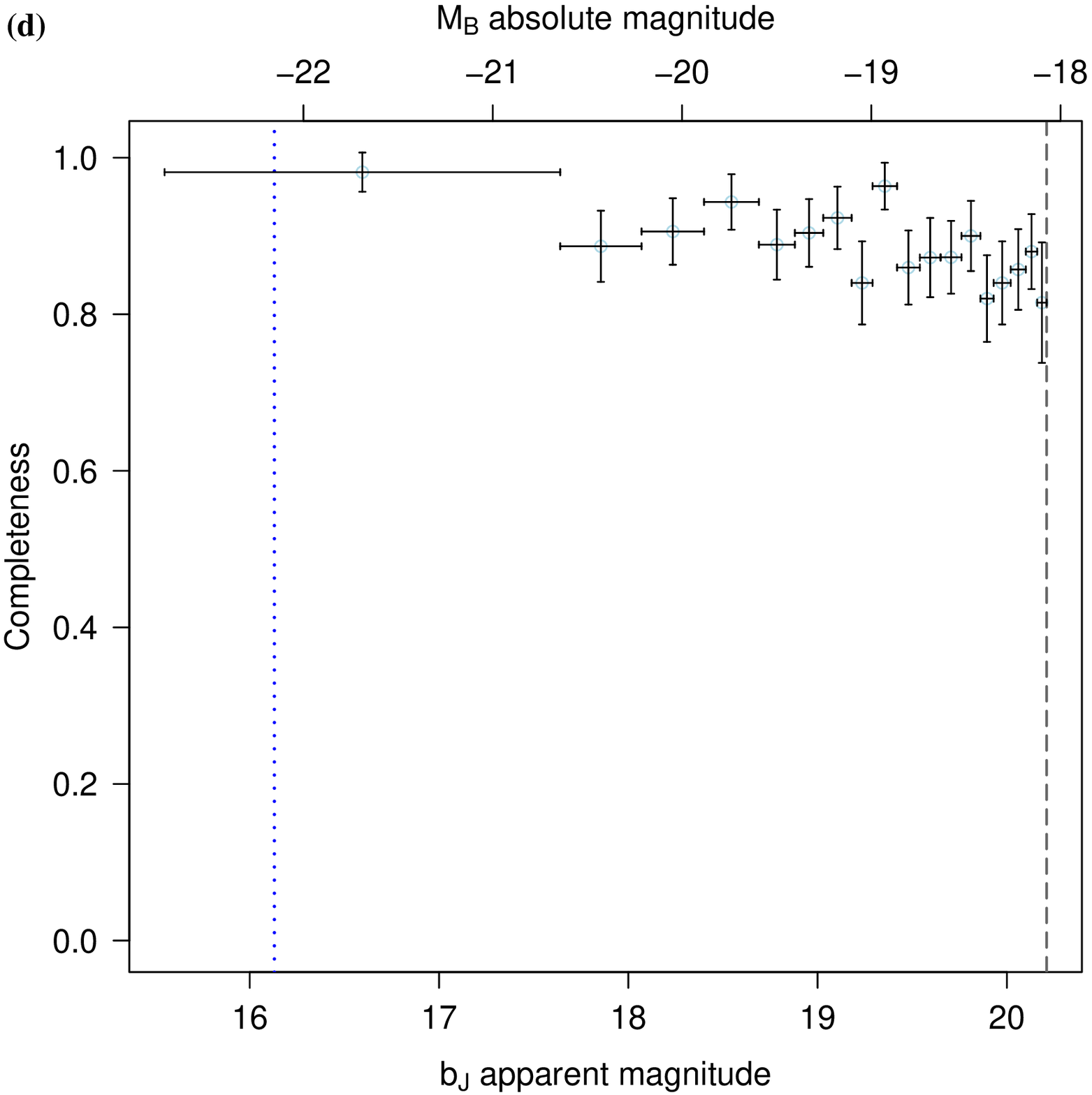}
\end{tabular}
\caption{(\subfigletter{a}) Distribution of galaxies around J0459--528; a 20 arcmin angular scale corresponds to a linear size of 2.10 Mpc. Plus symbols represent galaxies within $\pm0.003$ of the host redshift at z = 0.0957. Radio contours are shown at (6, 12, 24, 48, 96, 192, and 384) $\times$ 0.15 mJy beam$^{-1}$. (\subfigletter{b}) An enlarged view of the optical field, centred at the host galaxy, with dotted circles at 1, 2 and 3-Mpc radii. (\subfigletter{c}) A completeness plot showing the fraction of observed SuperCOSMOS targets against radius, measured in intervals of 0.1$^{\circ}$ from the field centre. (\subfigletter{d}) A completeness plot showing the fraction of observed targets against $B$-band magnitude with the $b_{\mathrm{J}}$ apparent magnitude (16.13) of the host galaxy indicated by a blue dotted line.}
\label{fig:J0459}
\end{minipage}
\end{figure*}

\begin{figure*}
\begin{minipage}[t][\textheight]{\textwidth}
  \centering
  \begin{tabular}{ll}
      \includegraphics[width=0.5\hsize]{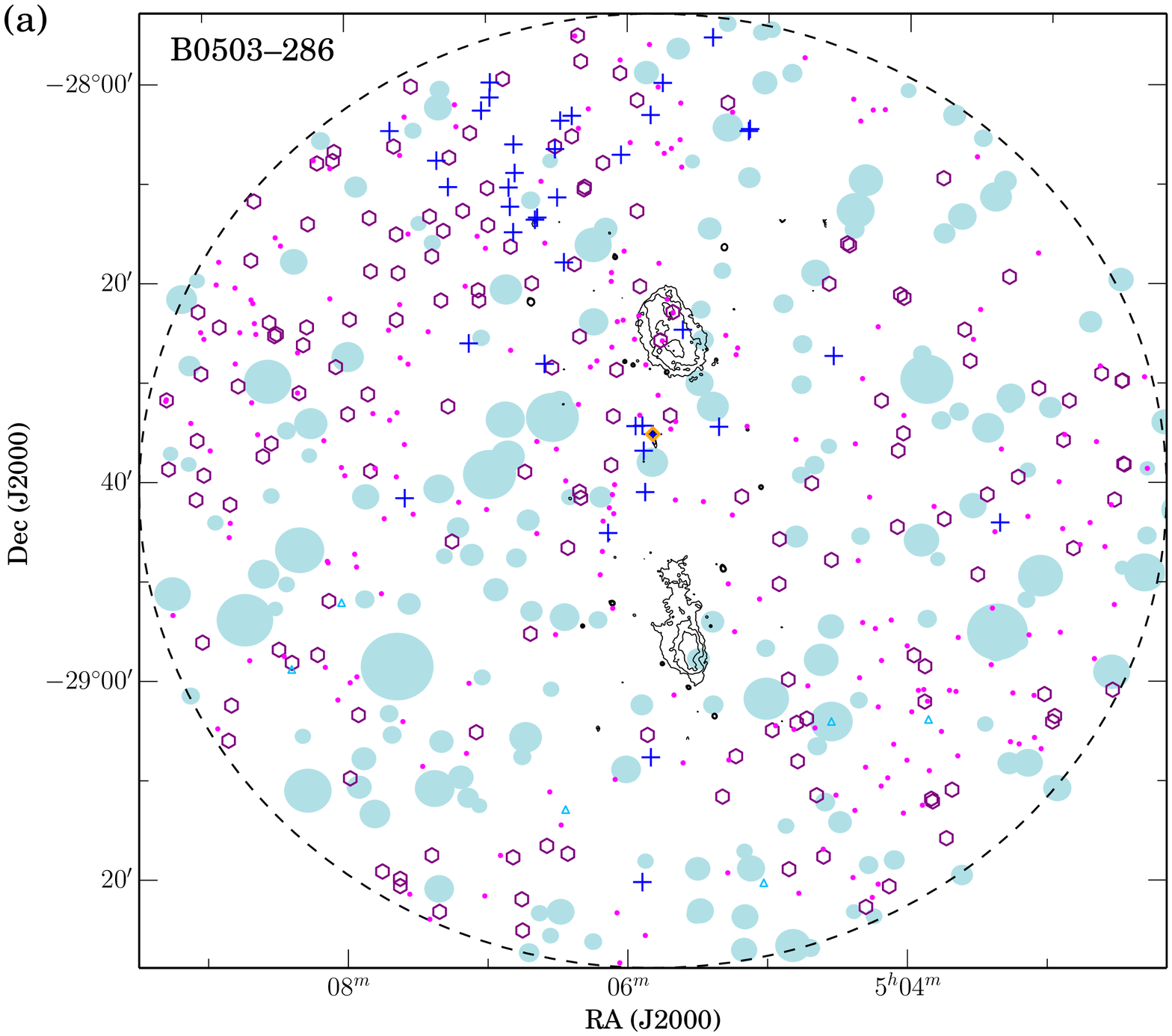} &
      \includegraphics[width=0.5\hsize]{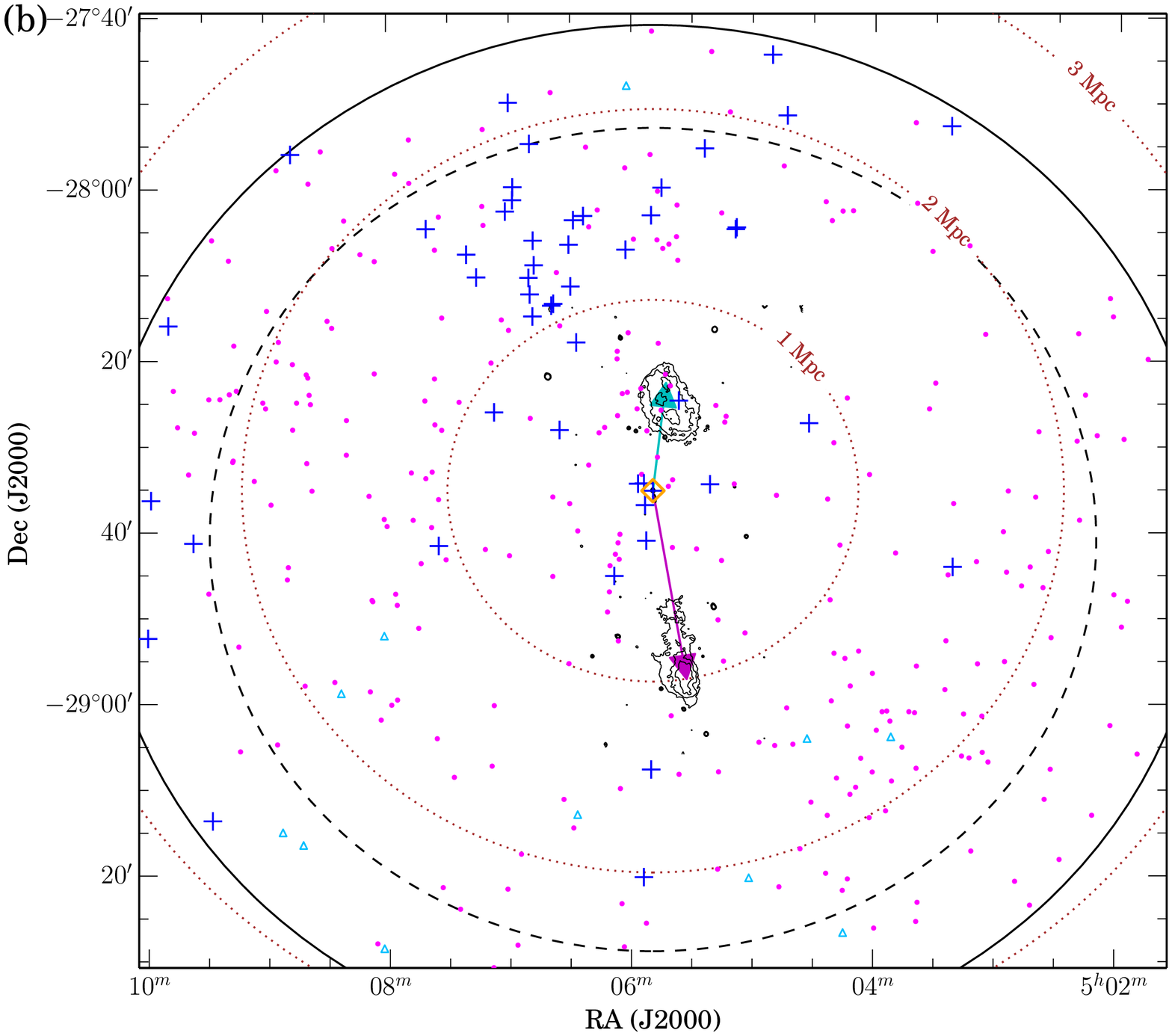}\\
      \includegraphics[width=0.5\hsize]{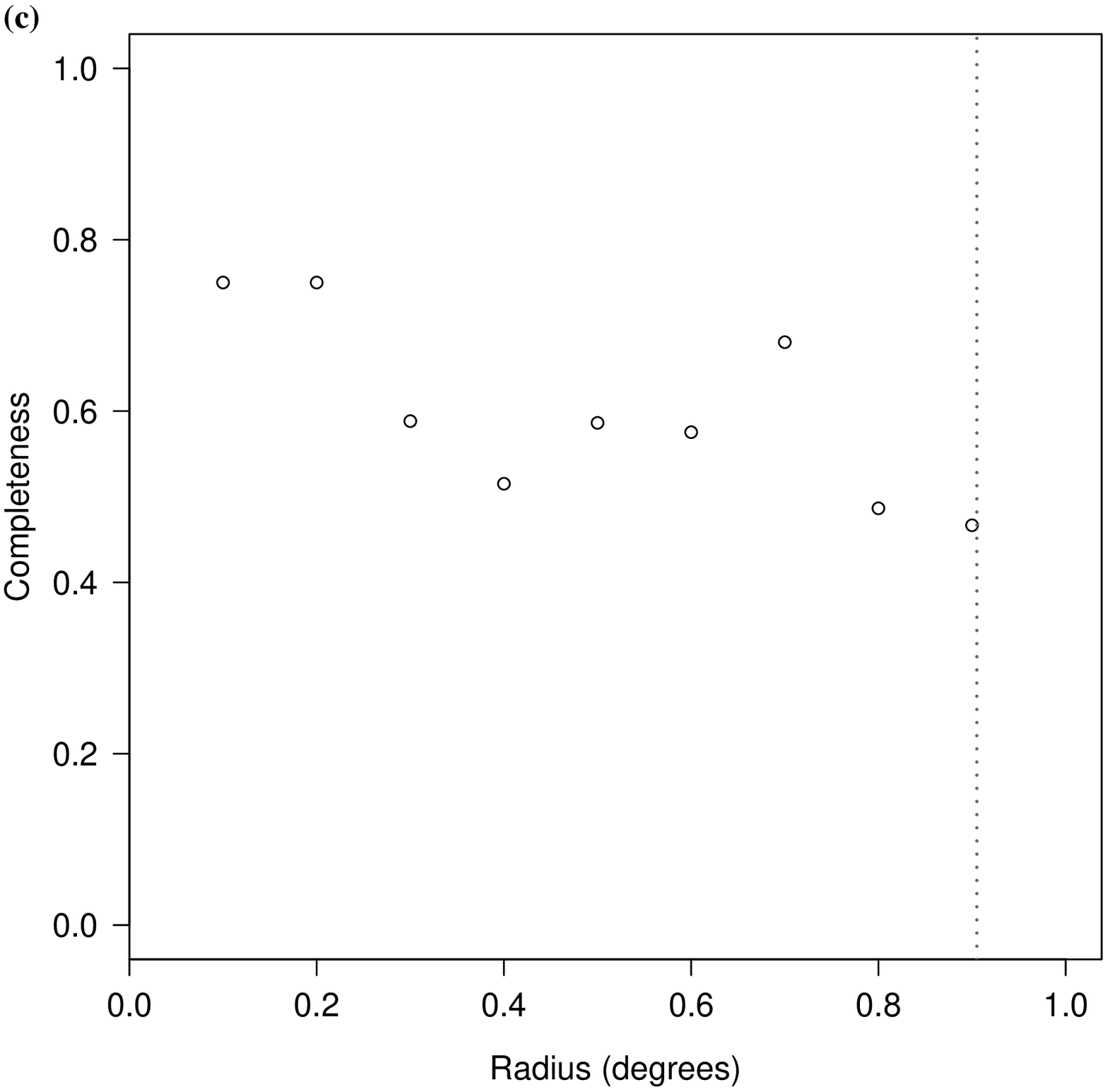} &
      \includegraphics[width=0.5\hsize]{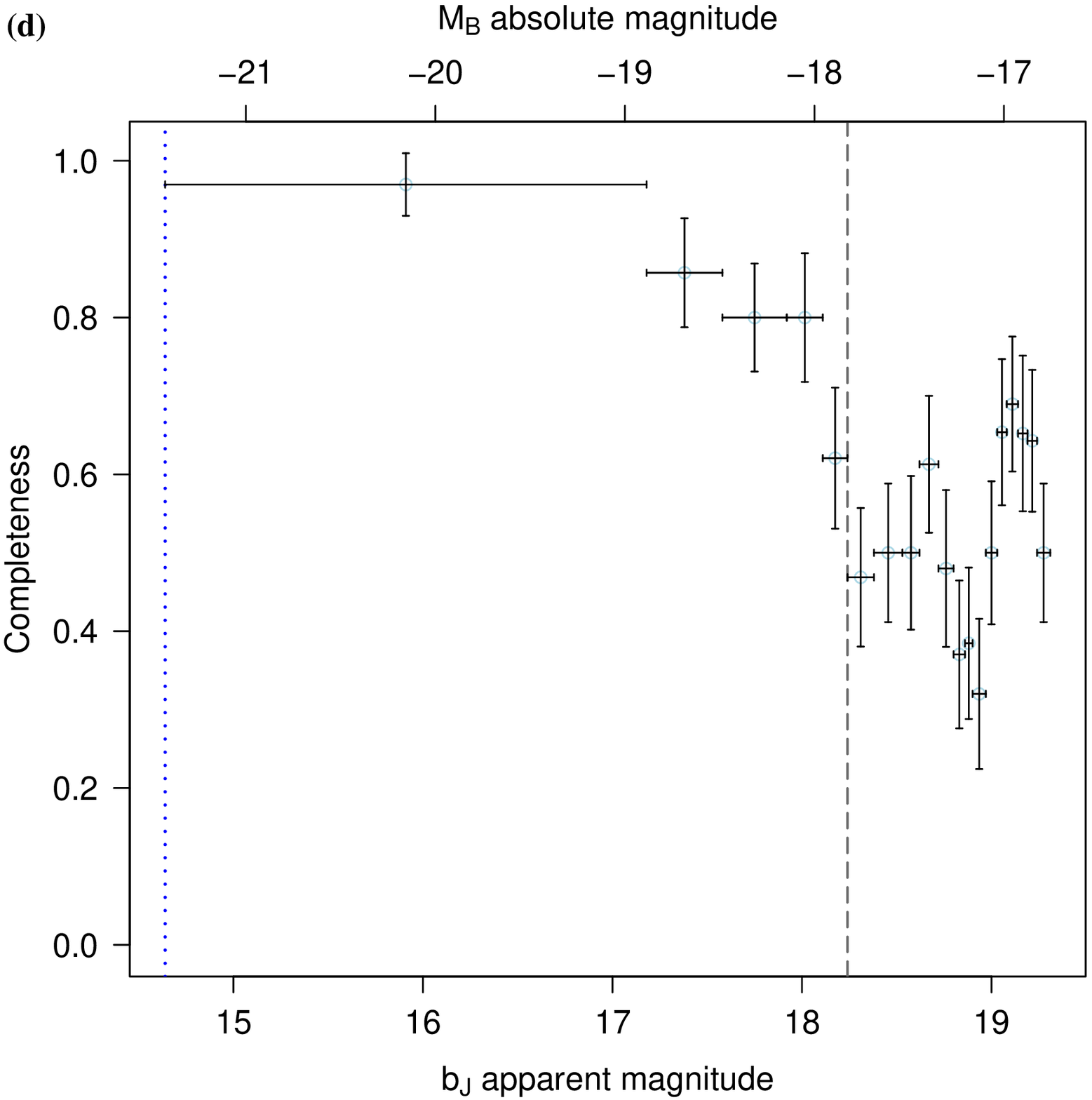}
\end{tabular}
\caption{(\subfigletter{a}) Distribution of galaxies around B0503--286 within the 2$^{\circ}$ field of the AAT/AAOmega (using redshift data from~\citet{Subrahmanyanetal2008}); a 20 arcmin angular scale corresponds to a linear size of 0.90 Mpc. Plus symbols represent galaxies within $\pm0.003$ of the host redshift at z = 0.0383. Radio contours are shown at (2, 4, 8, and 16) $\times$ 0.4 mJy beam$^{-1}$ for a 1520 MHz, total intensity, VLA image with a beam of FWHM $15.5\arcsec\times14.2\arcsec$ at a P.A. of 8.6$^{\circ}$. (\subfigletter{b}) An alternate view of the optical field, centred at the host galaxy, with dotted circles at 1, 2 and 3-Mpc radii. The solid, black circle represents the 1.0$^{\circ}$ field boundary. (\subfigletter{c}) A completeness plot showing the fraction of observed SuperCOSMOS targets against radius, measured in intervals of 0.1$^{\circ}$ from the position of the GRG host optical galaxy. The dotted line represents the radius of the largest circular region centred at the host that will fit within the observed 2$^{\circ}$ field. (\subfigletter{d}) A completeness plot showing the fraction of observed targets against $B$-band magnitude with the $b_{\mathrm{J}}$ apparent magnitude (14.64) of the host galaxy indicated by a blue dotted line.}
\label{fig:B0503}
\end{minipage}
\end{figure*}

\begin{figure*}
\begin{minipage}[t][\textheight]{\textwidth}
  \centering
  \begin{tabular}{ll}
      \includegraphics[width=0.5\hsize]{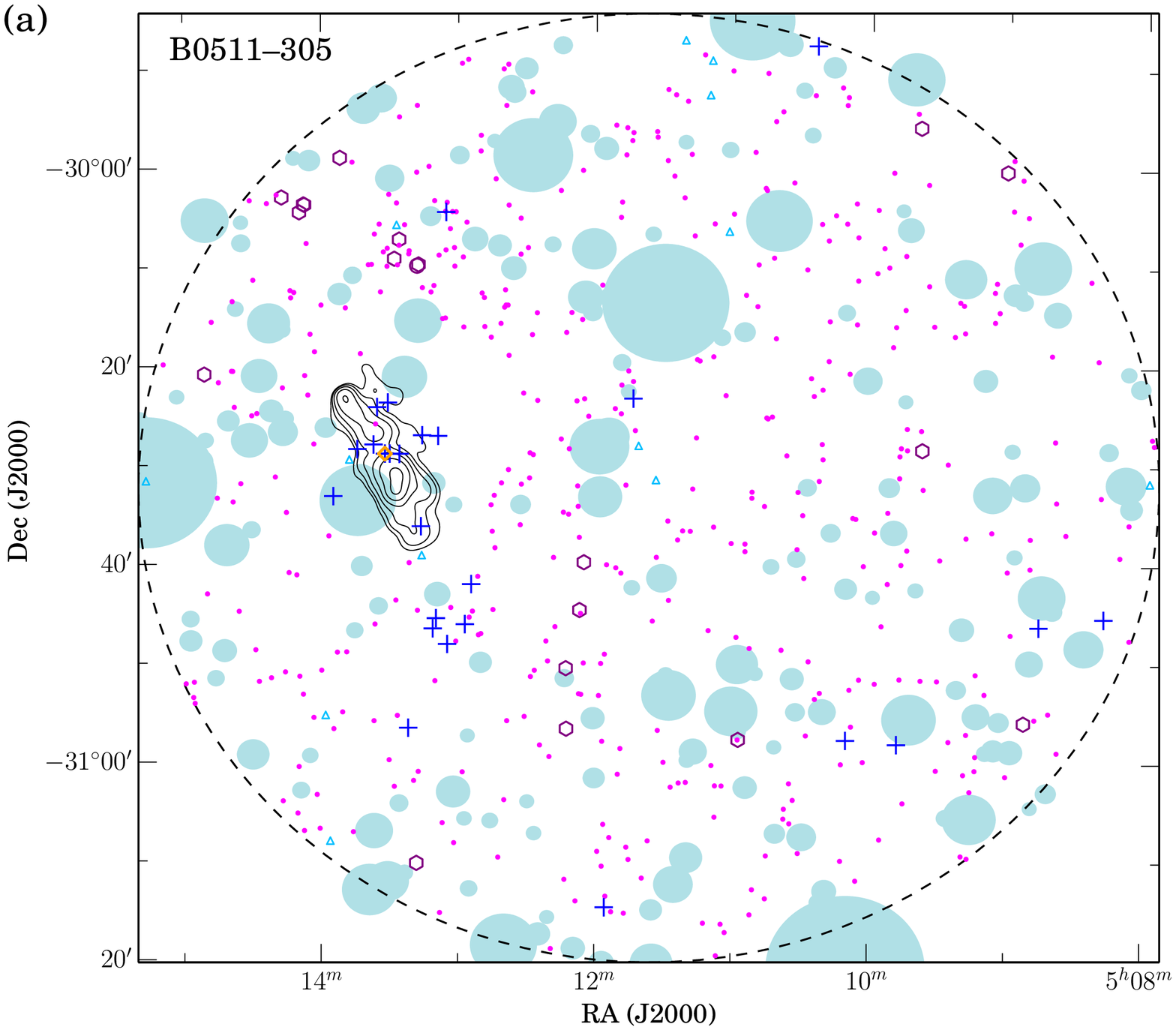} &
      \includegraphics[width=0.5\hsize]{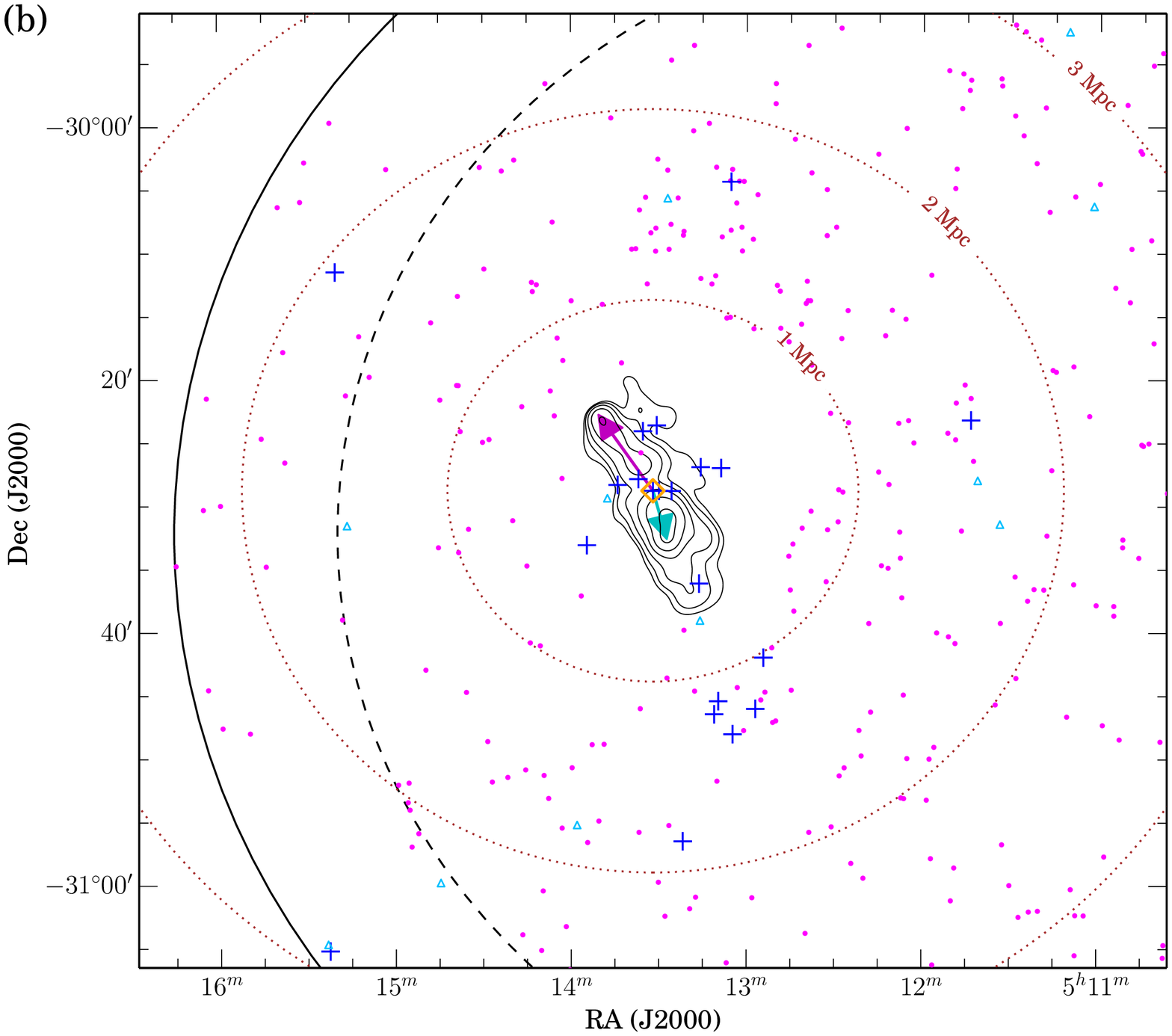}\\
      \includegraphics[width=0.5\hsize]{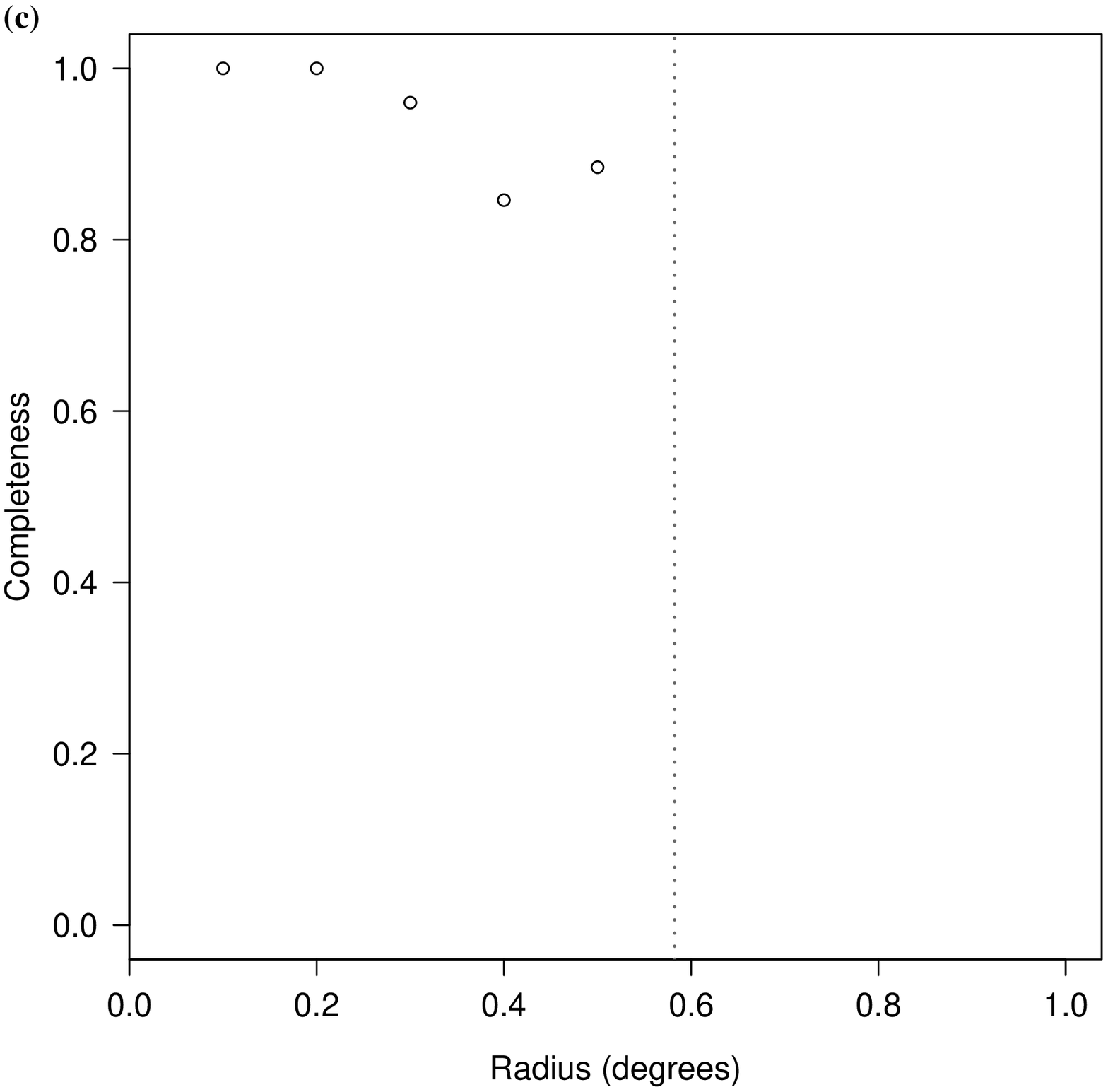} &
      \includegraphics[width=0.5\hsize]{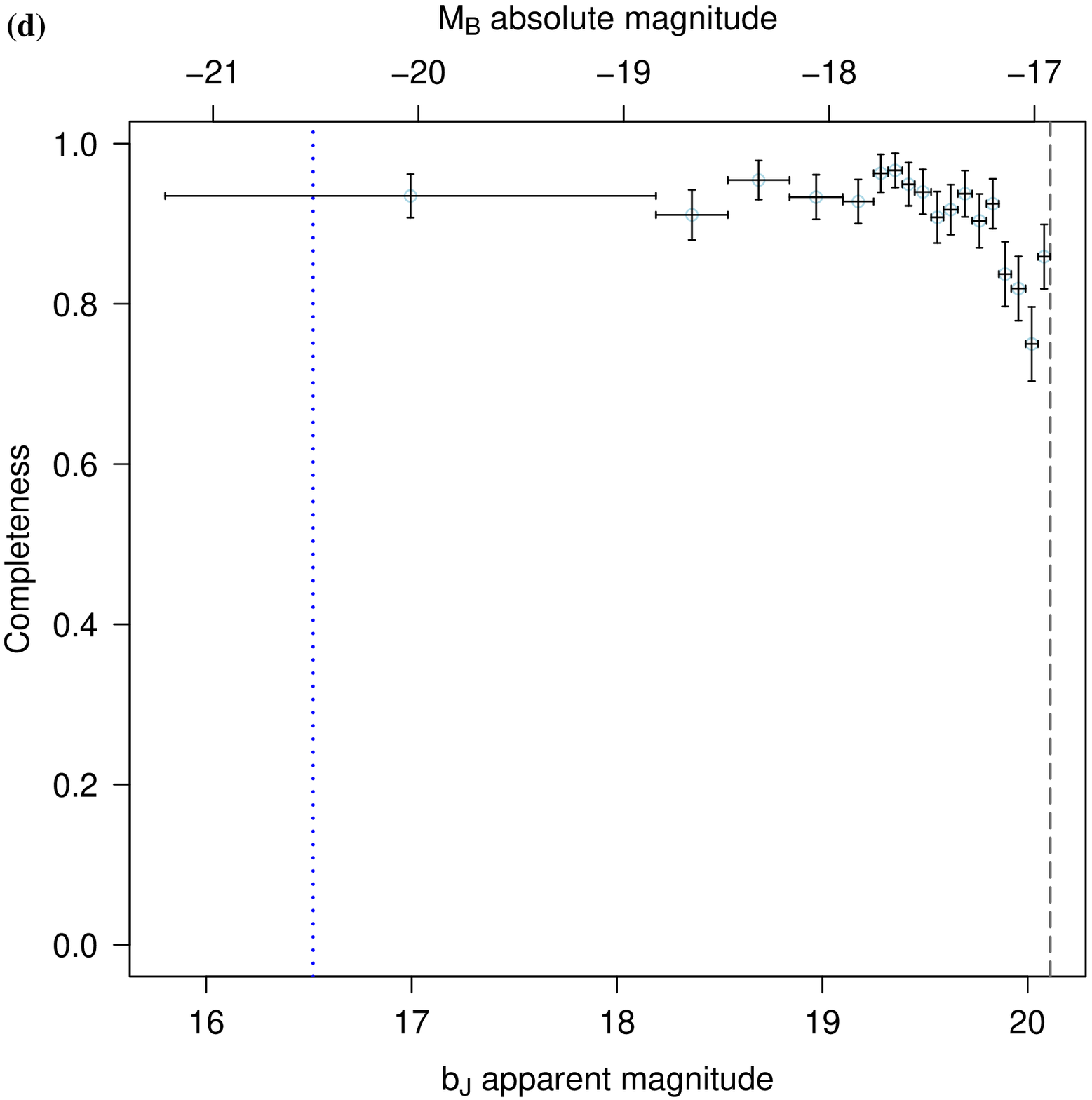}
\end{tabular}
\caption{(\subfigletter{a}) Distribution of galaxies around B0511--305; a 20 arcmin angular scale corresponds to a linear size of 1.32 Mpc. Plus symbols represent galaxies within $\pm0.003$ of the host redshift at z = 0.0577. Radio contours are shown at (6, 12, 24, 48, 96, and 192) $\times$ 1.25 mJy beam$^{-1}$ with a symmetric beam of FWHM 90 arcsec. (\subfigletter{b}) An enlarged view of the optical field, centred at the host galaxy, with dotted circles at 1, 2 and 3-Mpc radii. The solid, black circle indicates the 1.0$^{\circ}$ field boundary. (\subfigletter{c}) A completeness plot showing the fraction of observed SuperCOSMOS targets against radius, measured in intervals of 0.1$^{\circ}$ from the position of the GRG host optical galaxy. The dotted line represents the radius of the largest circular region centred at the host that will fit within the observed 2$^{\circ}$ field. (\subfigletter{d}) A completeness plot showing the fraction of observed targets against $B$-band magnitude with the $b_{\mathrm{J}}$ apparent magnitude (16.52) of the host galaxy indicated by a blue dotted line.}
\label{fig:B0511}
\end{minipage}
\end{figure*}

\begin{figure*}
\begin{minipage}[t][\textheight]{\textwidth}
  \centering
  \begin{tabular}{ll}
      \includegraphics[width=0.5\hsize]{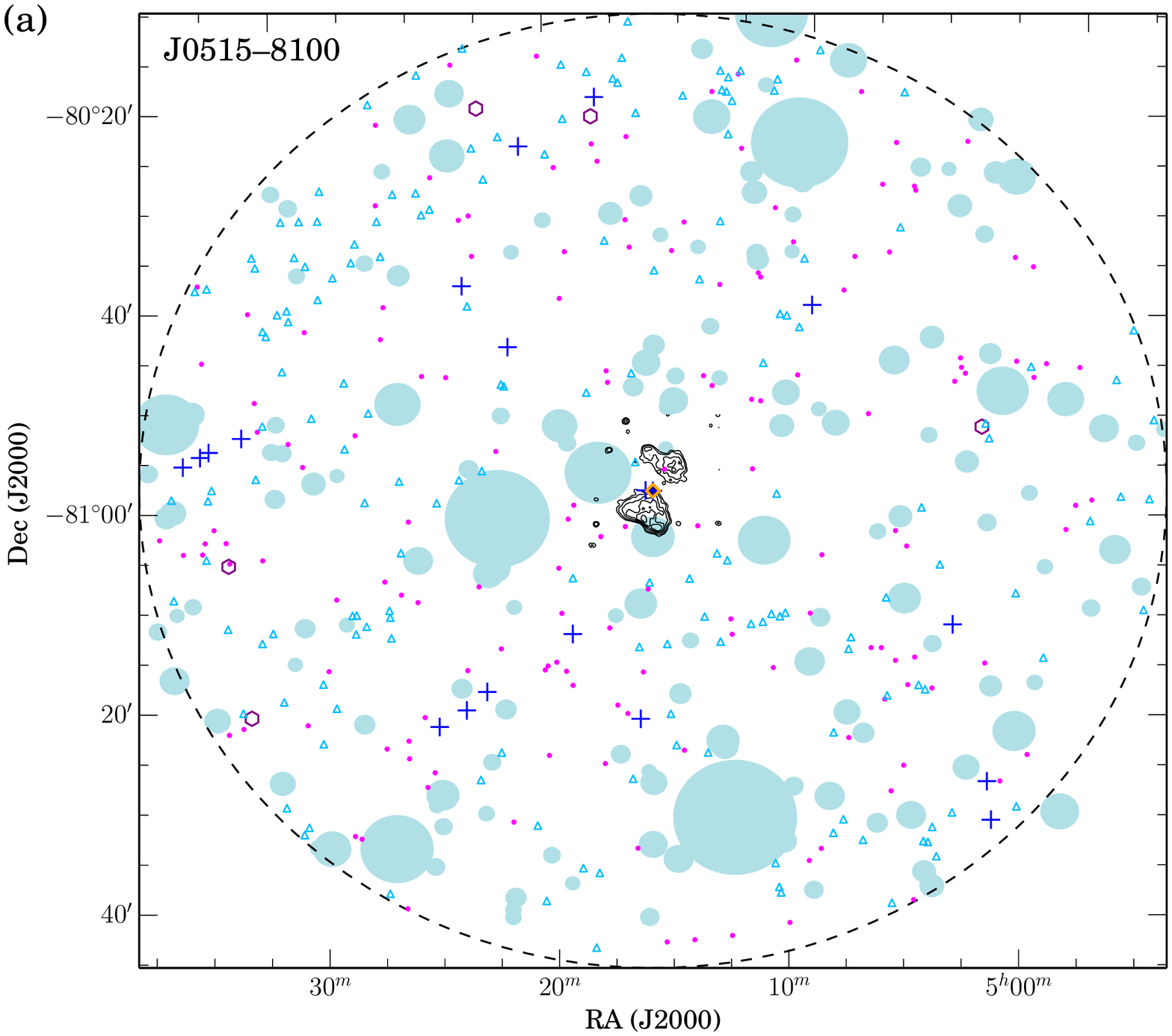} &
      \includegraphics[width=0.5\hsize]{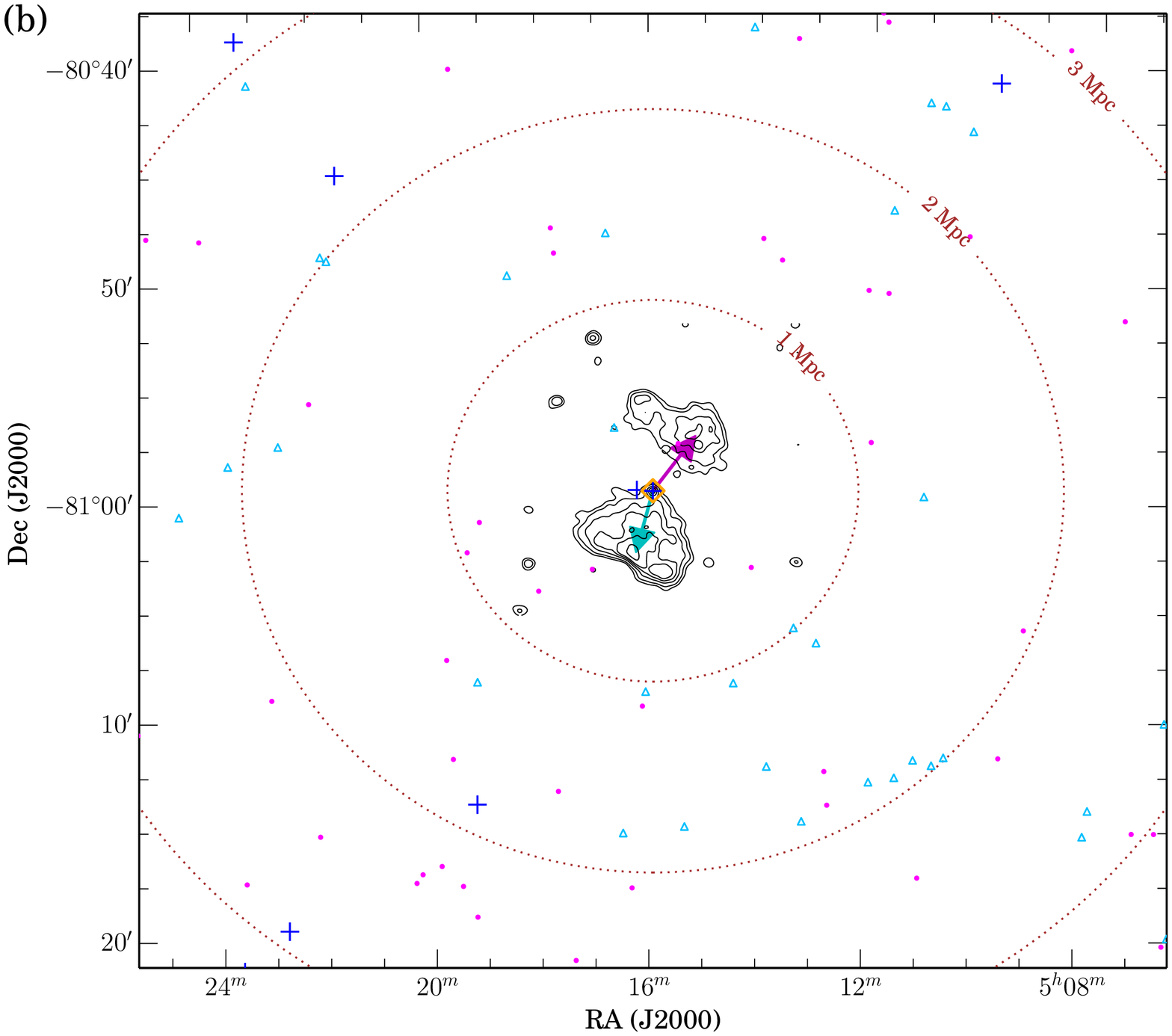}\\
      \includegraphics[width=0.5\hsize]{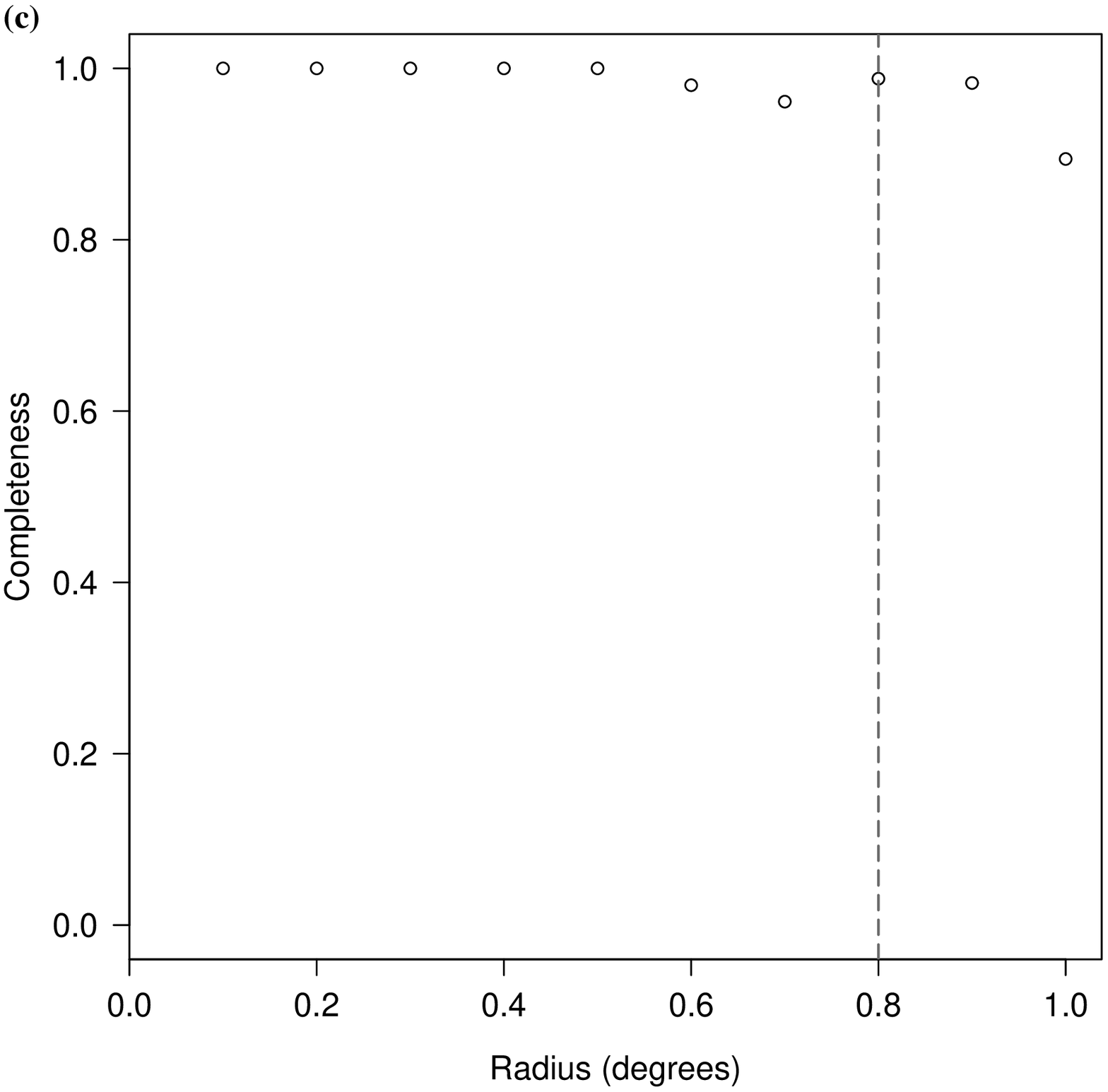} &
      \includegraphics[width=0.5\hsize]{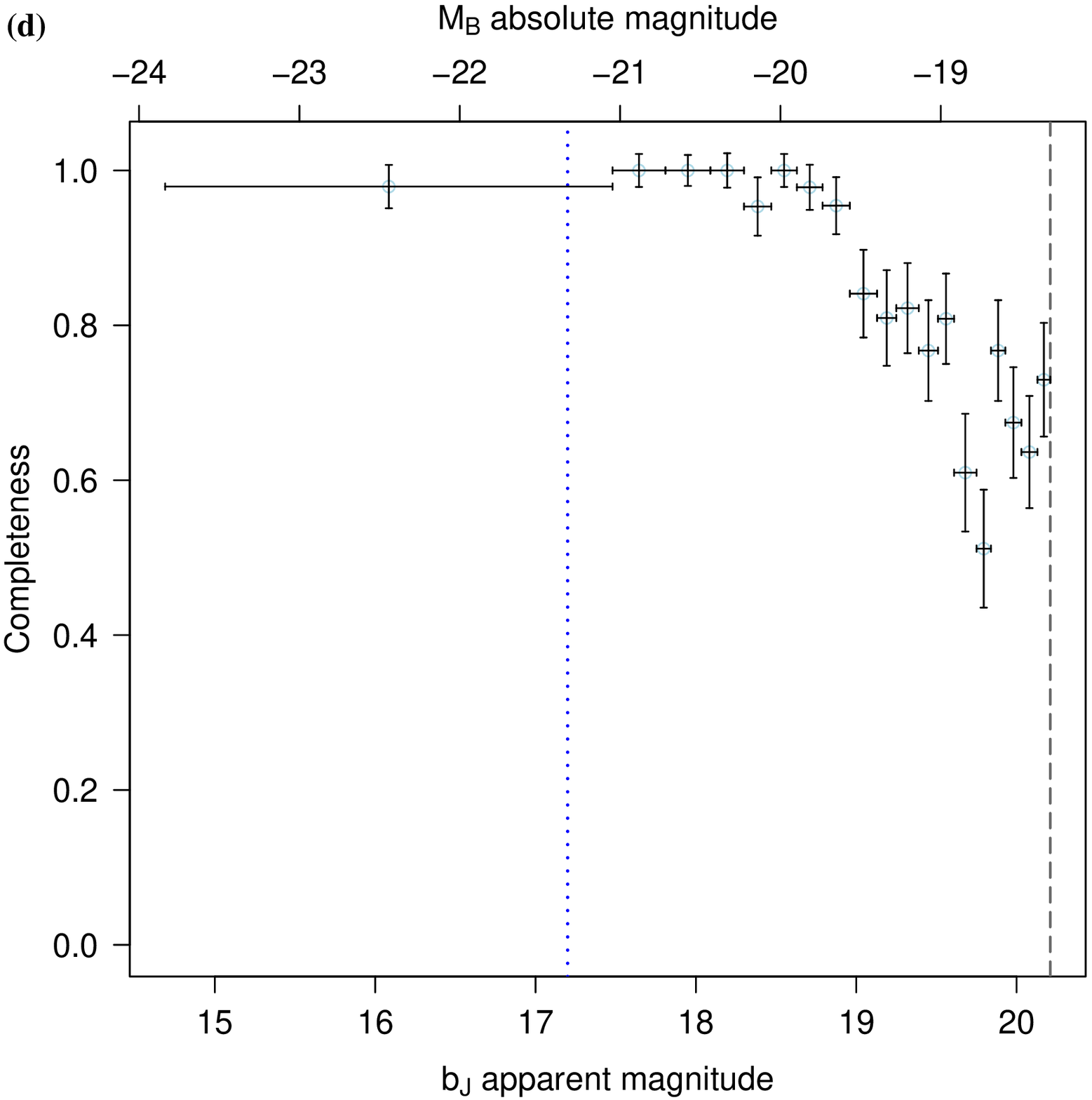}
\end{tabular}
\caption{(\subfigletter{a}) Distribution of galaxies around J0515--8100; a 20 arcmin angular scale corresponds to a linear size of 2.28 Mpc. Plus symbols represent galaxies within $\pm0.015$ of the host redshift at z = 0.105. Radio contours are shown at (2, 4, 8, 12, 16, and 24) $\times$ 0.15 mJy beam$^{-1}$ with a symmetric beam of FWHM 25 arcsec. (\subfigletter{b}) An enlarged view of the optical field, centred at the host galaxy, with dotted circles at 1, 2 and 3-Mpc radii. (\subfigletter{c}) A completeness plot showing the fraction of observed SuperCOSMOS targets against radius, measured in intervals of 0.1$^{\circ}$ from the field centre. (\subfigletter{d}) A completeness plot showing the fraction of observed targets against $B$-band magnitude with the $b_{\mathrm{J}}$ apparent magnitude (17.20) of the host galaxy indicated by a blue dotted line.}
\label{fig:J0515}
\end{minipage}
\end{figure*}

\begin{figure*}
\begin{minipage}[t][\textheight]{\textwidth}
  \centering
  \begin{tabular}{ll}
      \includegraphics[width=0.5\hsize]{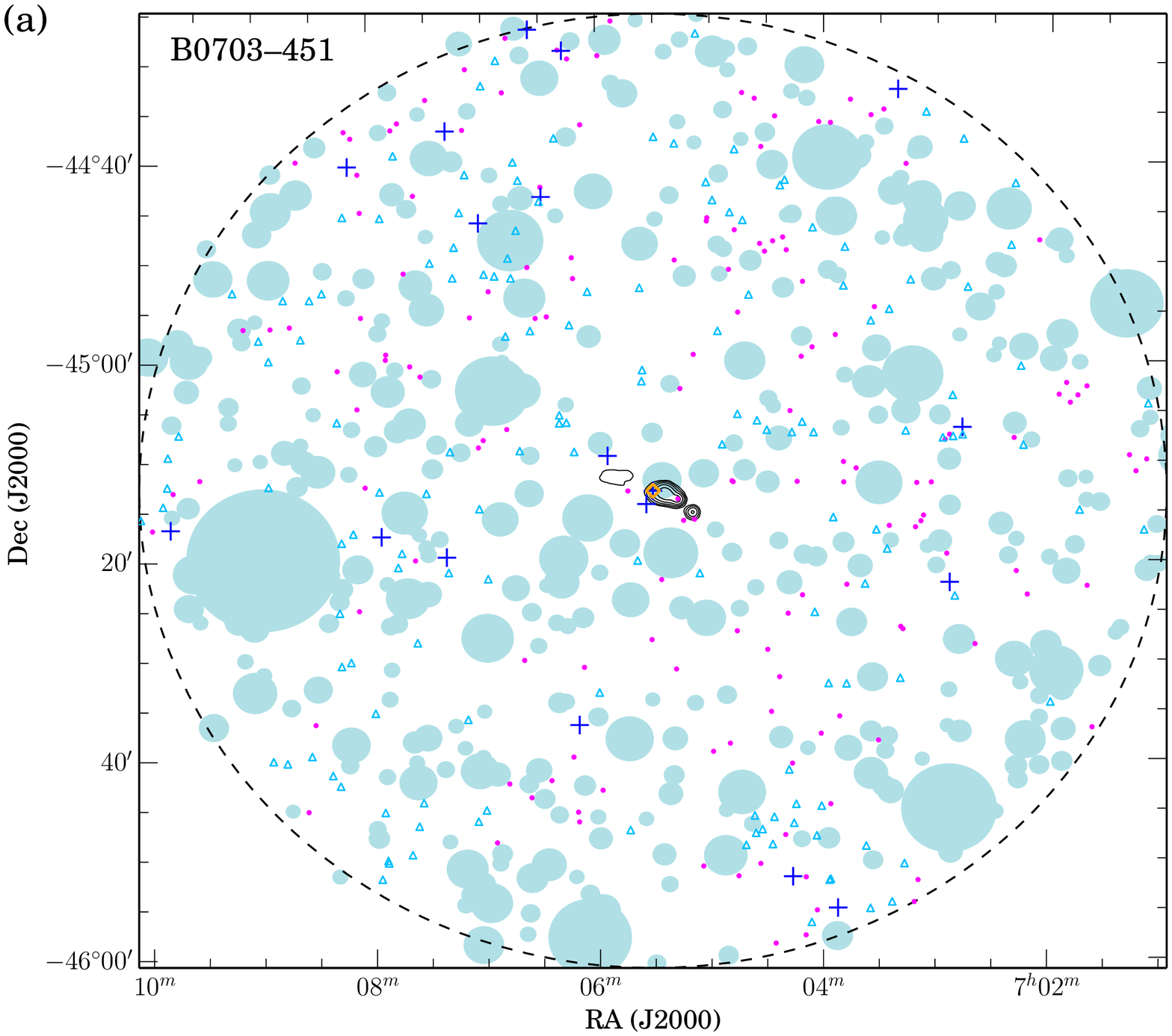} &
      \includegraphics[width=0.5\hsize]{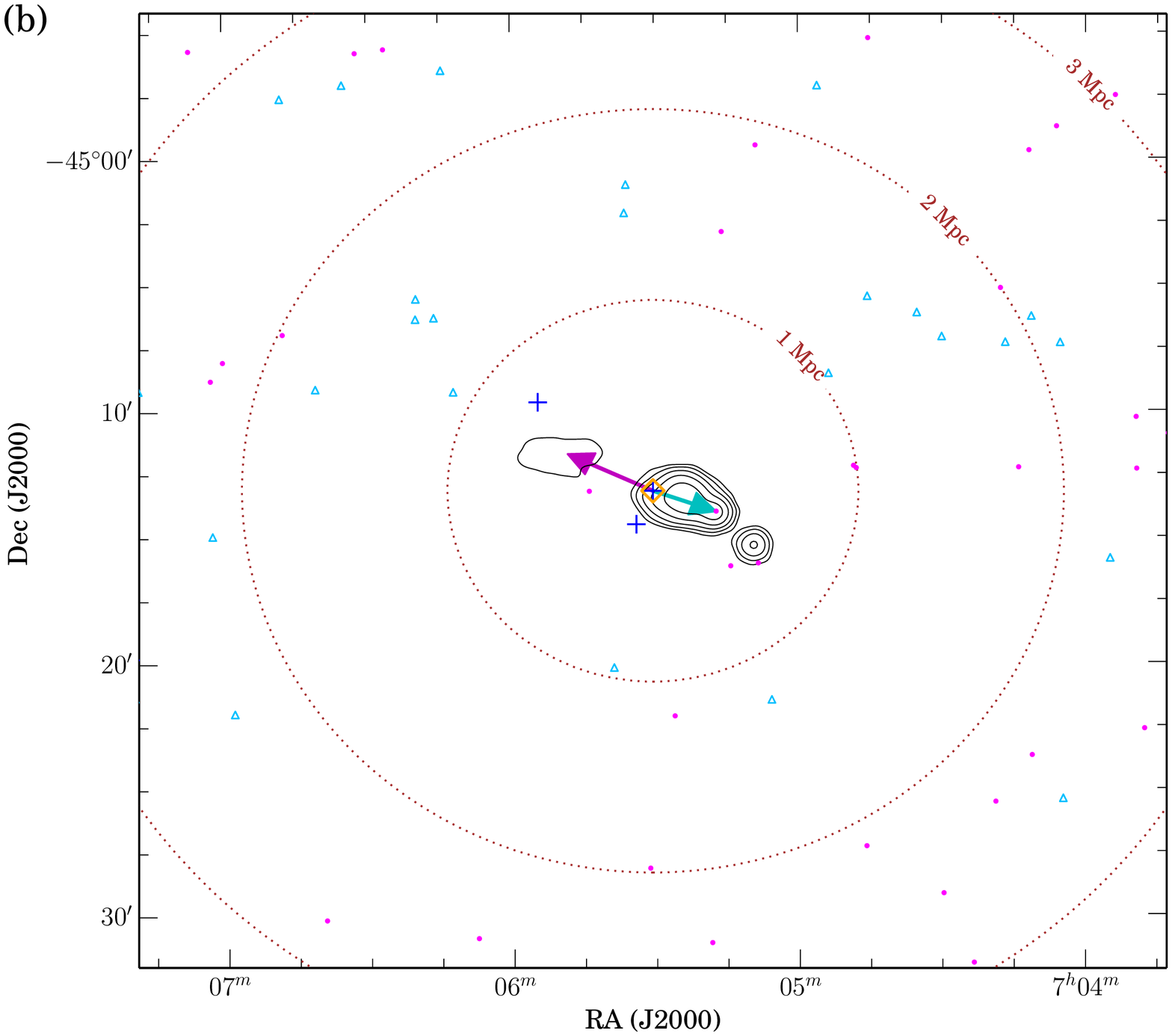}\\
      \includegraphics[width=0.5\hsize]{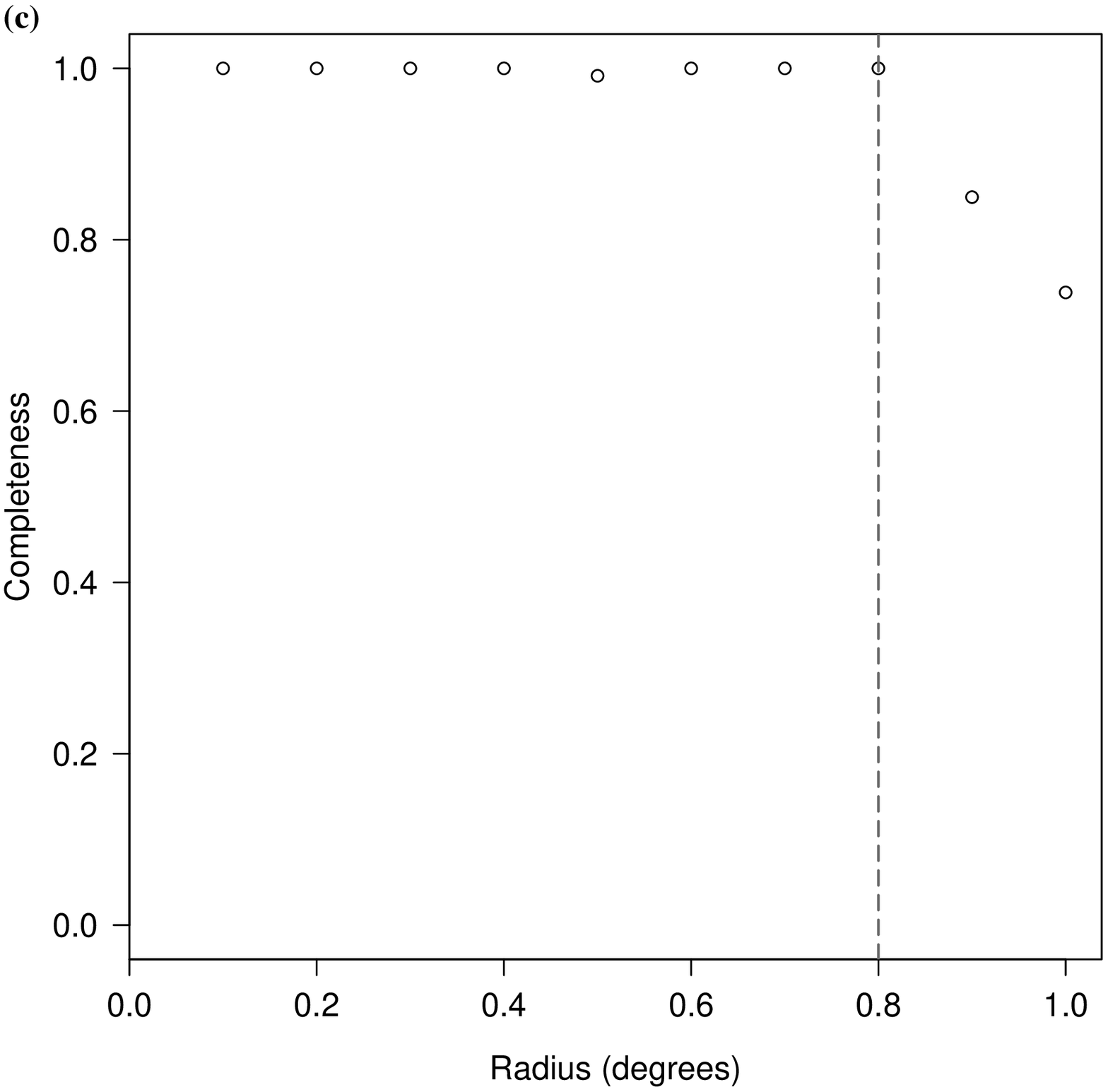} &
      \includegraphics[width=0.5\hsize]{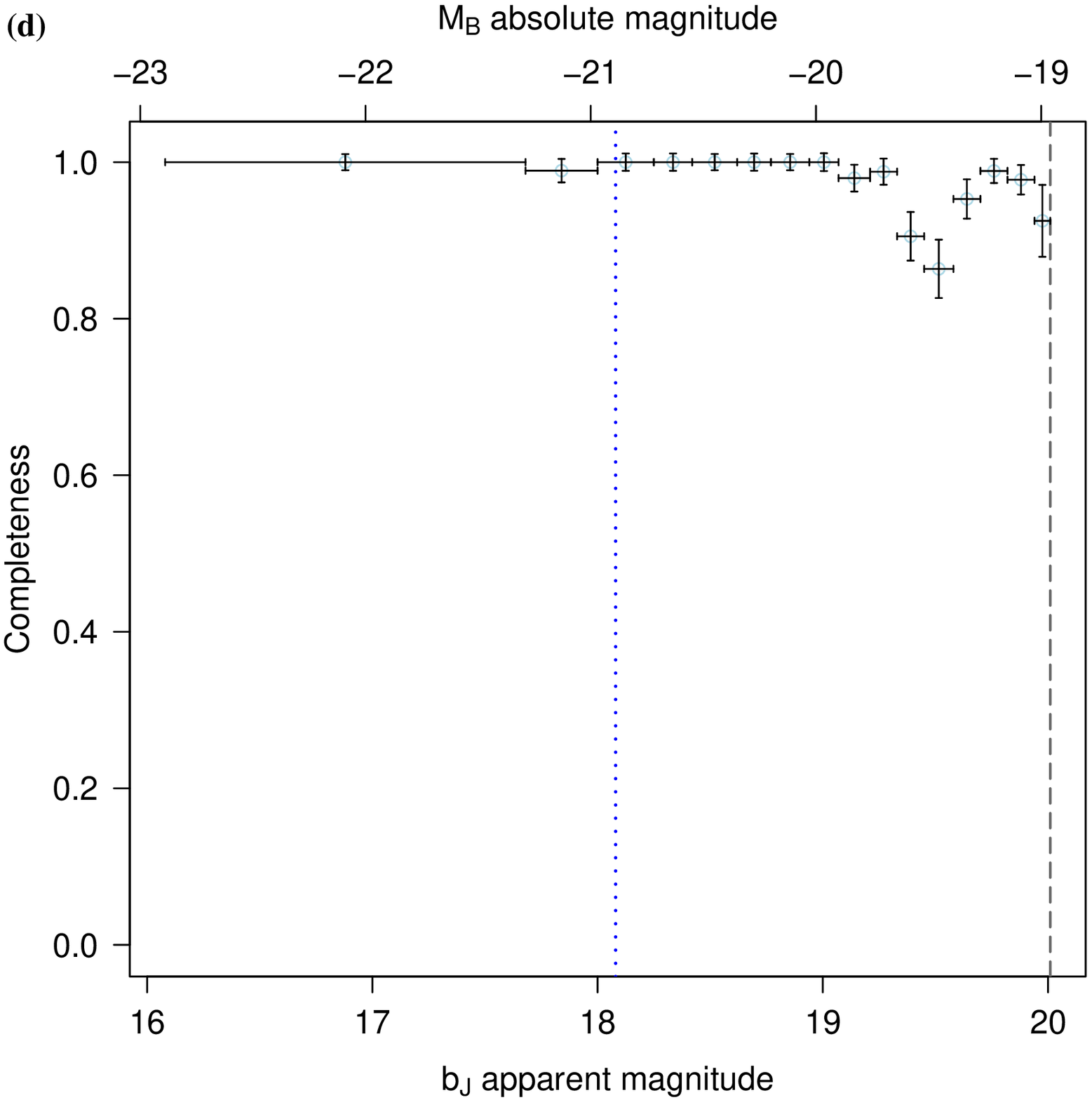}
\end{tabular}
\caption{(\subfigletter{a}) Distribution of galaxies around B0703--451; a 20 arcmin angular scale corresponds to a linear size of 2.64 Mpc. Plus symbols represent galaxies within $\pm0.003$ of the host redshift at z = 0.1241. Radio contours are shown at (6, 12, 24, 48, and 96) $\times$ 0.25 mJy beam$^{-1}$. (\subfigletter{b}) An enlarged view of the optical field, centred at the host galaxy, with dotted circles at 1, 2 and 3-Mpc radii. (\subfigletter{c}) A completeness plot showing the fraction of observed SuperCOSMOS targets against radius, measured in intervals of 0.1$^{\circ}$ from the field centre. (\subfigletter{d}) A completeness plot showing the fraction of observed targets against $B$-band magnitude with the $b_{\mathrm{J}}$ apparent magnitude (18.08) of the host galaxy indicated by a blue dotted line.}
\label{fig:B0703}
\end{minipage}
\end{figure*}

\begin{figure*}
\begin{minipage}[t][\textheight]{\textwidth}
  \centering
  \begin{tabular}{ll}
      \includegraphics[width=0.5\hsize]{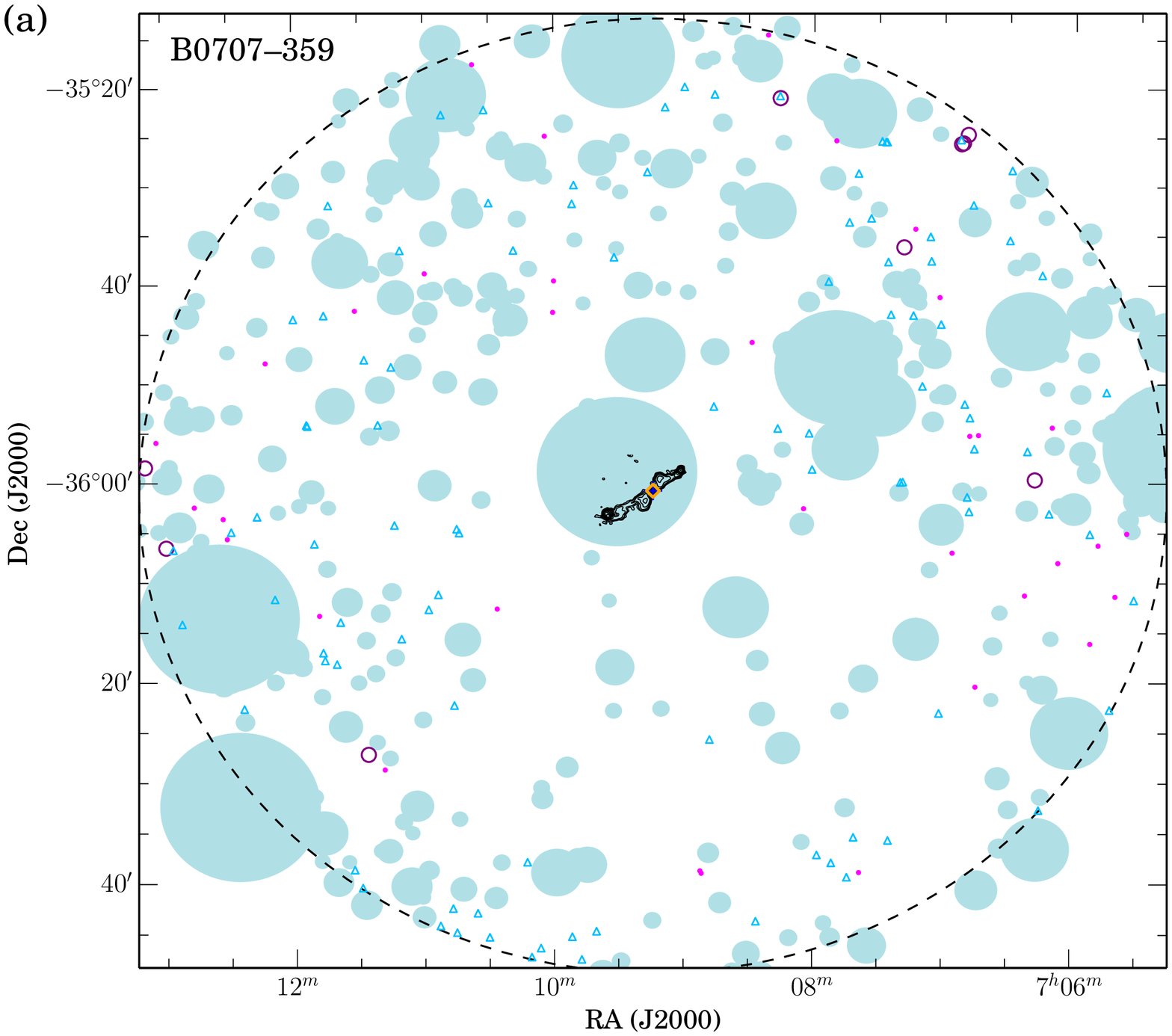} &
      \includegraphics[width=0.5\hsize]{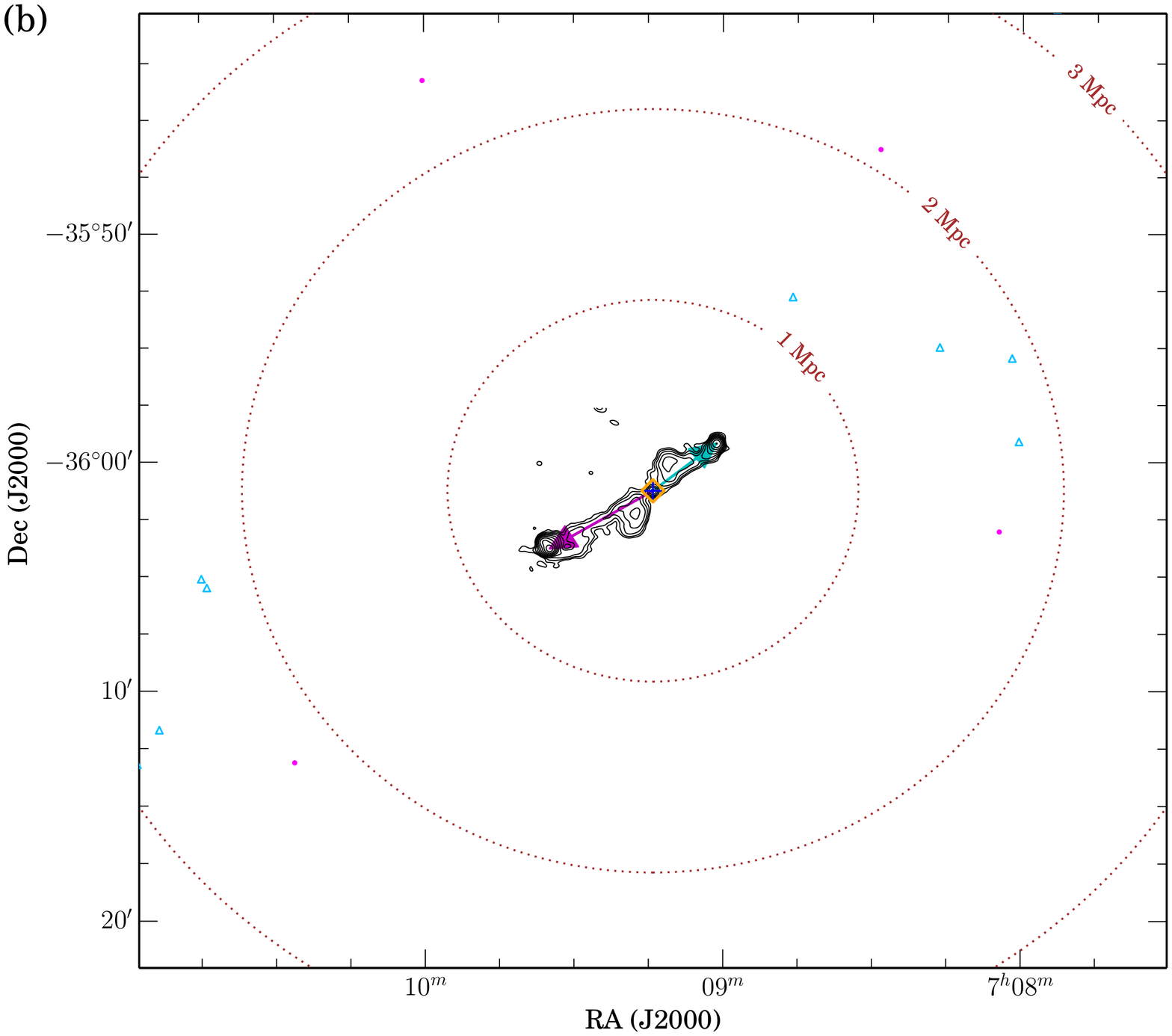}\\
      \includegraphics[width=0.5\hsize]{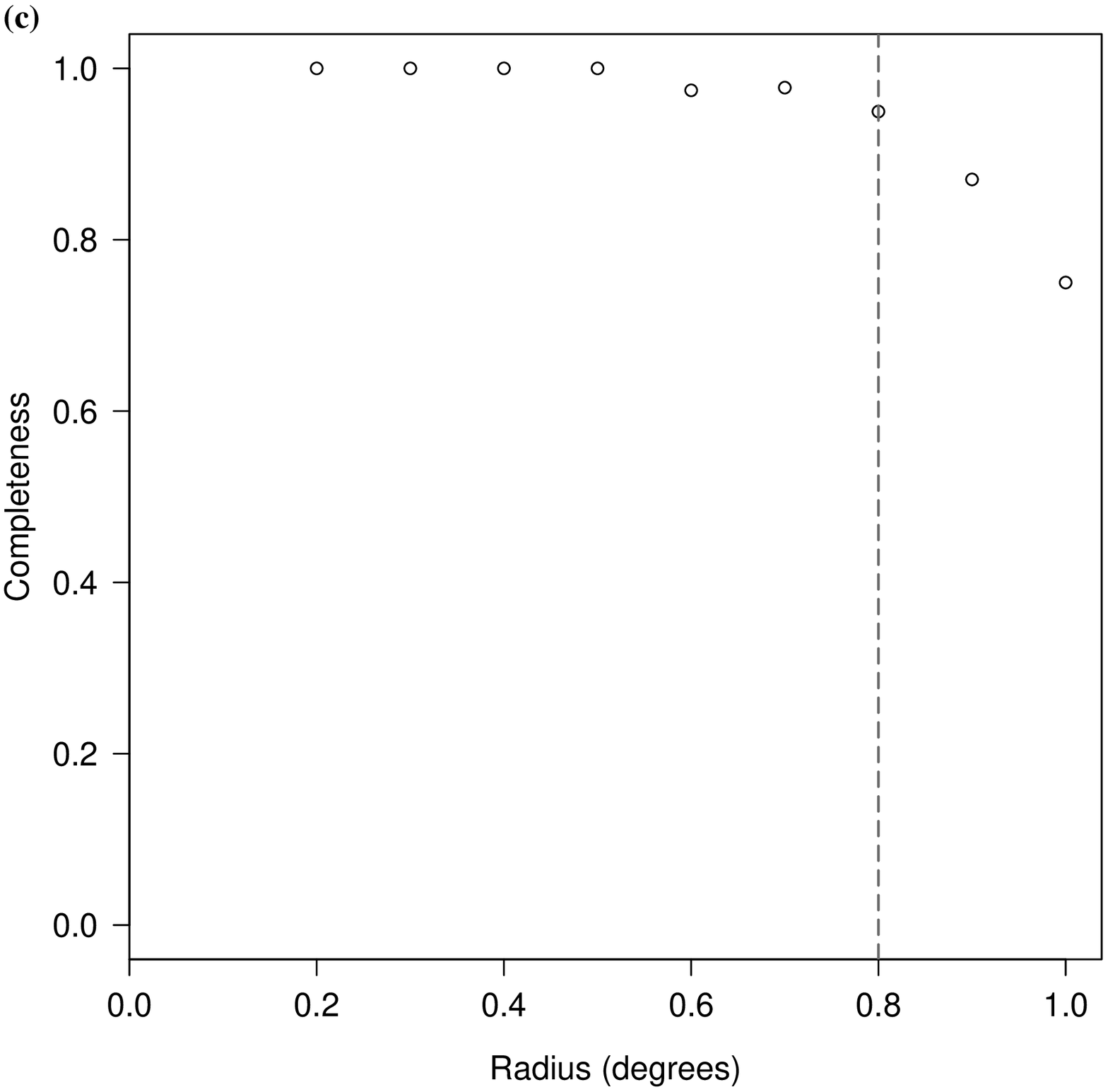} &
      \includegraphics[width=0.5\hsize]{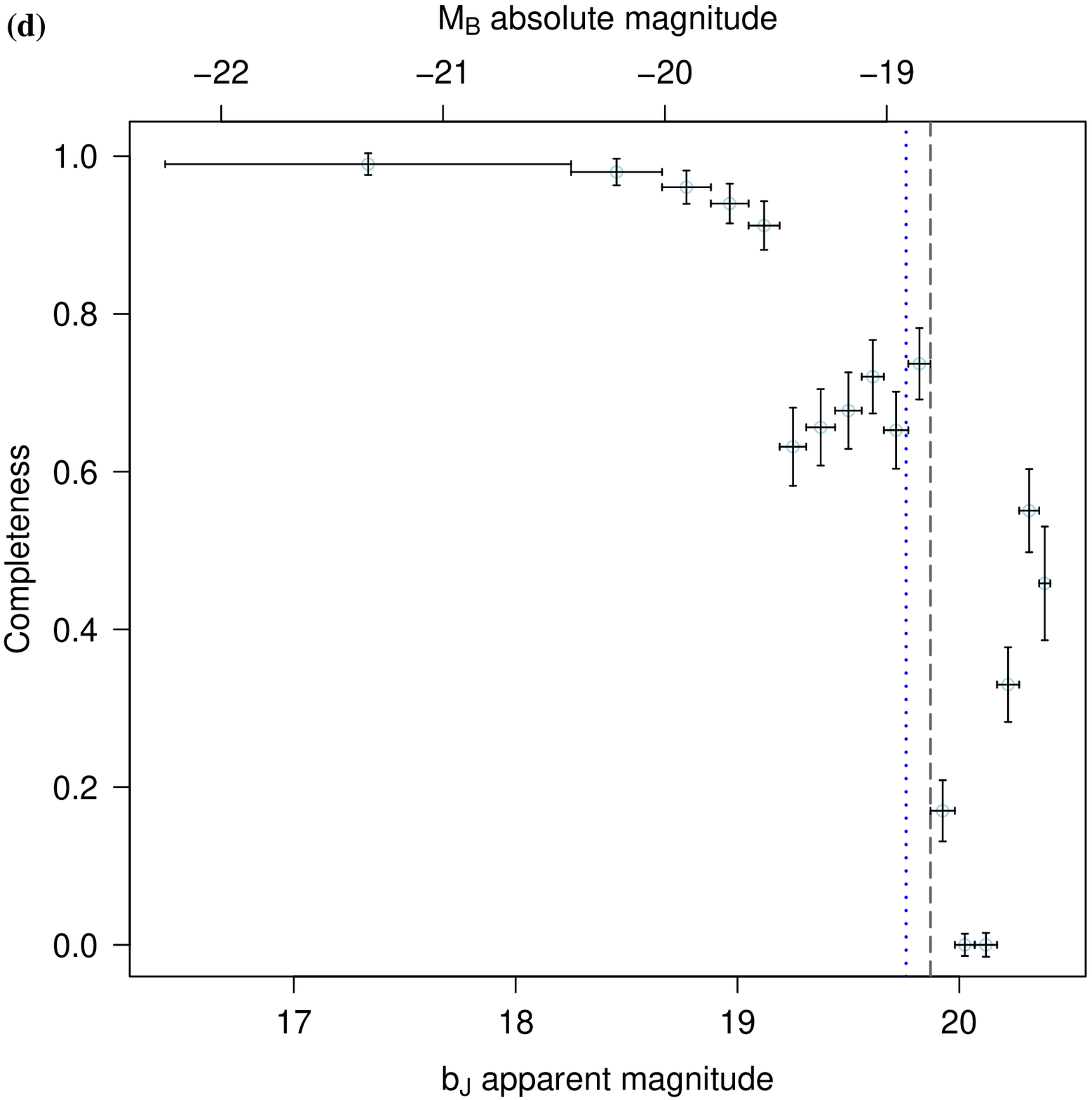}
\end{tabular}
\caption{(\subfigletter{a}) Distribution of galaxies around B0707--359; a 20 arcmin angular scale corresponds to a linear size of 2.40 Mpc. Plus symbols represent galaxies within $\pm0.003$ of the host redshift at z = 0.1109. Radio contours are shown at (2, 4, 8, 16, 32, 64, 128, and 256) $\times$ 0.25 mJy beam$^{-1}$ with a symmetric beam of FWHM 16 arcsec. (\subfigletter{b}) An enlarged view of the optical field, centred at the host galaxy, with dotted circles at 1, 2 and 3-Mpc radii. (\subfigletter{c}) A completeness plot showing the fraction of observed SuperCOSMOS targets against radius, measured in intervals of 0.1$^{\circ}$ from the field centre. (\subfigletter{d}) A completeness plot showing the fraction of observed targets against $B$-band magnitude with the $b_{\mathrm{J}}$ apparent magnitude (19.76) of the host galaxy indicated by a blue dotted line.}
\label{fig:B0707}
\end{minipage}
\end{figure*}

\begin{figure*}
\begin{minipage}[t][\textheight]{\textwidth}
  \centering
  \begin{tabular}{ll}
      \includegraphics[width=0.5\hsize]{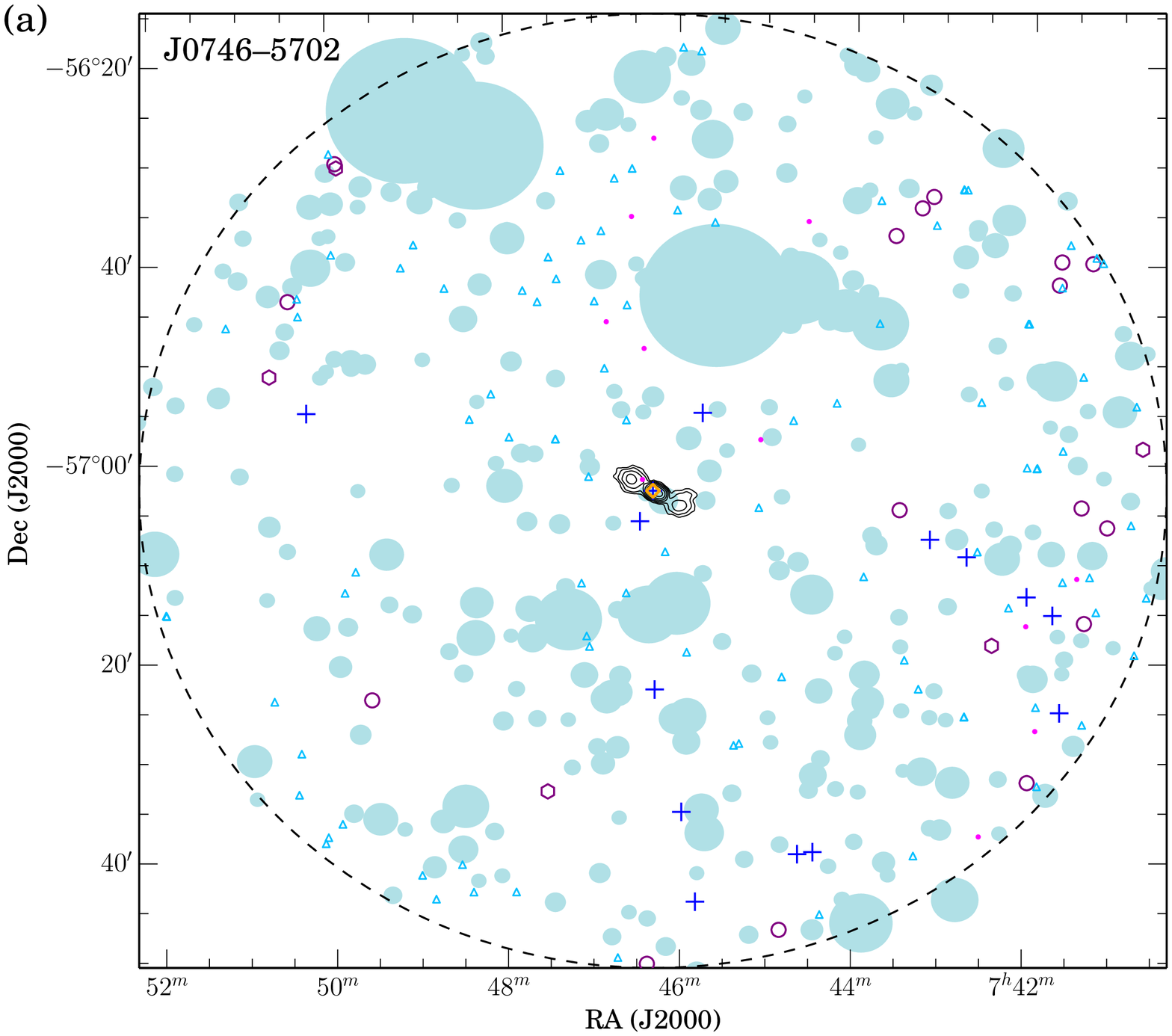} &
      \includegraphics[width=0.5\hsize]{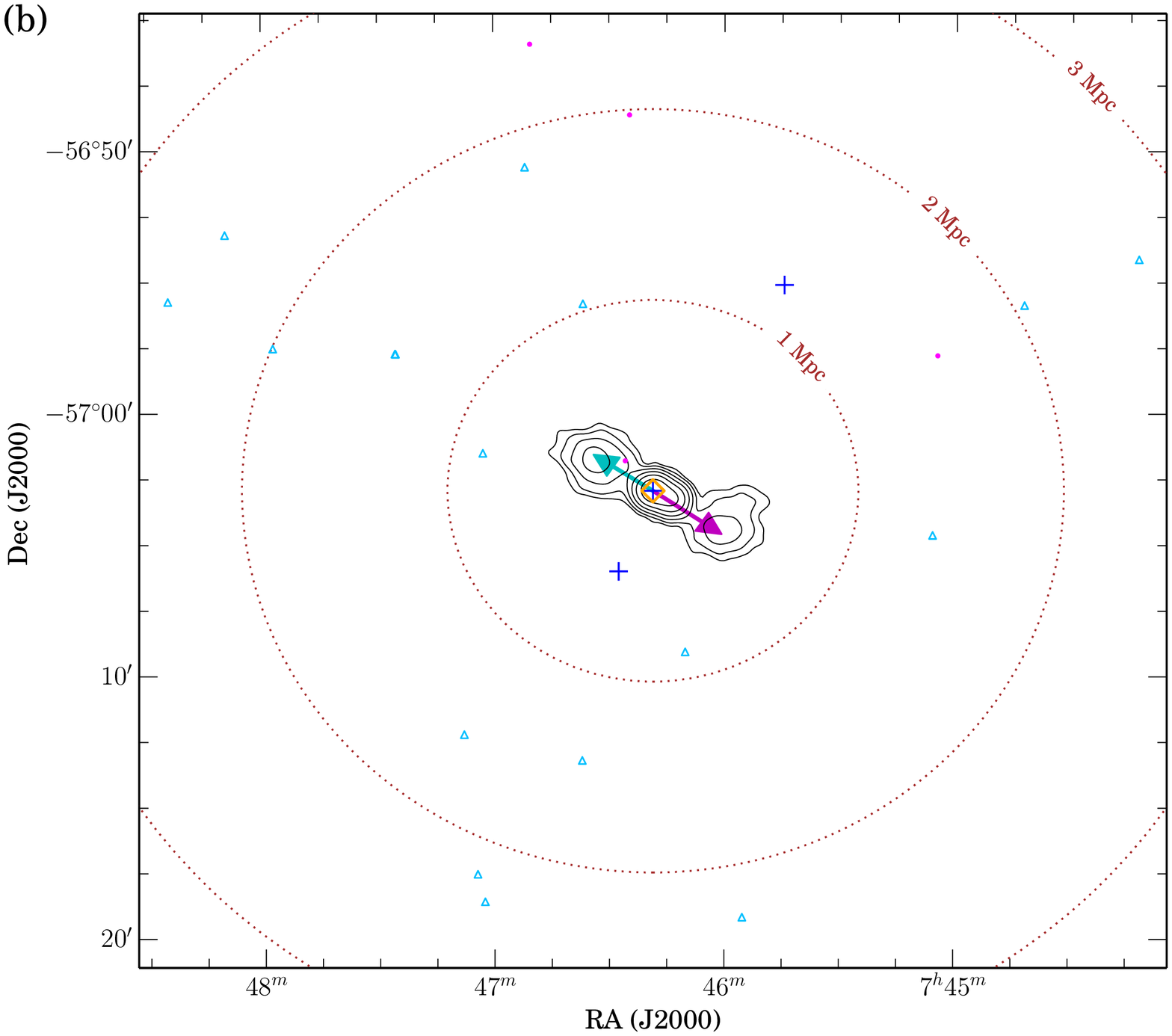}\\
      \includegraphics[width=0.5\hsize]{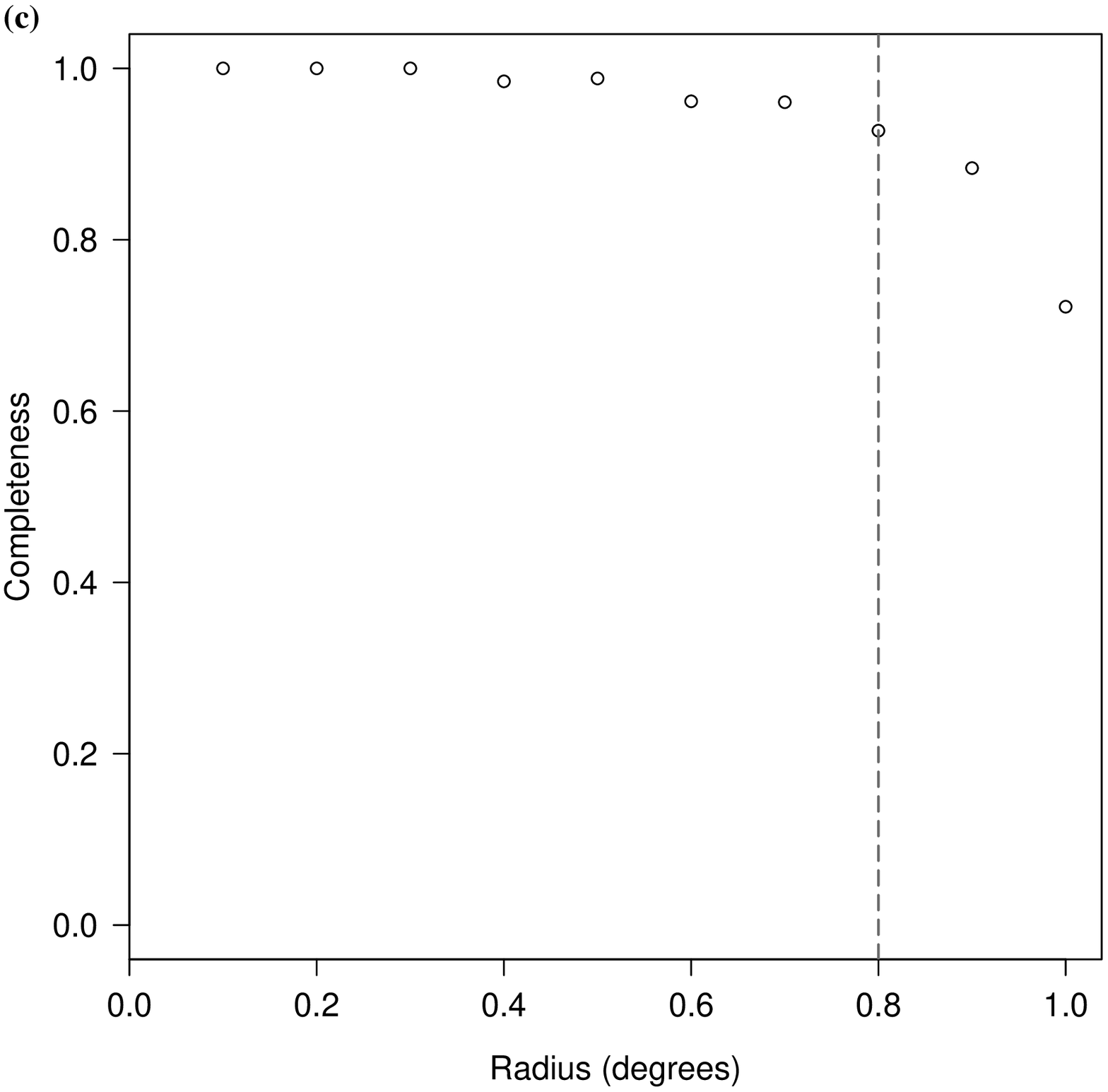} &
      \includegraphics[width=0.5\hsize]{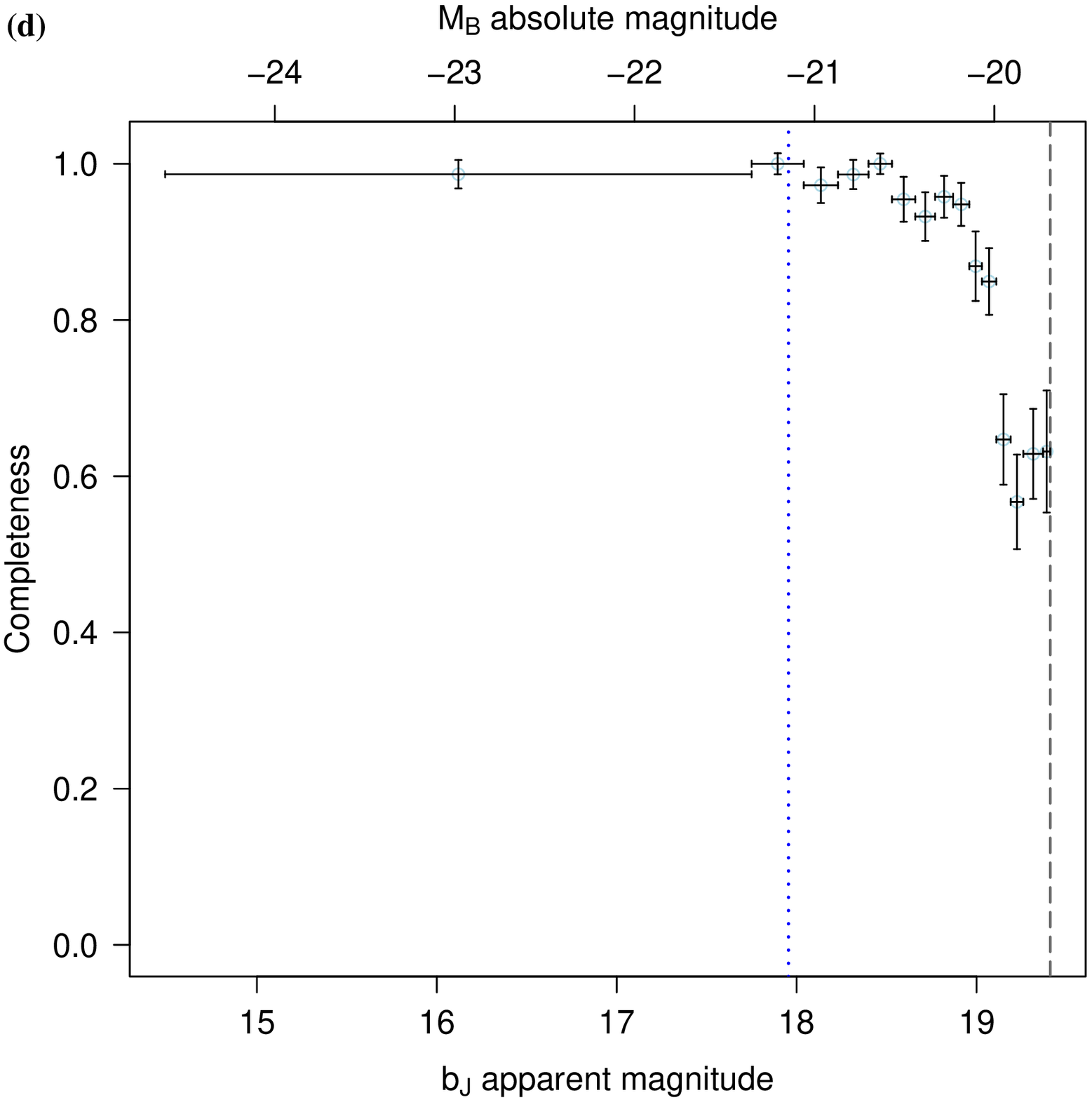}
\end{tabular}
\caption{(\subfigletter{a}) Distribution of galaxies around J0746--5702; a 20 arcmin angular scale corresponds to a linear size of 2.75 Mpc. Plus symbols represent galaxies within $\pm0.003$ of the host redshift at z = 0.1301. Radio contours are shown at (6, 12, 24, 48, 96, 192, and 384) $\times$ 0.12 mJy beam$^{-1}$. (\subfigletter{b}) An enlarged view of the optical field, centred at the host galaxy, with dotted circles at 1, 2 and 3-Mpc radii. (\subfigletter{c}) A completeness plot showing the fraction of observed SuperCOSMOS targets against radius, measured in intervals of 0.1$^{\circ}$ from the field centre. (\subfigletter{d}) A completeness plot showing the fraction of observed targets against $B$-band magnitude with the $b_{\mathrm{J}}$ apparent magnitude (17.96) of the host galaxy indicated by a blue dotted line.}
\label{fig:J0746}
\end{minipage}
\end{figure*}

\begin{figure*}
\begin{minipage}[t][\textheight]{\textwidth}
  \centering
  \begin{tabular}{ll}
      \includegraphics[width=0.5\hsize]{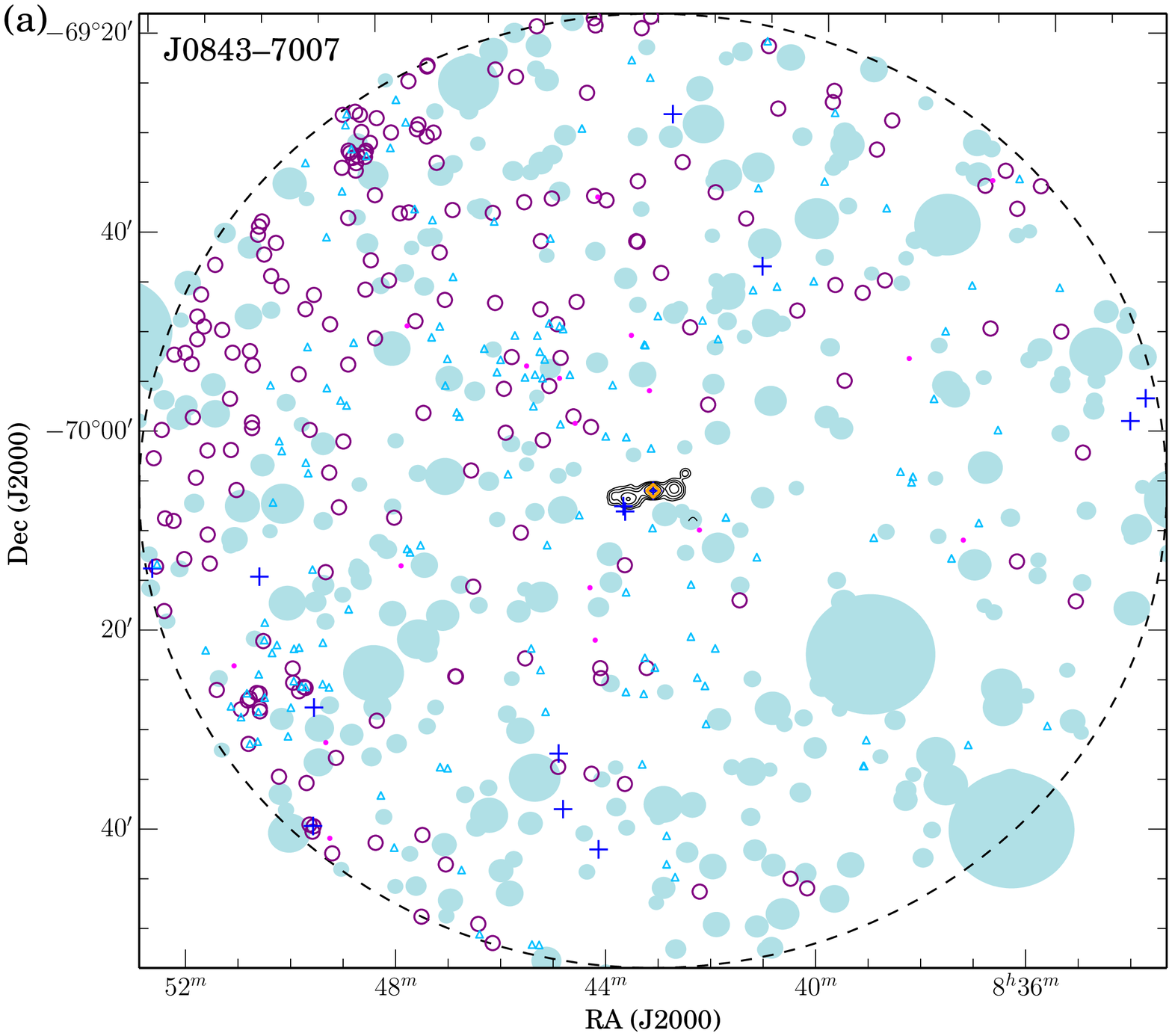} &
      \includegraphics[width=0.5\hsize]{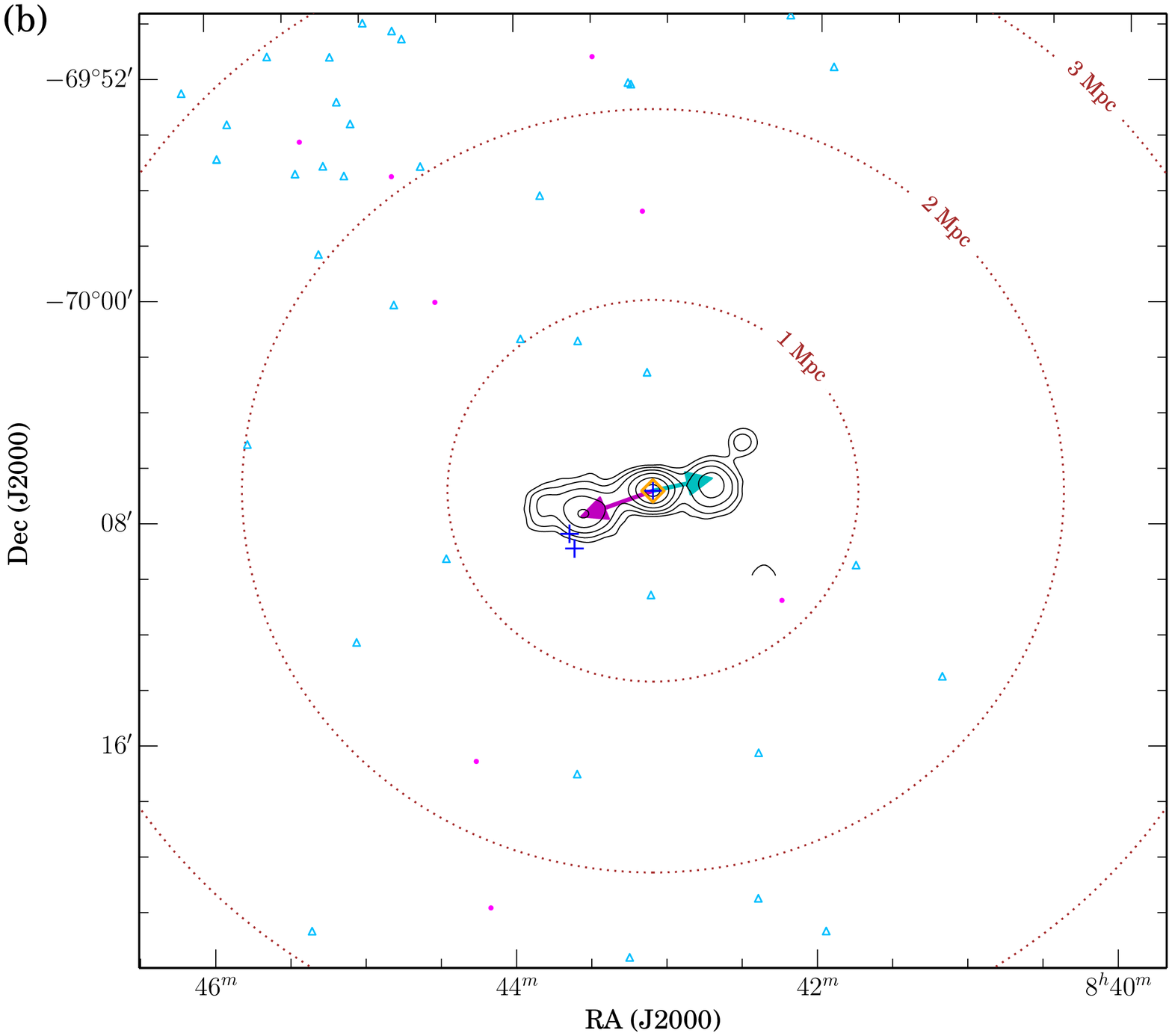}\\
      \includegraphics[width=0.5\hsize]{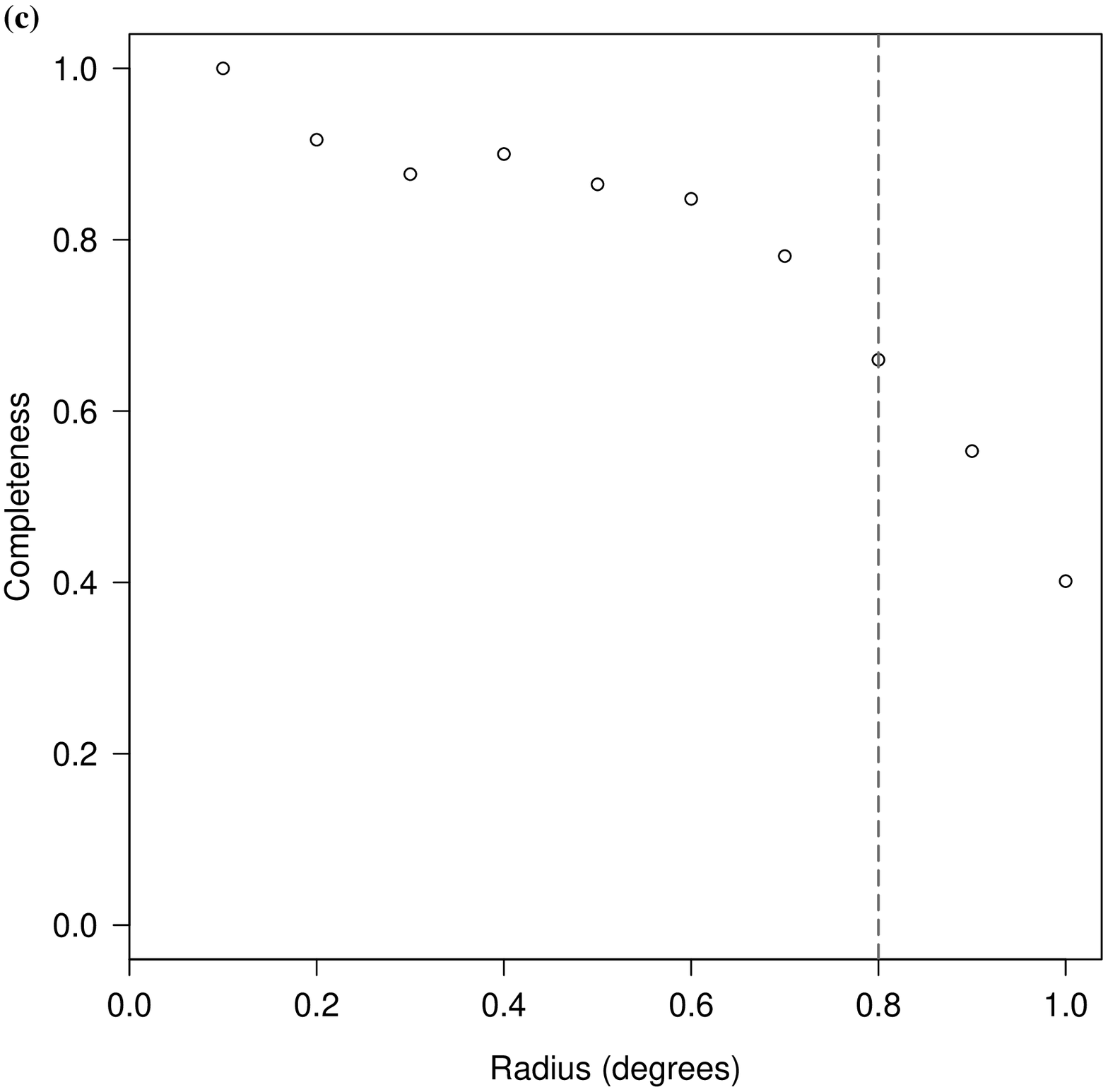} &
      \includegraphics[width=0.5\hsize]{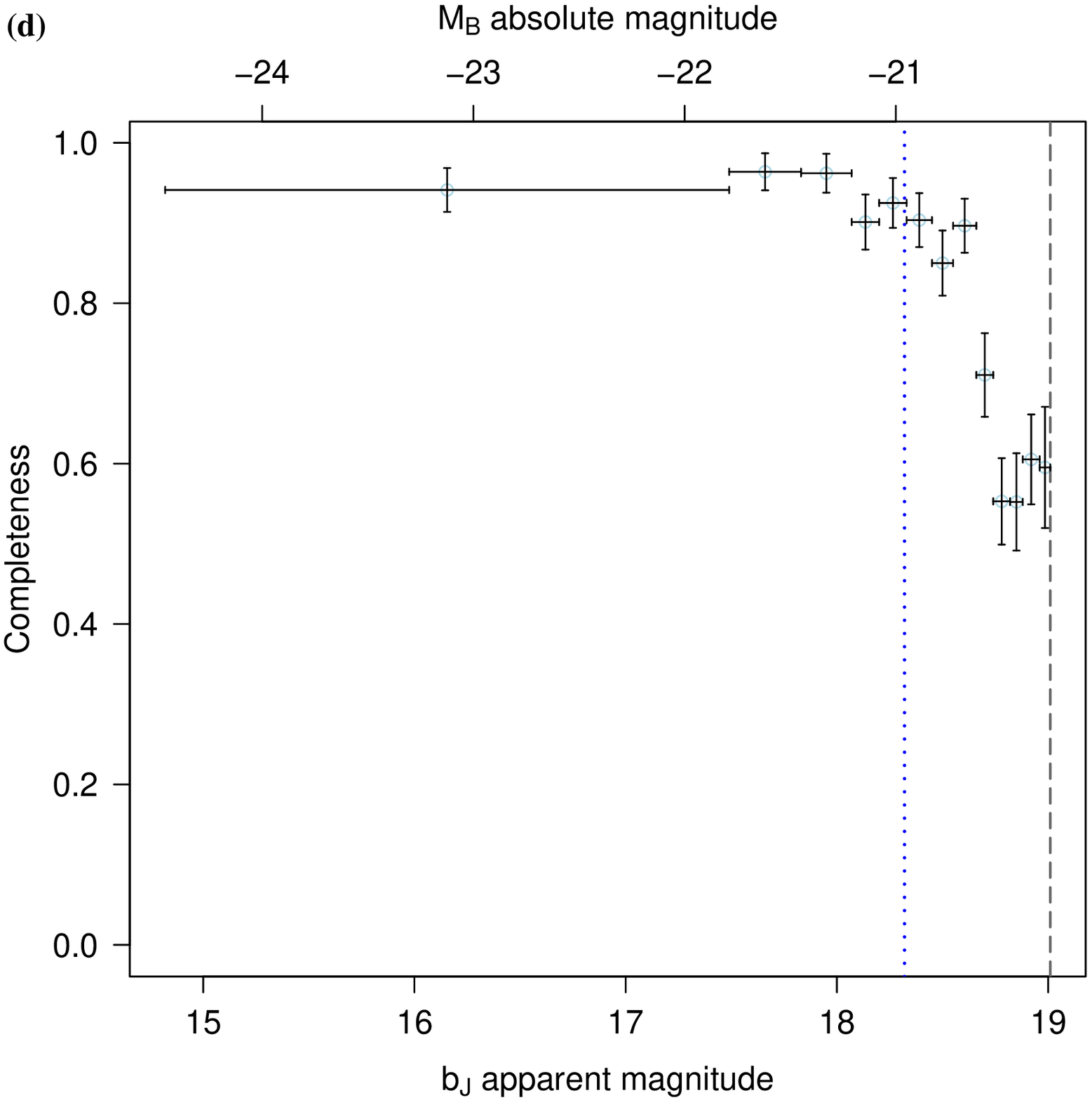}
\end{tabular}
\caption{(\subfigletter{a}) Distribution of galaxies around J0843--7007; a 20 arcmin angular scale corresponds to a linear size of 2.91 Mpc. Plus symbols represent galaxies within $\pm0.003$ of the host redshift at z = 0.1389. Radio contours are shown at (6, 12, 24, 48, 96, and 192) $\times$ 95 $\mu$Jy beam$^{-1}$. (\subfigletter{b}) An enlarged view of the optical field, centred at the host galaxy, with dotted circles at 1, 2 and 3-Mpc radii. (\subfigletter{c}) A completeness plot showing the fraction of observed SuperCOSMOS targets against radius, measured in intervals of 0.1$^{\circ}$ from the field centre. (\subfigletter{d}) A completeness plot showing the fraction of observed targets against $B$-band magnitude with the $b_{\mathrm{J}}$ apparent magnitude (18.32) of the host galaxy indicated by a blue dotted line.}
\label{fig:J0843}
\end{minipage}
\end{figure*}

\begin{figure*}
\begin{minipage}[t][\textheight]{\textwidth}
  \centering
  \begin{tabular}{ll}
      \includegraphics[width=0.5\hsize]{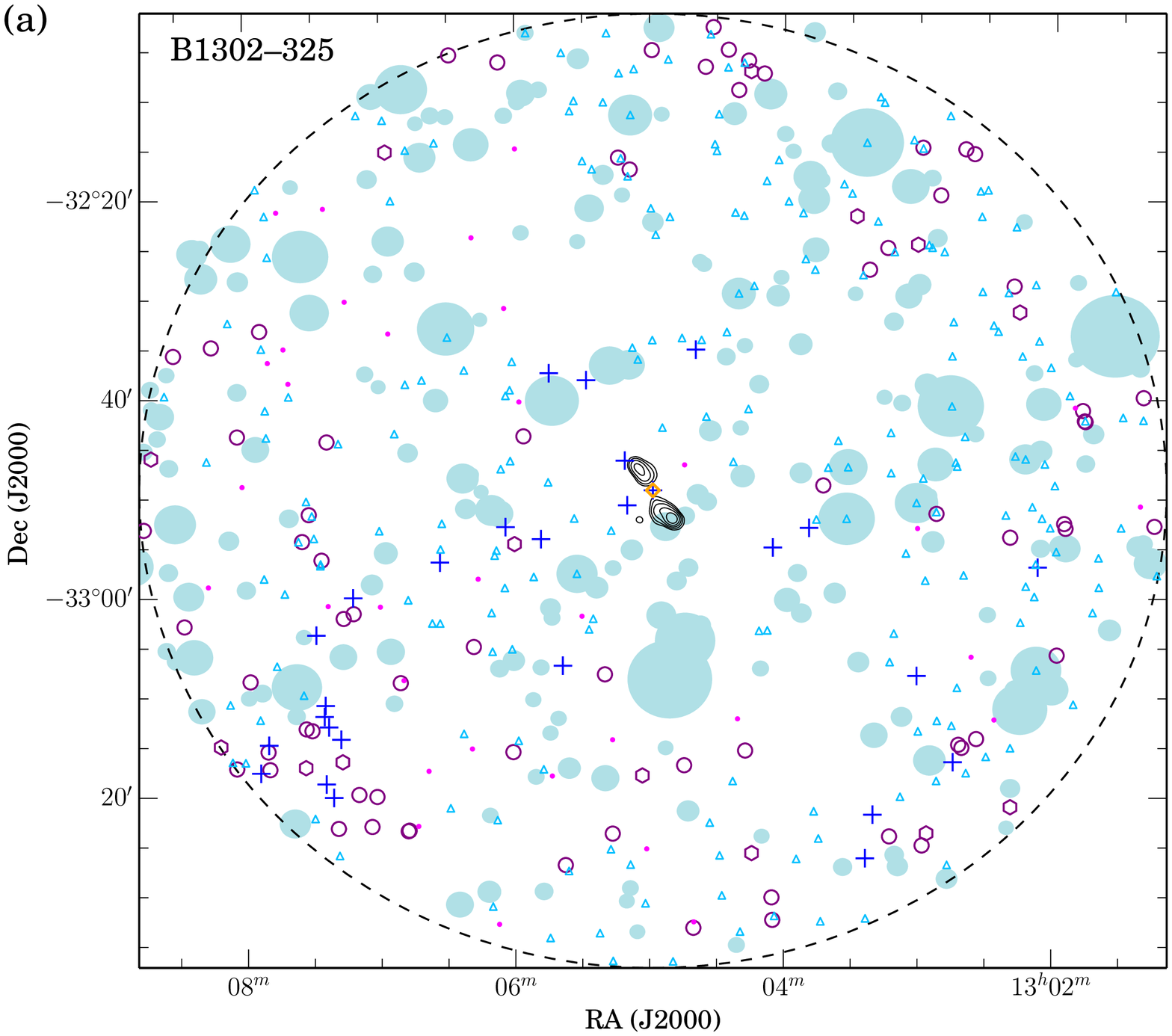} &
      \includegraphics[width=0.5\hsize]{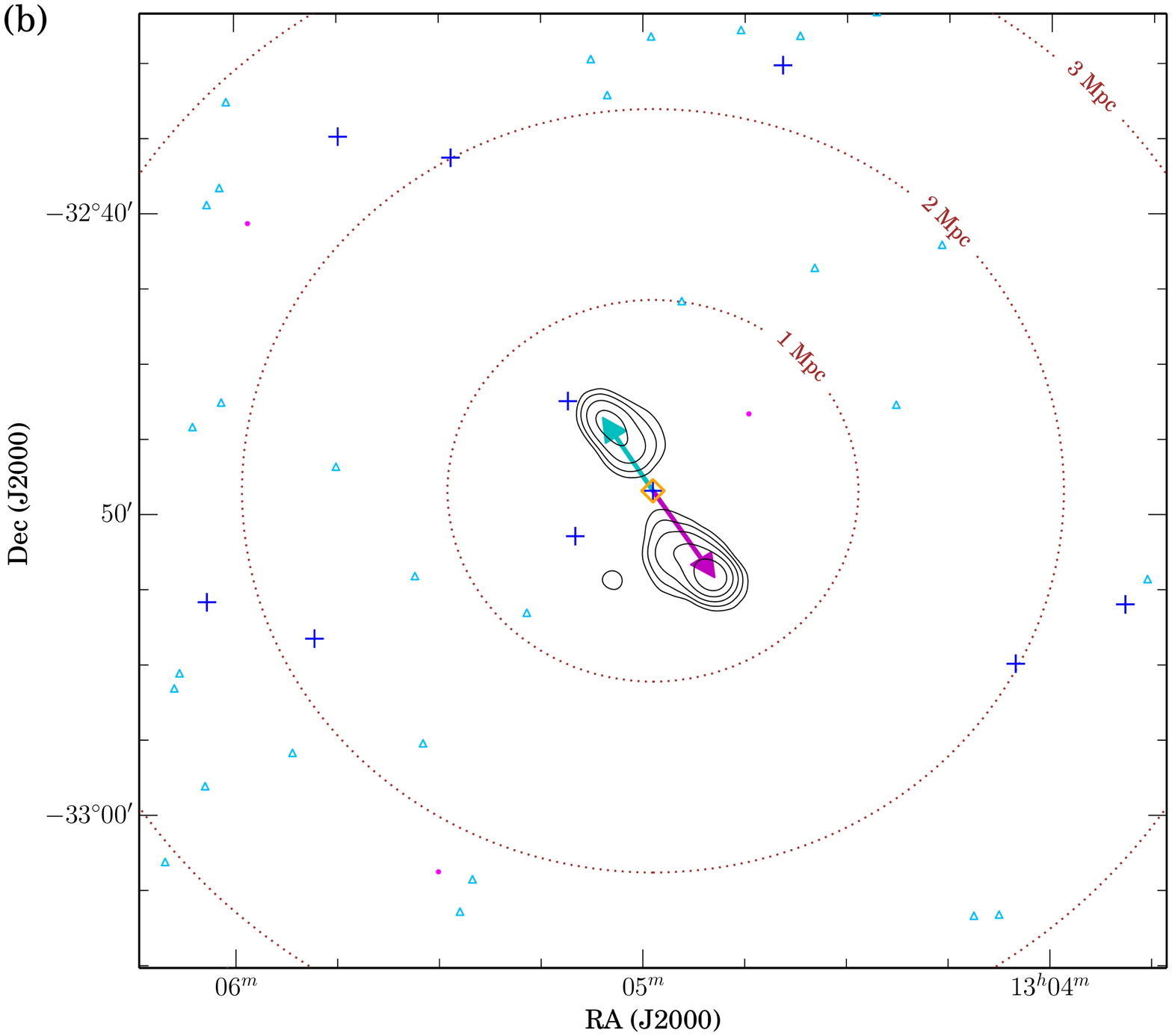}\\
      \includegraphics[width=0.5\hsize]{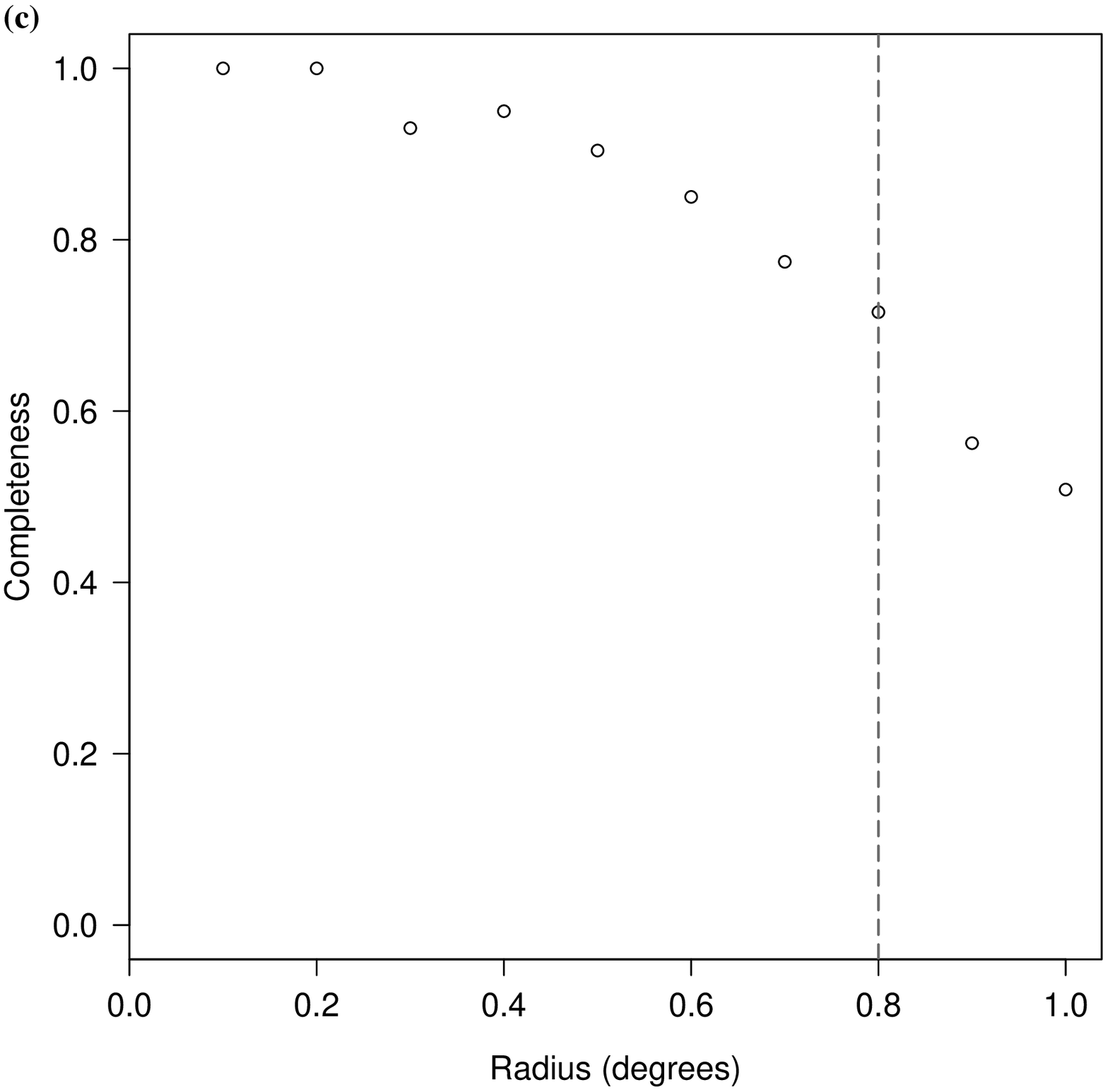} &
      \includegraphics[width=0.5\hsize]{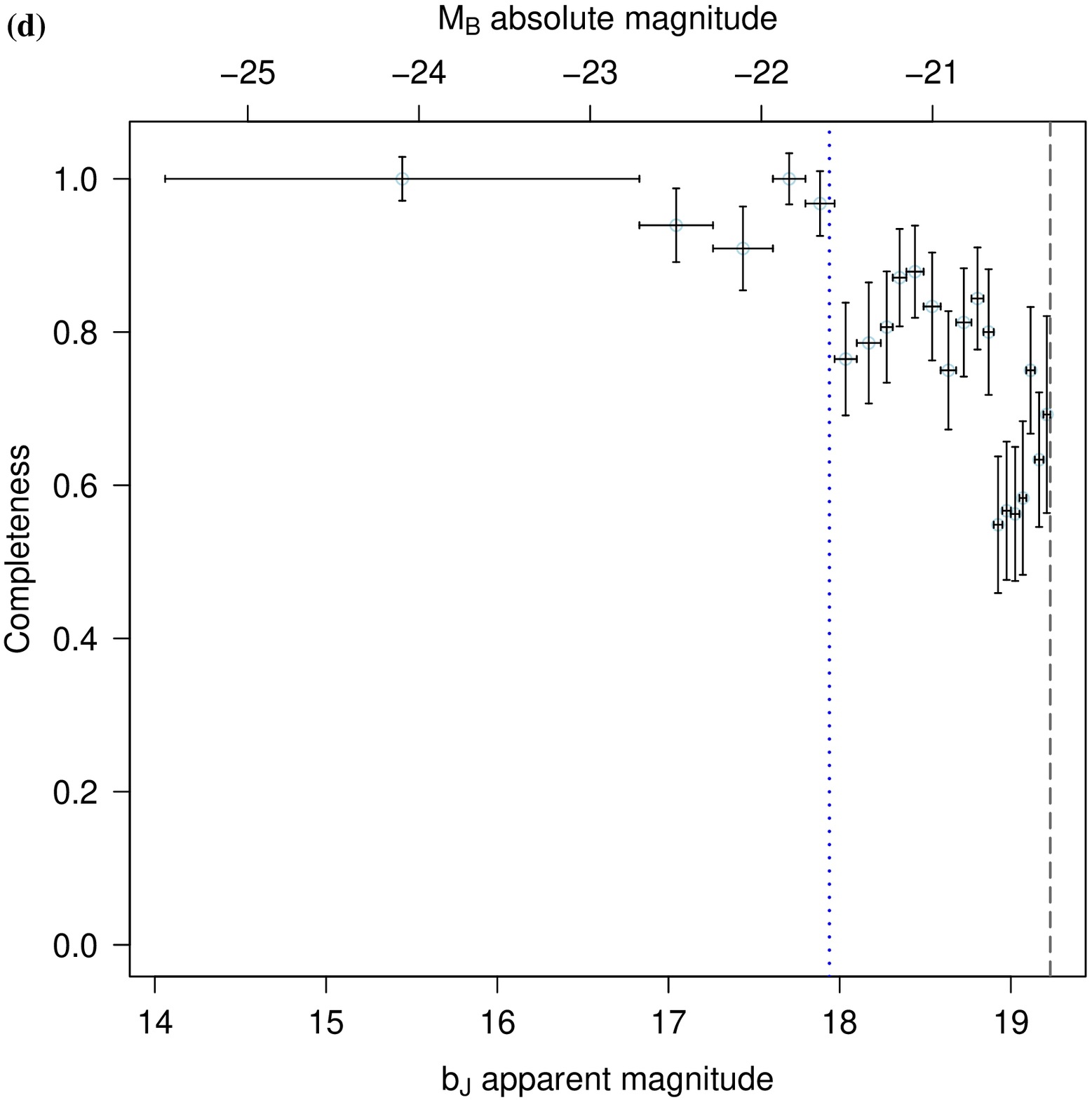}
\end{tabular}
\caption{(\subfigletter{a}) Distribution of galaxies around B1302--325; a 20 arcmin angular scale corresponds to a linear size of 3.15 Mpc. Plus symbols represent galaxies within $\pm0.003$ of the host redshift at z = 0.1528. Radio contours are shown at (6, 12, 24, 48, 96, and 192) $\times$ 0.8 mJy beam$^{-1}$. (\subfigletter{b}) An enlarged view of the optical field, centred at the host galaxy, with dotted circles at 1, 2 and 3-Mpc radii. (\subfigletter{c}) A completeness plot showing the fraction of observed SuperCOSMOS targets against radius, measured in intervals of 0.1$^{\circ}$ from the field centre. (\subfigletter{d}) A completeness plot showing the fraction of observed targets against $B$-band magnitude with the $b_{\mathrm{J}}$ apparent magnitude (17.94) of the host galaxy indicated by a blue dotted line.}
\label{fig:B1302}
\end{minipage}
\end{figure*}

\begin{figure*}
\begin{minipage}[t][\textheight]{\textwidth}
  \centering
  \begin{tabular}{ll}
      \includegraphics[width=0.5\hsize]{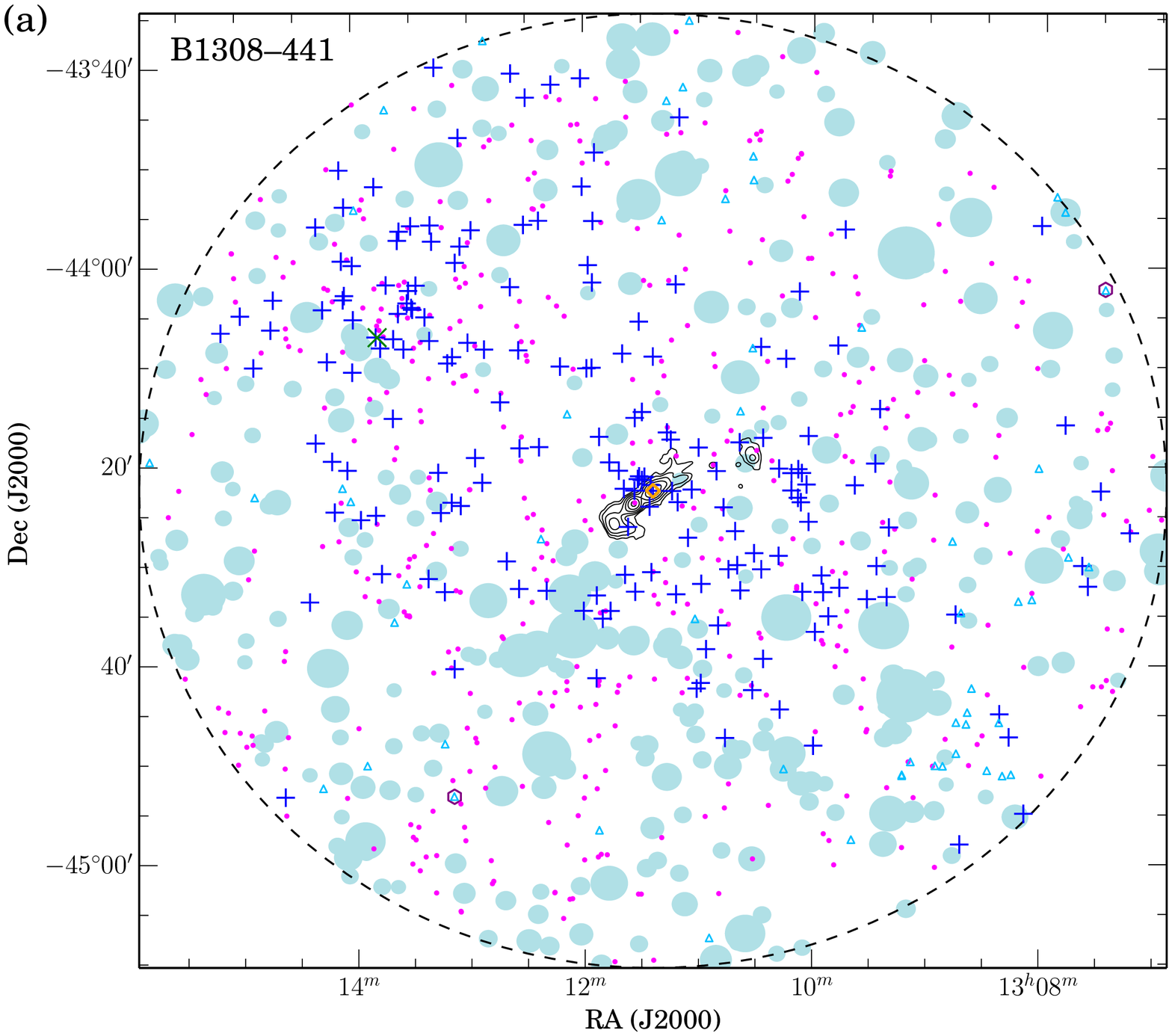} &
      \includegraphics[width=0.5\hsize]{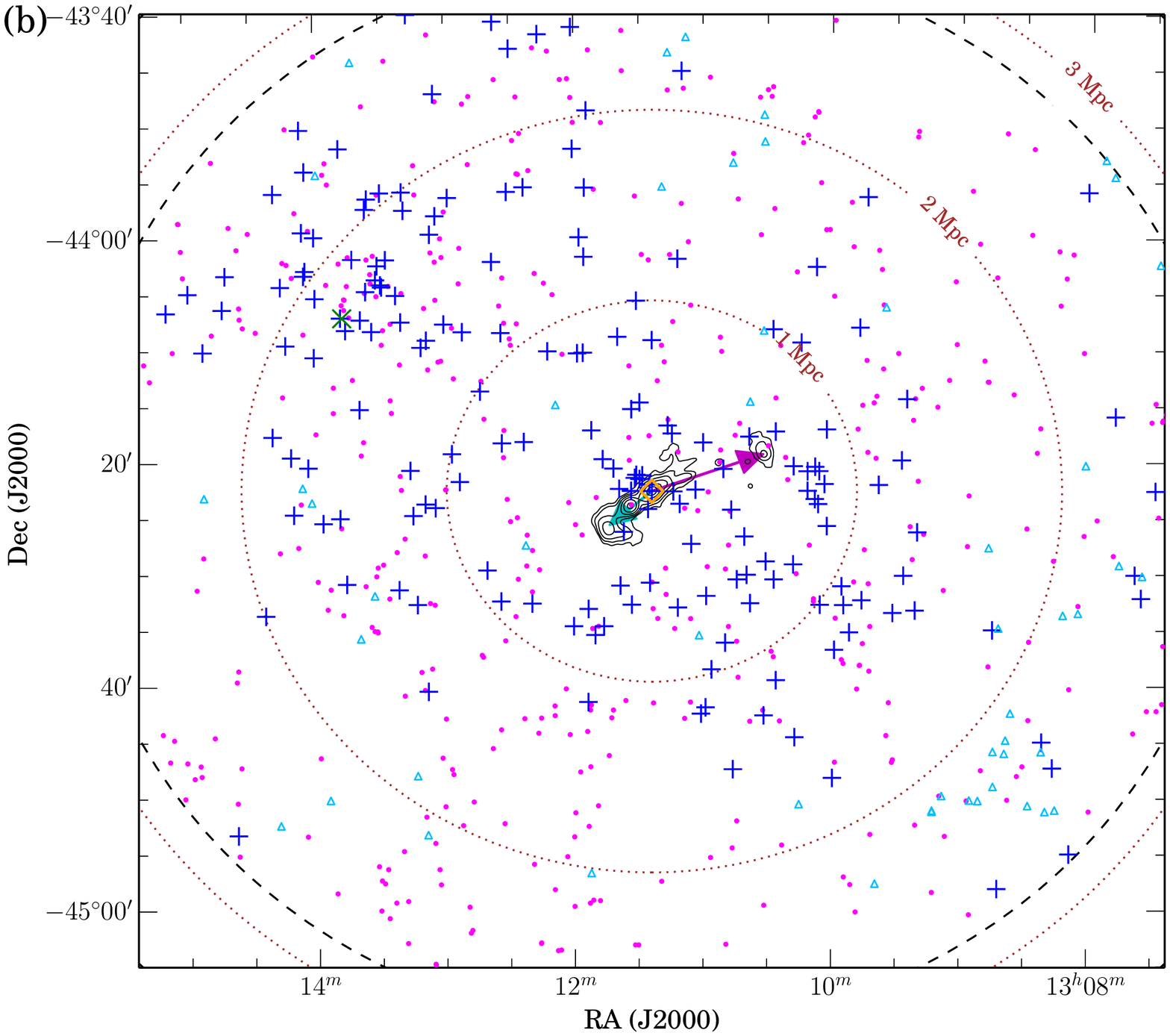}\\
      \includegraphics[width=0.5\hsize]{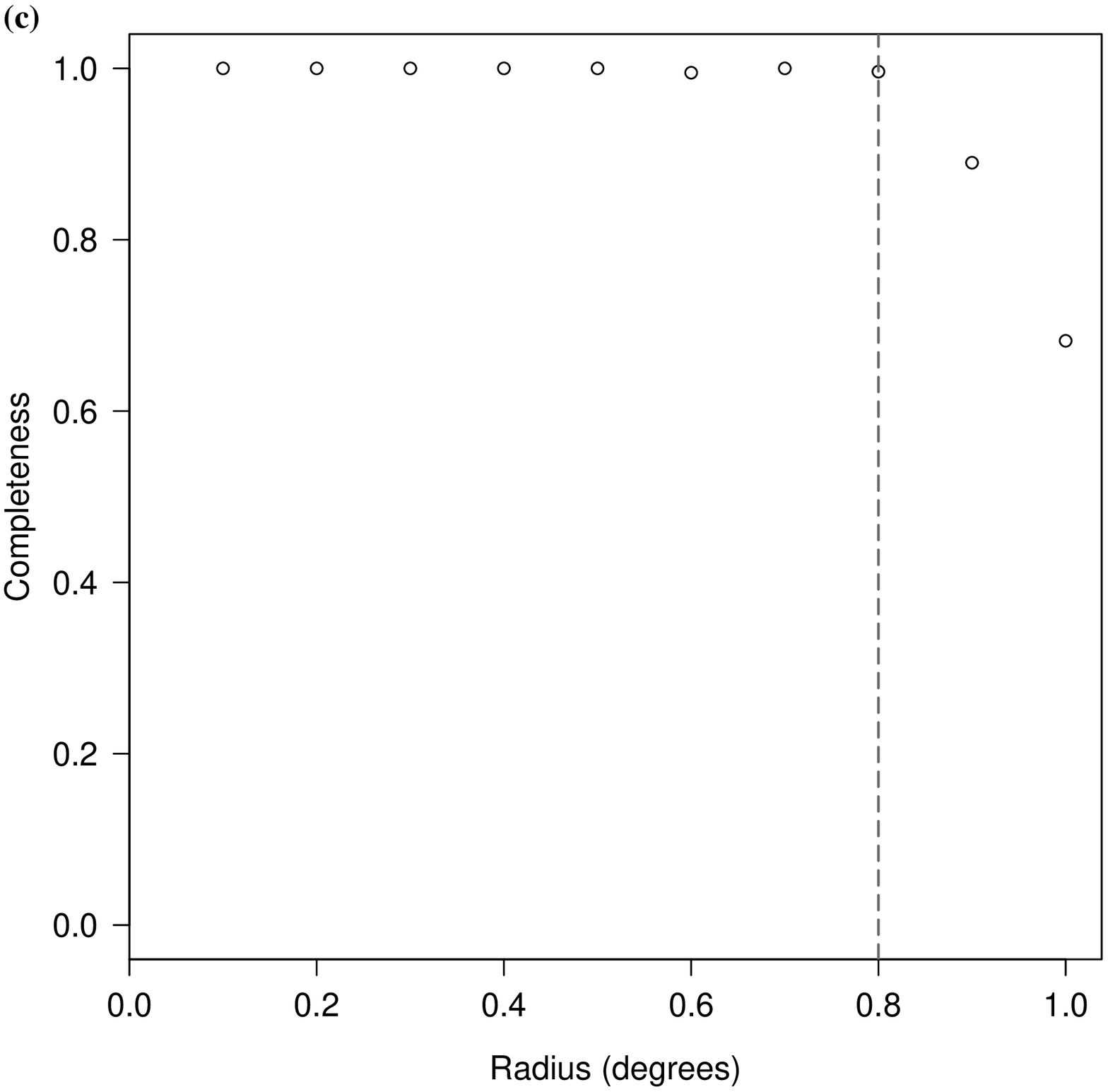} &
      \includegraphics[width=0.5\hsize]{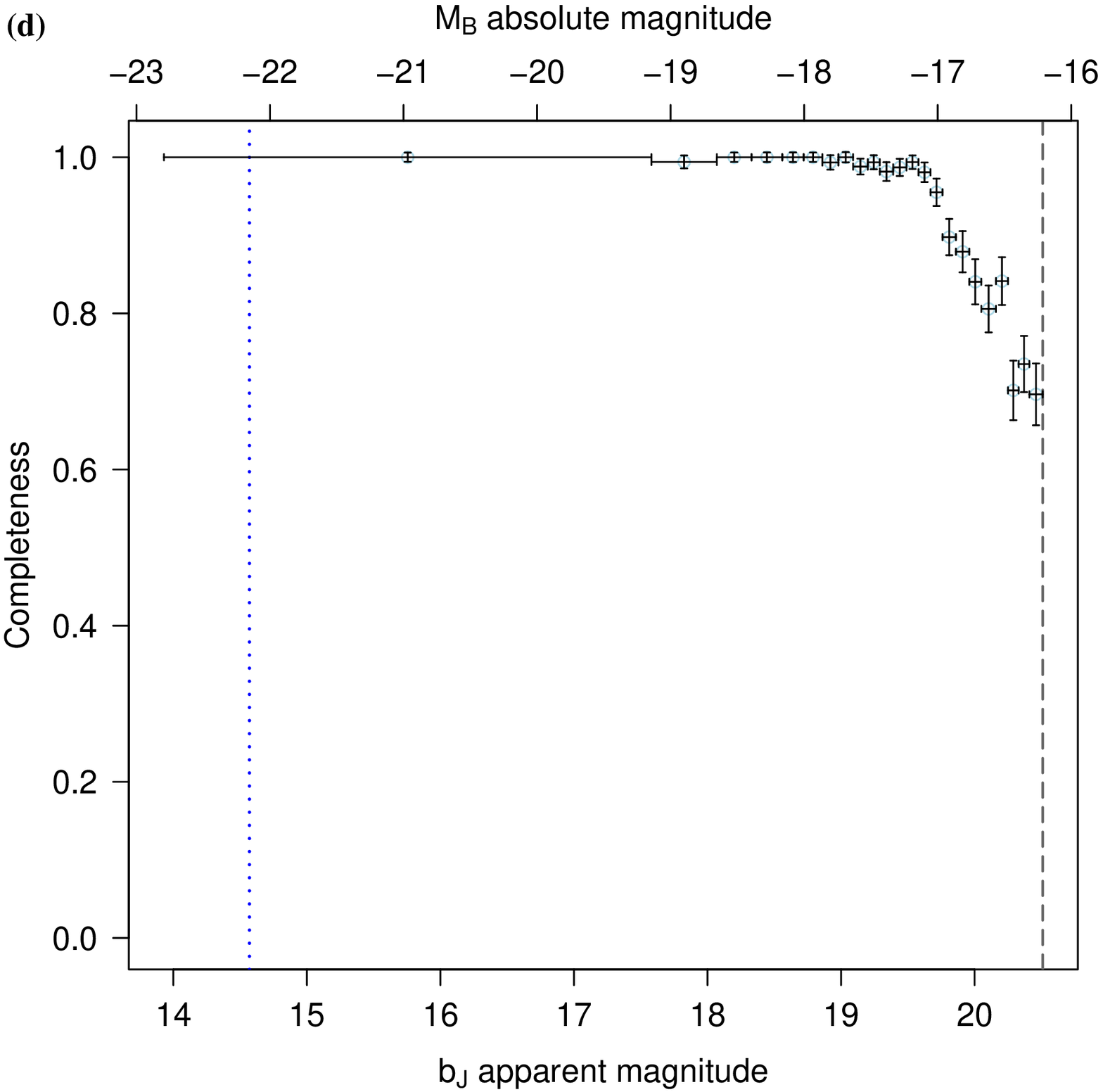}
\end{tabular}
\caption{(\subfigletter{a}) Distribution of galaxies around B1308--441; a 20 arcmin angular scale corresponds to a linear size of 1.17 Mpc. Plus symbols represent galaxies within $\pm0.003$ of the host redshift at z = 0.0507. Radio contours are shown at (6, 12, 24, 48, 96, and 192) $\times$ 0.4 mJy beam$^{-1}$. (\subfigletter{b}) An enlarged view of the optical field, centred at the host galaxy, with dotted circles at 1, 2 and 3-Mpc radii. (\subfigletter{c}) A completeness plot showing the fraction of observed SuperCOSMOS targets against radius, measured in intervals of 0.1$^{\circ}$ from the field centre. (\subfigletter{d}) A completeness plot showing the fraction of observed targets against $B$-band magnitude with the $b_{\mathrm{J}}$ apparent magnitude (14.57) of the host galaxy indicated by a blue dotted line.}
\label{fig:B1308}
\end{minipage}
\end{figure*}

\begin{figure*}
\begin{minipage}[t][\textheight]{\textwidth}
  \centering
  \begin{tabular}{ll}
      \includegraphics[width=0.5\hsize]{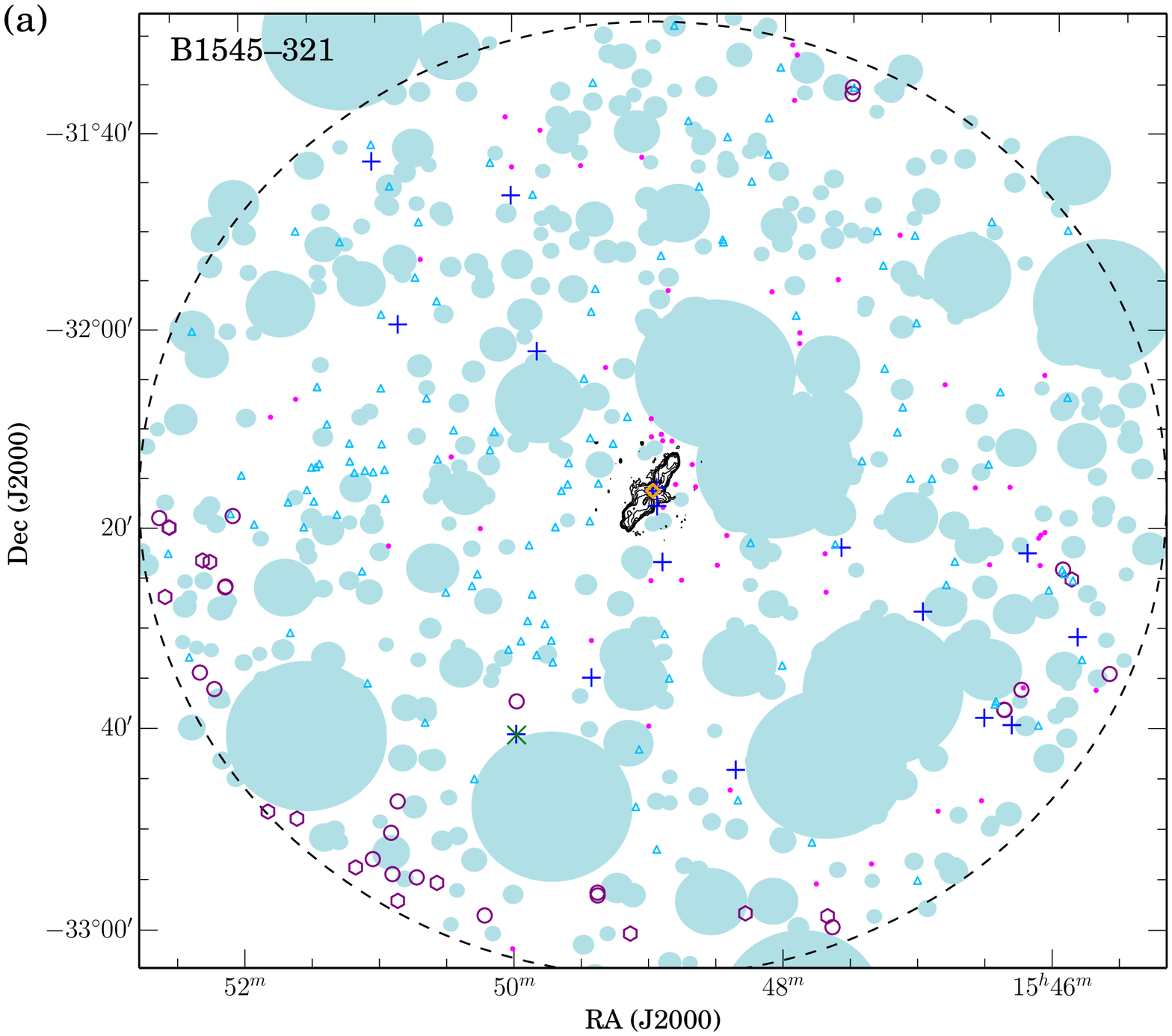} &
      \includegraphics[width=0.5\hsize]{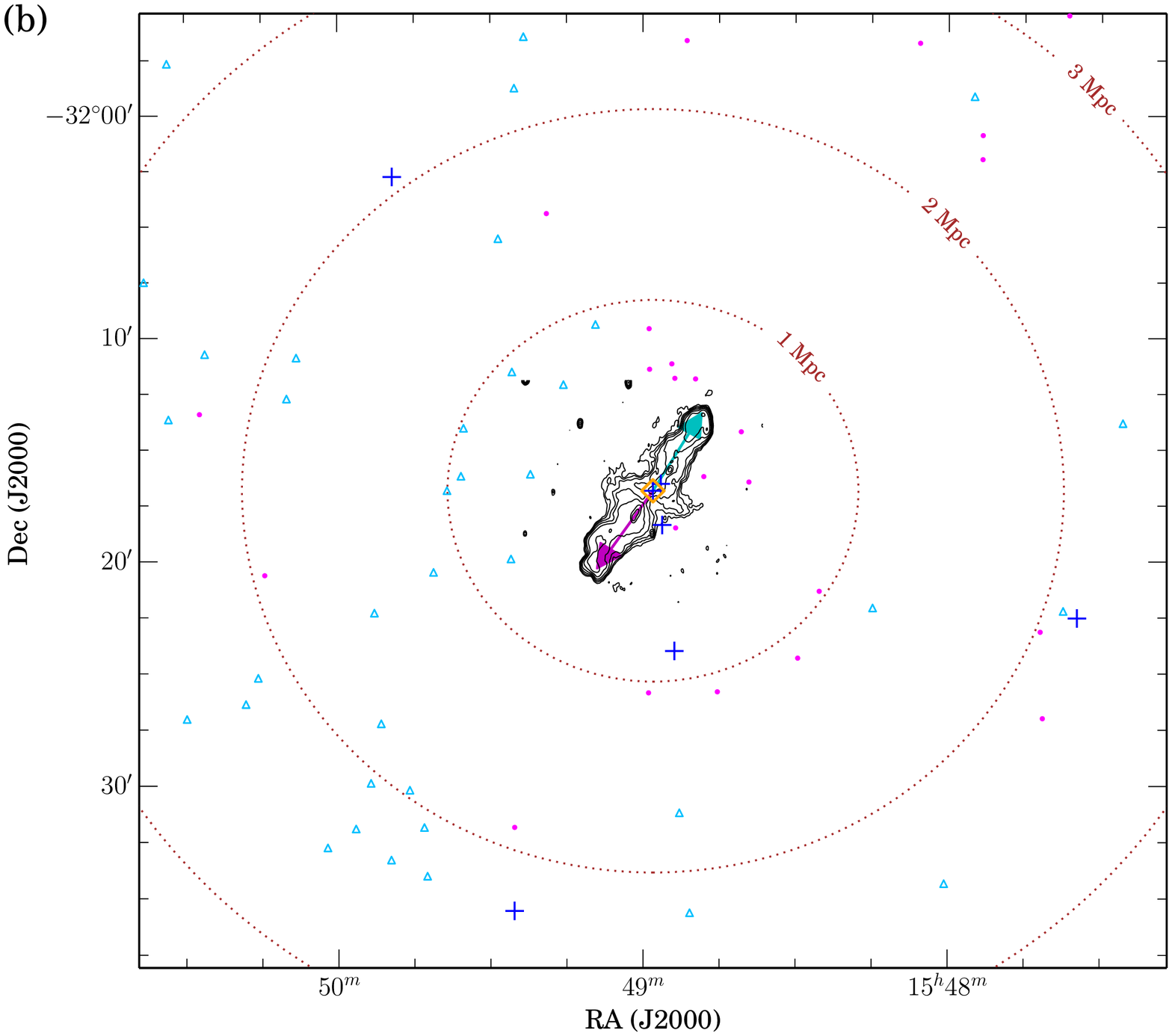}\\
      \includegraphics[width=0.5\hsize]{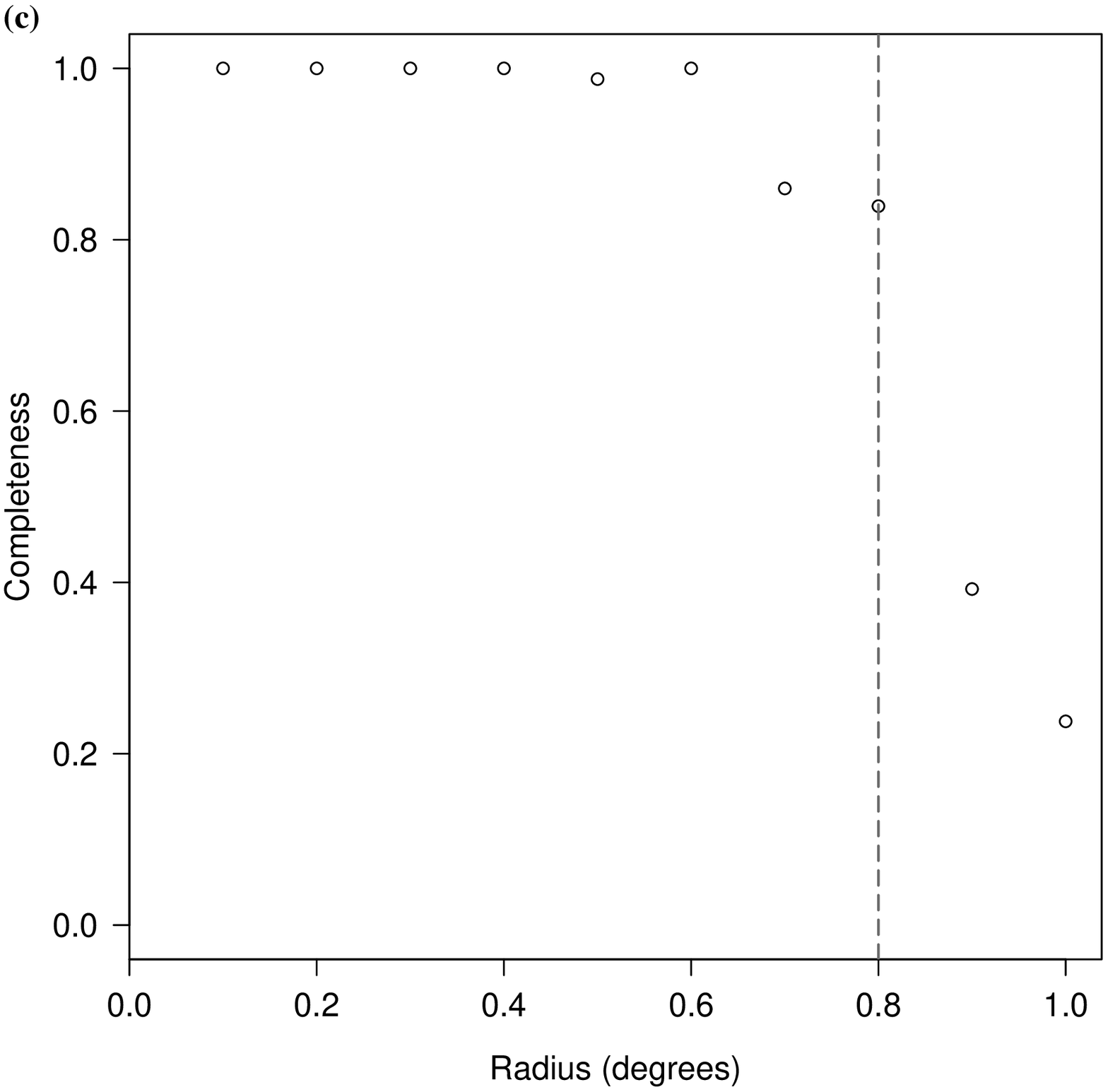} &
      \includegraphics[width=0.5\hsize]{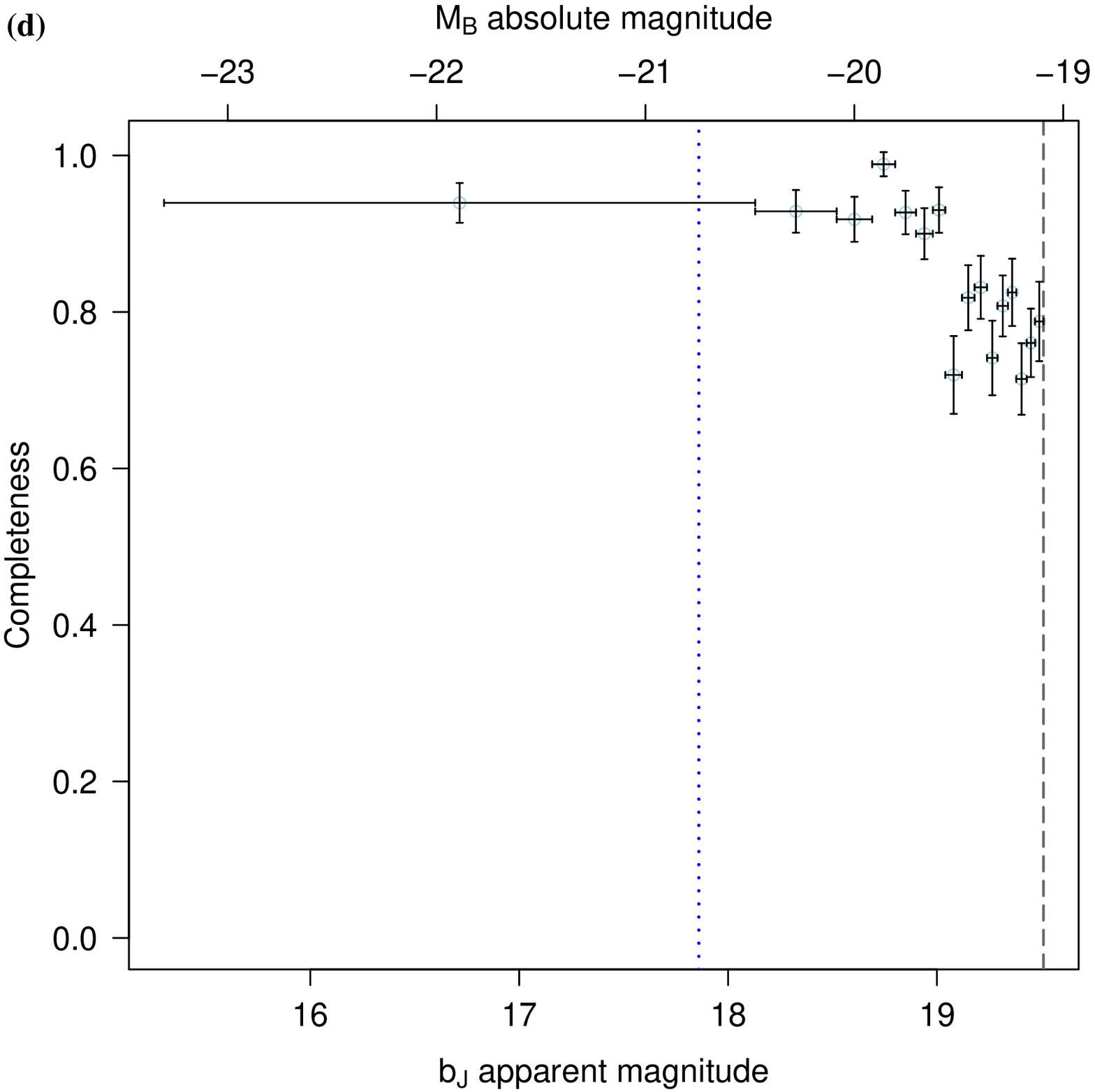}
\end{tabular}
\caption{(\subfigletter{a}) Distribution of galaxies around B1545--321; a 20 arcmin angular scale corresponds to a linear size of 2.34 Mpc. Plus symbols represent galaxies within $\pm0.003$ of the host redshift at z = 0.1081. Radio contours are shown at (1, 2, 4, 8, 16, 32, 64, and 128) $\times$ 0.2 mJy beam$^{-1}$ with a beam of FWHM $11.6\arcsec\times5.7\arcsec$ at a P.A. of $1.6^{\circ}$. (\subfigletter{b}) An enlarged view of the optical field, centred at the host galaxy, with dotted circles at 1, 2 and 3-Mpc radii. (\subfigletter{c}) A completeness plot showing the fraction of observed SuperCOSMOS targets against radius, measured in intervals of 0.1$^{\circ}$ from the field centre. (\subfigletter{d}) A completeness plot showing the fraction of observed targets against $B$-band magnitude with the $b_{\mathrm{J}}$ apparent magnitude (17.86) of the host galaxy indicated by a blue dotted line.}
\label{fig:B1545}
\end{minipage}
\end{figure*}

\begin{figure*}
\begin{minipage}[t][\textheight]{\textwidth}
  \centering
  \begin{tabular}{ll}
      \includegraphics[width=0.5\hsize]{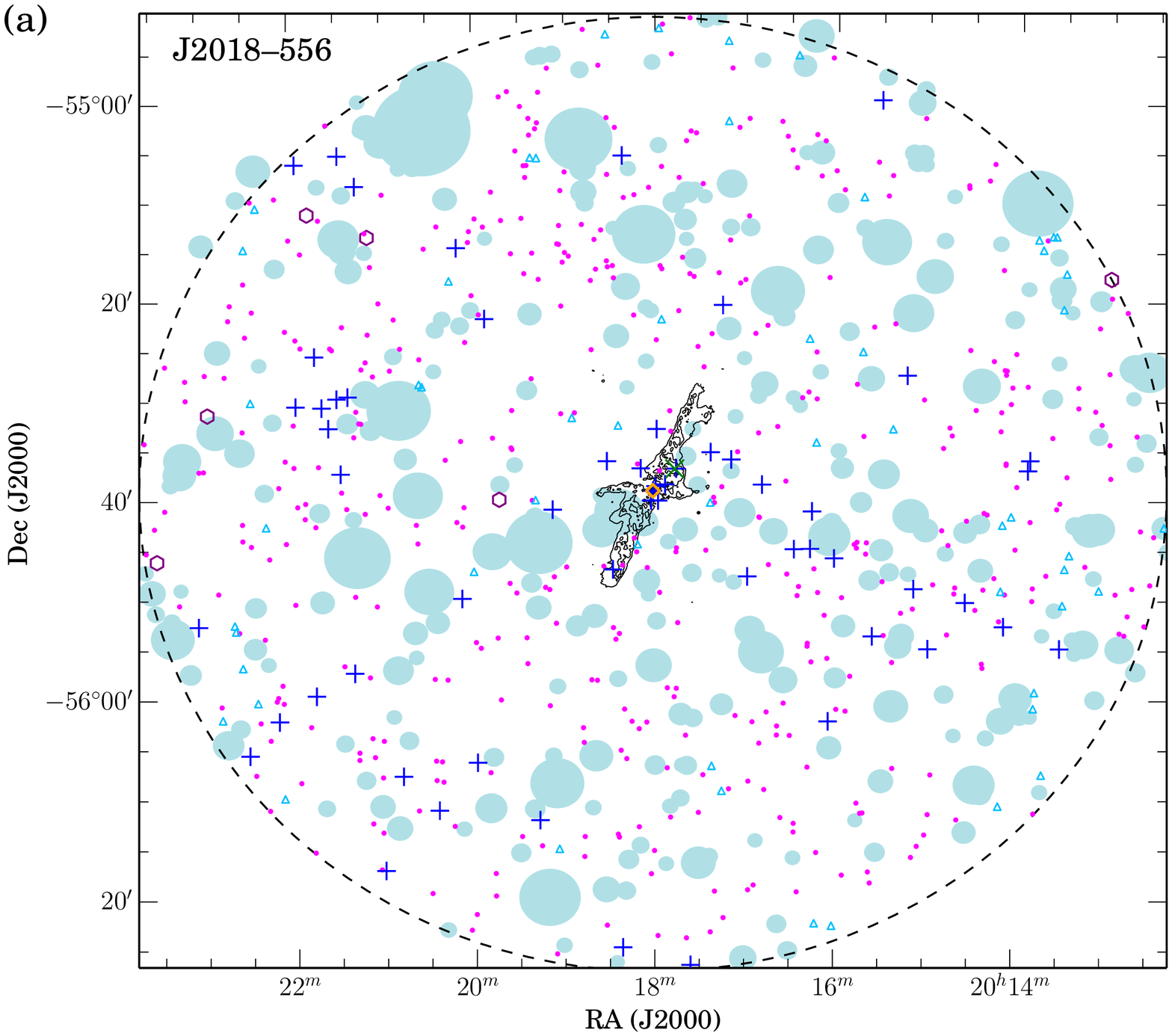} &
      \includegraphics[width=0.5\hsize]{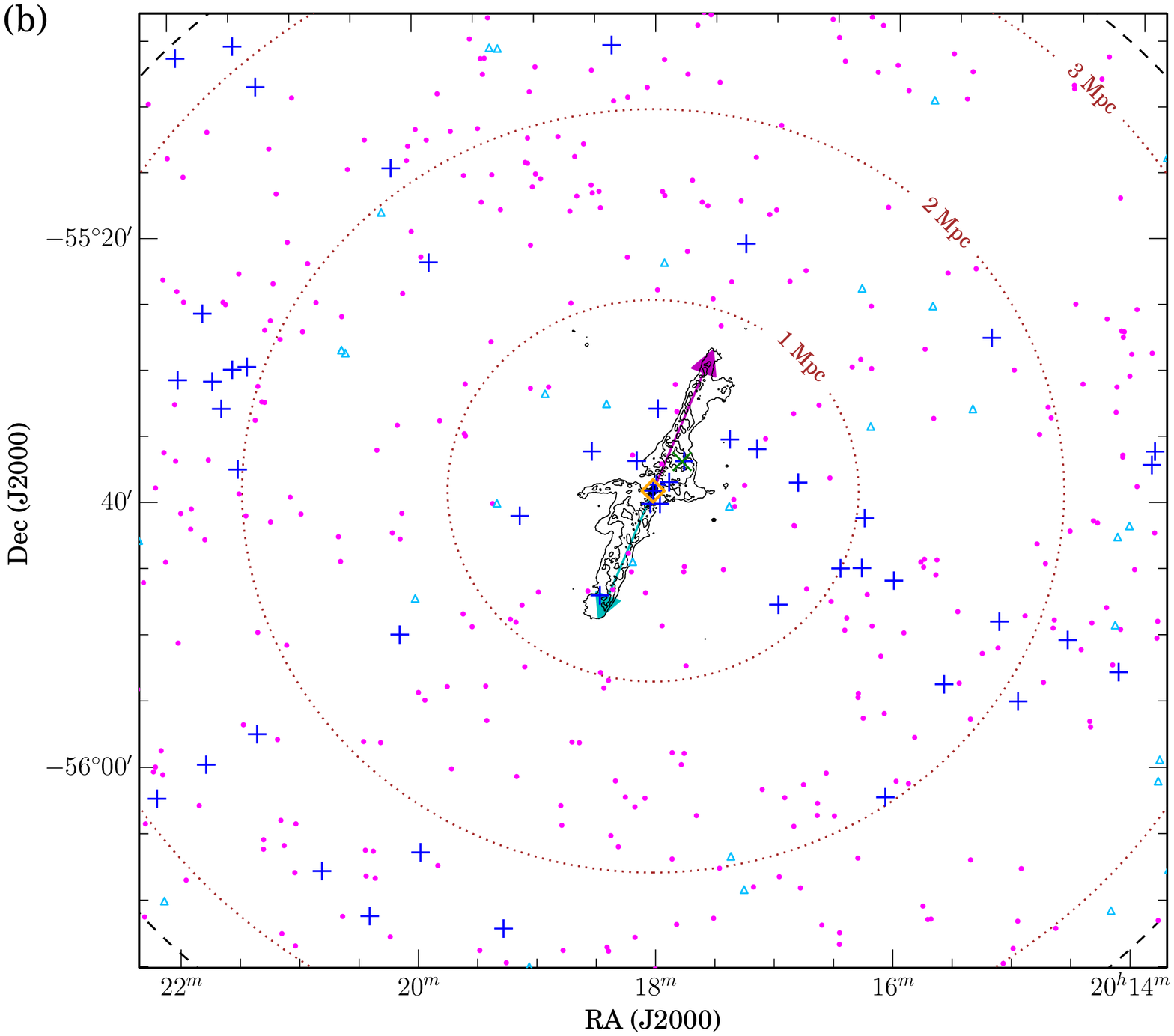}\\
      \includegraphics[width=0.5\hsize]{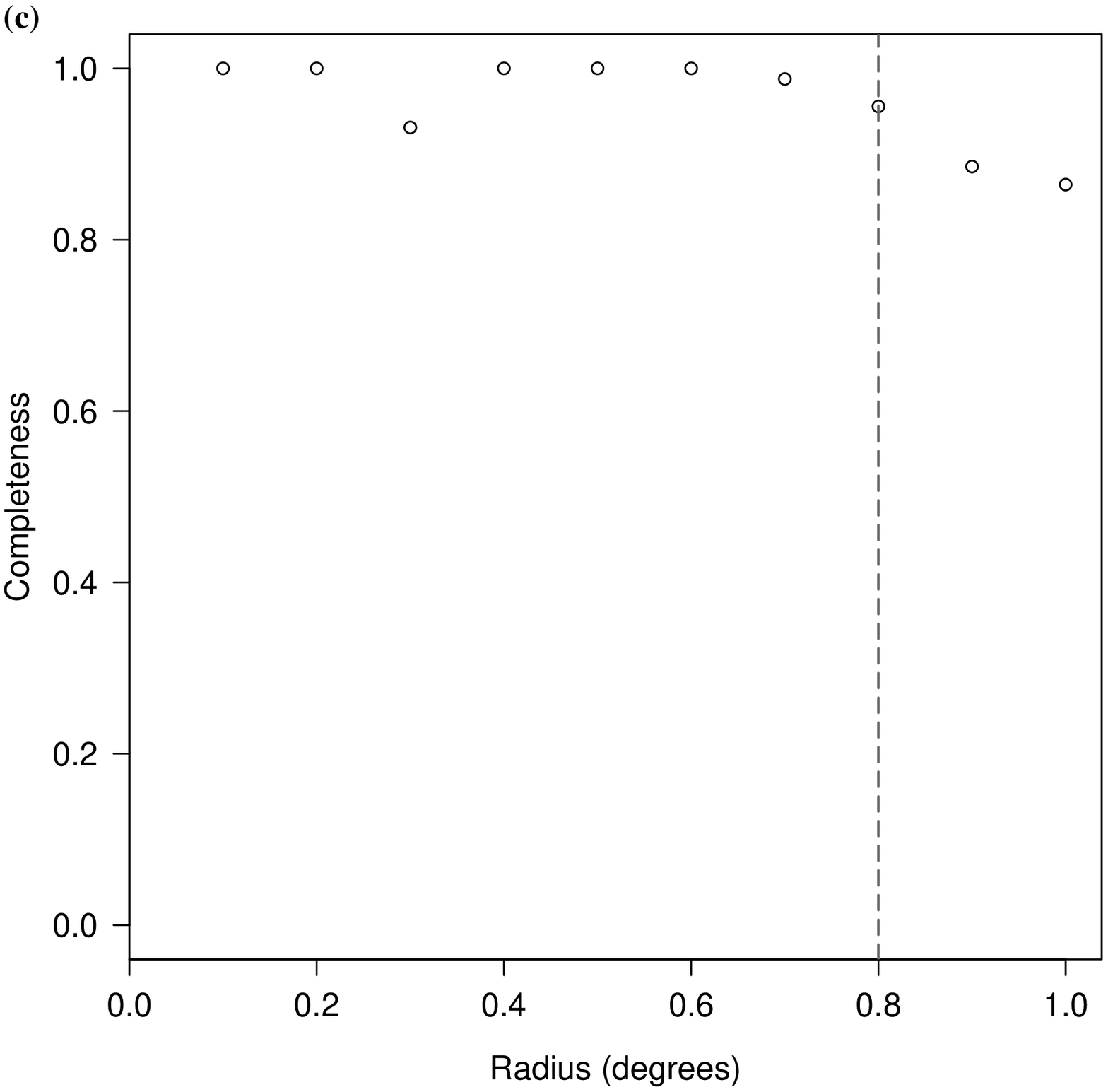} &
      \includegraphics[width=0.5\hsize]{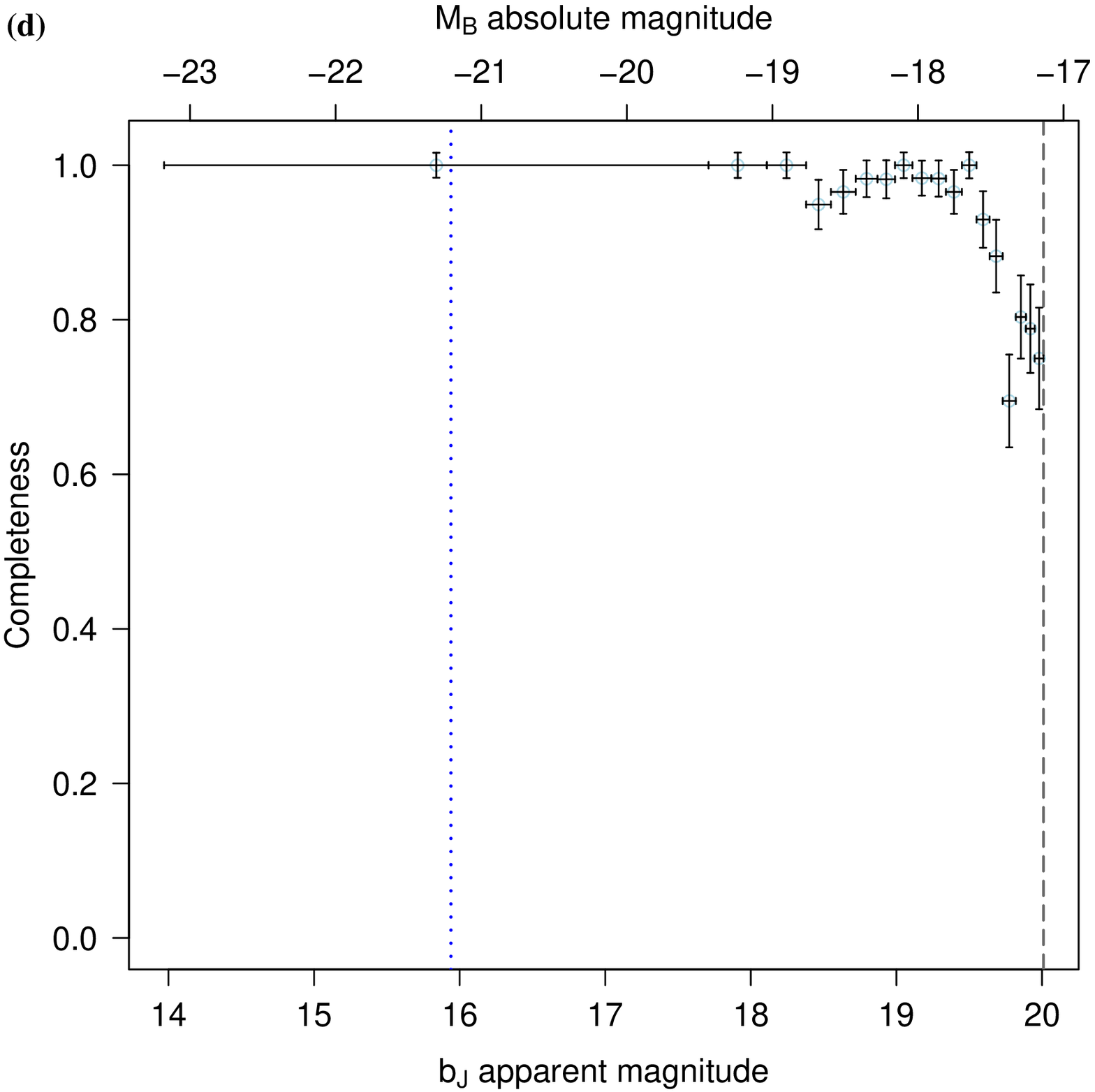}
\end{tabular}
\caption{(\subfigletter{a}) Distribution of galaxies around J2018--556; a 20 arcmin angular scale corresponds to a linear size of 1.38 Mpc. Plus symbols represent galaxies within $\pm0.003$ of the host redshift at z = 0.0605. Radio contours are shown at (1, 2, 4, 8, 16, 32, 64, and 128) $\times$ 0.7 mJy beam$^{-1}$. (\subfigletter{b}) An enlarged view of the optical field, centred at the host galaxy, with dotted circles at 1, 2 and 3-Mpc radii. (\subfigletter{c}) A completeness plot showing the fraction of observed SuperCOSMOS targets against radius, measured in intervals of 0.1$^{\circ}$ from the field centre. (\subfigletter{d}) A completeness plot showing the fraction of observed targets against $B$-band magnitude with the $b_{\mathrm{J}}$ apparent magnitude (15.94) of the host galaxy indicated by a blue dotted line.}
\label{fig:J2018}
\end{minipage}
\end{figure*}

\begin{figure*}
\begin{minipage}[t][\textheight]{\textwidth}
  \centering
  \begin{tabular}{ll}
      \includegraphics[width=0.5\hsize]{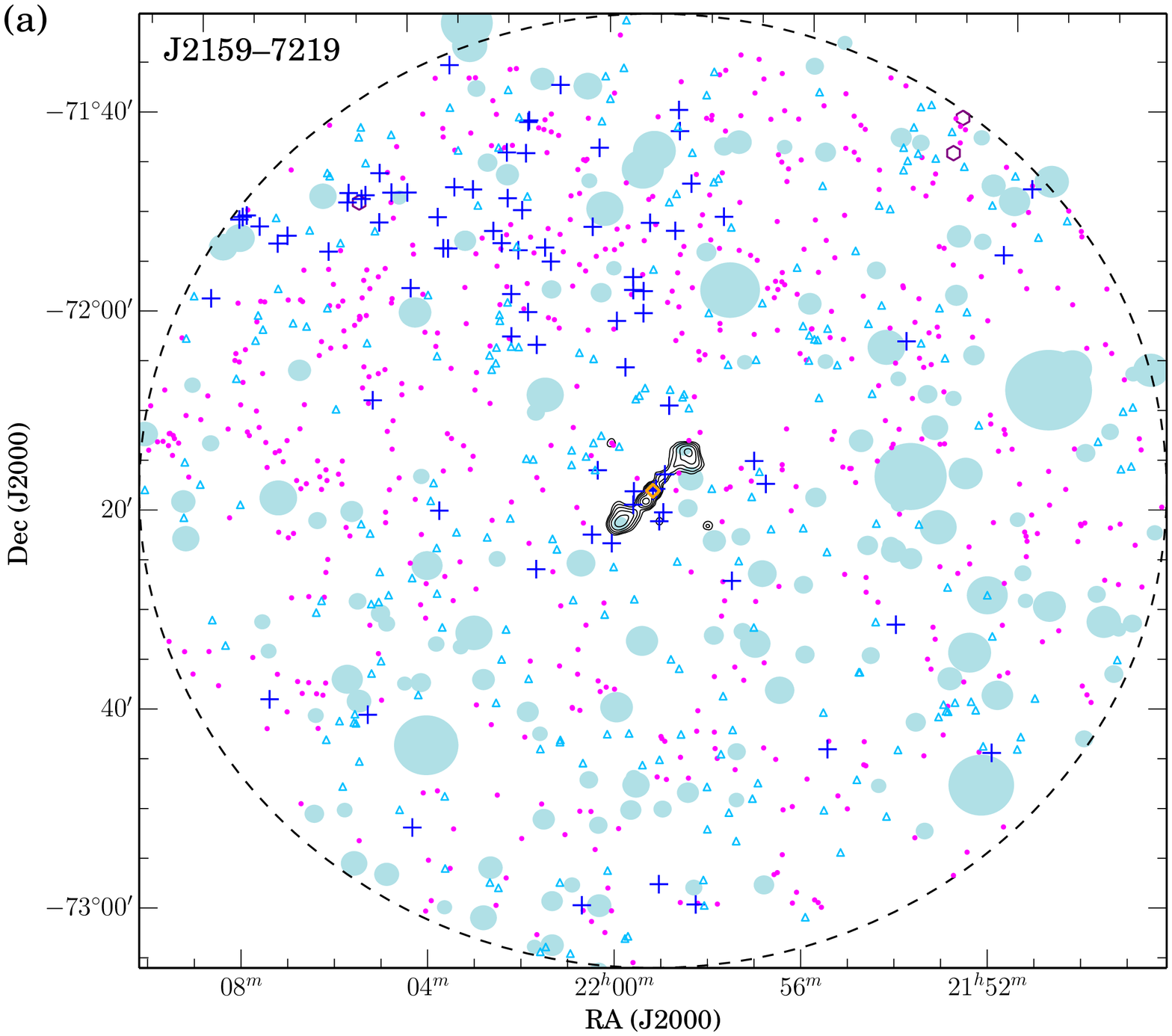} &
      \includegraphics[width=0.5\hsize]{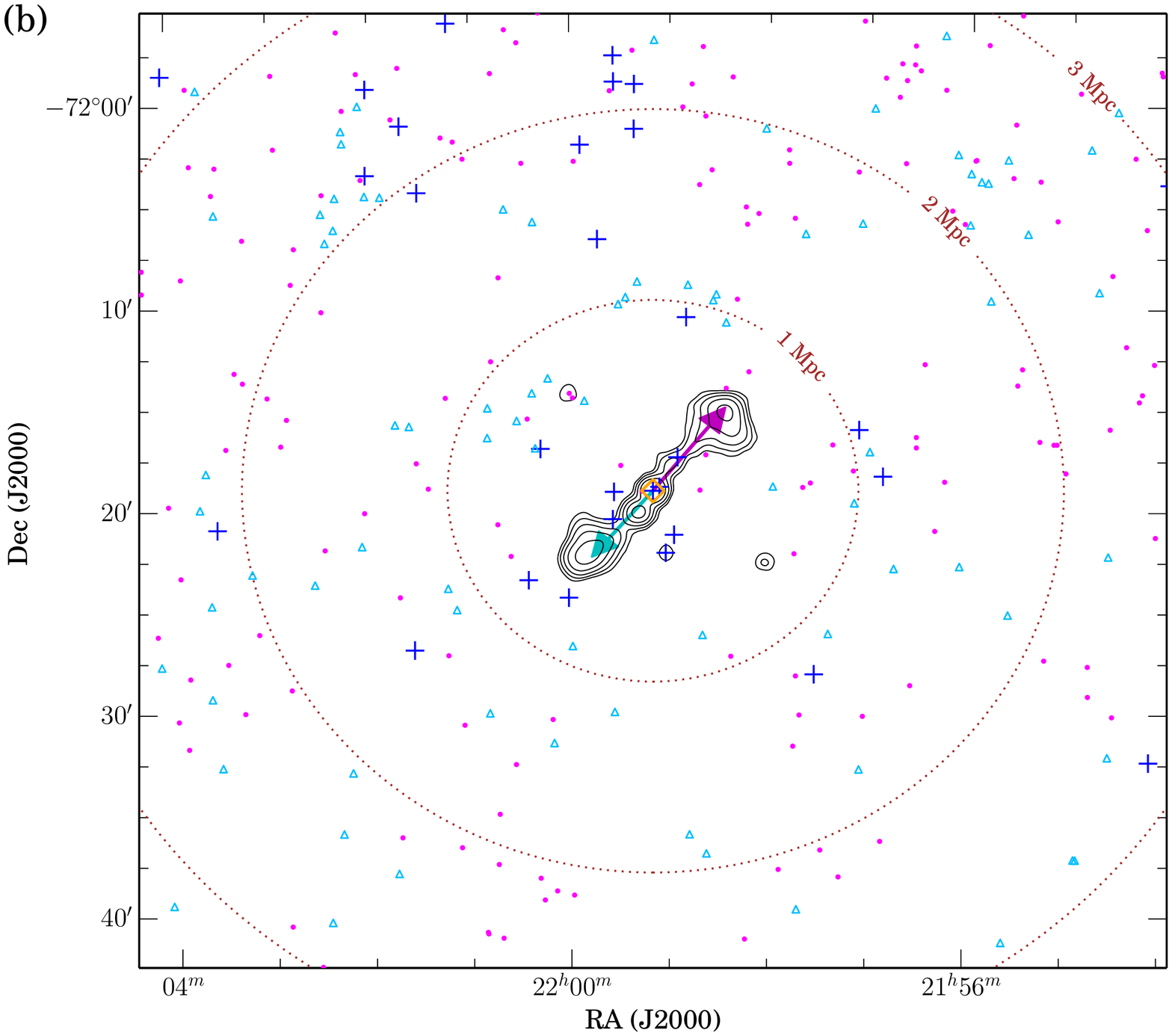}\\
      \includegraphics[width=0.5\hsize]{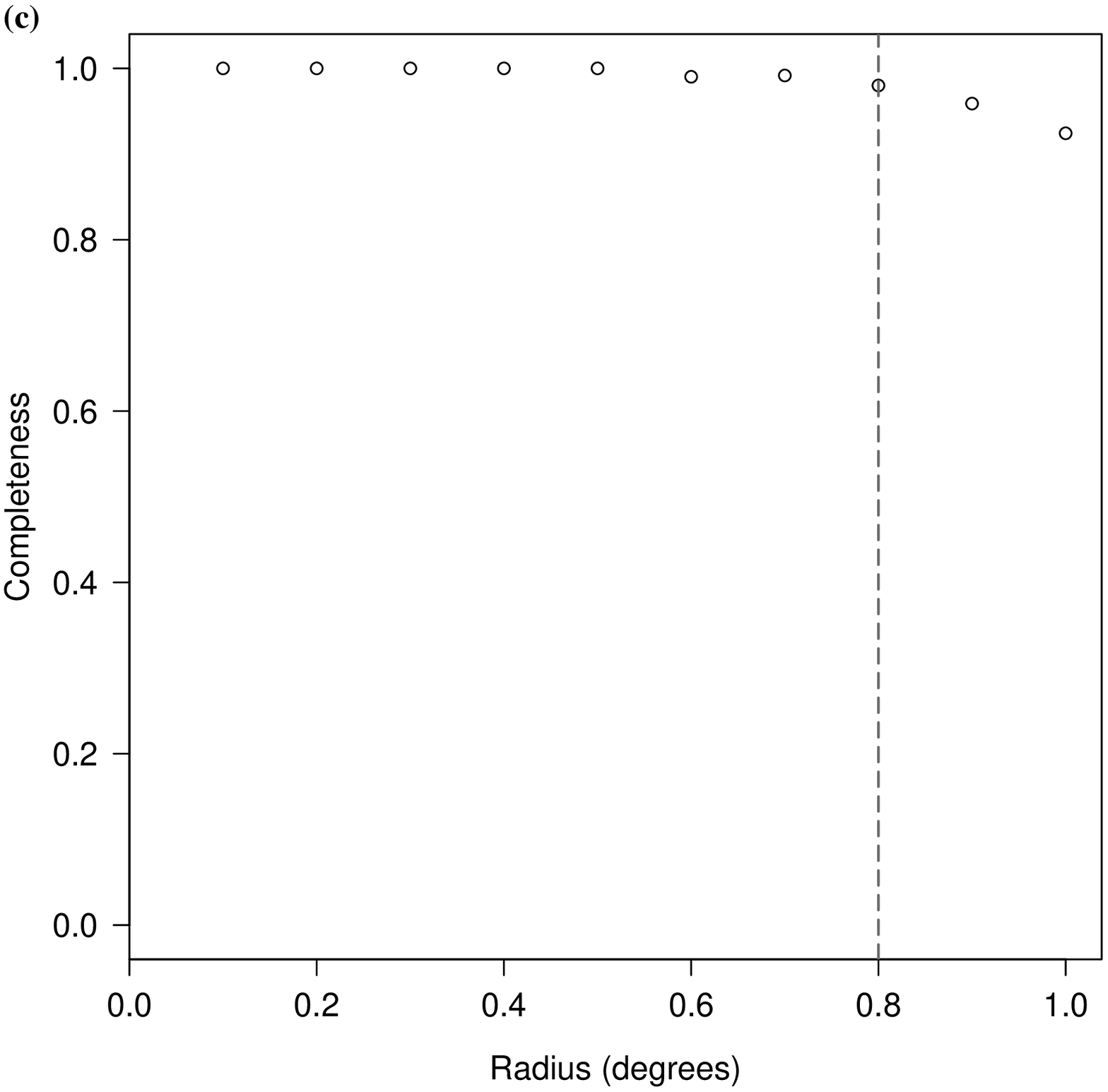} &
      \includegraphics[width=0.5\hsize]{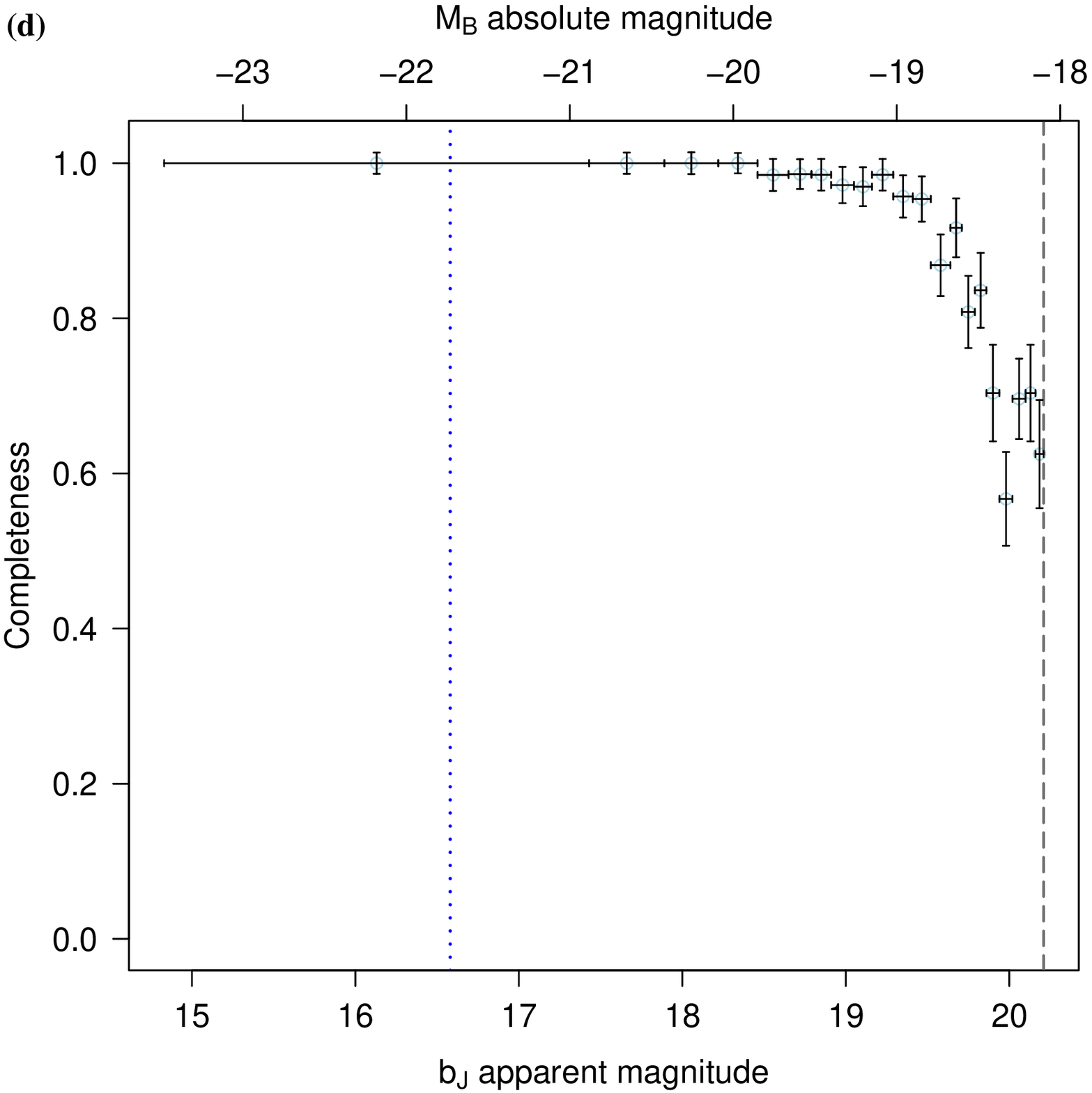}
\end{tabular}
\caption{(\subfigletter{a}) Distribution of galaxies around J2159--7219; a 20 arcmin angular scale corresponds to a linear size of 2.12 Mpc. Plus symbols represent galaxies within $\pm0.003$ of the host redshift at z = 0.0967. Radio contours are shown at (6, 12, 24, 48, and 96) $\times$ 75 $\mu$Jy beam$^{-1}$. (\subfigletter{b}) An enlarged view of the optical field, centred at the host galaxy, with dotted circles at 1, 2 and 3-Mpc radii. (\subfigletter{c}) A completeness plot showing the fraction of observed SuperCOSMOS targets against radius, measured in intervals of 0.1$^{\circ}$ from the field centre. (\subfigletter{d}) A completeness plot showing the fraction of observed targets against $B$-band magnitude with the $b_{\mathrm{J}}$ apparent magnitude (16.58) of the host galaxy indicated by a blue dotted line.}
\label{fig:J2159}
\end{minipage}
\end{figure*}

\begin{figure*}
\begin{minipage}[t][\textheight]{\textwidth}
  \centering
  \begin{tabular}{ll}
      \includegraphics[width=0.5\hsize]{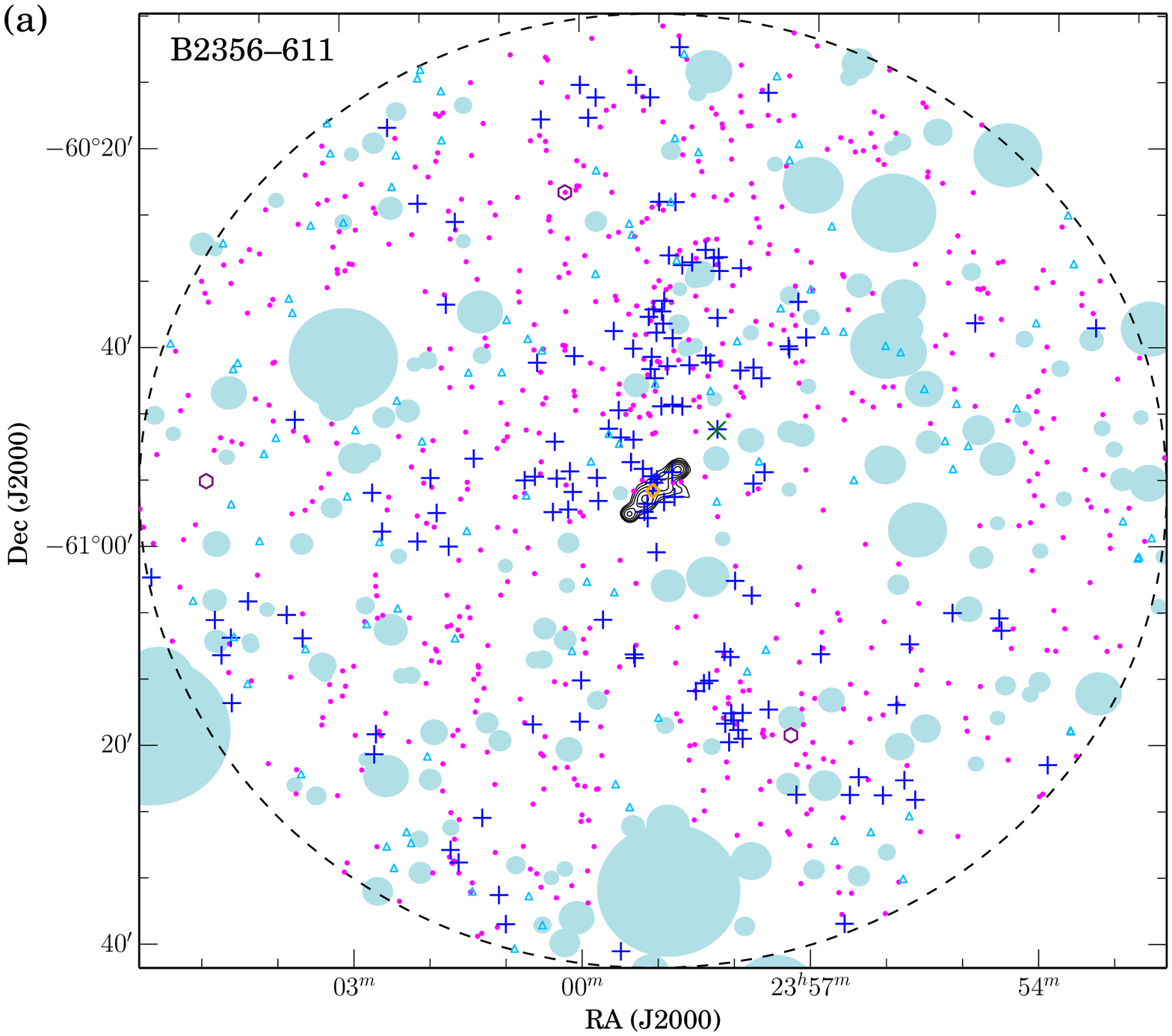} &
      \includegraphics[width=0.5\hsize]{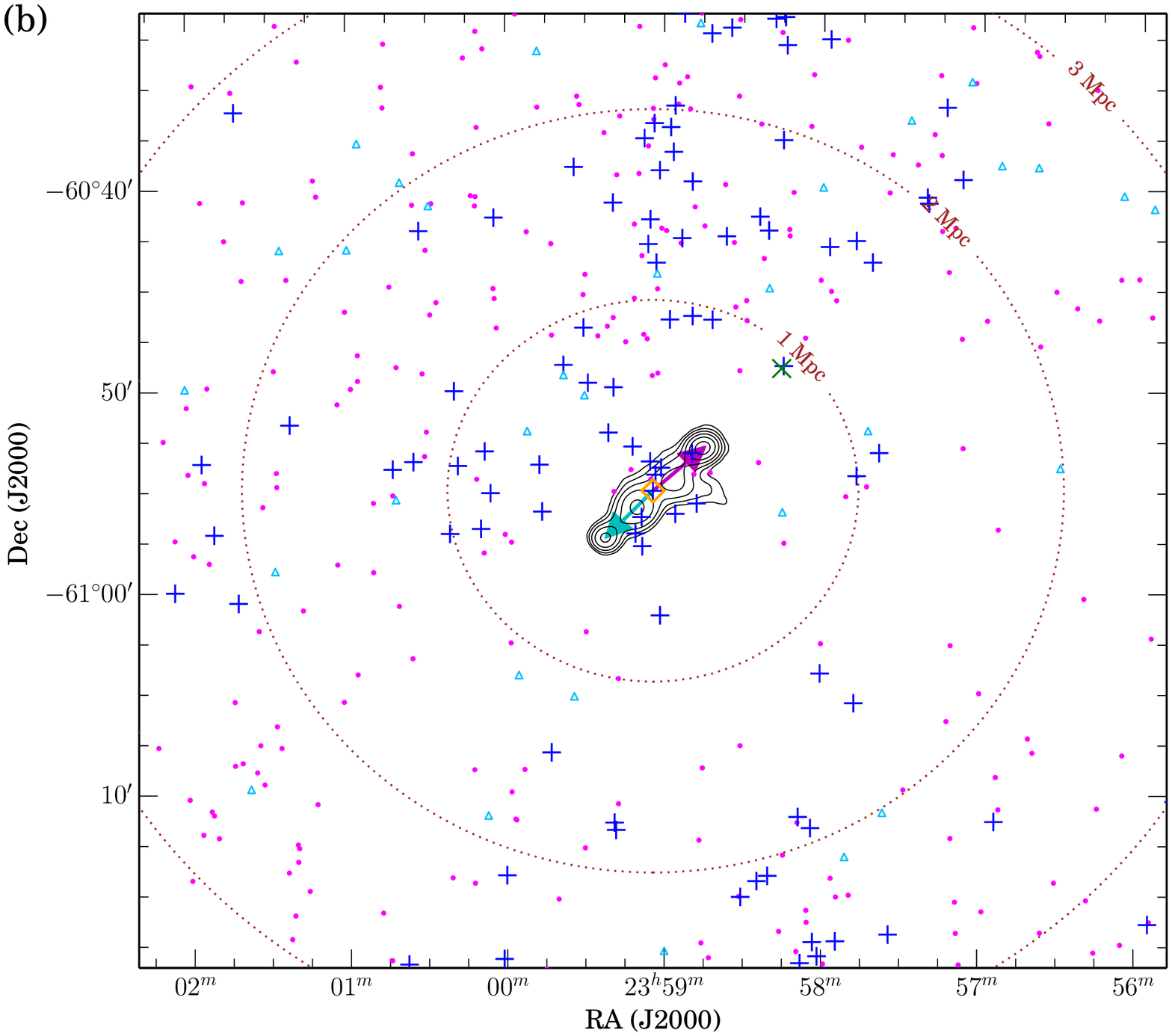}\\
      \includegraphics[width=0.5\hsize]{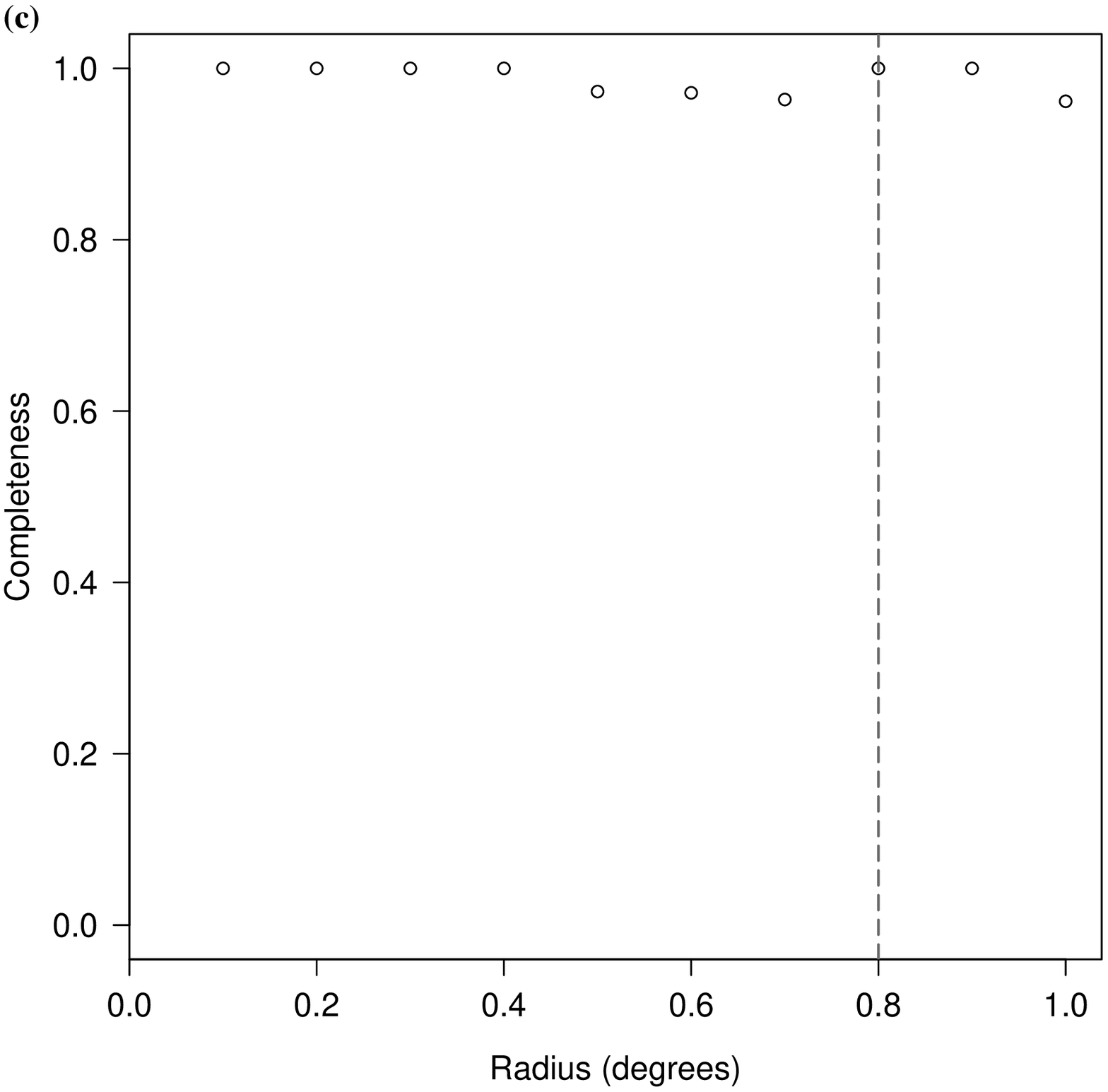} &
      \includegraphics[width=0.5\hsize]{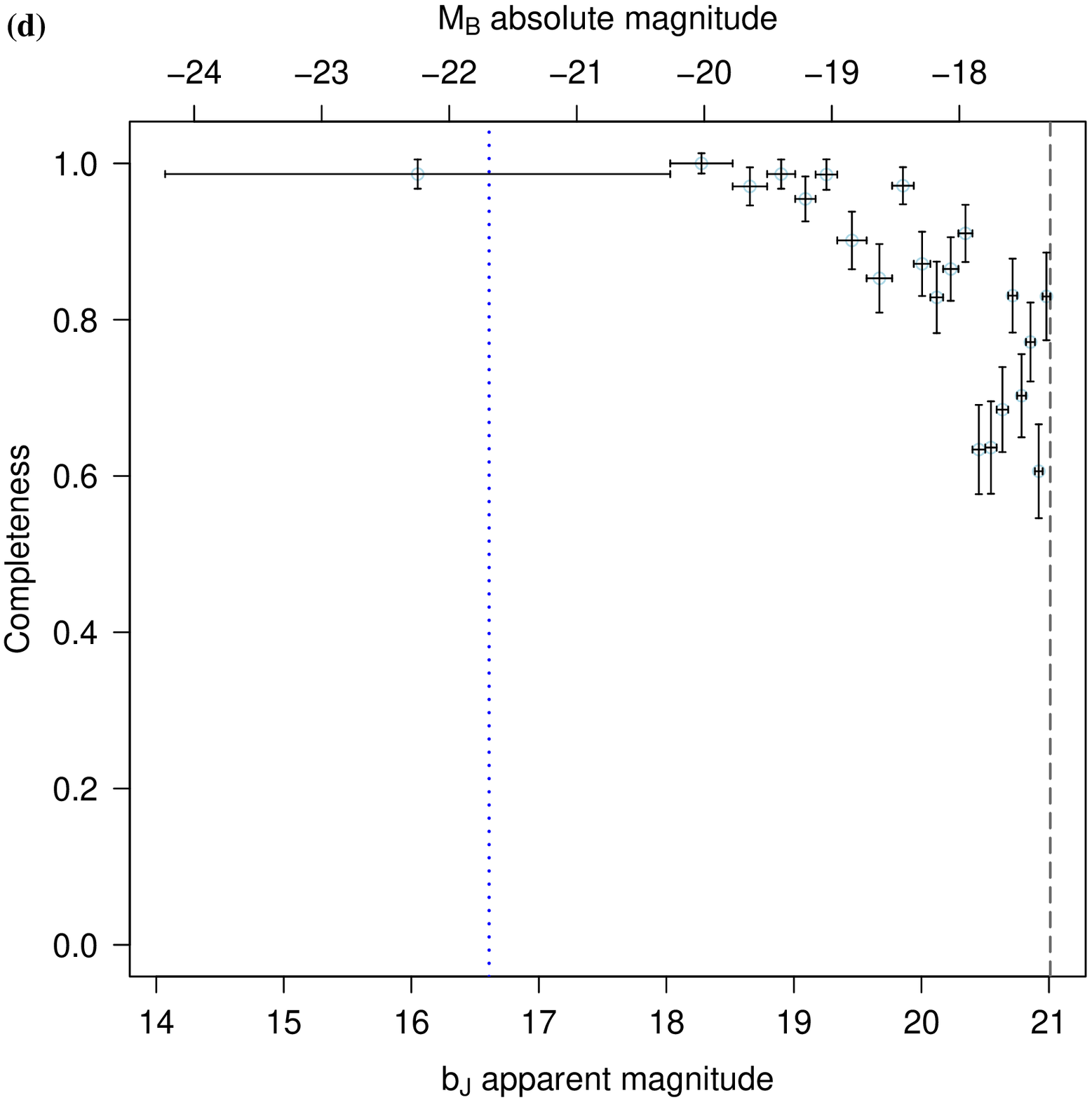}
\end{tabular}
\caption{(\subfigletter{a}) Distribution of galaxies around B2356--611; a 20 arcmin angular scale corresponds to a linear size of 2.11 Mpc. Plus symbols represent galaxies within $\pm0.003$ of the host redshift at z = 0.0962. Radio contours are shown at (6, 12, 24, 48, 96, and 192) $\times$ 15 mJy beam$^{-1}$. (\subfigletter{b}) An enlarged view of the optical field, centred at the host galaxy, with dotted circles at 1, 2 and 3-Mpc radii. (\subfigletter{c}) A completeness plot showing the fraction of observed SuperCOSMOS targets against radius, measured in intervals of 0.1$^{\circ}$ from the field centre. (\subfigletter{d}) A completeness plot showing the fraction of observed targets against $B$-band magnitude with the $b_{\mathrm{J}}$ magnitude (16.61) of the host galaxy indicated by a blue dotted line.}
\label{fig:B2356}
\end{minipage}
\end{figure*}

\begin{figure}
\includegraphics[width=\hsize]{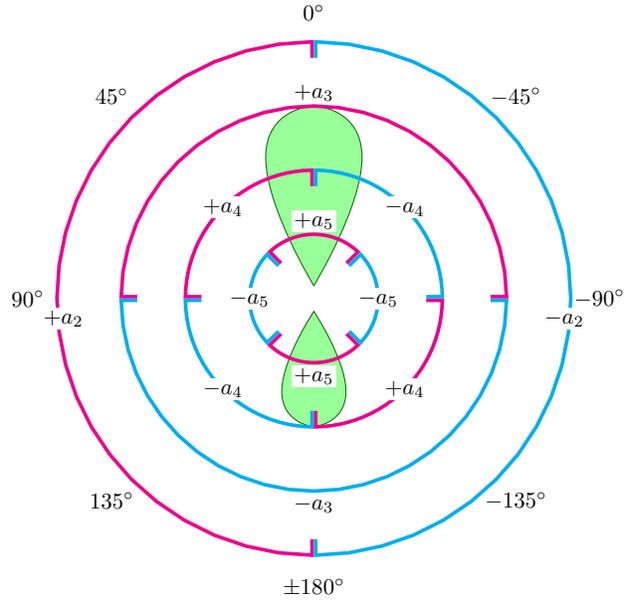}
\caption{The Fourier components ($a_1$, $a_2$, \ldots, $a_5$) used to measure anisotropy in the galaxy distribution surrounding the GRG host galaxy. The longer radio lobe designates $0^{\circ}$ with the sectors associated with the dipole moments ($a_2$ and $a_3$) and the quadrupole moments ($a_4$ and $a_5$) shown as coloured arcs -- positive in magenta, negative in cyan. The radio lobes are represented in green. The remaining Fourier component $a_1$ (not shown) is a measure of the overdensity of a volume containing the GRG compared to volumes placed randomly in the rest of the observationally complete field.}
\label{fig:Fourier_components_schematic}
\end{figure}

\begin{figure}
\centering
\includegraphics[width=\hsize]{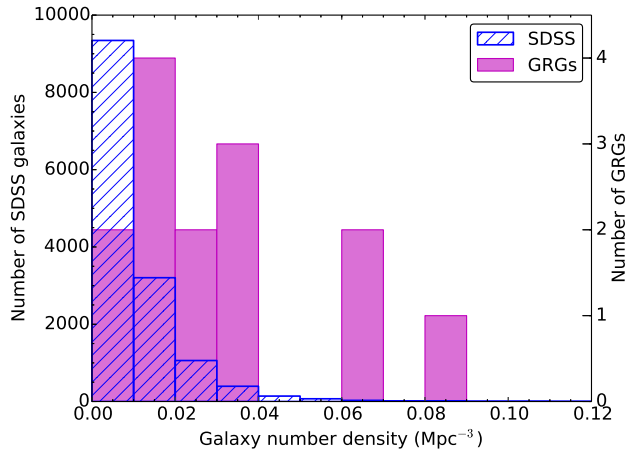}
\caption{Distributions of galaxy number density for cylindrical volumes centred on each GRG host and for equally sized volumes surrounding Sloan Digital Sky Survey (SDSS) galaxies with magnitudes within 0.5 of the GRG host median $B$-band absolute magnitude, $-21.4$~mag. Each cylinder is 2~Mpc in radius and 24 Mpc in length. Galaxy number density in the SDSS volumes varies from 0.003 to 0.123 Mpc$^{-3}$. We exclude the GRG fields J0746--5702, J0843--7007 and B1302--325 as their limiting magnitudes are brighter than the common absolute magnitude limit of $-19.49$~mag. Also, those centred on B0703--451 and B0707--359 are excluded due to uncertainty in the host galaxy and excess stellar contamination at low Galactic latitude (including masking of the region containing the host), respectively.}
\label{fig:GRG_SDSS_histogram}
\end{figure}

\begin{figure*}
  \centering
      \includegraphics[width=\hsize]{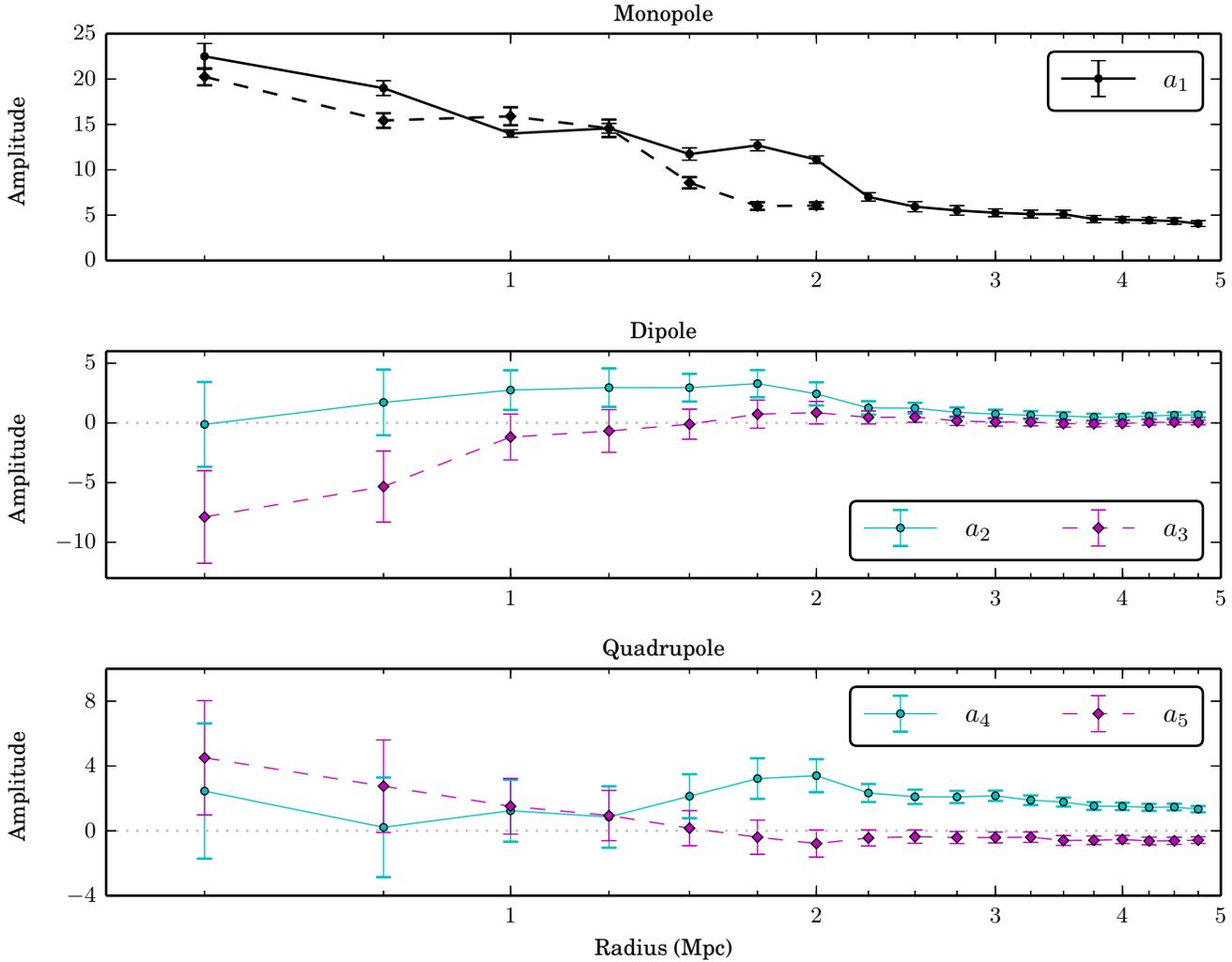}
\caption{Fourier component values against reference volume (cylinder) radius in Mpc for the stacked fields in Table~\ref{tbl:Fourier_components_equal_weight}, listed as entry (c), excluding the smallest fields (B0319--454, B0503--286, B0511--305, B1308--441 and J2018--556). The black dashed line in the upper panel represents the $a_1$ values for the full set of stacked fields from entry (c) in Table~\ref{tbl:Fourier_components_equal_weight} and so is limited to the completeness radius of the smallest field, B0511--305. A fixed number of reference volumes (3) is used for consistency.}
\label{fig:Stacked_Fourier_component_parameters_vs_radius}
\end{figure*}

\begin{figure*}
\centering
\begin{minipage}[t]{\columnwidth}
\centering
\includegraphics[width=\hsize]{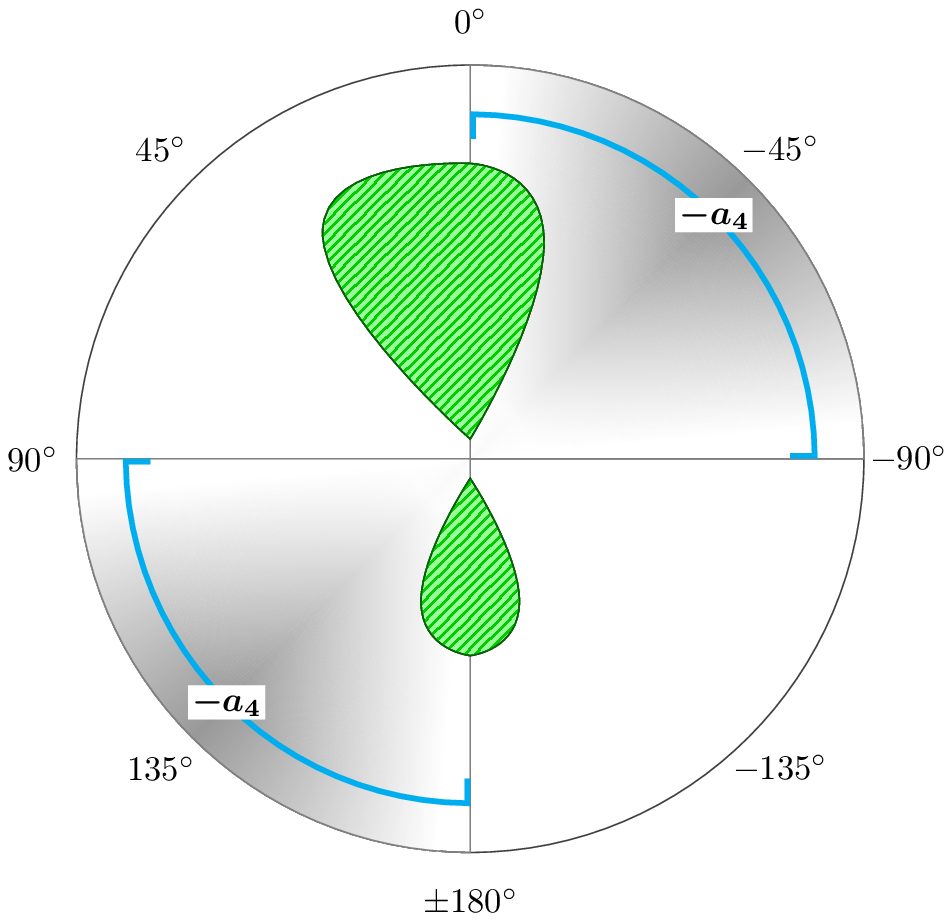}
  \caption{The significant quadrupole moment in the stacked galaxy distribution around GRGs with an extension on only one side of the jet axis, see entry (e) in Table~\ref{tbl:stacked}. The two fields (B0319--454 and B2356--611$\dagger$) are co-added, centred on each GRG host optical galaxy and aligned longer lobe to $0^{\circ}$. The field indicated by $\dagger$ was inverted about PA $=0^{\circ}$ to align the offset radio components. Grey-shaded sectors, which are labelled with the corresponding Fourier component parameter, indicate overdensity and the radio lobes are represented in hatched green.}\label{fig:Stacked_Fourier_components_wing_like}
\end{minipage}\hfill
\begin{minipage}[t]{\columnwidth}
\centering
\includegraphics[width=\hsize]{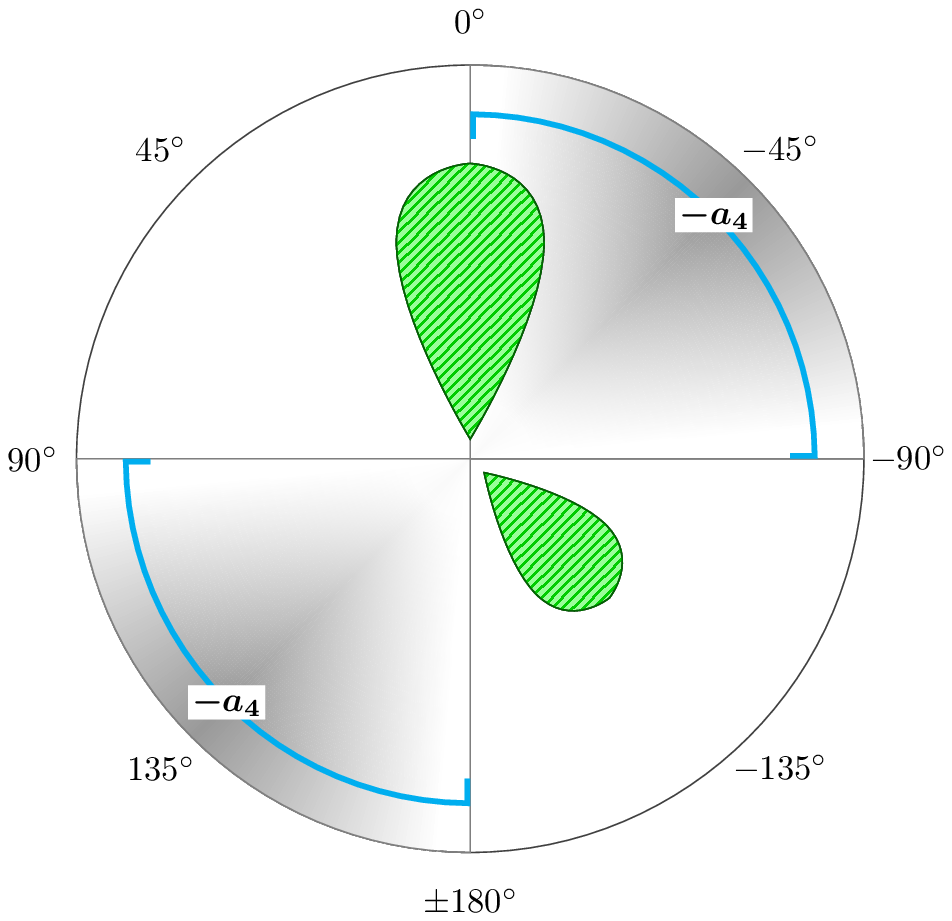}
  \caption{The significant quadrupole moment in the stacked galaxy distribution around GRGs with non-collinear lobes, see entry (f) in Table~\ref{tbl:stacked}. The five fields (J0116--473, B0503--286$\dagger$, B0511--305$\dagger$, J0515--8100 and B1308--441) are co-added, centred on each GRG host optical galaxy and aligned longer lobe to $0^{\circ}$. Fields indicated by $\dagger$ were inverted about \mbox{PA $=0^{\circ}$} to align the offset, shorter radio lobes. Grey-shaded sectors, which are labelled with the corresponding Fourier component parameter, indicate overdensity and the radio lobes are represented in hatched green.}\label{fig:Stacked_Fourier_components_noncollinear}
\end{minipage}
\end{figure*}

We now consider individual GRGs and attempt to understand the relation between galaxy distribution and GRG structure. Our comprehensive optical spectroscopic data is presented in Fig.~\ref{fig:J0034} to~\ref{fig:B2356} as radio-optical overlays in addition to completeness plots for $0.8^{\circ}$ radial regions. The Fourier component analysis complements these representations by highlighting anisotropy in the galaxy distribution within 2-Mpc radii around each host as projected on the plane of the sky.

The following radio-optical overlays feature projected distributions of galaxies within $\pm0.003$ of each host redshift (indicated by plus symbols). These distributions vary in the number of galaxy members from rich distributions, such as the field of B1308--441, to those consisting of few members, such as J0116--473. In some cases the distributions resemble large galaxy filaments (e.g. B1308--441, J2018--556) and in others the galaxy distributions are more localized to the vicinity of the GRG hosts (e.g. B0511--305, B2356--611). In most of these cases the GRGs are oriented nearly orthogonal or at large angles to the galaxy distributions.\\

\centerline{\emph{J0034--6639 (Fig.~\ref{fig:J0034})}}
The GRG lobes appear to have evolved into local voids. In the 2-Mpc radial region surrounding this symmetric GRG, there appears to be a bi-conical region across the radio axis within which all the galaxies are located. Within the 1-Mpc radius circle it is even more pronounced; the four galaxies with redshift offsets of $|\Delta z| \leq 0.003$ from the host are distributed largely perpendicular to the radio axis. In Table~\ref{tbl:Fourier_components_equal_weight}, the negative $a_5$ value strongly reflects this division in the galaxy overdensity and that the GRG lobe position angles avoid populated regions about the radio axis.\\

\centerline{\emph{J0116--473 (Fig.~\ref{fig:J0116})}}
This asymmetric GRG host lies in a sparse environment. There are only three galaxies within the 2-Mpc radius circle about the host galaxy. Over a larger region about the host, within a radius of 4--5 Mpc, more galaxies are seen with redshifts in the $|\Delta z| \leq 0.003$ range. Once again these galaxies avoid a wide conical region about the GRG radio axis. The GRG lobes appear to have expanded into regions that avoid these galaxies.\\

\centerline{\emph{B0319--454 (Fig.~\ref{fig:B0319})}}
The galaxy redshift data used in this case is that obtained by~\citet{Safourisetal2009}. The authors had reported an increased concentration of galaxies ($0.0022<\Delta z <0.0038$) located on the side of the shorter NE bright lobe in contrast to the sparser environment in which the longer SW lobe is located~\citep[see Fig. 9,][]{Safourisetal2009}. The host galaxy, clearly seen to be surrounded by galaxies, was inferred to belong to a galaxy group and the prospect of higher gas concentration associated with the group was reasoned to account for the shorter extent and higher brightness of the NE lobe as compared to the SW lobe.

On examining the SuperCOSMOS data we find several bright galaxies, particularly those in the vicinity of bright stars, have not been catalogued and so were not observed. There are as many as 6 on the side of the NE lobe, (with 4 of them located within the lobe) and 5 on the SW lobe side well removed from any radio emission. The jet on the NE side appears to have run into a larger concentration of galaxies in its path than the jet on the SW side. If we take these bright galaxies into consideration the conclusion arrived at by~\citet{Safourisetal2009} is strengthened further. In Table~\ref{tbl:Fourier_components_equal_weight}, the $a_3$ value does not reflect the correlation however and may be affected by the missed galaxies (also we have used galaxies with $|\Delta z| \leq 0.003$ instead, which excludes the galaxy to the extreme NE tip of the lobe). Moreover the Fourier components have all been calculated for a region of 2-Mpc radius around the host galaxy and it appears that environment more local to the host (over 600--700~kpc distance) has a greater influence than the environment on a 2-Mpc radius region.\\

\centerline{\emph{J0331--7710 (Fig.~\ref{fig:J0331})}}
This highly asymmetric GRG is in a very sparse region. There are only two other galaxies, within $|\Delta z|\leq0.003$ of the host redshift, in a region of radius 3 Mpc about the host. Both these galaxies are to the north of the northern lobe, which is characterized by much shorter extent and much brighter flux as compared to the southern lobe. The strongly significant, negative $a_3$ parameter reflects the location of the neighbouring galaxies on the side of the shorter lobe. Similarly, the positive $a_5$ parameter may be attributed to these few neighbours. The data on this source appears inadequate to reveal the cause of the severe asymmetry on the basis of galaxy distribution around the GRG. With the shorter lobe also being the brighter as expected if gas associated with a richer galaxy environment were confining the lobe, it would be interesting to investigate whether deeper observations may reveal a larger galaxy concentration on the side of the shorter lobe.\\

\centerline{\emph{J0400--8456 (Fig.~\ref{fig:J0400})}}
A striking characteristic of this GRG is the non-collinearity of its radio structure. The curved inner structure leading up to the hotspots (the high resolution image 4(b) in \citetalias{Malareckietal2013} shows this more clearly) and billowing lobes thereafter are strongly suggestive of a Wide-Angle Tail-like structure. However, the two lobe features are at nearly orthogonal orientations with respect to the local axis and extend out in opposing directions. Although its rather symmetric twin-jet~\citep{Saripallietal2005} as well as symmetric twin-hotspot structure may suggest that this GRG lies in the plane of the sky, the deflected NW and SE lobes hint at either a rapid rotation of the beam axis or opposing environmental effects at play at large distances from the host galaxy. Whatever the scenario the two lobes are sampling environment far from the host galaxy where environmental effects may be stronger.

The field within 2~Mpc of the GRG is quite sparse. A band of galaxies exists to the south (at ${\sim}2$--$3$~Mpc from the host) at a position angle of 120$^{\circ}$. In the same position angle another stream exists to the north but it weakens in the vicinity of the GRG host (both streams are seen more clearly when the redshift range is increased to $\pm0.015$). While the northern stream has almost all galaxies with positive $\Delta z$, both streams have predominantly positive $\Delta z$ with relatively high values (${>}0.003$ and hence only marked as magenta dots) to the west. To the NE are two galaxies with $|\Delta z|\leq0.003$ at a distance of about 500 kpc from the northern lobe. The southern stream flows past the host galaxy where there is a group of 8 galaxies with $|\Delta z| \leq 0.003$ (Fig.~\ref{fig:J0400}a) at a separation of 1.5--2~Mpc to the south of the southern lobe with negative as well as a few positive $\Delta z$ values. The locations (although less so the distances) of the galaxies suggest influence of the environment on the radio galaxy lobes. The GRG appears to have formed in a region that is relatively sparse and clearly avoids the dense parts galaxy streams. A galaxy only few arcsec to the SE of the host could not be observed due to its close proximity. This is reflected in the dip in the radial completeness plot shown in Fig.~\ref{fig:J0400}(c). None of the Fourier component parameters shows a significant value for this GRG.\\

\centerline{\emph{J0459--528 (Fig.~\ref{fig:J0459})}}
Fig.~\ref{fig:J0459}(a) clearly shows a swathe of galaxies forming a band in a NW-SE direction of which the host is a member. This GRG appears to have grown in a direction that avoids this band of galaxies, which is reflected in the $a_5$ parameter at $2.2\sigma$.\\

\centerline{\emph{B0503--286 (Fig.~\ref{fig:B0503})}}
The galaxy redshift data used in this field is that obtained by~\citet{Subrahmanyanetal2008}. These authors studied the environment of this GRG is some detail and showed a strong link between the radio source asymmetries and variation in the galaxy distribution around the GRG. The $a_3$ parameter we calculate clearly reflects the correlation between the lobe extent asymmetry and the difference in the galaxy distribution on the two sides of the radio source (Fig.~\ref{fig:B0503}b). Although the host is in a group it is in a relatively sparse environment except for the prominent concentration of galaxies more than 1~Mpc to the NE (in Table~\ref{tbl:Fourier_components_equal_weight}, this source has the greatest $a_1$ value among the GRGs). The positive $a_5$ parameter reflects this north-south distribution of galaxies, which is relatively close to the radio axis position angle. Given the strong north-south asymmetry in the galaxy distribution about the host, the jet on the sparse southern side has grown to a larger extent than on the opposite northern side.\\

\centerline{\emph{B0511--305 (Fig.~\ref{fig:B0511})}}
Our observations of this field have missed 12 SuperCOSMOS targets within the 2-Mpc region with magnitudes between 16.7 and 19.0 due to their proximity to one another. This is seen in Fig.~\ref{fig:B0511}(c) as a drop at 30 arcsec in the fraction of SuperCOSMOS targets observed. However, within the 1-Mpc radius region there is 100 per cent completeness. Close to the host is seen a band of galaxies stretching nearly 800~kpc from NW to SE. The GRG radio axis is nearly perpendicular to this galaxy chain. If we turn our attention to the targets within the larger 2-Mpc region, there appear to be two separate galaxy condensations in a north-south direction: a group of 5 galaxies to the south and a group of 10 galaxies at the centre (including the host). We do not know if they are related as a galaxy filament. If these galaxy concentrations form a filament, the GRG appears to have grown nearly along the filament although the northern lobe is oriented away. With the clear orthogonality however between the GRG axis and the chain of 8 galaxies -- closer to the host in velocity space than the pair just to the north -- it may also be that GRG growth is guided more by the local distribution of galaxies.

This GRG is an example of a restarted AGN where its southern lobe, which has a bright hotspot at its end, is embedded in diffuse synchrotron plasma. Such a feature surrounding the northern lobe is absent. A possible explanation may lie in the galaxy group to the south, which is only 500 kpc south of the southern lobe. The intra-group gas associated with these galaxies may provide an ambient medium that `preserves' the lobe, preventing it from expanding and hence from suffering substantial losses due to adiabatic expansion. The asymmetry in the lobe lengths (where the southern hotspot is significantly closer than the northern hotspot) may also be related to the group to the south. We note here that the host galaxy has a prominent feature that stretches nearly 150~kpc to the west and turning south at the end.\\

\centerline{\emph{J0515--8100 (Fig.~\ref{fig:J0515})}}

This rather symmetric GRG is in a very sparse field. A fainter, close companion to the east of the host is seen, earlier studied by~\citet{Subrahmanyanetal2006} to be in close interaction with the GRG host. There is only one other galaxy (1.6 Mpc to the south) in a region of 2-Mpc radius around the host. It is interesting to note the orientation of the GRG jets that are nearly orthogonally to the line formed by the host and its companion.\\

\centerline{\emph{B0703--451 (Fig.~\ref{fig:B0703})}}
This source suffers from a problem of host galaxy identification. As of yet we are not certain of the host galaxy, whether it is the relatively bright galaxy at the GRG centroid or the fainter object located at the centre of the western peak component. We nevertheless observed the field. The observations are suitable to observe galaxies in the redshift range of the brighter host candidate.\\

\centerline{\emph{B0707--359 (Fig.~\ref{fig:B0707})}}
This source has the lowest absolute Galactic latitude of any GRG in the sample ($b=-12.3^{\circ}$) and consequently the surrounding field has significant stellar contamination. The field is heavily masked due to bright stars, including a mask encapsulating the entirety of the radio structure and the region to the NE. Our observations of the GRG field were constrained by the limited reliable SuperCOSMOS data available for this source. There are no galaxies detected within a redshift offset of $|\Delta z| \leq 0.003$ from the host and so this GRG appears to reside in a very sparse region.\\

\centerline{\emph{J0746--5702 (Fig.~\ref{fig:J0746})}}
This GRG is in a sparse environment and one of the most isolated GRGs in the sample. Several SuperCOSMOS targets have remained unobserved, however these are well outside the 3-Mpc radius circle. The field is fairly well sampled within the 2-Mpc radius region surrounding the host. In this region, the only two galaxies in the $|\Delta z| \leq 0.003$ range are along a position angle perpendicular to the GRG axis. This is reflected in the negative $a_5$ parameter.\\

\centerline{\emph{J0843--7007 (Fig.~\ref{fig:J0843})}}
This GRG is in a sparse environment as indicated by its low galaxy number overdensity compared to the other GRG fields (see Table~\ref{tbl:host_volume_overdensity}). Although the field suffers from incompleteness (Fig.~\ref{fig:J0843}d), with several SuperCOSMOS targets in the magnitude range 18--18.5 unobserved, the 1-Mpc radius region is 100 per cent complete and in this region there are only two galaxies with a redshift offset of $|\Delta z| \leq 0.003$ about the host.\\

\centerline{\emph{B1302--325 (Fig.~\ref{fig:B1302})}}
This GRG has relatively low observational completeness (see Table~\ref{tbl:Optical_field_properties}), which is seen in Fig.~\ref{fig:B1302}(c). However, most of the unobserved objects are on the peripheries of the field. The region within 2-Mpc radius has 100 per cent completeness. From the overlay in Fig.~\ref{fig:B1302}(a) we see that this GRG is in a sparse region. Within the 2-Mpc radius region surrounding the host there are only 5 galaxies in the redshift range of $|\Delta z| \leq 0.003$ about the host redshift. There is a hint of a string of galaxies mostly to the east of the host and out to a distance of at least 4~Mpc (including the galaxy at the NE tip of the GRG), which lie in a position angle nearly orthogonal to the radio galaxy axis. Although the negative sign of the $a_5$ parameter suggests such a galaxy distribution, the significance is affected due to the sparseness of the field within the 2-Mpc region for which the parameter is computed.\\

\centerline{\emph{B1308--441 (Fig.~\ref{fig:B1308})}}
A broad, ${\sim}2$-Mpc wide NE-SW galaxy filament is seen across the field. The GRG host galaxy is situated close to the centre. A concentration of galaxies is seen close to the host within a few hundred kpc. This group of galaxies is spread along the same direction as the larger filament. The GRG jets appear to have grown in a direction nearly orthogonal to the filament. This is reflected in the $a_5$ parameter, although only at low significance. The significant $a_4$ value reflects that this is a source with non-collinear lobes. Once again, local environments in the vicinity of the host galaxy may have a greater influence on the advance of the jets, which in the case of B1308--441 have a direction nearly orthogonal to the ${\sim}500$-kpc extent galaxy chain of which the host is a member. The rather high asymmetry in lobe extents may also be related to the local galaxy distribution, where there is a higher galaxy concentration on the shorter, SE side as compared to the longer, NW side. The $a_0$ parameter has the highest value among all GRGs indicating its environment as being the most dense among all GRGs in our sample.\\

\centerline{\emph{B1545--321 (Fig.~\ref{fig:B1545})}}
Despite masking of the input SuperCOSMOS catalogue due to bright stars to the NW of the host (Fig.~\ref{fig:B1545}a), the GRG likely resides in a relatively sparse region of the sky. The remaining 2-Mpc radial region is relatively unaffected by stellar contamination but appears devoid of galaxies with only 3 others seen with redshift offsets from the host of $|\Delta z| \leq 0.003$.\\

\centerline{\emph{J2018--556 (Fig.~\ref{fig:J2018})}}
In Fig.~\ref{fig:J2018}(b) a rather sparse field is seen within the 2-Mpc radius region with the exception of a 500-kpc wide, linear distribution of galaxies to the west extending just beyond the GRG host. On a larger scale (Fig.~\ref{fig:J2018}a) a few groups of galaxies are seen to the east at a distance of more than 2 Mpc from the host, spread nearly north-south over a large region nearly 3 Mpc in extent. The GRG is in a fairly dense region of galaxies although these are clearly not uniformly distributed. The GRG jets appear to have grown into sparse regions nearly orthogonal to the linear distribution of galaxies. The $a_5$ parameter, although indicating this with its negative sign, has a large error. The galaxy distribution seen predominantly on one side, west of the host galaxy, is reflected by the negative $a_2$ parameter.\\

\centerline{\emph{J2159--7219 (Fig.~\ref{fig:J2159})}}
In the 2-Mpc radial region surrounding the host there appear to be several galaxies in the vicinity of the GRG host. If we look at a larger region (Fig.~\ref{fig:J2159}a), up to more than 3 Mpc out, a rich concentration of galaxies (more than 2 Mpc wide) is seen to the NE. It is not clear if the galaxies in the vicinity of the host form a continuum with this rich group to the NE but it is interesting to note the near orthogonality of the GRG radio axis with respect to it. This is not reflected in the $a_5$ parameter, which is calculated only for a region of 2-Mpc radius.

Within the 1-Mpc region the two galaxies with the smallest redshift offsets relative to the host ($|\Delta z| \leq 0.0005$) are seen to the east, which also form an axis with the host at a position angle nearly orthogonal to the GRG radio axis. There are two other galaxies with even smaller redshift offsets that lie along the same orthogonal position angle further to the SW of the GRG host. Interestingly, there appears to be a band of galaxies all with redshift offsets in this narrow range including these four galaxies, the host, one more to the NW of the host (at ${\sim}1$~Mpc) and 8 galaxies in the concentration to the NE (at a distance of 2--3 Mpc), which all lie along a position angle nearly orthogonal to the GRG radio axis.\\

\centerline{\emph{B2356--611 (Fig.~\ref{fig:B2356})}}

Fig.~\ref{fig:B2356}(a) shows a large-scale view of the field surrounding this GRG. Galaxies lie in a larger concentration to the north and east of the host galaxy, forming a linear feature nearly 1.5 Mpc wide and more than 3 Mpc long. This is reflected in the positive $a_2$ parameter. From this feature a string of galaxies is seen extending towards the GRG host over a linear extent of ${\sim}600$~kpc and orthogonal to the GRG radio axis. The GRG appears to have grown into local void regions orthogonal to the galaxy chain. Although the $a_5$ parameter reflects this in its negative sign the errors are once again large. Interestingly, the few-hundred kpc radio lobe feature seen to the west of the host is extended in a direction orthogonal to the galaxy concentration surrounding the GRG, which is reflected in the highly significant $a_4$ value. This may indicate higher gas pressure and density in the region of the galaxy concentration that causes the radio lobe to extend away from it.\\

In summary, for the individual sources, the correlations between GRG morphology and large-scale structure agree with the Fourier component analysis. In a number of the radio-optical overlays, the GRG lobes are seen extending into local voids (e.g. J0034--6639, J0116--473, J0459--528, B2356--611). In others, radio components are deflected away from concentrations of galaxies (e.g. B0503--286) or the expansion of one lobe is truncated compared to the other (e.g. B0319--454, J0331--7710, B0503--286, B0511--305). Often neighbouring galaxies are seen perpendicular to the radio axis (e.g. J0515--8100, J0746--5702) and in several cases these form chains of galaxies across the radio axis (e.g. J0034--6639, B0319--454, B0511--305, B1308--441, B2356--611). Hence, it appears that GRG jets often develop in directions perpendicular to the position angles at which nearby galaxies are found. Furthermore, the presence of galaxy concentrations in their path correspond with the GRG lobes becoming asymmetric due to radio components being deflected or limited in their expansion.

\section{Discussion}
\label{sec:Discussion}

\subsection{Growth to giant sizes}
\begin{figure}
  \centering
      \includegraphics[width=\hsize]{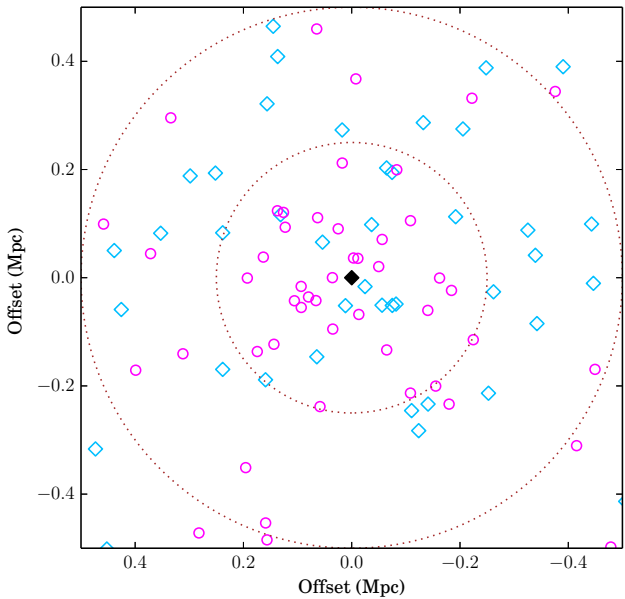}
\caption{Projected galaxy positions relative to the host (indicated by a filled diamond) stacked for all GRG fields, excluding B0319--454 and B0503--286 due to insufficient spectrum quality, within a redshift offset of $|\Delta z| \leq 0.003$. Each field is rotated about the host to align the longer radio lobes to PA $=0^{\circ}$. Galaxies with emission lines are marked with open diamonds and galaxies without emission lines are marked with open circles (in the online version these symbols are shown in cyan and magenta, respectively). The dotted circles represent radii in increments of 0.25 Mpc.}
\label{fig:Galaxy_spectra_features}
\end{figure}

A longstanding puzzle has been the factors that may be responsible for the growth to extraordinary linear sizes of some radio-loud AGN. While it is recognized that GRGs rarely reside in rich cluster environments, little else could be inferred about their surroundings due to lack of detailed spectroscopic studies of the galaxy distributions around GRG hosts. Our spectroscopic observations of the fields surrounding 17 GRG hosts allow us to investigate, for the first time, the kind of environments these largest of radio galaxies occupy. From the radio-optical overlays as well as the Fourier component analysis for individual fields, it appears that while GRGs are often found in sparse regions, they may also be in the vicinity of or members of galaxy groups, chains, or larger-scale filaments. Remarkably, in the latter case, their jets almost invariably have grown in directions that often avoid galaxy concentrations. The GRG jets instead preferentially grow into void regions. The jets are seen to grow nearly orthogonal or at large angles to the galaxy neighbourhood. With the exception of the hosts of J0459--528 and B1308--441, in four out of seven GRGs where a galaxy filament is seen the host galaxy is located at the edge (B0503--286, J2018--556, J2159--7219, B2356--611) or is displaced from the filament (J0400--8456). This is partially consistent with active galaxies lying at the edge of overdense environments and increased clustering around radio-loud AGN~\citep[e.g.][]{Stockton1986,Lietzenetal2009,Worpeletal2013}. We further explore the environments of the GRG hosts by examining the optical spectra of neighbouring galaxies and find that 63 per cent of those greater than 0.25 Mpc from the hosts have emission features whilst only 30 per cent of those within this radius exhibit emission features (see Fig.~\ref{fig:Galaxy_spectra_features}).

For a separate set of three northern-declination GRGs, NGC 6251, NGC 315 and 4C 73.08, a closer inspection of the surrounding optical fields~\citep[shown in figs 5, 5, 3 in][respectively]{Chenetal2011a,Chenetal2012a,Chenetal2012b} once again reveals compact galaxy chains and sheets (galaxies within $\pm0.002$ of host redshifts) to which the hosts belong and again demonstrates an avoidance of these galaxy concentrations by the GRG jets. The three GRGs have axes that are at large angles to the galaxy distribution about their hosts. In NGC 315 it is even more striking with the GRG having grown nearly orthogonal to the large ${\sim}25$-Mpc Perseus-Pisces galaxy filament. The giant sizes of GRGs may therefore be attributed to the sparse environments in which their hosts reside or, in the case when the host is a member of a chain or filament of galaxies, to the fortuitous orientation of the jets that avoids galaxy concentrations.

It is likely therefore that when jets happen to be oriented such that they propagate into the galaxy concentrations (formed by the galaxy chains or filaments) the gas associated with the galaxy chains or filaments may be strongly impeding jet propagation. This resistance to jet growth appears to occur even when the chains or filaments contain relatively few members as in J0116--473, J0746--5702 and B1302--325. The effect of the environment is therefore seen even in sparse environments where the jets grow in directions that avoid the sparse galaxy chain or distribution about the host.

\begin{figure*}
  \centering
      \includegraphics[width=\hsize]{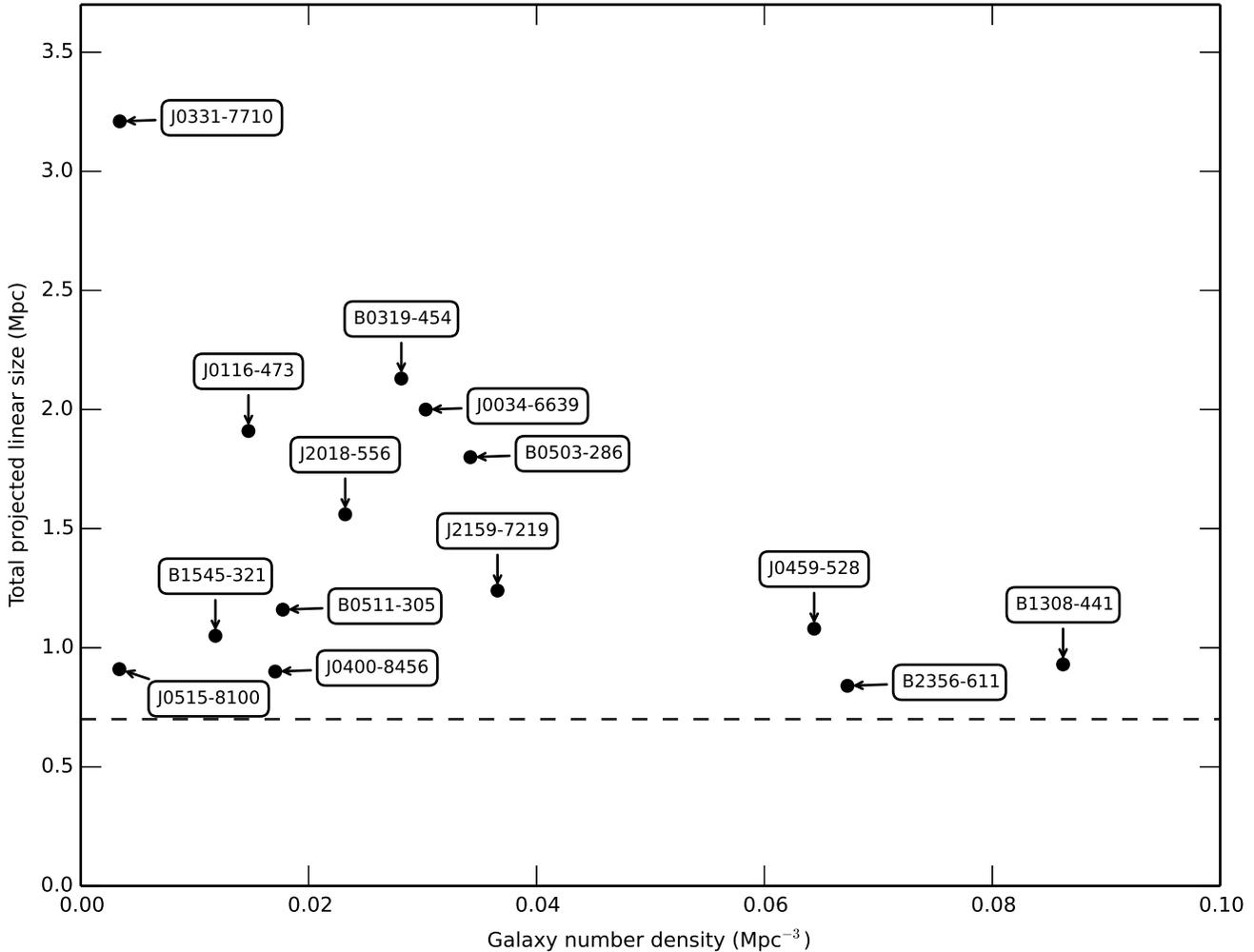}
\caption{GRG projected linear size against galaxy number density in a cylindrical volume (2 Mpc in radius and 24 Mpc in length along the line of sight) centred on the host optical galaxy. A faint absolute $B$-band magnitude threshold of $-19.49$~mag is applied to all fields before computing the galaxy number density. The three fields with limiting magnitudes brighter than this absolute magnitude limit (J0746--5702, J0843--7007 and B1302--325) are excluded, as well as B0703--451 due to uncertainty in the host ID and B0707--359 due to excess masking of bright stars at low Galactic latitude. The dashed line represents the minimum projected length threshold for GRGs ($0.7$~Mpc).}
\label{fig:Linear_size_vs_density}
\end{figure*}

Fig.~\ref{fig:Linear_size_vs_density} shows a plot of the projected linear size of each GRG and the galaxy number density in a 2-Mpc radius, 24-Mpc length cylindrical region about the host optical galaxy, which is similar to the cylinders used in the Fourier component analysis. It should be noted that a faint absolute magnitude limit of $-19.49$~mag is applied to all constituent fields to account for differences in observational completeness. A weak correlation is seen between the projected linear size of each GRG and the density of its immediate environment. In particular, no sources occupy the upper right quadrant of the plot. We note several extreme cases: J0331--7710, which is located in a sparse environment and has grown to the greatest projected linear size in the sample; while the hosts of J0459--528, B1308--441 and B2356--611 are in densely populated regions but to a lesser extent in the path of the jets and have still grown to giant sizes.

\subsection{GRG orientations with respect to host galaxies and large-scale structure}
\label{subsec:GRG_orientations}
In a study of relative orientations of radio axes and host galaxy axes,~\citet{SaripalliSubrahmanyan2009} reported a strong tendency for GRGs to have radio axes that are at large angles to the major axes of their host elliptical galaxies. Separately, for the 3C sample, they reported that while FR-II radio galaxies have radio axes distributed over a wide range of angles with respect to their host major axes, there was a significant difference in the physical sizes of major-axis radio galaxies and minor-axis radio galaxies. Those with radio axes closer to the host minor axes were nearly twice as large. We have measured position angles of the major axes of 12 GRGs in our sample using the {\sc ellipse} task in {\sc iraf}. Among the remaining seven, we use information available in the literature for six (J0116--473, B0511--305, B1545--321, J2018--556 and B2356--611 from~\citet{SaripalliSubrahmanyan2009}; B0707--359 from~\citet{Saripallietal2013}) and exclude one due to uncertain host ID (B0703--451). For our sample of 18 GRGs, four hosts are classified circular (J0331--7710, B0511--305, J0843--7007, B1545--321). Among the remaining 14 GRGs, 9 have radio axes within $30^\circ$ of the host minor axes (within errors of $3$--$5^\circ$) and B0707--359 has a radio axis within $35^\circ$. We note that for B0511--305 the host is accompanied by a prominent east-west feature; the jets have an orientation nearly orthogonal to this feature. The remaining four GRGs (J0116--473, J0400--8456, J2018--556, B2356--611) have jets within close angles to the host major axes (interestingly these four are also the only GRGs with prominent off-axis lobe deviations). A similar tendency for minor axis growth among GRGs is shown by our GRG sample (no galaxies with minor-axis proximity are in common with the GRG sample of~\citet{SaripalliSubrahmanyan2009}; three of the four major-axis GRGs are common to the two samples).

GRGs may be said to have grown to giant linear sizes aided by environment (minor axis growth and avoiding galaxy chains and filaments) both local and large-scale. Their lobes can be further shaped by the large-scale galaxy distribution that also affects their relative extents.

In this study, we have examined galaxy distributions in narrow redshift slices over cylindrical regions of 2-Mpc radius and in many cases well beyond. We have reported relationships between the large-scale structure and GRG morphological properties, essentially between galaxy distributions in close redshift space of the GRG host, jet orientations, and the extent and morphologies of synchrotron-plasma lobes. This requires the probed environments to be permeated with gas with which the synchrotron jets and lobes interact. What is the nature of the galaxy environments we have probed on these length scales? For regions on scales of 2.5 Mpc (corresponding to the volumes considered for the Fourier component study), the RMS mass fluctuation in the local Universe is ${\sim}1.8$. Hence, these regions are on the average below the threshold for turnaround in linear theory. The mean overdensity of ${\sim}70$ surrounding GRGs is consistent with environments that are detached from the Hubble flow but non-virialized since virialization would be expected only at an overdensity of ${\sim}178$~\citep{Silketal2013}.

The tendency for GRGs to grow at large angles with respect to their host major axes as well as galaxy chains or filaments to which their hosts belong suggests a correlation between the orientations of GRG host ellipticals (all being among the brightest in the vicinity) and large-scale galaxy distribution. This could simply imply that there is preferential growth of radio lobes associated with galaxies oriented in a favourable direction. However, correspondence between axes of bright elliptical galaxies and larger galaxy structures like groups, clusters, filaments and sheets or between galaxy clusters themselves (independent of richness) has been reported elsewhere~\citep[][and references therein]{Binggeli1982,Pazetal2011,Zhangetal2013}. The alignments are found to be strongest for redder and more massive galaxies. The correlations have been found to extend to large scales of several tens of Mpc. Alignments are also reported between dark matter halos and the galaxy filaments and sheets~\citep{Hahnetal2007,Tempeletal2013,TempelLibeskind2013,Zhangetal2013}. Such large-scale alignments could arise from galaxy formation processes with gas flows along the filaments. Further study is required to resolve which process, alignment selection or large-scale spin alignment, dominates for GRGs.

\subsection{The role of significant dipole and quadrupole moments in radio morphology}
\label{subsec:QuadrupoleMoment}

For GRGs grouped together purely on the basis of lobe non-collinearities or offset features, the only significant Fourier component parameter is $a_4$ and its negative sign indicates overdensity along $-45^{\circ}$ as opposed to $+45^{\circ}$ (see Table~\ref{tbl:stacked}). This indicates the lack of a dipole asymmetry in the surrounding galaxy distribution and only a quadrupole moment at angles of $135^{\circ}$ and $315^{\circ}$ to the radio axis (as defined by the longer GRG lobe). Whether it is sources with non-collinear lobes or sources that are otherwise collinear but have offset extensions (e.g. B0319--454 or B2356--611), $a_4$ is negative and highly significant. Asymmetries in sources, in terms of non-collinearity of the lobes and offset features, appear to be accompanied by asymmetries in the environment.

The fields in several GRGs have been rotated to align the longer lobe and inverted along this axis such that all GRGs in this subset have off-axis deviations in the same quadrant (see Section~\ref{subsubsec:GRG_subsets}). It is therefore highly significant that the overdensity excess is found in quadrants along an axis orthogonal to the quadrants in which the off-axis features lie rather than along the $+45^{\circ}$ and $+225^{\circ}$ axis that shares the quadrants of the offsets. The observed orientation of lobes away from high density regions is consistent with a model in which synchrotron lobes are deflected away from high densities and down the gradient in density. GRGs grouped on the basis of collinear lobes (category \emph{g}) instead show a sharply contrasting behaviour with only a 1.2$\sigma$ and positive $a_4$ parameter.

Considering symmetric galaxies as a control sample, for which the $a_3$ parameter is found to have the opposite (positive) sign, we find that the distributions of asymmetric and symmetric GRGs differ significantly based on the relative proportions of positive and negative $a_3$ parameters, which is consistent with the model where the lobe is shorter on the side with greater density due to ram pressure.

Besides the previously reported detailed case studies of the GRGs B0503--286~\citep{Subrahmanyanetal2008} and B0319--454~\citep{Safourisetal2009} there have been a few recent studies of the environments of some well known and nearby GRGs (NGC 6251,~\citet{Chenetal2011a}; NGC 315,~\citet{Chenetal2012a}; 4C 73.08,~\citet{Chenetal2012b} and 3C 326,~\citet{Piryaetal2012}). These recent spectroscopic studies all report the same finding of the shorter lobe lying on the side of larger galaxy number density as first reported by~\citet{Saripallietal1986} for a modest sample of GRGs and~\citet{Subrahmanyanetal2008} and~\citet{Safourisetal2009}.

\subsection{Previous work}

The aim of our study has been to use GRGs to enable us to derive IGM properties. This, we argued, was possible only because the GRGs are among the largest of radio galaxies and their lobes sample a medium well outside of host corona or any group to which their hosts belong. In \citetalias{Malareckietal2013} we presented radio maps and computed pressures in the diffuse parts of the GRG lobes. Comparing with the OWL simulations of~\citet{Schayeetal2010} we placed limits on the IGM temperature and density in the vicinity of the GRG lobes. However to what galaxy environments do these IGM properties correspond? The IGM is known to be a multi-layered medium and its properties vary with distance from galaxy filaments. The IGM may therefore not be a uniform medium and its pressure and density can vary from location to location.

The mean galaxy overdensity we find (${\sim}70$; see Table~\ref{tbl:host_volume_overdensity}), computed over a larger radius of 3 Mpc, is within the range of the particle overdensities ($50$--$500$) that we inferred in \citetalias{Malareckietal2013}. Galaxy overdensities within smaller volumes, closer in size to the GRGs, are a factor of ${\sim}2$ higher (Table~\ref{tbl:host_volume_density}). As noted in our previous work, particle overdensities on this order correspond to a small fraction of the expected WHIM by mass and volume. However, these estimates provide an improved upper limit on the density of the WHIM compared to the higher densities inferred from X-ray observations of galaxy filaments.

\section*{Summary}
We have measured optical spectroscopic redshifts for $9{,}076$ galaxies in fields surrounding a complete sample of 19 GRGs and have shown a correlation between the morphologies of these giant radio sources and their environments. In particular, overdensities in the galaxy distributions surrounding the GRGs, which reflect greater pressure and density in the ambient IGM, often correspond with asymmetries in lobe length or deflections of radio components.

Two complementary methods were used to study the GRG fields: (1) a Fourier component technique to quantify anisotropy in the surrounding field, limited to cylindrical volumes around each GRG; (2) a visual analysis of the individual fields using radio-optical overlays. Both methods independently show that the asymmetric radio components of GRGs have been influenced by their environments.

The major findings of this paper are as follows:
\begin{itemize}
\item The environments of GRGs have a mean galaxy number overdensity of ${\sim}70$ consistent with galaxy groups (using cylindrical volumes centred on each host with radius 3 Mpc and length 24 Mpc).
\item GRG jets tend to form in directions orthogonal to galaxy distributions around the host, which can often be seen as a chain across the jet axis.
\item Lobe length asymmetries can be attributed to greater galaxy density in the path of one jet compared to the other.
\item Non-collinear radio components are deflected in directions orthogonal to the surrounding overdensity reflected by significant quadrupole moments.
\end{itemize}

We find a clear correlation between GRG growth and the environments in which they form. GRGs form in sparse environments or in galaxy concentrations where the jets and lobes evolve in directions away from (or even nearly orthogonal to) the surrounding galaxy distribution. The galaxy distribution therefore affects the GRG morphologies: the extent and the orientation are affected by higher galaxy densities. While not strongly shown by the Fourier component analysis, the noted evidence of shorter lobes being on the side of higher galaxy density supports the finding that a GRG's growth to large linear sizes is facilitated by reduced galaxy densities.

The Fourier component method we have described here for quantifying anisotropy in the fields surrounding the GRGs would benefit from additional observations to improve completeness across all fields to a fainter absolute magnitude and also extending the sample to include northern GRGs. These would help to better distinguish correlations with the environment for asymmetric and symmetric GRGs, and sources with evidence of restarted activity.

\section*{Acknowledgements}
The Australia Telescope Compact Array is part of the Australia Telescope National Facility which is funded by the Commonwealth of Australia for operation as a National Facility managed by CSIRO. We gratefully acknowledge the Australian Astronomical Observatory for use of the Anglo-Australian Telescope and the AAOmega spectrograph. We thank R.~W.~Hunstead for bringing the GRG J0459--528 to our attention, S.~Brough for advice regarding optical data reduction and M.~Read for assistance with the SuperCOSMOS database. This work makes use of the {\sc runz} redshifting code developed by Will Sutherland, Will Saunders, Russell Cannon and Scott Croom. SuperCOSMOS was used for identifications of optical objects. This research has made use of NASA's Astrophysics Data System and the NASA/IPAC Extragalactic Database (NED), which is operated by the Jet Propulsion Laboratory, California Institute of Technology, under contract with the National Aeronautics and Space Administration. The \RadioAppName{casa} software is developed by the National Radio Astronomy Observatory (NRAO). Luminosity distance values were computed with E.~L.~Wright's Javascript Cosmology Calculator~\citep[][]{Wright2006}. Parts of this research were conducted by the Australian Research Council Centre of Excellence for All-sky Astrophysics (CAASTRO), through project number CE110001020. This research made use of APLpy, an open-source plotting package for Python hosted at http://aplpy.github.com. We thank the editors and referee for their helpful comments.

\bibliographystyle{mn2e}
\bibliography{references,grgsnorthern}

\label{lastpage}

\end{document}